\documentclass[12pt]{iopart}

\pdfoutput=1
\usepackage{iopams}  
\usepackage{graphicx}
\usepackage{amssymb}
\usepackage{bm,bbm}    
\usepackage{subfigure}
\usepackage{multirow} 
\usepackage{array} 
\usepackage[pdftex]{color}
\usepackage{times}
\usepackage{indentfirst}
\usepackage{layout}
\usepackage{hyperref}
\usepackage{subfigure}
\usepackage{multirow}
\usepackage{srcltx}
\usepackage{mathrsfs}
\expandafter\let\csname equation*\endcsname\relax
\expandafter\let\csname endequation*\endcsname\relax
\usepackage{amsmath} 
\usepackage{amssymb}

\begin{document}

\title{Symmetry-protected entangling boundary zero modes
in crystalline topological insulators
      }
\date{\today}

\author{Po-Yao Chang$^1$,
            Christopher Mudry$^2$,
            Shinsei Ryu$^1$}
\address{$^1$Department of Physics, University of Illinois at Urbana-Champaign, Urbana, IL 61801, USA}
\address{$^2$Condensed Matter Theory Group, 
Paul Scherrer Institute, 
CH-5232 Villigen PSI, Switzerland}

\ead{
\mailto{chang153@illinois.edu}}

\begin{abstract}
Crystalline topological insulators owe their topological character
to the protection that certain boundary states acquire because of
certain point-group symmetries. 
We first show that a Hermitian operator obeying 
supersymmetric quantum mechanisms (SUSY QM)
delivers the entanglement spectrum.
We then show that such an entanglement spectrum that 
is compatible with a certain point-group symmetry
obeys a certain local spectral symmetry. 
The latter result is applied to the stability analysis
of four fermionic non-interacting Hamiltonians, the last of which
describes graphene with a Kekule distortion. All examples have the remarkable 
property that their entanglement spectra inherit
a local spectral symmetry 
from either an inversion or reflection symmetry that 
guarantees the stability of gapless boundary entangling states,
even though all examples fail to support 
protected gapless boundary states at their physical boundaries.
\end{abstract}

\maketitle
\tableofcontents

\numberwithin{equation}{section}

\newpage

\noindent
\section{Introduction}
\label{sec:intro}

One of the fundamental distinctive feature of phases of matter
is the spontaneous breaking of symmetries. As a corollary,
phases of matter are gapless under very general
conditions when the broken symmetry group is continuous.
There are, however, incompressible phases of matter at zero temperature
that are featureless from the point of view
of spontaneous symmetry breaking. Examples 
thereof are the integer quantum Hall effect (IQHE)
and topological insulators (superconductors).%
~\cite{Hasan2010,Qi2011}
Such incompressible phases of matter are inherently quantum mechanical.
They have no classical counterparts, unlike phases breaking
spontaneously a symmetry. They are characterized by topological attributes 
such as a quantized response function 
or the existence of (symmetry protected) gapless modes
that propagate along a physical boundary, while they are 
exponentially localized away from the physical boundary
(in short gapless boundary modes or edge states),
when the ground state is non-degenerate in the thermodynamic limit.
Another probe of their topological character is the
entanglement of their incompressible ground state,%
\cite{Ryu2006,Levin2006,Kitaev2006,Li2008,Ivan2009,Fidkowski2010,Chen2010,ZC2010,Legner}
whether short ranged
when the ground state is non-degenerate (as in the IQHE), 
or long ranged when the ground states becomes degenerate
in the thermodynamic limit 
[as in the fractional quantum Hall effect (FQHE)].

One probe that measures the entanglement
of an incompressible  ground state $|\Psi\rangle$
is the entanglement entropy (von Neumann entropy) defined by
\begin{subequations}
\begin{equation}
S^{\,}_{A}:=
-\mathrm{tr}^{\,}_{A}\,
\left(
\hat{\rho}^{\,}_{A}\,\ln\,\hat{\rho}^{\,}_{A}
\right).
\end{equation}
Here, the total system is divided into two subsystems
$A$ and $B$,
\begin{equation}
\hat{\rho}^{\,}_{A}:=
\mathrm{tr}^{\,}_{B}\,|\Psi\rangle \langle\Psi|
\end{equation}
is the reduced density matrix obtained
by tracing over the states in subsystem $B$ 
of the total density matrix
\begin{equation}
\hat{\rho}:=
|\Psi\rangle\,\langle\Psi|
\end{equation}
\end{subequations}
in the incompressible ground state $|\Psi\rangle$.
In this paper, we will almost exclusively consider
single-particle Hamiltonians, 
their non-degenerate incompressible ground states $|\Psi\rangle$,
and partitionings into $A$ and $B$ 
with respect to a referred basis
for which locality is manifest.
We will also assume that
it is possible to associate with
$A$ and $B$ two regions of $d$-dimensional position space 
sharing a $(d-1)$-dimensional boundary. This boundary is called
the entangling boundary of the partition into $A$ and $B$.
This entangling boundary is unrelated to any physical boundary
selected by the choice of open boundary conditions.

Another probe for the entanglement contained in an incompressible
ground state $|\Psi\rangle$ is the entanglement spectrum.% 
~\cite{Ryu2006,Li2008,Ivan2009,Fidkowski2010}
On the one hand, topological phases that are characterized 
by the presence of gapless boundary states
in the energy spectrum must support 
gapless modes that propagate along the entangling boundary
but decay exponentially fast away from the entangling boundary
(in short entangling boundary states or entangling edge modes)
in the entanglement spectrum.%
~\cite{Ryu2006,Turner2010,Fidkowski2010,Hughes2011}
On the other hand, there are
symmetry-protected topological phases that
do not show gapless boundary modes in the energy spectrum,
while they do show gapless boundary modes in the entanglement spectrum. 
The entanglement spectrum can thus be thought of as  
a more refined diagnostic for identifying and 
classifying topological incompressible ground states
than the energy spectrum.
For example, non-interacting insulators that owe their topological character
to the existence of an inversion symmetry,
may support mid-gap states in the entanglement spectrum that are protected
by the inversion symmetry,
while they need not support gapless boundary modes in the energy spectrum.
These mid-gap modes in the entanglement spectrum
cannot be removed by an adiabatic and local deformation of the single-particle
Hamiltonian. As such there existence may be used as a mean
to quantify a topological invariant.
Symmetry protected topological phases can also arise
from other discrete symmetries, such as 
reflection symmetry,%
~\cite{Yao2012,Chingkai2012,Morimoto13a, Shiozaki, Tim, SY_Xu,Tanaka,Dziawa}
or more general point-group symmetries.%
~\cite{Liang2011,Fang2012,Chen2012,Chen-12012,Jadaun2012}
Moreover, it is known that
the eigenvalues of discrete symmetry operators
at the symmetric points in the Brillouin zone (BZ)
are related to the number of the mid-gap states  
in the entanglement spectrum
for inversion- and point-group-symmetric topological 
ground states.%
~\cite{Hughes2011,Fang2012,Chen2012,Chen-12012}.

In this paper, we explain how crystalline symmetries
of single-particle Hamiltonians with an incompressible ground state
manifest themselves in the entanglement spectrum and how they
can protect gapless boundary states in the entanglement spectrum,
when no such protection is operative for the boundary states
in the energy spectrum due to the non-local character of 
crystalline symmetries.

This paper is organized as follows.
In Sec.~\ref{sec: Symmetries and entanglement Spectrum}, 
after some preliminary definitions in
Secs.~\ref{subsubsec: Hamiltonian}
and~\ref{subsubsec: Entanglement spectrum},
we show that the entanglement Hamiltonian
is an example of supersymmetric quantum mechanics
in Sec.~\ref{Equal-time one-point correlation matrix and SUSY QM}.
The main result of this paper follows in
Sec.~\ref{Chiral symmetry of the entanglement spectrum},
where we show under what conditions a crystalline symmetry
can induce a spectral symmetry of the entanglement spectrum
that is absent from the energy spectrum of a single-particle Hamiltonian
with an incompressible ground state.%
~\cite{Turner2010,Hughes2011,Ari2012}
The subtle interplay between the geometry imposed by the boundary conditions,
the non-local crystalline symmetry, and its (local) realization on the physical
(entangling) boundaries is explained in
Sec.%
~\ref{sec: Physical versus entangling boundaries, spectral gap, and locality}.
We then apply our main result,
Eqs.~(\ref{eq: main result on how to get chiral sym})
and~(\ref{eq: induced symmetry}),
to the stability analysis of gapless edge states in the energy and
entanglement spectra for
four examples 
of single-particle Hamiltonians
with incompressible ground states 
in Secs.%
\ref{sec: example A},
\ref{sec: example B},
\ref{sec: example C},
and
\ref{sec: example D},
respectively.
This stability analysis is the most intricate when
treating a one-dimensional tight-binding model
with inversion symmetry in Sec.~\ref{sec: example A}.
Two copies of a pair of Chern insulators 
differing by the sign of their Chern numbers
that accommodate a reflection symmetry are treated 
in Sec.~\ref{sec: example B}
as the first two-dimensional example.
By considering two copies of the filled lowest Landau level
with opposite Chern numbers, we treat the example
of two reflection symmetries in two dimensions
in Sec.~\ref{sec: example C}.
We close the applications of
Eqs.~(\ref{eq: main result on how to get chiral sym})
and~(\ref{eq: induced symmetry})
by demonstrating that the Kekule distortion of graphene
(see Ref.~\cite{Hou2007}) 
should be thought of as a crystalline topological insulator.
We conclude in Sec.\ \ref{sec:conclusion}. 

\section{Symmetries and entanglement spectrum}
\label{sec: Symmetries and entanglement Spectrum}

The goal of this section is to study the relationship 
between the symmetries shared by a fermionic Hamiltonian
and its ground state and the symmetries of the
reduced density matrix of the pure-state density matrix 
constructed from the ground state, whereby the 
reduced density matrix presumes
a partitioning of the fermionic Fock space
into the tensor product of two Fock subspaces, i.e.,
the decomposition of the single-particle Hilbert space
into the direct sum of two subspaces.

We shall begin by studying the non-interacting case
in Sec.~\ref{subsec: Noninteracting fermions}.
The non-interacting limit is defined in 
Sec.~\ref{subsubsec: Hamiltonian}.
The ground state for
$N^{\,}_{\mathrm{f}}$ fermions hopping between the sites 
and orbitals of a lattice is a Slater determinant
(Fermi sea). By Wick's theorem, all the information
contained in the density matrix for the Fermi sea
can be retrieved from the equal-time one-point correlation
function (matrix) and conversely, as is reviewed in
Sec.~\ref{subsubsec: Entanglement spectrum}.
The direct sum decomposition of the single-particle Hilbert space is
an example of a graded vector space. We show 
in Sec.~\ref{Equal-time one-point correlation matrix and SUSY QM}
that this grading allows to construct from the 
equal-time one-point correlation matrix
a supersymmetric single-particle Hamiltonian.
This supersymmetry has consequences for the spectral
properties of the reduced density matrix (entanglement spectrum)
as is shown
in Sec.~\ref{Chiral symmetry of the entanglement spectrum},
where we demonstrate that certain symmetries
of the Hamiltonian induce symmetries of the entanglement spectrum.
Conversely, we show 
in Sec.~\ref{subsubsec: ... and PCT symmetry}
that a spectral symmetry of the Hamiltonian 
can turn into a symmetry of the reduced density matrix
in view of the hidden supersymmetry of the latter.
Spectral symmetry of the Hamiltonian can thus turn into
degeneracies of the entanglement spectrum.

The interacting case is considered in
Sec.~\ref{subsec: Interacting fermions}.
We show how a symmetry of the Hamiltonian and of its ground state
that interchanges the partition is realized on the reduced density 
matrix.

\subsection{Non-interacting fermions}
\label{subsec: Noninteracting fermions}

\subsubsection{Hamiltonian}
\label{subsubsec: Hamiltonian}

We consider a lattice $\Lambda\subset\mathbb{Z}^{d}$ 
whose $N$ sites are labeled by the vector 
$\bm{r}:=(r^{\,}_{1},\cdots,r^{\,}_{N})^{\mathsf{T}}$.
Each site $\bm{r}$ is also associated with $N^{\,}_{\mathrm{orb}}$
orbital (flavors) labeled by the Greek letter
$\alpha=1,\cdots,N^{\,}_{\mathrm{orb}}$.
We define the non-interacting second-quantized Hamiltonian
\begin{subequations}
\label{eq: def noninteractin fermion Hamiltonian}
\begin{equation}
\hat{H}:=
\sum_{\bm{r},\bm{r}'=1}^{N} 
\sum_{\alpha,\alpha'=1}^{N^{\,}_{\mathrm{orb}}}
\hat{\psi}^{\dag}_{\alpha,\bm{r}}\, 
\mathcal{H}^{\,}_{\alpha,\bm{r}; \alpha',\bm{r}'}\, 
\hat{\psi}^{\ }_{\alpha',\bm{r}'}.
\label{eq: def noninteractin fermion Hamiltonian a}
\end{equation}
The pair of creation 
($\hat{\psi}^{\dag}_{\alpha,\bm{r}}$)
and annihilation operators
($\hat{\psi}^{\ }_{\alpha',\bm{r}'}$)
obey the fermion algebra
\begin{equation}
\begin{split}
&
\left\{
\hat{\psi}^{\dag}_{\alpha,\bm{r}},
\hat{\psi}^{\ }_{\alpha',\bm{r}'}
\right\}=
\delta^{\,}_{\alpha,\alpha'}\,\delta^{\,}_{\bm{r},\bm{r}'},
\\
&
\left\{
\hat{\psi}^{\dag}_{\alpha,\bm{r}},
\hat{\psi}^{\dag}_{\alpha',\bm{r}'}
\right\}=
\left\{
\hat{\psi}^{\,}_{\alpha,\bm{r}},
\hat{\psi}^{\,}_{\alpha',\bm{r}'}
\right\}=0.
\end{split}
\label{eq: def noninteractin fermion Hamiltonian b}
\end{equation}
Hermiticity $\hat{H}=\hat{H}^{\dag}$ is imposed
by demanding that the single-particle matrix elements
obey
\begin{equation}
\mathcal{H}^{\,}_{\alpha,\bm{r}; \alpha',\bm{r}'}=
\mathcal{H}^{*}_{\alpha',\bm{r}'; \alpha,\bm{r}}.
\label{eq: def noninteractin fermion Hamiltonian c}
\end{equation}
Locality, in the orbital-lattice basis, is imposed by demanding that
\begin{equation}
\lim_{|\bm{r}-\bm{r}'|\to\infty}
\left|\mathcal{H}^{\,}_{\alpha,\bm{r}; \alpha',\bm{r}'}\right|
\leqslant
\lim_{|\bm{r}-\bm{r}'|\to\infty}
c\times
e^{-|\bm{r}-\bm{r}'|/\ell}
\label{eq: def noninteractin fermion Hamiltonian d}
\end{equation}
for some positive constant $c$ and for some characteristic length
scale $\ell$ independent of the lattice sites
$\bm{r},\bm{r}'=1,\cdots,N$ 
and of the orbital indices 
$\alpha,\alpha'=1,\cdots,N^{\,}_{\mathrm{orb}}$.
If we collect the orbital ($\alpha$)
and lattice ($\bm{r}$) indices into
a single collective index $I\equiv(\alpha,\bm{r})$, we may
introduce the short-hand notation
\begin{equation}
\hat{H}\equiv
\sum_{I,I'\in\Omega}
\hat{\psi}^{\dag}_{I}\, 
\mathcal{H}^{\,}_{II'}
\hat{\psi}^{\,}_{I'}\equiv
\hat{\psi}^{\dag}\, 
\mathcal{H}\,
\hat{\psi},
\label{eq: def noninteractin fermion Hamiltonian e}
\end{equation}
where $\mathcal{H}$ is a Hermitian
$N^{\,}_{\mathrm{tot}}\times N^{\,}_{\mathrm{tot}}$
matrix with $N^{\,}_{\mathrm{tot}}=N^{\,}_{\mathrm{orb}}\times N$
and $\Omega$ is the set
\begin{equation}
\Omega:=\{1,\cdots,N^{\,}_{\mathrm{orb}}\}\times\Lambda
\label{eq: def noninteractin fermion Hamiltonian f}
\end{equation}
\end{subequations}
obtained by taking the Cartesian product of the set of orbitals with
the set of lattice points.

The Hermitian matrix $\mathcal{H}$ has
$N^{\,}_{\mathrm{tot}}$ pairs of 
eigenvalues and eigenvectors
$(\varepsilon^{\,}_{I},\upsilon^{\,}_{I})$
that are defined by demanding that
\begin{subequations}
\label{eq: def noninteractin fermion Hamiltonian bis}
\begin{equation}
\mathcal{H}\,\upsilon^{\,}_{I}=
\varepsilon^{\,}_{I}\,
\upsilon^{\,}_{I},
\qquad
\upsilon^{\dag}_{I}\,\upsilon^{\vphantom{\dag}}_{I^{\prime}}=\delta^{\,}_{I,I^{\prime}},
\end{equation}
for $I,I'=1,\cdots,N^{\,}_{\mathrm{tot}}$.
With the help of the unitary matrix
\begin{equation}
\mathcal{U}=(\upsilon^{\,}_{1},\ldots,\upsilon^{\,}_{N^{\,}_{\mathrm{tot}}}),
\end{equation}
we may represent the single-particle Hamiltonian $\mathcal{H}$
as the diagonal matrix
\begin{equation}
\mathcal{U}^{\dag}\,\mathcal{H}\,\mathcal{U}=
\mathrm{diag}\, \left(
\varepsilon^{\, }_{1}, \ldots, 
\varepsilon^{\,}_{N^{\,}_{\mathrm{tot}}}
\right)
\end{equation}
The canonical transformation
\begin{equation}
\hat{\psi}^{\,}_{I}=: 
\sum_{J=1}^{N^{\,}_{\mathrm{tot}}} 
\mathcal{U}^{\,}_{IJ}\,
\hat{\chi}^{\,}_{J},
\label{eq: def chi basis noninteracting fermions}
\end{equation}
gives the representation
\begin{equation}
\hat{H}=
\sum_{I=1}^{N^{\,}_{\mathrm{tot}}} 
\hat{\chi}^{\dag}_{I}\, 
\varepsilon^{\,}_{I}\,
\hat{\chi}^{\,}_{I}.
\end{equation}
\end{subequations}
The ground state of $N^{\,}_{\mathrm{f}}$ fermions is then the Fermi sea 
\begin{subequations}
\label{eq: def ground state noninteracting fermions}
\begin{equation}
|\Psi^{\,}_{\mathrm{FS}}\rangle:=
\prod_{I=1}^{N^{\,}_{\mathrm{f}}} 
\hat{\chi}^{\dag}_{I}\, 
|0\rangle, 
\qquad
\hat{\chi}^{\,}_{I}\, 
|0\rangle=0,
\label{eq: def ground state noninteracting fermions a}
\end{equation}
whereby we have assumed that
\begin{equation}
I<I'\Longrightarrow
\varepsilon^{\,}_{I}\leqslant\varepsilon^{\,}_{I'}.
\label{eq: def ground state noninteracting fermions b}
\end{equation}
\end{subequations}

\subsubsection{Entanglement spectrum of
the equal-time one-point correlation matrix} 
\label{subsubsec: Entanglement spectrum} 

We seek to partition the Fock space $\mathfrak{F}$,
on which the non-interacting Hamiltonian
$\hat{H}$ defined by 
Eq.~(\ref{eq: def noninteractin fermion Hamiltonian bis})
acts, 
into two Fock subspace, which we denote by 
$\mathfrak{F}^{\,}_{A}$ and $\mathfrak{F}^{\,}_{B}$,
according to the tensorial decomposition
\begin{equation}
\mathfrak{F}=
\mathfrak{F}^{\,}_{A}
\otimes
\mathfrak{F}^{\,}_{B}.
\label{eq: tensorial decomposition mathfrak F}
\end{equation}
Two ingredients are necessary to define the subspaces
$\mathfrak{F}^{\,}_{A}$ and $\mathfrak{F}^{\,}_{B}$.
We need a state from $\mathfrak{F}$
that is a single Slater determinant.
It is for this quality that
we choose the ground state%
~(\ref{eq: def ground state noninteracting fermions}).
We need a basis, that we choose to be the orbital-lattice basis
defined by the representation%
~(\ref{eq: def noninteractin fermion Hamiltonian bis}).

Following Ref.~\cite{Ingo2003}, we start from the
equal-time one-point correlation function (matrix)
\begin{equation}
\mathcal{C}^{\,}_{IJ}:=
\left\langle\Psi^{\,}_{\mathrm{FS}}\left|
\hat{\psi}^{\dag}_{I}\, 
\hat{\psi}^{\,}_{J} 
\right|\Psi^{\,}_{\mathrm{FS}}\right\rangle,
\qquad
I,J=1,\cdots,N^{\,}_{\mathrm{tot}}.
\label{eq: def matrix elements C}
\end{equation}
Insertion of Eqs.%
~(\ref{eq: def ground state noninteracting fermions})
and~(\ref{eq: def chi basis noninteracting fermions})
delivers
\begin{equation}
\begin{split}
\mathcal{C}^{\,}_{IJ}=&\,
\sum_{I'=1}^{N^{\,}_{\mathrm{f}}}
\mathcal{U}^{*}_{II'}\,
\mathcal{U}^{\,}_{JI'}
\equiv
\sum_{I'=1}^{N^{\,}_{\mathrm{f}}}
\langle
\upsilon^{\,}_{I'}|
I\rangle
\langle
J|
\upsilon^{\,}_{I'}\rangle,
\end{split}
\label{eq: matrix elements C}
\end{equation}
where $I,J=1,\cdots,N^{\,}_{\mathrm{tot}}$.
We define the $N^{\,}_{\mathrm{tot}}\times N^{\,}_{\mathrm{tot}}$
correlation matrix $\mathcal{C}$ 
by its matrix elements~(\ref{eq: matrix elements C}).
One then verifies that
\begin{equation}
\mathcal{C}^{\dag}=\mathcal{C},
\qquad
\mathcal{C}^{2}=\mathcal{C},
\label{eq: cal C is Hermitian projector}
\end{equation}
i.e., the correlation matrix $\mathcal{C}$ is a Hermitian projector.
The last equality of Eq.~(\ref{eq: matrix elements C})
introduces the bra and ket notation of Dirac for the 
single-particle eigenstates
of $\mathcal{H}$ defined by 
Eq.~(\ref{eq: def noninteractin fermion Hamiltonian bis})
to emphasize that this projector is nothing but the sum over
all the single-particle eigenstates of $\mathcal{H}$ that are
occupied in the ground state.
Thus, all $N^{\,}_{\mathrm{tot}}$
eigenvalues of the correlation matrix
$\mathcal{C}$ are either the numbers 0 or 1. 
When it is convenient to shift the eigenvalues of the
correlation matrix from the numbers 0 or 1 to the numbers $\pm1$,
this is achieved through the linear transformation
\begin{equation}
\mathcal{Q}^{\,}_{IJ}:=
%\openone
\mathbb{I}
-
2\,
\mathcal{C}^{\,}_{IJ},
\qquad
I,J=1,\cdots,N^{\,}_{\mathrm{tot}}.
\label{eq: def cal Q}
\end{equation}
The occupied single-particle eigenstates of
$\mathcal{H}$ in the ground state of $\hat{H}$
are all eigenstates of $\mathcal{Q}$ 
with eigenvalue $-1$.
The unoccupied single-particle eigenstates of
$\mathcal{H}$ in the ground state of $\hat{H}$
are all eigenstates of $\mathcal{Q}$ 
with eigenvalue $+1$. In other words,
$\mathcal{Q}$ is the difference between the projector
onto the unoccupied single-particle eigenstates of
$\mathcal{H}$ in the ground state of $\hat{H}$
and the projector 
onto the occupied single-particle eigenstates of
$\mathcal{H}$ in the ground state of $\hat{H}$.
As such $\mathcal{Q}$ possesses all the symmetries
of $\mathcal{H}$ and all the spectral symmetries of
$\mathcal{H}$.

We denote the $N^{\,}_{\mathrm{tot}}$-dimensional
single-particle Hilbert space 
on which the correlation function $\mathcal{C}$
acts by $\mathfrak{H}$. The labels
$A$ and $B$ for the partition are introduced through
the direct sum decomposition
\begin{subequations}
\label{eq: partitioning single particles}
\begin{equation}
\mathfrak{H}=
\mathfrak{H}^{\,}_{A}
\oplus
\mathfrak{H}^{\,}_{B},
\label{eq: partitioning single particles a}
\end{equation}
whereby $A$ and $B$ are two non-intersecting subsets of the set
$\Omega$ defined by Eq.%
~(\ref{eq: def noninteractin fermion Hamiltonian f}) 
such that
\begin{equation}
\Omega=A\cup B,
\qquad
A\cap B=\emptyset,
\end{equation}
and
\begin{equation}
\mathcal{C}=
\begin{pmatrix}
C^{\,}_{A} & 
C^{\,}_{AB} 
\\
C^{\dag}_{AB} 
& 
C^{\,}_{B}
\end{pmatrix},
\quad
\mathcal{Q}=
\begin{pmatrix}
%\openone
\mathbb{I}-2\,C^{\,}_{A} & 
-2\,C^{\,}_{AB} 
\\
-2\,C^{\dag}_{AB} 
& 
%\openone
\mathbb{I}-2\,C^{\,}_{B}
\end{pmatrix}. 
\label{eq: partitioning single particles b}
\end{equation}
\end{subequations}
By construction, 
the $N^{\,}_{A}\times N^{\,}_{A}$ block $C^{\,}_{A}$
and 
the $N^{\,}_{B}\times N^{\,}_{B}$ block $C^{\,}_{B}$
are Hermitian matrices.
These blocks inherit the property 
that their eigenvalues are real numbers bounded 
between the numbers 0 and 1 from $\mathcal{C}$
being a Hermitian projector.
We call the set
\begin{subequations}
\label{eq: def sigma CA}
\begin{equation}
\sigma(C^{\,}_{A}):=
\left\{
\zeta^{\,}_{\iota}\left|
\exists v^{\,}_{\iota}\in\mathbb{C}^{N^{\,}_{\mathrm{A}}},\
C^{\,}_{A}\,v^{\,}_{\iota}=\zeta^{\,}_{\iota}\,v^{\,}_{\iota},\
\iota=1,\cdots,N^{\,}_{\mathrm{A}}
\right.
\right\}
\label{eq: def sigma CA a}
\end{equation}
of single-particle eigenvalues of the block $C^{\,}_{A}$
the entanglement spectrum of the correlation matrix $\mathcal{C}$.
Any eigenvalue from $\sigma(C^{\,}_{A})$ obeys
\begin{equation}
0\leqslant\zeta^{\,}_{\iota}\leqslant1,
\qquad
\iota=1,\cdots,N^{\,}_{A}.
\label{eq: def sigma CA b}
\end{equation}
\end{subequations}
The single-particle eigenvalues 
of the $N^{\,}_{A}\times N^{\,}_{A}$ Hermitian matrix $C^{\,}_{A}$
can be shifted from their support in the interval $[0,1]$
to the interval $[-1,+1]$ through the linear transformation
\begin{equation}
Q^{\,}_{A}:=
%\openone
\mathbb{I}
-
2\,
C^{\,}_{A}.
\label{eq: def QA}
\end{equation}
We refer to the set $\sigma(Q^{\,}_{A})$
of single-particle eigenvalues of
the $N^{\,}_{A}\times N^{\,}_{A}$ Hermitian matrix $Q^{\,}_{A}$
as the entanglement spectrum of the correlation matrix $\mathcal{Q}$,
which we shall abbreviate as \textit{the entanglement spectrum}.

It is shown in Ref.~\cite{Ingo2003}
that there exists a $N^{\,}_{A}\times N^{\,}_{A}$ block 
Hermitian matrix (entanglement Hamiltonian) 
$H^{E}\equiv(H^{E}_{K,L})^{\,}_{K,L\in A}$
with the positive definite operator
\begin{subequations}
\begin{equation}
\hat{\rho}^{\,}_{A}:= 
\frac{
e^{-\sum_{K',L'\in A}\hat{\psi}^{\dag}_{K'}\,H^{E}_{K'L'}\,\hat{\psi}^{\,}_{L'}}
     }
     {
\mathrm{tr}^{\,}_{\mathfrak{F}^{\,}_{A}}\,
e^{-\sum_{K',L'\in A}\hat{\psi}^{\dag}_{K'}\,H^{E}_{K'L'}\,\hat{\psi}^{\,}_{L'}}
     }
\end{equation}
whose domain of definition defines the Fock space
$\mathfrak{F}^{\,}_{A}$
such that the block $C^{\,}_{A}$ from the correlation matrix%
~(\ref{eq: partitioning single particles b})
is
\begin{equation}
C^{\,}_{A}=
\left(
\mathrm{tr}^{\,}_{\mathfrak{F}^{\,}_{A}}\,
\hat{\rho}^{\,}_{A}\,\hat{\psi}^{\dag}_{K}\,\hat{\psi}^{\,}_{L}\,
\right)^{\,}_{K,L\in A}.
\end{equation}
The positive definite matrix $\hat{\rho}^{\,}_{A}$ is the reduced density
matrix acting on the Fock space $\mathfrak{F}^{\,}_{A}$
obtained by tracing the degrees of freedom from the 
Fock space $\mathfrak{F}^{\,}_{B}$
in the density matrix 
\begin{equation}
\hat{\rho}^{\,}_{\mathrm{FS}}:=
\left|\Psi^{\,}_{\mathrm{FS}}\left\rangle
\right\langle\Psi^{\,}_{\mathrm{FS}}\right|
\end{equation}
\end{subequations}
whose domain of definition is the Fock space $\mathfrak{F}$.

It is also shown in Ref.~\cite{Ingo2003} that
the single-particle spectrum $\sigma(C^{\,}_{A})$
of the $N^{\,}_{A}\times N^{\,}_{A}$ Hermitian matrix $C^{\,}_{A}$
is related to the single-particle spectrum 
$\sigma(H)$ of the $N^{\,}_{A}\times N^{\,}_{A}$ Hermitian matrix 
$H^{E}$ by
\begin{equation}
\zeta^{\,}_{\iota}= 
\frac{1}{e^{\varpi^{\,}_{\iota}}+1},
\qquad
\iota=1,\cdots,N^{\,}_{A}.
\label{eq: relation sigma(CA) to sigma(H)}
\end{equation}
Equation~(\ref{eq: relation sigma(CA) to sigma(H)})
states that the dependence of the eigenvalue
$\zeta^{\,}_{\iota}$ of $C^{\,}_{A}$
on the eigenvalue $\varpi^{\,}_{\iota}$ of $H$ 
is the same as that of the Fermi-Dirac function on the single-particle
energy $\varpi^{\,}_{\iota}$
when the inverse temperature is unity and the chemical potential
is vanishing in units for which the Boltzmann constant is unity. 
Equation~(\ref{eq: relation sigma(CA) to sigma(H)})
is a one-to-one mapping between
$\sigma(C^{\,}_{A})$ and $\sigma(H)$.

Equation~(\ref{eq: relation sigma(CA) to sigma(H)})
allows to express the entanglement entropy 
\begin{equation}
S^{\mathrm{ee}}_{A}:=
- 
\mathrm{tr}^{\,}_{\mathfrak{F}^{\,}_{A}}\,
\left( 
\hat{\rho}_{A}\,\ln\hat{\rho}^{\,}_{A}
\right)
\end{equation}
of the reduced density matrix $\hat{\rho}^{\,}_{A}$ 
in terms of $\sigma(C^{\,}_{A})$ through
\begin{equation}
S^{\mathrm{ee}}_{A}=
-
\sum_{\iota=1}^{N^{\,}_{A}} 
\left[ 
\zeta^{\,}_{\iota}\,\ln\zeta^{\,}_{\iota}
+
\left(1-\zeta^{\,}_{\iota}\right)\, 
\ln(1-\zeta^{\,}_{\iota})
\right].
\end{equation}
Zero modes of $C^{\,}_{A}$ are eigenstates of
$C^{\,}_{A}$ with vanishing eigenvalues.
The vector space spanned by the zero modes of $C^{\,}_{A}$ 
is denoted by
$\mathrm{ker}\,(C^{\,}_{A})$
in linear algebra.
We conclude that the zero modes of $C^{\,}_{A}$ 
do not contribute to the entanglement entropy.

\subsubsection{Equal-time one-point correlation matrix and SUSY QM}
\label{Equal-time one-point correlation matrix and SUSY QM}

It was observed in Refs.~\cite{Turner2010}and 
\cite{Hughes2011}
that the $2\times2$ block structure
(\ref{eq: partitioning single particles b})
on the Hermitian correlation matrix $\mathcal{C}$
defined by its matrix elements~(\ref{eq: def matrix elements C})
is compatible with the condition~(\ref{eq: cal C is Hermitian projector})
that the correlation matrix
is a projector if and only if the four conditions
\begin{subequations}
\label{eq: hidden SUSY QM}
\begin{align}
&
C^{2}_{A}-C^{\,}_{A}=
-C^{\ }_{AB}\,C^{\dag}_{AB},
\label{eq: hidden SUSY QM a}
\\
&
Q^{\,}_{A}\, 
C^{\,}_{AB}= 
-
C^{\,}_{AB}\, 
Q^{\,}_{B}, 
\label{eq: hidden SUSY QM b}
\\
&
C^{\dag}_{AB}\, 
Q^{\,}_{A}= 
- 
Q^{\,}_{B}\,
C^{\dag}_{AB},
\label{eq: hidden SUSY QM c}
\\
&
C^{2}_{B}
-
C^{\,}_{B} 
=
-
C^{\dag}_{AB}\, 
C^{\,}_{AB},
\label{eq: hidden SUSY QM d}
\end{align} 
\end{subequations}
hold. Here, the $N^{\,}_{A}\times N^{\,}_{A}$ matrix
$Q^{\,}_{A}$ was defined in Eq.~(\ref{eq: def QA})
and we have introduced the $N^{\,}_{B}\times N^{\,}_{B}$ matrix
$Q^{\,}_{B}:=%\openone
\mathbb{I}-2\,C^{\,}_{B}$.

The family of $2N^{\,}_{\mathrm{susy}}+1$ operators 
\begin{subequations}
\label{eq: def SUSY QM}
\begin{equation}
\left(
\widehat{\mathcal{Q}}^{\dag}_{1},
\widehat{\mathcal{Q}}^{\,}_{1},
\cdots,
\widehat{\mathcal{Q}}^{\dag}_{N^{\,}_{\mathrm{susy}}},
\widehat{\mathcal{Q}}^{\,}_{N^{\,}_{\mathrm{susy}}},
\widehat{\mathcal{H}}
\right)
\label{eq: def SUSY QM a}
\end{equation}
acting on a common Hilbert space
realizes the graded Lie algebra 
of supersymmetric quantum mechanics (SUSY QM)
if and only if
\begin{eqnarray}
&&
\left[
\widehat{\mathcal{Q}}^{\dag}_{i},
\widehat{\mathcal{H}}
\right]=
\left[
\widehat{\mathcal{Q}}^{\,}_{i},
\widehat{\mathcal{H}}
\right]=0,
\label{eq: def SUSY QM b}
\\
&&
\left\{
\widehat{\mathcal{Q}}^{\,}_{i},
\widehat{\mathcal{Q}}^{\dag}_{j}
\right\}=
\delta^{\,}_{i,j}\,
\widehat{\mathcal{H}},
\label{eq: def SUSY QM c}
\end{eqnarray}
\end{subequations}
holds for
$i,j=1,\cdots,N^{\,}_{\mathrm{susy}}$. 
SUSY QM is realized when $\widehat{\mathcal{H}}$, 
which is Hermitian by construction 
because of Eq.~(\ref{eq: def SUSY QM c}),
can be identified with a Hamiltonian.
Conversely, a Hamiltonian in quantum mechanics
is supersymmetric if there exists a factorization
of the form~(\ref{eq: def SUSY QM}).
It is then convention to call the 
operators 
$\widehat{\mathcal{Q}}^{\dag}_{i}$ 
and
$\widehat{\mathcal{Q}}^{\,}_{i}$ 
with $i=1,\cdots,N^{\,}_{\mathrm{susy}}$ 
supercharges owing to
Eq.~(\ref{eq: def SUSY QM b}).
We are going to show that the algebra~(\ref{eq: hidden SUSY QM})
is an example of SUSY QM with $N^{\,}_{\mathrm{susy}}=1$ in disguise.

To this end, we define the 
$N^{\,}_{A}\times N^{\,}_{A}$,
$N^{\,}_{B}\times N^{\,}_{B}$,
$N^{\,}_{A}\times N^{\,}_{B}$,
and
$N^{\,}_{B}\times N^{\,}_{A}$
block matrices
\begin{subequations}
\label{eq: def Mpm SA SB}
\begin{align}
&
S^{\,}_{A}:= 
%\openone
\mathbb{I}
- 
Q^{2}_{A},
\qquad
S^{\,}_{B}:= 
%\openone
\mathbb{I}
- 
Q^{2}_{B},
\\
&
M^{+}\equiv M^{\dag}:= 
2\,C^{\,}_{AB},
\qquad
M^{-}\equiv M:= 
2\,C^{\dag}_{AB},
\end{align}
\end{subequations}
respectively. By construction, the spectrum of 
the Hermitian $N^{\,}_{A}\times N^{\,}_{A}$ matrix $S^{\,}_{A}$ 
and that of the Hermitian $N^{\,}_{B}\times N^{\,}_{B}$ matrix $S^{\,}_{B}$ 
belong to the interval $[0,1]$. One verifies with the help
of Eq.~(\ref{eq: hidden SUSY QM}) that
\begin{subequations}
\label{eq: hidden SUSY QM bis}
\begin{align}
&
M^{+}\, 
S^{\,}_{B}
-
S^{\,}_{A}\, 
M^{+}=0, 
\qquad
M^{-}\,
S^{\,}_{A}
-
S^{\,}_{B}\, 
M^{-}=0, 
\label{eq: hidden SUSY QM bis a}
\\
&
M^{+}\, 
M^{-}=
S^{\,}_{A},
\qquad
M^{-}\, 
M^{+}=
S^{\,}_{B}. 
\label{eq: hidden SUSY QM bis b}
\end{align}
\end{subequations}
We cannot close the graded Lie algebra~(\ref{eq: def SUSY QM})
with the four block matrices~(\ref{eq: def Mpm SA SB}).
However, we still have the possibility to define
the triplet of $N^{\,}_{\mathrm{tot}}\times N^{\,}_{\mathrm{tot}}$ matrices
\begin{equation}
\mathcal{S}^{\,}_{\mathrm{susy}}=
\begin{pmatrix}
S^{\,}_{A}  
& 
0 
\\
0 
& 
S^{\,}_{B}
\end{pmatrix},
\qquad
\mathcal{Q}^{\,}_{\mathrm{susy}}:=
\begin{pmatrix}
0 
& 
0 
\\
M 
& 
0 
\end{pmatrix},
\label{eq: def triplet cal S cal Q cal Qdag}
\end{equation}
and $\mathcal{Q}^{\dag}$. One verifies that they satisfy the graded
Lie algebra
\begin{subequations}
\label{eq: SUSY algebra from C matrix}
\begin{align}
&
[\mathcal{Q}^{\,}_{\mathrm{susy}},\mathcal{S}^{\,}_{\mathrm{susy}}]= 
[\mathcal{Q}^{\dag}_{\mathrm{susy}},\mathcal{S}^{\,}_{\mathrm{susy}}]=0,
\\
&
\{\mathcal{Q}^{\,}_{\mathrm{susy}},\mathcal{Q}^{\,}_{\mathrm{susy}}\}=
\{\mathcal{Q}^{\dag}_{\mathrm{susy}},\mathcal{Q}^{\dag}_{\mathrm{susy}}\}=0,
\\
&
\{\mathcal{Q}^{\,}_{\mathrm{susy}},\mathcal{Q}^{\dag}_{\mathrm{susy}}\}= 
\mathcal{S}^{\,}_{\mathrm{susy}},
\qquad
\end{align}
\end{subequations}
i.e., the pair of supercharge matrices 
$\mathcal{Q}^{\dag}_{\mathrm{susy}}$ and $\mathcal{Q}^{\,}_{\mathrm{susy}}$
and the Hamiltonian matrix 
$\mathcal{S}^{\,}_{\mathrm{susy}}$ 
realize SUSY QM with 
$N^{\,}_{\mathrm{tot}}=1$.

The grading labeled by $A$ and $B$ in the definition
of the matrices in Eq.%
~(\ref{eq: SUSY algebra from C matrix})
originates from the decomposition of the single
particle Hilbert space $\mathcal{H}$ into the direct
sum~(\ref{eq: partitioning single particles a}).
The matrix $\mathcal{S}^{\,}_{\mathrm{susy}}$ is block diagonal,
i.e., it does not mix the subspaces
$\mathfrak{H}^{\,}_{A}$
and
$\mathfrak{H}^{\,}_{B}$.
The pair of matrices
$\mathcal{Q}^{\dag}_{\mathrm{susy}}$ and $\mathcal{Q}^{\,}_{\mathrm{susy}}$
are off-diagonal with respect to the labels $A$ and $B$,
i.e., they mix the subspaces
$\mathfrak{H}^{\,}_{A}$
and
$\mathfrak{H}^{\,}_{B}$.
The matrix $M^{-}\equiv M$ 
maps $\mathfrak{H}^{\,}_{A}$ into $\mathfrak{H}^{\,}_{B}$.
Its adjoint $M^{+}\equiv M^{\dag}$ 
maps  $\mathfrak{H}^{\,}_{B}$ into $\mathfrak{H}^{\,}_{A}$.
In the context of SUSY QM, the pair $M^{-}$ and $M^{+}$
are called intertwiner. 

We are now going to show that the 
Eqs.~(\ref{eq: def Mpm SA SB}),
(\ref{eq: def triplet cal S cal Q cal Qdag}),
and (\ref{eq: SUSY algebra from C matrix})
imply that the number of linearly independent eigenstates
of $\mathcal{S}^{\,}_{\mathrm{susy}}$ with vanishing eigenvalues, i.e., the
number of  linearly independent zero modes is larger or equal to
\begin{equation}
\left|
\mathrm{dim}\,\mathfrak{H}^{\,}_{A}
-
\mathrm{dim}\,\mathfrak{H}^{\,}_{B}
\right|\equiv
\left|
N^{\,}_{A}-N^{\,}_{B}
\right|.
\end{equation}

To see this, we assume without loss of generality that
$N^{\,}_{A}>N^{\,}_{B}$.
When the dimension 
$N^{\,}_{A}$ of $\mathfrak{H}^{\,}_{A}$
is larger than the dimension
$N^{\,}_{B}$ of $\mathfrak{H}^{\,}_{B}$,
$M^{+}$ is a rectangular matrix with more rows than columns
while $M^{-}$ is a rectangular matrix with more columns than rows.
There follows two consequences. On the one hand, the condition
\begin{equation}
M^{-}\, \mathfrak{H}^{\,}_{A}=0
\end{equation}
that defines the null space of $M^{-}$
necessarily admits $N^{\,}_{A}-N^{\,}_{B}>0$ linearly independent solutions
in $\mathfrak{H}^{\,}_{A}$, for we must solve $N^{\,}_{B}$ equations
for $N^{\,}_{A}$ unknowns. 
On the other hand, the condition
\begin{equation}
M^{+}\, \mathfrak{H}^{\,}_{B}=0
\end{equation}
that defines the null space of $M^{+}$
is overdetermined, for we must solve $N^{\,}_{A}$ equations
for $N^{\,}_{B}$ unknowns. The condition
\begin{equation}
\mathcal{S}^{\,}_{\mathrm{susy}}\,\mathfrak{H}=0
\end{equation}
that defines the null space of $\mathcal{S}^{\,}_{\mathrm{susy}}$
delivers at least 
$N^{\,}_{A}-N^{\,}_{B}>0$ linearly independent solutions
of the form
\begin{equation}
\upsilon=
\begin{pmatrix}
\upsilon^{\,}_{A}
\\
0
\end{pmatrix}
\end{equation}
where
\begin{equation}
M^{-}\,\upsilon^{\,}_{A}=0.
\end{equation}
As observed by Witten in Ref.~\cite{Witten1982},
the number of zero modes plays an important role
in SUSY QM. The number of zero modes of $\mathcal{S}^{\,}_{\mathrm{susy}}$
is given by the Witten index
\begin{equation}
\Delta^{\,}_{\mathrm{w}}:=
\left|
\mathrm{dim}\,\mathrm{ker}(M^{-})
-
\mathrm{dim}\,\mathrm{ker}(M^{+})
\right|
\geqslant
\left|
N^{\,}_{A}-N^{\,}_{B}
\right|.
\label{eq: def Witten index}
\end{equation}
The relevance of  
Eq.~(\ref{eq: def Witten index})
to the entanglement spectrum of non-interacting fermions
is the main result of 
Sec.~\ref{Equal-time one-point correlation matrix and SUSY QM}.

\subsubsection{Chiral symmetry of the entanglement spectrum $\sigma(Q^{\,}_{A})$}
\label{Chiral symmetry of the entanglement spectrum}

We have seen that the existence of zero modes 
in the spectrum $\sigma(\mathcal{S}^{\,}_{\mathrm{susy}})$
of $\mathcal{S}^{\,}_{\mathrm{susy}}$
defined in Eq.~(\ref{eq: def triplet cal S cal Q cal Qdag})
is guaranteed when the dimensionalities 
$N^{\,}_{A}$
and 
$N^{\,}_{B}$
of the single-particle Hilbert spaces
$\mathfrak{H}^{\,}_{A}$
and 
$\mathfrak{H}^{\,}_{B}$,
respectively, are unequal.
We have shown that this property 
is a consequence of a hidden supersymmetry.
When $N^{\,}_{A}=N^{\,}_{B}$,
we cannot rely on SUSY QM to decide
if $0\in\sigma(\mathcal{S}^{\,}_{\mathrm{susy}})$.

The non-interacting Hamiltonian $\hat{H}$ defined by Eq.%
~(\ref{eq: def noninteractin fermion Hamiltonian a})
in the orbital-lattice basis acts on the Fock space $\mathfrak{F}$.
Let $\mathscr{O}$ denote an operation, i.e., 
an invertible mapping of time, of the lattice, 
of the orbital degrees of freedom, or of compositions thereof.
We represent this operation
by either a unitary or an anti-unitary transformation
on the Fock space $\mathfrak{F}$. In turn,
it suffices to specify how the creation and annihilation
operators defined by their algebra%
~(\ref{eq: def noninteractin fermion Hamiltonian b})
transform under the operation $\mathscr{O}$ to
represent $\mathscr{O}$ on the Fock space $\mathfrak{F}$.

For example, the transformation law
\begin{equation}
\hat{\psi}^{\,}_{I}\to
\hat{O}\,
\hat{\psi}^{\,}_{I}\,
\hat{O}^{\dag}\equiv
\sum_{I'=1}^{N^{\,}_{\mathrm{tot}}}
\mathcal{O}^{\,}_{I' I}\,
\hat{\psi}^{\,}_{I'},
\qquad
I=1,\cdots,N^{\,}_{\mathrm{tot}},
\end{equation}
where $\mathcal{O}=(\mathcal{O}^{\,}_{IJ})\in U(N^{\,}_{\mathrm{tot}})$
is a unitary matrix, realizes in a unitary fashion
the operation $\mathscr{O}$. 
The non-interacting Hamiltonian $\hat{H}$
defined in Eq.~(\ref{eq: def noninteractin fermion Hamiltonian})
and the correlation matrix $\mathcal{C}$ 
defined in Eq.~(\ref{eq: def matrix elements C})
obey the 
transformation laws
\begin{equation}
\hat{H}\to
\hat{O}\,
\hat{H}\,
\hat{O}^{\dag}
\Rightarrow
\mathcal{H}^{\mathsf{T}}\to
\mathcal{O}^{\dag}\,
\mathcal{H}^{\mathsf{T}}\,
\mathcal{O},
\end{equation}
and
\begin{equation}
\mathcal{C}\to
\mathcal{O}^{\dag}\,
\mathcal{C}\,
\mathcal{O},
\end{equation}
respectively. 

For comparison, the operations of time reversal
$(\mathscr{O}^{\,}_{\mathrm{tr}},\hat{O}^{\,}_{\mathrm{tr}},\mathcal{O}^{\,}_{\mathrm{tr}})$
and charge conjugation (that exchanges particle and holes)
$(\mathscr{O}^{\,}_{\mathrm{ph}},\hat{O}^{\,}_{\mathrm{ph}},\mathcal{O}^{\,}_{\mathrm{ph}})$
are anti-unitary transformations for which
\begin{equation}
\mathcal{C}\to
\mathcal{O}^{\dag}_{\mathrm{tr}}\,
\mathcal{C}\,
\mathcal{O}^{\,}_{{\mathrm{tr}}},
\quad
\mathcal{O}^{\,}_{{\mathrm{tr}}}\equiv
\mathcal{T}^{\,}_{{\mathrm{tr}}}\,K,
\quad
\mathcal{T}^{-1}_{{\mathrm{tr}}}=\mathcal{T}^{\dag}_{{\mathrm{tr}}},
\end{equation}
and
\begin{equation}
\mathcal{C}\to
\mathcal{O}^{\dag}_{\mathrm{ph}}\,
\mathcal{C}\,
\mathcal{O}^{\,}_{{\mathrm{ph}}},
\quad
\mathcal{O}^{\,}_{{\mathrm{ph}}}\equiv
\mathcal{T}^{\,}_{{\mathrm{ph}}}\,K,
\quad
\mathcal{T}^{-1}_{{\mathrm{ph}}}=\mathcal{T}^{\dag}_{{\mathrm{ph}}},
\end{equation}
respectively. Here, $K$ is the anti-linear operation of complex conjugation.

The non-interacting Hamiltonian $\hat{H}$ has
$\mathscr{O}$ as a symmetry if and only if
\begin{equation}
\hat{O}\,
\hat{H}\,
\hat{O}^{\dag}=
\hat{H}
\Longleftrightarrow
\mathcal{H}^{\mathsf{T}}=
\mathcal{O}^{\dag}\,
\mathcal{H}^{\mathsf{T}}\,
\mathcal{O}.
\label{eq: unitary symmetry H}
\end{equation}
Moreover, if we assume the transformation law
\begin{equation}
|\Psi^{\,}_{\mathrm{FS}}\rangle\to
e^{\mathrm{i}\Theta}\,
|\Psi^{\,}_{\mathrm{FS}}\rangle,
\qquad
0\leqslant\Theta<2\pi,
\end{equation}
for the ground state%
~(\ref{eq: def ground state noninteracting fermions}),
i.e., if we assume that the ground state does not break
spontaneously the symmetry~(\ref{eq: unitary symmetry H}),
then 
\begin{subequations}
\label{eq: condition cal O dag cal C cal O = cal C}
\begin{equation}
\mathcal{C}=
\mathcal{O}^{\dag}\,
\mathcal{C}\,
\mathcal{O}
\Longleftrightarrow
\mathcal{Q}=
\mathcal{O}^{\dag}\,
\mathcal{Q}\,
\mathcal{O}.
\label{eq: condition cal O dag cal C cal O = cal C a}
\end{equation}
We want to derive what effect condition%
~(\ref{eq: condition cal O dag cal C cal O = cal C a})
has on the entanglement spectrum $\sigma(Q^{\,}_{A})$, i.e.,
when the direct sum decompositions%
~(\ref{eq: partitioning single particles})
and
\begin{equation}
\mathcal{O}=
\begin{pmatrix}
O^{\,}_{A} 
& 
O^{\,}_{AB} 
\\
O^{\,}_{BA} 
& 
O^{\,}_{B} 
\end{pmatrix}\in U(N^{\,}_{\mathrm{tot}}).
\label{eq: condition cal O dag cal C cal O = cal C b}
\end{equation}
\end{subequations}
hold for two special cases.

First, we assume that 
\begin{subequations}
\label{eq: assumption cal O block diagonal}
\begin{equation}
\mathcal{O}=
\begin{pmatrix}
O^{\,}_{A} 
& 
0
\\
0
& 
O^{\,}_{B} 
\end{pmatrix},
\quad
O^{\,}_{A}\in U(N^{\,}_{A}),
\quad
O^{\,}_{B}\in U(N^{\,}_{B}).
\label{eq: assumption cal O block diagonal a}
\end{equation}
This situation arises when the operation $\mathscr{O}$
is compatible with the partitioning encoded by the two subsets $A$ and $B$
of $\Omega$ in the sense that 
\begin{equation}
A=\mathscr{O}A, 
\qquad
B=\mathscr{O}B.
\label{eq: assumption cal O block diagonal b}
\end{equation}
\end{subequations}
If so, condition~(\ref{eq: condition cal O dag cal C cal O = cal C})
simplifies to
\begin{subequations}
\label{eq: O operates on QA if block diagonal}
\begin{align}
&
Q^{\,}_{A}=
O^{\dag}_{A}\,
Q^{\,}_{A}\,
O^{\,}_{A},
\\
&
C^{\,}_{AB}=
O^{\dag}_{A}\,
C^{\,}_{AB}\,
O^{\,}_{B},
\\
&
Q^{\,}_{B}=
O^{\dag}_{B}\,
Q^{\,}_{B}\,
O^{\,}_{B}.
\end{align}
\end{subequations}
Thus, the $N^{\,}_{A}\times N^{\,}_{A}$ Hermitian block matrices
$C^{\,}_{A}$ and $Q^{\,}_{A}=%\openone
\mathbb{I}-2\,C^{\,}_{A}$
inherit the symmetry obeyed by the
$N^{\,}_{\mathrm{tot}}\times N^{\,}_{\mathrm{tot}}$ 
Hermitian matrix $\mathcal{H}$
when Eq.~(\ref{eq: assumption cal O block diagonal}) 
holds.

Second, we assume that 
\begin{subequations}
\label{eq: cal O off block diagonal}
\begin{equation}
N^{\,}_{A}=N^{\,}_{B}=N^{\,}_{\mathrm{tot}}/2
\label{eq: cal O off block diagonal a}
\end{equation}
and
\begin{equation}
\mathcal{O}=
\begin{pmatrix}
0
& 
O^{\,}_{AB} 
\\
O^{\,}_{BA} 
& 
0
\end{pmatrix},
\quad
O^{\,}_{AB},O^{\,}_{BA}\in U(N^{\,}_{\mathrm{tot}}/2).
\label{eq: cal O off block diagonal b}
\end{equation}
This situation arises when the operation $\mathscr{O}$
interchange the two subsets $A$ and $B$ of $\Omega$,
\begin{equation}
A=\mathscr{O}B,
\qquad
B=\mathscr{O}A.
\label{eq: cal O off block diagonal c}
\end{equation}
\end{subequations}
If so, condition~(\ref{eq: condition cal O dag cal C cal O = cal C})
simplifies to
\begin{subequations}
\label{eq: condition cal O dag cal C cal O= cal c}
\begin{align}
&
Q^{\,}_{A}=
O^{\dag}_{BA}\,
Q^{\,}_{B}\, 
O^{\,}_{BA},
\label{eq: condition cal O dag cal C cal O= cal c a}
\\ 
&
C^{\,}_{AB}=
O^{\dag}_{BA}\, 
C^{\,}_{BA}\, 
O^{\,}_{AB}
\label{eq: condition cal O dag cal C cal O= cal c b}
\\
&
Q^{\,}_{B}=
O^{\dag}_{AB}\,
Q^{\,}_{A}\, 
O^{\,}_{AB}.
\label{eq: condition cal O dag cal C cal O= cal c c}
\end{align}
\end{subequations}
We are going to combine Eq.%
~(\ref{eq: condition cal O dag cal C cal O= cal c})
with Eq.~(\ref{eq: hidden SUSY QM}).
Multiplication of
Eq.~(\ref{eq: condition cal O dag cal C cal O= cal c a})
from the right by 
Eq.~(\ref{eq: condition cal O dag cal C cal O= cal c b})
gives the relation
\begin{equation}
Q^{\,}_{A}\,C^{\,}_{AB}=
O^{\dag}_{BA}\,Q^{\,}_{B}\,C^{\,}_{BA}\,O^{\,}_{AB},
\end{equation}
while multiplication of
Eq.~(\ref{eq: condition cal O dag cal C cal O= cal c b})
from the right by
Eq.~(\ref{eq: condition cal O dag cal C cal O= cal c c})
gives the relation
\begin{equation}
C^{\,}_{AB}\,Q^{\,}_{B}=
O^{\dag}_{BA}\,C^{\,}_{BA}\,Q^{\,}_{A}\,O^{\,}_{AB}.
\end{equation}
We may then use 
Eq.~(\ref{eq: hidden SUSY QM b}) to infer that
\begin{equation}
Q^{\,}_{A}\,C^{\,}_{AB}=
-
C^{\,}_{AB}\,Q^{\,}_{B}.
\label{eq: mapping QA to QB}
\end{equation}
A second use of
Eq.~(\ref{eq: condition cal O dag cal C cal O= cal c c})
delivers
\begin{equation}
Q^{\,}_{A}\,C^{\,}_{AB}=
-
C^{\,}_{AB}\,O^{\dag}_{AB}\,Q^{\,}_{A}\,O^{\,}_{AB}.
\label{eq: cal Q A times cal C AB nearly ready}
\end{equation}
We introduce the auxiliary 
$(N^{\,}_{\mathrm{tot}}/2)\times(N^{\,}_{\mathrm{tot}}/2)$
matrix
\begin{subequations}
\label{eq: main result on how to get chiral sym}
\begin{equation} 
\Gamma^{\ }_{\mathscr{O}A}:= 
C^{\,}_{AB}\,O^{\dag}_{AB},
\label{eq: def Gamma mathsf O} 
\end{equation}
whose domain and co-domain are 
the single-particle Hilbert space $\mathfrak{H}^{\,}_{A}$.
We may then rewrite
Eq.~(\ref{eq: cal Q A times cal C AB nearly ready})
as the vanishing anti-commutator
\begin{equation}
\left\{Q^{\,}_{A},\Gamma^{\ }_{\mathscr{O}A}\right\}=0.
\label{eq: def Gamma mathsf O anticommutes with Q A} 
\end{equation}
The same manipulations with the substitution $A\leftrightarrow B$
deliver
\begin{equation}
\left\{Q^{\,}_{B},\Gamma^{\ }_{\mathscr{O}B}\right\}=0,
\label{eq: def Gamma mathsf O anticommutes with Q B} 
\end{equation}
where
\begin{equation}
\Gamma^{\ }_{\mathscr{O}B}:= 
C^{\,}_{BA}\,O^{\dag}_{BA},
\label{eq: def Gamma mathsf O} 
\end{equation}
\end{subequations}
Equations~(\ref{eq: def Gamma mathsf O anticommutes with Q A})
and~(\ref{eq: def Gamma mathsf O anticommutes with Q B})
are the main result of
Sec.~\ref{Chiral symmetry of the entanglement spectrum}.

An important consequence of
Eq.~(\ref{eq: def Gamma mathsf O anticommutes with Q A})
is that the spectra 
$\sigma(Q^{\,}_{A})$
and 
$\sigma(Q^{\,}_{B})$,
are endowed with a symmetry not necessarily present
in the spectrum $\sigma(\mathcal{H})$ 
when Eqs.~(\ref{eq: unitary symmetry H}),
(\ref{eq: condition cal O dag cal C cal O = cal C}),
and~(\ref{eq: cal O off block diagonal})
hold. This spectral symmetry is reminiscent of
the so-called chiral symmetry, the property
that a single-particle Hamiltonian anti-commutes with
a unitary operator, although, here, the operator
$\Gamma^{\,}_{\mathscr{O}}$ is not necessarily unitary.
If we assume that $C^{\,}_{AB}$ is an invertible
$(N^{\,}_{\mathrm{tot}}/2)\times(N^{\,}_{\mathrm{tot}}/2)$
matrix, 
to any pair 
$(1-2\,\zeta^{\,}_{\iota}\neq0,\upsilon^{\,}_{\iota})$
that belongs to $\sigma(Q^{\,}_{A})$,
the image pair
$(-1+2\,\zeta^{\,}_{\iota}\neq0,\Gamma^{\,}_{\mathscr{O}}\,\upsilon^{\,}_{\iota})$
also belongs to $\sigma(Q^{\,}_{A})$, and conversely.
We shall say that the entanglement spectrum $\sigma(Q^{\,}_{A})$
is chiral symmetric, when Eqs.~(\ref{eq: unitary symmetry H}),
(\ref{eq: condition cal O dag cal C cal O = cal C}),
(\ref{eq: cal O off block diagonal})
hold with $\Gamma^{\,}_{\mathscr{O}}$ defined in 
Eq.~(\ref{eq: def Gamma mathsf O})
invertible.

In this paper, we shall consider a regular $d$-dimensional
lattice $\Lambda$, say a Bravais lattice. 
We denote with $P^{\,}_{\parallel}\subset\mathbb{R}^{d}$ the plane with
the coordinates
\begin{subequations}
\label{eq: def coordinates in Rd to define R and P trsf}
\begin{equation}
\begin{pmatrix}
\bm{r}^{\,}_{\parallel}
\\
\bm{0}
\end{pmatrix},
\qquad
\bm{r}^{\,}_{\parallel}\in\mathbb{R}^{n}, 
\qquad
n=1,\cdots,d-1.
\label{eq: def coordinates in Rd to define R and P trsf b}
\end{equation}
We denote with $P^{\,}_{\perp}\subset\mathbb{R}^{d}$ the plane with
the coordinates
\begin{equation}
\begin{pmatrix}
\bm{0}
\\
\bm{r}^{\,}_{\perp}
\end{pmatrix},
\qquad
\bm{r}^{\,}_{\perp}\in\mathbb{R}^{d-n},
\qquad
n=1,\cdots,d-1.
\label{eq: def coordinates in Rd to define R and P trsf c}
\end{equation}
Any lattice point $\bm{r}$ from $\Lambda$ can be written as 
\begin{equation}
\bm{r}\equiv
\begin{pmatrix}
\bm{r}^{\,}_{\parallel}
\\
\bm{r}^{\,}_{\perp}
\end{pmatrix}
\in\mathbb{R}^{d}.
\end{equation}
\end{subequations}
We define the operation of reflection $\mathscr{R}$ 
about the plane $P^{\,}_{\perp}\subset\mathbb{R}^{d}$ by
\begin{subequations}
\label{eq: def reflection}
\begin{align}
&
\bm{r}\equiv
\begin{pmatrix}
\bm{r}^{\,}_{\parallel}
\\
\bm{r}^{\,}_{\perp}
\end{pmatrix}
\to
\mathscr{R}\,\bm{r}:=
\begin{pmatrix}
-\bm{r}^{\,}_{\parallel}
\\
+\bm{r}^{\,}_{\perp}
\end{pmatrix},
\\
&
\hat{\psi}^{\,}_{\alpha,\bm{r}}\to
\hat{R}\,\hat{\psi}^{\,}_{\alpha,\bm{r}}\,\hat{R}^{\dag}\equiv
\sum_{\beta=1}^{N^{\,}_{\mathrm{orb}}}
\mathtt{R}^{\,}_{\alpha\beta}\,
\hat{\psi}^{\,}_{\beta,\mathscr{R}\bm{r}},
\end{align}
\label{eq: reflection}
\end{subequations}
for any $\alpha=1,\cdots,N^{\,}_{\mathrm{orb}}$ and $\bm{r}\in\Lambda$.
We define the operation of parity (inversion) $\mathscr{P}$ by
\begin{subequations}
\label{eq: def inversion}
\begin{align}
&
\bm{r}\to
\mathscr{P}\,\bm{r}:=
-\bm{r},
\\
&
\hat{\psi}^{\,}_{\alpha,\bm{r}}\to
\hat{P}\,\hat{\psi}^{\,}_{\alpha,\bm{r}}\,\hat{P}^{\dag}\equiv
\sum_{\beta=1}^{N^{\,}_{\mathrm{orb}}}
\mathtt{P}^{\,}_{\alpha\beta}\,
\hat{\psi}^{\,}_{\beta,\mathscr{P}\bm{r}},
\label{eq: inversion}
\end{align}
\end{subequations}
for any $\alpha=1,\cdots,N^{\,}_{\mathrm{orb}}$ and $\bm{r}\in\Lambda$.

Even though we may impose on the non-interacting
Hamiltonian~(\ref{eq: def noninteractin fermion Hamiltonian})
the conditions of translation and point-group symmetries
when the lattice $\Lambda$ is regular, 
a partitioning~(\ref{eq: partitioning single particles})
might break these symmetries, for it involves lattice degrees of freedom.
We seek a partitioning that
preserves translation invariance within the ``plane''%
~(\ref{eq: def coordinates in Rd to define R and P trsf c})
that is normal to the ``plane''%
~(\ref{eq: def coordinates in Rd to define R and P trsf b}).
With this partition in mind,
we first perform the Fourier transformations
\begin{subequations}
\label{eq: def Fourier Trsf if partition}
\begin{align}
\mathcal{H}^{\,}_{I,I'}(\bm{k}^{\,}_{\perp}):=
\frac{1}{N^{\,}_{\perp}}
\sum_{\bm{r}^{\,}_{\perp},\bm{r}^{\prime}_{\perp}}
e^{\mathrm{i}\bm{k}^{\,}_{\perp}\cdot(\bm{r}^{\,}_{\perp}-\bm{r}^{\prime}_{\perp})}
\mathcal{H}^{\,}_{
I,\bm{r}^{\,}_{\perp};
I^{\prime},\bm{r}^{\prime}_{\perp}
                },
\label{eq: def Fourier Trsf if partition a}
\end{align}
and
\begin{align}
\mathcal{C}^{\,}_{I,I'}(\bm{k}^{\,}_{\perp}):=
\frac{1}{N^{\,}_{\perp}}
\sum_{\bm{r}^{\,}_{\perp},\bm{r}^{\prime}_{\perp}}
e^{\mathrm{i}\bm{k}^{\,}_{\perp}\cdot(\bm{r}^{\,}_{\perp}-\bm{r}^{\prime}_{\perp})}
\mathcal{C}^{\,}_{
I,\bm{r}^{\,}_{\perp};
I^{\prime},\bm{r}^{\prime}_{\perp}
                },
\label{eq: def Fourier Trsf if partition b}
\end{align}
on the single-particle Hamiltonian%
~(\ref{eq: def noninteractin fermion Hamiltonian c})
and equal-time one-point correlation matrix%
~(\ref{eq: def matrix elements C}),
respectively,
where
\begin{equation}
I\equiv(\alpha,\bm{r}^{\,}_{\parallel}),
\qquad
I^{\prime}\equiv
(\alpha^{\prime},\bm{r}^{\prime}_{\parallel}),
\label{eq: def Fourier Trsf if partition c}
\end{equation}
and $N^{\,}_{\perp}$ is the number of lattice sites in $\Lambda$
holding a suitably chosen $\bm{r}^{\,}_{\parallel}$ fixed. 
For some suitably chosen $\bm{r}^{\,}_{\perp}$ held fixed, 
we may then define
the non-intersecting partitioning of the set
\begin{equation}
\Omega:=
\left\{
(\alpha,\bm{r}^{\,}_{\parallel})\left|
\alpha=1,\cdots,N^{\,}_{\mathrm{orb}},\
\begin{pmatrix}
\bm{r}^{\,}_{\parallel}
\\
\bm{r}^{\,}_{\perp}
\end{pmatrix}
\in\Lambda
\right.
\right\}
\end{equation}
into 
\begin{equation}
A=\mathscr{R}B=\mathscr{P}B
\end{equation}
and 
\begin{equation}
B=\mathscr{R}A=\mathscr{P}A.
\end{equation}
\end{subequations}
The discussion in Sec.%
\ref{Equal-time one-point correlation matrix and SUSY QM}
is now applicable
for each $\bm{k}^{\,}_{\perp}$ separately, i.e., 
Eq.~(\ref{eq: hidden SUSY QM}) becomes
\begin{subequations}
\label{eq: hidden SUSY QM k resolved}
\begin{align}
&
C^{2}_{A}(\bm{k}^{\,}_{\perp})-C^{\,}_{A}(\bm{k}^{\,}_{\perp})=
-C^{\ }_{AB}(\bm{k}^{\,}_{\perp})\,C^{\dag}_{AB}(\bm{k}^{\,}_{\perp}),
\label{eq: hidden SUSY QM k resolved a}
\\
&
Q^{\,}_{A}(\bm{k}^{\,}_{\perp})\, 
C^{\,}_{AB}(\bm{k}^{\,}_{\perp})= 
-
C^{\,}_{AB}(\bm{k}^{\,}_{\perp})\, 
Q^{\,}_{B}(\bm{k}^{\,}_{\perp}), 
\label{eq: hidden SUSY QM k resolved b}
\\
&
C^{\dag}_{AB}(\bm{k}^{\,}_{\perp})\, 
Q^{\,}_{A}(\bm{k}^{\,}_{\perp})= 
- 
Q^{\,}_{B}(\bm{k}^{\,}_{\perp})\,
C^{\dag}_{AB}(\bm{k}^{\,}_{\perp}),
\label{eq: hidden SUSY QM k resolved c}
\\
&
C^{2}_{B}(\bm{k}^{\,}_{\perp})
-
C^{\,}_{B}(\bm{k}^{\,}_{\perp})
=
-
C^{\dag}_{AB}(\bm{k}^{\,}_{\perp})\, 
C^{\,}_{AB}(\bm{k}^{\,}_{\perp}).
\label{eq: hidden SUSY QM k resolved d}
\end{align} 
\end{subequations}
We impose the symmetry conditions
\begin{subequations}
\begin{align}
&
\mathcal{R}^{\dag}\, 
C(\bm{k}^{\,}_{\perp})\, 
\mathcal{R}=
C(+\bm{k}^{\,}_{\perp}),
\\
&
\mathcal{P}^{\dag}\, 
C(\bm{k}^{\,}_{\perp})\, 
\mathcal{P}=
C(-\bm{k}^{\,}_{\perp}),
\end{align}
\end{subequations}
where $\mathcal{R}$ and $\mathcal{P}$ are unitary representations
of the actions of the operations of reflection $\mathscr{R}$ 
and parity (inversion) $\mathscr{P}$ 
on the labels~(\ref{eq: def Fourier Trsf if partition c})
for each $\bm{k}^{\,}_{\perp}$ separately, respectively. We assume that
$N^{\,}_{A}=N^{\,}_{B}$,
for which it is necessary that the dimensionality $d$ of the lattice is even
with the pair of orthogonal planes spanned by $\bm{r}^{\,}_{\parallel}$ and
$\bm{r}^{\,}_{\perp}$ each $(d/2)$-dimensional, respectively.
We then conclude Sec.~\ref{Chiral symmetry of the entanglement spectrum} 
with the identities
\begin{subequations}
\label{eq: induced symmetry}
\begin{align}
&
\left\{
Q^{\,}_{A}(\bm{k}^{\,}_{\perp}),
\Gamma^{\,}_{\mathscr{R}}(\bm{k}^{\,}_{\perp})
\right\}=0,
\label{eq: induced symmetry a}
\\
&
Q^{\,}_{A}(\bm{k}^{\,}_{\perp})\,
\Gamma^{\,}_{\mathscr{P}}(\bm{k}^{\,}_{\perp})
+
\Gamma^{\,}_{\mathscr{P}}(\bm{k}^{\,}_{\perp})\,
Q^{\,}_{A}(-\bm{k}^{\,}_{\perp})=0,
\label{eq: induced symmetry b}
\end{align}
\end{subequations}
where $\Gamma^{\,}_{\mathscr{R}}(\bm{k}^{\,}_{\perp})$
and $\Gamma^{\,}_{\mathscr{P}}(\bm{k}^{\,}_{\perp})$
are not assumed invertible for each $\bm{k}^{\,}_{\perp}$ separately.

\subsubsection{Equal-time one-point correlation matrix
and $\mathscr{P}\,\mathscr{C}\,\mathscr{T}$ symmetry}
\label{subsubsec: ... and PCT symmetry}

Any local quantum field theory with a Hermitian Hamiltonian 
for which Lorentz invariance is neither explicitly nor spontaneously broken
must preserve the composition ($\mathscr{P}\,\mathscr{C}\,\mathscr{T}$) 
of parity ($\mathscr{P}$), 
charge conjugation ($\mathscr{C}$), 
and time-reversal ($\mathscr{T}$),
even though neither $\mathscr{P}$, nor $\mathscr{C}$, nor $\mathscr{T}$
need to be separately conserved.%
~\cite{streater1978pct} 
The $\mathscr{P}\,\mathscr{C}\,\mathscr{T}$
theorem implies the existence of antiparticles.
It also implies that any composition of two out 
of the triplet of transformations
$\mathscr{P}$, $\mathscr{C}$, and $\mathscr{T}$
is equivalent to the third. 
[
Both the $\mathscr{P}$ and the $\mathscr{C}\,\mathscr{P}$ 
symmetries are violated by the weak interactions.%
~\cite{Wu57,Garwin57,Christenson64}
Consequently, the $\mathscr{P}\,\mathscr{C}\,\mathscr{T}$
theorem predicts that the $\mathscr{T}$
symmetry is violated by the weak interactions. 
A direct observation for the violation of the $\mathscr{T}$ symmetry
by the weak interaction has been reported in
Ref.~\cite{Lees12}.]

The $\mathscr{P}\,\mathscr{C}\,\mathscr{T}$
theorem does not hold anymore after relaxing
the condition of Lorentz invariance. The effective
Hamiltonians used in condensed matter physics
generically break Lorentz invariance. 
However, the extend to which
$\mathscr{P}$, $\mathscr{C}$, and $\mathscr{T}$ are separately
conserved is not to be decided as a matter of principle but 
depends on the material and the relevant energy scales in 
condensed matter physics. For example,
the effects of the earth magnetic field or those of
cosmic radiation are for most practical purposes irrelevant to
the properties of materials in condensed matter physics.   
A non-relativistic counterpart to
charge conjugation symmetry holds in a mean-field treatment
of superconductivity. Inversion symmetry, 
the non-relativistic counterpart to symmetry under parity, 
is common to many crystalline states of matter.
The effective Hamiltonian of electrons in magnetically inert materials 
preserves time-reversal symmetry in the absence of an external magnetic field.
Conversely, the effective Hamiltonian of electrons in the presence of
magnetic impurities in metals or if the crystalline host is magnetic
break explicitly time-reversal symmetry.

It has become clear in the last eight years that 
the notion of symmetry protected phases of matter 
is useful both theoretically and experimentally 
in condensed matter physics. It has first lead 
to a classification for the possible insulating phases 
of electrons that are adiabatically connected to Slater determinants.%
~\cite{Schnyder2008,Schnyder2009,Kitaev2009,Ryu2010,Stone2011,Ryu12}
Some of the predicted insulating phases have then been observed in
suitable materials. For each insulating phase, 
there exists a non-interacting many-body
Hamiltonian $\hat{H}$ that is smoothly connected to all Hamiltonians
describing this insulating phase by local unitary transformations.%
~\cite{Chen2010}
This classification depends on the dimensionality of space and
on the presence or absence of the following three discrete 
(i.e., involutive) symmetries. 
There is the symmetry of $\hat{H}$ 
under time reversal $\hat{O}^{\,}_{\mathrm{tr}}$,
\begin{equation}
[\hat{H},\hat{O}^{\,}_{\mathrm{tr}}]=0,
\qquad 
\hat{O}^{2}_{\mathrm{tr}}=\pm1.
\end{equation}
There is the spectral symmetry of
$\hat{H}$ under an anti-unitary transformation 
(particle-hole)
$\hat{O}^{\,}_{\mathrm{ph}}$,
\begin{equation}
\{\hat{H},\hat{O}^{\,}_{\mathrm{ph}}\}=0,
\qquad 
\hat{O}^{2}_{\mathrm{ph}}=\pm1.
\end{equation}
There is the spectral symmetry of
$\hat{H}$ under a unitary transformation 
(chiral)
$\hat{O}^{\,}_{\mathrm{ch}}$,
\begin{equation}
\{\hat{H},\hat{O}^{\,}_{\mathrm{ch}}\}=0.
\end{equation}
There are two distinct insulating phases characterized by the
presence or absence of the chiral spectral symmetry when
both time reversal and particle-hole symmetry are broken.
There are another eight distinct insulating phases when
both time reversal and particle-hole symmetry are satisfied.
For a fixed dimension of space, five of the ten insulating
phases support gapless extended boundary states in geometries with open 
boundaries. Moreover, these three discrete symmetries can 
be augmented by discrete (involutive)
symmetries that enforce reflection or mirror symmetries
resulting in a rearrangement of which of the five insulating
phases supporting gapless extended boundary states.

The single-particle matrix $\mathcal{Q}$
defined in Eq.~(\ref{eq: def cal Q})
inherits from the single-particle matrix $\mathcal{H}$
defined by 
Eq.~(\ref{eq: def noninteractin fermion Hamiltonian bis})
any one of the three symmetries
\begin{subequations}
\label{eq: cal Q inherits tr ph ch from cal H}
\begin{align}
&
\left[\mathcal{Q},\mathcal{O}^{\,}_{\mathrm{tr}}\right]=0,
\qquad
\mathcal{O}^{2}_{\mathrm{tr}}=\pm1,
\label{eq: cal Q inherits tr ph ch from cal H a}
\\
&
\left\{\mathcal{Q},\mathcal{O}^{\,}_{\mathrm{ph}}\right\}=0,
\qquad
\mathcal{O}^{2}_{\mathrm{ph}}=\pm1,
\label{eq: cal Q inherits tr ph ch from cal H b}
\\
&
\left\{\mathcal{Q},\mathcal{O}^{\,}_{\mathrm{ch}}\right\}=0,
\label{eq: cal Q inherits tr ph ch from cal H c}
\end{align}
with $\mathscr{O}^{\,}_{\mathrm{ch}}$ 
implementing the chiral transformation
and $\mathcal{O}^{\,}_{\mathrm{ch}}$ its single-particle representation, 
$\mathscr{O}^{\,}_{\mathrm{ph}}$  
implementing the particle-hole transformation
and $\mathcal{O}^{\,}_{\mathrm{ph}}$ its single-particle representation, and 
$\mathscr{O}^{\,}_{\mathrm{tr}}$ 
implementing reversal of time
and $\mathcal{O}^{\,}_{\mathrm{tr}}$ its single-particle representation, 
respectively. Here, we are assuming that there exists a partition
\begin{equation}
\Omega= A\cup B,
\qquad
A\cap B=\emptyset,
\label{eq: cal Q inherits tr ph ch from cal H d}
\end{equation}
that obeys
\begin{equation}
A=\mathscr{O}\,A,
\qquad
B=\mathscr{O}\,B,
\label{eq: cal Q inherits tr ph ch from cal H e}
\end{equation}
\end{subequations}
with 
$\mathscr{O}=
\mathscr{O}^{\,}_{\mathrm{tr}},
\mathscr{O}^{\,}_{\mathrm{ph}},
\mathscr{O}^{\,}_{\mathrm{ch}},$
when the time-reversal symmetry 
or the particle-hole 
or the chiral
spectral symmetries 
of $\mathcal{Q}$ are inherited from those of $\mathcal{H}$.

Conservation of $\mathscr{P}\,\mathscr{C}\,\mathscr{T}$,
as occurs in a relativistic local quantum-field theory,
implies that the operations of 
parity $\mathscr{P}$, 
charge conjugation $\mathscr{C}$,  
and time-reversal $\mathscr{T}$
are not independent. The composition of any two of them is
equivalent to the third one.

Even though $\mathscr{P}\,\mathscr{C}\,\mathscr{T}$ may not be conserved
for the non-interacting Hamiltonian $\mathcal{H}$ defined in Eq.%
~(\ref{eq: def noninteractin fermion Hamiltonian bis}),
its correlation matrix $\mathcal{Q}$
defined in Eq.~(\ref{eq: def cal Q})
or its entanglement matrix $Q^{\,}_{A}$ 
defined in Eq.~(\ref{eq: def QA})
may obey a weaker form of a $\mathscr{P}\,\mathscr{C}\,\mathscr{T}$-like
relation as we now illustrate by way of two examples.

\textit{Case when $\mathscr{P}\,\mathscr{C}$ is equivalent to $\mathscr{T}$:}
When the parity (inversion) transformation $\mathscr{P}$ is a symmetry of
$\mathcal{H}$, there exists a $\Gamma^{\ }_{\mathscr{P}A}$ 
such that, by Eq. (\ref{eq: def Gamma mathsf O anticommutes with Q A}), 
it anti-commutes with $Q^{\,}_{A}$, i.e.,
\begin{align}
\{Q^{\,}_{A},\Gamma^{\ }_{\mathscr{P}A}\}=0.
\end{align}
If we also assume that there exists a particle-hole transformation 
$O^{\,}_{\mathrm{ph}}$ (an anti-unitary transformation)
that anti-commutes with with $Q^{\,}_{A}$
(e.g., if $\mathcal{H}$ is the Bogoliubov-de-Gennes Hamiltonian
of a superconductor), i.e.,
\begin{equation}
\{Q^{\,}_{A},O^{\ }_{\mathrm{ph}}\}=0,
\label{eq: QA anticommutes with ph}
\end{equation}
it then follows that the composition
\begin{equation}
O^{\,}_{\mathrm{tr}}:=
O^{\,}_{\mathrm{ph}}\,
\Gamma^{\ }_{\mathscr{P}A}
\end{equation}
is an anti-linear transformation that commutes with $Q^{\,}_{A}$, i.e.,
\begin{align}
[Q^{\,}_{A},O^{\,}_{\mathrm{tr}}]=0.
\end{align}
The symmetry of $\mathcal{H}$ under parity and
the spectral symmetry of $\mathcal{H}$ under particle-hole transformation
have delivered an effective anti-linear symmetry of $Q^{\,}_{A}$.

\textit{Case when $\mathscr{C}\,\mathscr{T}$ is equivalent to $\mathscr{P}$:}
We assume that
$\mathcal{H}$ defined in Eq.%
~(\ref{eq: def noninteractin fermion Hamiltonian bis})
anti-commutes with the conjugation of charge $\mathscr{C}$ 
represented by $\mathcal{O}^{\,}_{\mathrm{ph}}$,
while it commutes with reversal of time $\mathscr{T}$
represented by $\mathcal{O}^{\,}_{\mathrm{tr}}$.
We also assume that the partition%
~(\ref{eq: cal Q inherits tr ph ch from cal H d})
obeys
Eq.~(\ref{eq: cal Q inherits tr ph ch from cal H e})
when $\mathscr{O}$ is $\mathscr{O}^{\,}_{\mathrm{ph}}$ or 
$\mathcal{O}^{\,}_{\mathrm{tr}}$, i.e.,
\begin{equation}
\mathcal{O}^{\,}_{\mathrm{ph}}=
\begin{pmatrix}
O^{\,}_{\mathrm{ph}A}
&
0
\\
0
&
O^{\,}_{\mathrm{ph}B}
\end{pmatrix},
\qquad
\mathcal{O}^{\,}_{\mathrm{tr}}=
\begin{pmatrix}
O^{\,}_{\mathrm{tr}A}
&
0
\\
0
&
O^{\,}_{\mathrm{tr}B}
\end{pmatrix}.
\end{equation}
We define
\begin{equation}
\mathcal{O}^{\,}_{\mathrm{ch}}:=
\mathcal{O}^{\,}_{\mathrm{ph}}\,
\mathcal{O}^{\,}_{\mathrm{tr}},
\qquad
\mathcal{O}^{\,}_{\mathrm{ch}}=
\begin{pmatrix}
O^{\,}_{\mathrm{ch}A}
&
0
\\
0
&
O^{\,}_{\mathrm{ch}B}
\end{pmatrix}.
\end{equation}
It follows that
$\mathcal{O}^{\,}_{\mathrm{ch}}$
is unitary,
obeys Eq.~(\ref{eq: cal Q inherits tr ph ch from cal H e}),
and anti-commutes with $\mathcal{H}$.
Hence, we may interpret
$\mathcal{O}^{\,}_{\mathrm{ch}}$
as an effective chiral transformation.
The symmetries and spectral symmetries of $\mathcal{H}$
are passed on to $\mathcal{Q}$
defined in Eq.~(\ref{eq: def cal Q}).
In particular, 
\begin{equation}
\mathcal{O}^{\,}_{\mathrm{ch}}\,
\mathcal{Q}=
-
\mathcal{Q}\,
\mathcal{O}^{\,}_{\mathrm{ch}}.
\label{eq: case ii) intermediary anticommutator}
\end{equation}
We make the additional assumption that
$\mathfrak{h}^{\,}_{A}$ 
and 
$\mathfrak{h}^{\,}_{B}$ 
are isomorphic and that the action of
$O^{\,}_{\mathrm{ch}A}$ on $\mathfrak{h}^{\,}_{A}$
is isomorphic to that of
$O^{\,}_{\mathrm{ch}B}$ on $\mathfrak{h}^{\,}_{B}$.
If so, we write
\begin{equation}
O^{\,}_{\mathrm{ch}A}\cong
O^{\,}_{\mathrm{ch}B}\cong
O^{\,}_{\mathrm{ch}},
\qquad
\mathcal{O}^{\,}_{\mathrm{ch}}=
\begin{pmatrix}
O^{\,}_{\mathrm{ch}}
&
0
\\
0
&
O^{\,}_{\mathrm{ch}}
\end{pmatrix}.
\end{equation}
Equation~(\ref{eq: case ii) intermediary anticommutator})
then gives four conditions,
\begin{subequations}
\begin{align}
&
O^{\,}_{\mathrm{ch}}\,  
Q^{\,}_{A} = 
- 
Q^{\,}_{A}\, 
O^{\,}_{\mathrm{ch}}, 
\\
&
O^{\,}_{\mathrm{ch}}\,  
C^{\,}_{AB} = 
- 
C^{\,}_{AB}\, 
O^{\,}_{\mathrm{ch}} , 
\\
&
O^{\,}_{\mathrm{ch}}\,  
C^{\,}_{BA} = 
- 
C^{\,}_{BA}\, 
O^{\,}_{\mathrm{ch}} , 
\\
&
O^{\,}_{\mathrm{ch}}\,   
Q^{\,}_{B} = 
- 
Q^{\,}_{B}\, 
O^{\,}_{\mathrm{ch}}. 
\end{align}
\end{subequations}
With the help of Eq.%
~(\ref{eq: hidden SUSY QM}),
one verifies that
\begin{subequations}
\begin{align}
&
(O^{\,}_{\mathrm{ch}}\,C^{\,}_{BA})\, Q^{\,}_{A}=
-O^{\,}_{\mathrm{ch}}\,  
Q^{\,}_{B}\, 
C^{\,}_{BA}=
Q^{\,}_{B}\, 
(O^{\,}_{\mathrm{ch}}\,C^{\,}_{BA}), 
\\
&
(O^{\,}_{\mathrm{ch}}\,C^{\,}_{BA})\, 
C^{\,}_{AB}=
C^{\,}_{BA}\, 
(C^{\,}_{AB}\,O^{\,}_{\mathrm{ch}}), 
\\
&
(C^{\,}_{AB}\,O^{\,}_{\mathrm{ch}})\, 
C^{\,}_{BA} 
=
C^{\,}_{AB}\,(O^{\,}_{\mathrm{ch}}\,C^{\,}_{BA}), 
\\
&
(C^{\,}_{AB}\,O^{\,}_{\mathrm{ch}})\,  
Q^{\,}_{B} =
- 
C^{\,}_{AB}\,
Q^{\,}_{B}\, 
O^{\,}_{\mathrm{ch}}=
Q^{\,}_{A}\, 
(C^{\,}_{AB}\,O^{\,}_{\mathrm{ch}}).  
\end{align}
\end{subequations}
This should be compared with
Eq.\ (\ref{eq: condition cal O dag cal C cal O= cal c}).
We conclude with the observation that there exists the
transformation
\begin{align}
\mathcal{O}^{\,}_{\mathrm{eff}}:=
\begin{pmatrix}
0 
& 
C^{\,}_{AB}\,
O^{\,}_{\mathrm{ch}}  
\\
O^{\,}_{\mathrm{ch}}\,  
C^{\,}_{BA}  
& 
0
\end{pmatrix}
\end{align}
that commutes with $\mathcal{Q}$,
\begin{align}
[\mathcal{Q},\mathcal{O}^{\,}_{\mathrm{eff}}]=0.
\end{align}
The spectral symmetry of $\mathcal{H}$ under a particle-hole transformation
and the symmetry of $\mathcal{H}$ under time reversal have conspired to
provide $\mathcal{Q}$ with a symmetry under the transformation
$\mathcal{O}^{\,}_{\mathrm{eff}}$.

\subsection{Interacting fermions}
\label{subsec: Interacting fermions}

We consider a many-body Hamiltonian 
acting on the Fock space $\mathfrak{F}$ introduced
in Eq.~(\ref{eq: def noninteractin fermion Hamiltonian bis})
that describes $N^{\,}_{\mathrm{f}}$ 
interacting fermions. 
Its normalized ground state is
\begin{subequations}
\label{eq: def ground state if interactions}
\begin{align}
|\Psi\rangle:=&\,
\sum_{n=1}^{2^{N^{\,}_{\mathrm{tot}}}}
\sum_{n^{(n)}_{1}=0,1}
\cdots
\sum_{n^{(n)}_{N^{\,}_{\mathrm{tot}}}=0,1}
\delta^{\,}_{n^{(n)}_{1}+\cdots+n^{(n)}_{N^{\,}_{\mathrm{tot}}},N^{\,}_{\mathrm{f}}}\,
%\\
%&\,
\times
c^{(\Psi)}_{n^{(n)}_{1},\cdots,n^{(n)}_{N^{\,}_{\mathrm{tot}}}}\,
\left|n^{(n)}_{1},\cdots,n^{(n)}_{N^{\,}_{\mathrm{tot}}}\right\rangle.
\label{eq: def ground state if interactions a}
\end{align}
The Slater determinant
\begin{equation}
\left\langle n^{(n)}_{1},\cdots,n^{(n)}_{N^{\,}_{\mathrm{tot}}}\right|:=
\prod_{I=1}^{N^{\,}_{\mathrm{tot}}} 
\langle0|
\left(\hat{\chi}^{\,}_{I}\right)^{n^{(n)}_{I}}, 
\label{eq: def ground state if interactions b}
\end{equation}
\end{subequations}
has
$c^{(\Psi)}_{n^{(n)}_{1},\cdots,n^{(n)}_{N^{\,}_{\mathrm{tot}}}}$
as its overlap with the ground state $|\Psi\rangle$.

In the presence of interactions,
the equal-time one-point correlation function (matrix)
\begin{equation}
\mathcal{C}^{\,}_{IJ}:=
\langle\Psi|
\hat{\psi}^{\dag}_{I}\,
\hat{\psi}^{\,}_{J}\,
|\Psi\rangle,
\qquad
I,J=1,\cdots,N^{\,}_{\mathrm{tot}},
\label{eq: def equal time one point fct with interactions} 
\end{equation}
does not convey anymore the same information as the density matrix
\begin{subequations}
\label{eq: interacting reduced density matrix}
\begin{equation}
\hat{\rho}^{\,}_{\Psi}:=
|\Psi\rangle
\,
\langle\Psi|,
\qquad
\langle\Psi|\Psi\rangle=1.
\label{eq: interacting density matrix}
\end{equation}
In particular, Eq.~(\ref{eq: def equal time one point fct with interactions})
does not encode the full information contained 
in the reduced density matrices
\begin{equation}
\hat{\rho}^{(A)}_{\Psi}:=
\mathrm{tr}^{\,}_{\mathfrak{F}^{\,}_{B}}\,
\hat{\rho}^{\,}_{\Psi}
\label{eq: interacting reduced density matrix A}
\end{equation}
and
\begin{equation}
\hat{\rho}^{(B)}_{\Psi}:=
\mathrm{tr}^{\,}_{\mathfrak{F}^{\,}_{A}}\,
\hat{\rho}^{\,}_{\Psi},
\label{eq: interacting reduced density matrix B}
\end{equation}
\end{subequations}
where we have defined the partition in
Eqs.~(\ref{eq: tensorial decomposition mathfrak F})
and~(\ref{eq: partitioning single particles}).

We seek a useful representation of
the reduced density matrices
(\ref{eq: interacting reduced density matrix A})
and 
(\ref{eq: interacting reduced density matrix B})
and how they might be related by symmetries
of the interacting Hamiltonian. To this end,
we assign the labels 
$\mu^{\,}_{A}=1,\cdots,\mathrm{dim}\,\mathfrak{F}^{\,}_{A}$ 
and
$\mu^{\,}_{B}=1,\cdots,\mathrm{dim}\,\mathfrak{F}^{\,}_{B}$ 
to any orthonormal basis 
$\{|\Psi^{(A)}_{\mu^{\,}_{A}}\rangle\}$ 
and
$\{|\Psi^{(B)}_{\mu^{\,}_{B}}\rangle\}$ 
that span the Fock spaces 
$\mathfrak{F}^{\,}_{A}$
and 
$\mathfrak{F}^{\,}_{B}$,
respectively. Without loss of generality, we assume
$\mathrm{dim}\,\mathfrak{F}^{\,}_{A}\geqslant\mathrm{dim}\,\mathfrak{F}^{\,}_{B}$.
We may write the expansion
\begin{equation}
|\Psi\rangle=
\sum_{\mu^{\,}_{A}=1}^{\mathrm{dim}\,\mathfrak{F}^{\,}_{A}}
\sum_{\mu^{\,}_{B}=1}^{\mathrm{dim}\,\mathfrak{F}^{\,}_{B}}
D^{\,}_{\mu^{\,}_{A}\mu^{\,}_{B}}\,
|\Psi^{(A)}_{\mu^{\,}_{A}}\rangle
\otimes
|\Psi^{(B)}_{\mu^{\,}_{B}}\rangle
\end{equation}
with the overlaps $D^{\,}_{\mu^{\,}_{A}\mu^{\,}_{B}}\in\mathbb{C}$
defining the matrix elements of the 
$\mathrm{dim}\,\mathfrak{F}^{\,}_{A}\times\mathrm{dim}\,\mathfrak{F}^{\,}_{B}$
matrix $D$. 

At this stage, we make use of the singular value decomposition
\begin{subequations}
\label{eq: def singular value decomposition}
\begin{equation}
D=
U\,
\Sigma\,
V^{\dag}
\label{eq: def singular value decomposition a}
\end{equation}
with 
\begin{equation}
U=(U^{\,}_{\mu^{\,}_{A}\mu^{\prime}_{A}})
\label{eq: def singular value decomposition b}
\end{equation}
a $\mathrm{dim}\,\mathfrak{F}^{\,}_{A}\times\mathrm{dim}\,\mathfrak{F}^{\,}_{A}$
unitary matrix,
\begin{equation}
\Sigma=
\begin{pmatrix}
\sigma^{\,}_{1}   % Column 1
&                
0               % Column 2
&
\cdots          % Column 3
&
{}              % Column 4
&
{}              % Column 5
&
0               % Column 6
\\  % END LINE 1
0
&
\ddots
&
\ddots
&
&
&
\vdots
\\ % END LINE 2
\vdots
&
\ddots
&
\sigma^{\,}_{\mathfrak{R}}
&
&
&
\\ % END LINE 3
&
&
&
0
&
\ddots
&
\\ % END LINE 4
&
&
&
\ddots
&
\ddots
&
0
\\ % END LINE 5
0
&
&
&
&
0
&
0
\\ % END LINE 6
\vdots
&
&
&
&
\vdots
&
\vdots
   % END LINE 7
\end{pmatrix},
\quad
0<\sigma^{\,}_{1}\leqslant\cdots\leqslant\sigma^{\,}_{\mathfrak{R}},
\label{eq: def singular value decomposition c}
\end{equation}
a $\mathrm{dim}\,\mathfrak{F}^{\,}_{A}\times\mathrm{dim}\,\mathfrak{F}^{\,}_{B}$
rectangular diagonal matrix 
of rank $\mathfrak{R}\leqslant\mathrm{dim}\,\mathfrak{F}^{\,}_{B}$,
and 
\begin{equation}
V=\left(V^{\,}_{\mu^{\,}_{B}\mu^{\prime}_{B}}\right)
\label{eq: def singular value decomposition d}
\end{equation}
\end{subequations}
a $\mathrm{dim}\,\mathfrak{F}^{\,}_{B}\times\mathrm{dim}\,\mathfrak{F}^{\,}_{B}$
unitary matrix.

With the help of Eq.~(\ref{eq: def singular value decomposition}), 
we have the {\it Schmidt decomposition}:
\begin{subequations}
\label{eq: SVD Psi}
\begin{equation}
|\Psi\rangle=
\sum_{\nu=1}^{\mathfrak{R}}
\sigma^{\,}_{\nu}\,
|\widetilde{\Psi}^{(A)}_{\nu}\rangle
\otimes
|\widetilde{\Psi}^{(B)}_{\nu}\rangle
\end{equation}
where
\begin{equation}
|\widetilde{\Psi}^{(A)}_{\nu}\rangle:=
\sum_{\mu^{\,}_{A}=1}^{\mathrm{dim}\,\mathfrak{F}^{\,}_{A}}
U^{\,}_{\mu^{\,}_{A}\nu}\,
|\Psi^{(A)}_{\mu^{\,}_{A}}\rangle
\end{equation}
and
\begin{equation}
|\widetilde{\Psi}^{(B)}_{\nu}\rangle:=
\sum_{\mu^{\,}_{B}=1}^{\mathrm{dim}\,\mathfrak{F}^{\,}_{B}}
V^{*}_{\mu^{\,}_{B}\nu}\,
|\Psi^{(B)}_{\mu^{\,}_{B}}\rangle.
\end{equation}
The singular values $\{\sigma^{\,}_{\nu}\}\equiv\sigma(\Sigma)$
of the rectangular matrix $\Sigma$ are non-negative and obey
the normalization condition
\begin{equation}
\sum_{\nu=1}^{\mathfrak{R}}
\sigma^{2}_{\nu}=1
\end{equation} 
\end{subequations}
owing to the facts that the ground state%
~(\ref{eq: def ground state if interactions})
is normalized to one and that the basis of
$\mathfrak{F}^{\,}_{A}$ and $\mathfrak{F}^{\,}_{B}$
are chosen orthonormal.

In the basis~(\ref{eq: SVD Psi}),
\begin{subequations}
\label{eq: interacting reduced density matrix bis}
\begin{equation}
\hat{\rho}^{\,}_{\Psi}:=
\sum_{\nu =1}^{\mathfrak{R}}
\sum_{\nu'=1}^{\mathfrak{R}}
\sigma^{\,}_{\nu }\,
\sigma^{\,}_{\nu'}\,
|\widetilde{\Psi}^{(A)}_{\nu }\rangle
\,
\langle\widetilde{\Psi}^{(A)}_{\nu'}|
\otimes
|\widetilde{\Psi}^{(B)}_{\nu }\rangle
\,
\langle\widetilde{\Psi}^{(B)}_{\nu'}|
\label{eq: rho in tilde psi basis}
\end{equation}
and the reduced density matrices%
~(\ref{eq: interacting reduced density matrix A})
and%
~(\ref{eq: interacting reduced density matrix B})
become the spectral decompositions 
\begin{equation}
\hat{\rho}^{(A)}_{\Psi}=
\sum_{\nu=1}^{\mathfrak{R}}
\sigma^{2}_{\nu}\,
|\widetilde{\Psi}^{(A)}_{\nu}\rangle
\,
\langle\widetilde{\Psi}^{(A)}_{\nu}|
\label{eq: interacting reduced density matrix A bis}
\end{equation}
and
\begin{equation}
\hat{\rho}^{(B)}_{\Psi}:=
\sum_{\nu=1}^{\mathfrak{R}}
\sigma^{2}_{\nu}\,
|\widetilde{\Psi}^{(B)}_{\nu}\rangle
\,
\langle\widetilde{\Psi}^{(B)}_{\nu}|,
\label{eq: interacting reduced density matrix B bis}
\end{equation}
respectively. The reduced density matrices 
$\hat{\rho}^{(A)}_{\Psi}$
and
$\hat{\rho}^{(B)}_{\Psi}$
are explicitly positive semi-definite. 
They share the same non-vanishing eigenvalues
\begin{equation}
0<\sigma^{2}_{\nu}\equiv 
e^{-\omega^{\,}_{\nu}},
\qquad\!
0\leqslant\omega^{\,}_{\nu}<\infty,
\qquad\!
\nu=1,\cdots,\mathfrak{R},
\label{eq: sigma are nonvanishing}
\end{equation}
\end{subequations}
each of which can be interpreted as the probability for 
the ground state~(\ref{eq: def ground state if interactions}) 
to be in the orthonormal basis state 
$
|\widetilde{\Psi}^{(A)}_{\nu}\rangle
\otimes
|\widetilde{\Psi}^{(B)}_{\nu}\rangle.
$

As we did with Eq.~(\ref{eq: cal O off block diagonal a}),
we assume that
\begin{subequations}
\begin{equation}
\mathrm{dim}\,\mathfrak{F}^{\,}_{A}=
\mathrm{dim}\,\mathfrak{F}^{\,}_{B}=
\mathfrak{D}.
\end{equation}
Hence,
$\mathrm{dim}\,\mathfrak{F}^{\,}_{A}$
and
$\mathrm{dim}\,\mathfrak{F}^{\,}_{B}$
are isomorphic. This means that
$\hat{\Gamma}^{\,}_{AB}:\mathfrak{F}^{\,}_{B}\mapsto\mathfrak{F}^{\,}_{A}$
defined by
\begin{equation}
\hat{\Gamma}^{\,}_{AB}:=
\sum_{\nu=1}^{\mathfrak{D}}
|\widetilde{\Psi}^{(A)}_{\nu}\rangle
\,
\langle\widetilde{\Psi}^{(B)}_{\nu}|
\end{equation}
and
$\hat{\Gamma}^{\,}_{BA}:\mathfrak{F}^{\,}_{A}\mapsto\mathfrak{F}^{\,}_{B}$
defined by
\begin{equation}
\hat{\Gamma}^{\,}_{BA}:=
\sum_{\nu=1}^{\mathfrak{D}}
|\widetilde{\Psi}^{(B)}_{\nu}\rangle
\,
\langle\widetilde{\Psi}^{(A)}_{\nu}|
\end{equation}
\end{subequations}
are linear one-to-one maps.
One verifies that
\begin{subequations}
\label{eq: action hat Gamma on reduced density matrix}
\begin{align}
&
\hat{\Gamma}^{-1}_{AB}=\hat{\Gamma}^{\,}_{BA}=\hat{\Gamma}^{\dag}_{AB},
\\
&
\hat{\rho}^{(A)}_{\Psi}=
\hat{\Gamma}^{\,}_{AB}\,
\hat{\rho}^{(B)}_{\Psi}\,
\hat{\Gamma}^{\dag}_{AB},
\quad
\hat{\rho}^{(B)}_{\Psi}=
\hat{\Gamma}^{\dag}_{AB}\,
\hat{\rho}^{(A)}_{\Psi}\,
\hat{\Gamma}^{\,}_{AB}.
\end{align}
\label{eq: interacting fermion mapping}
\end{subequations}

We conclude Sec.~\ref{subsec: Interacting fermions}
with a counterpart for interacting fermions  
of the spectral symmetry%
~(\ref{eq: main result on how to get chiral sym}).
We assume that the operation $\mathscr{O}$
under which the partition is interchanged,
\begin{equation}
A=\mathscr{O}B,
\qquad
B=\mathscr{O}A,
\end{equation}
is represented by the unitary map
$\mathcal{O}:\mathfrak{F}^{\,}_{A}\mapsto\mathfrak{F}^{\,}_{B}$
defined either by 
\begin{subequations}
\label{eq: def map from FA to FB tensorial basis}
\begin{equation}
|\widetilde{\Psi}^{(B)}_{\nu}\rangle^{\,}_{\mathscr{O}}:=
\sum_{\nu'=1}^{\mathfrak{R}}
\mathcal{O}^{*}_{\nu\nu'}
|\widetilde{\Psi}^{(A)}_{\nu'}\rangle,
\label{eq: def map from FA to FB tensorial basis a}
\end{equation}
or by its inverse
$\mathcal{O}^{\dag}:\mathfrak{F}^{\,}_{B}\mapsto\mathfrak{F}^{\,}_{A}$
defined by 
\begin{equation}
|\widetilde{\Psi}^{(A)}_{\nu'}\rangle=:
\sum_{\nu''=1}^{\mathfrak{R}}
\mathcal{O}^{\,}_{\nu''\nu'}
|\widetilde{\Psi}^{(B)}_{\nu''}\rangle^{\,}_{\mathscr{O}},
\label{eq: def map from FA to FB tensorial basis b}
\end{equation}
whereby
\begin{equation}
\sum_{\nu'=1}^{\mathfrak{R}}
\mathcal{O}^{*}_{\nu\nu'}
\mathcal{O}^{\,}_{\nu''\nu'}=
\delta^{\,}_{\nu,\nu''}
\label{eq: def map from FA to FB tensorial basis c}
\end{equation}
\end{subequations}
with 
$\nu,\nu',\nu''=1,\cdots,\mathfrak{R}$.

By Eq.~(\ref{eq: interacting reduced density matrix A bis}),
\begin{align}
\left(
\hat{\rho}^{(A)}_{\Psi}
\right)^{\,}_{\mathscr{O}}:=
\sum_{\nu,\nu',\nu''=1}^{\mathfrak{R}}
\sigma^{2}_{\nu}\,
\mathcal{O}^{*}_{\nu'\nu}\,
\mathcal{O}^{\,}_{\nu''\nu}\,
|\widetilde{\Psi}^{(B)}_{\nu'}{\rangle}^{\,}_{\mathscr{O}}
\,
{\vphantom{\rangle}}^{\,}_{\mathscr{O}}
\langle\widetilde{\Psi}^{(B)}_{\nu''}|.
\end{align}
%\end{subequations}
If we impose the symmetry constraint
\begin{subequations}
\label{eq: imposing symmetry so that rho A maps to rho B}
\begin{equation}
\sum_{\nu=1}^{\mathfrak{R}}
\sigma^{2}_{\nu}\,
\mathcal{O}^{*}_{\nu'\nu}\,
\mathcal{O}^{\,}_{\nu''\nu}=
\begin{cases}
\sigma^{2}_{\nu'}\,\delta^{\,}_{\nu',\nu''},
&
\hbox{if $\nu'=1,\cdots,\mathfrak{R}$,}
\\
0,
&
\hbox{otherwise,}
\end{cases}
\label{eq: imposing symmetry so that rho A maps to rho B a}
\end{equation}
(i.e., $\mathcal{O}^{\dag}\,\Sigma\,\mathcal{O}=\Sigma$),
we conclude that
\begin{align}
\left(
\hat{\rho}^{(A)}_{\Psi}
\right)^{\,}_{\mathscr{O}}=
\sum_{\nu=1}^{\mathfrak{R}}
\sigma^{2}_{\nu}\,
|\widetilde{\Psi}^{(B)}_{\nu}{\rangle}^{\,}_{\mathscr{O}}
\,
{\vphantom{\rangle}}^{\,}_{\mathscr{O}}
\langle\widetilde{\Psi}^{(B)}_{\nu}|
\equiv
\hat{\rho}^{(B)}_{\mathscr{O}\Psi}
\end{align}
\end{subequations}

We can combine Eqs.
~(\ref{eq: action hat Gamma on reduced density matrix})
and~(\ref{eq: imposing symmetry so that rho A maps to rho B})
as follows. Given are the interacting Hamiltonian
$\hat{H}$ acting on the Fock space $\mathfrak{F}$ defined in the 
orbital-lattice basis,
the density matrix $\hat{\rho}^{\,}_{\Psi}$ for the ground state 
$|\Psi\rangle$ of $\hat{H}$, the partition 
$\mathfrak{F}=\mathfrak{F}^{\,}_{A}\otimes\mathfrak{F}^{\,}_{B}$,
and the reduced density matrix 
$\hat{\rho}^{(A)}_{\Psi}$
and 
$\hat{\rho}^{(B)}_{\Psi}$
acting on the Fock spaces 
$\mathfrak{F}^{\,}_{A}$
and
$\mathfrak{F}^{\,}_{B}$
with $\mathrm{dim}\,\mathfrak{F}^{\,}_{A}=\mathrm{dim}\,\mathfrak{F}^{\,}_{B}$,
respectively.
Let $\hat{\Sigma}^{\,}$ represent 
the action by conjugation on the space of operators
of the matrix%
~(\ref{eq: def singular value decomposition c}).
Let $\hat{O}$ represent
the action by conjugation on the space of operators
of the matrix $\mathcal{O}$ with the matrix elements%
~(\ref{eq: def map from FA to FB tensorial basis}).
Let 
\begin{equation}
\hat{\Xi}:=\hat{\Gamma}^{\,}_{BA}\,\hat{O}
\end{equation}
and assume the symmetry
\begin{equation}
\hat{O}^{\dag}\,\hat{\Sigma}\,\hat{O}=\hat{\Sigma}.
\end{equation}
The symmetry 
\begin{equation}
\hat{\rho}^{(A)}_{\Psi}=
\hat{\Xi}^{\dag}\,
\hat{\rho}^{(A)}_{\Psi}\,
\hat{\Xi}
\end{equation}
obeyed by the reduced density matrix then follows.

\section{Physical versus entangling boundaries, spectral gap, and locality}
\label{sec: Physical versus entangling boundaries, spectral gap, and locality}

Topological band insulators are insulators with two equivalent
properties in the thermodynamic limit. 
If the discrete symmetry of time reversal, 
or the spectral symmetries of
charge conjugation or chirality
are imposed together with
periodic boundary conditions, the bundle
of Bloch states making up a ground state defines
a quantized index, i.e., a number that does not change if the 
bundle of Bloch states is changed smoothly without violating the
discrete symmetries or closing the band gap.
The quantization of this index is a topological attribute of the 
occupied bands that is protected by certain discrete symmetries.
If open boundary conditions are imposed instead of periodic ones,
mid-gap single-particle states 
that are exponentially localized on the boundaries
descend from the continuum of occupied and unoccupied Bloch states. 
These boundary states are protected, i.e.,
robust to perturbations that are compliant with the discrete symmetries,
change smoothly Bloch states, and do not close the band gap.
This equivalence between a topological index when 
periodic boundary conditions are imposed and protected boundary states 
when open  boundary conditions are imposed is an example of a 
bulk-edge correspondence. 

We shall call the boundaries in position space (i.e., in the orbital basis)
that are selected by imposing open boundary conditions on the
single-particle Hamiltonian $\mathcal{H}$ defined in
Eq.~(\ref{eq: def noninteractin fermion Hamiltonian})
the physical boundaries.
It is possible to define another class of boundaries in position space
(i.e., in the orbital basis) by performing the partitioning%
~(\ref{eq: partitioning single particles}). We call the boundaries
that separate the lattice labels of the orbital basis
in set $A$ from those in set $B$ the entangling boundaries.
The number of entangling boundaries depends on whether
periodic, open, or mixed boundary conditions are imposed.
For a one-dimensional lattice such as the one shown in
Fig.~\ref{fig: dimerized chain}(a),
choosing periodic boundary
conditions selects no physical boundaries as is shown in
the left panel of Fig.~\ref{fig: dimerized chain}(b),
while it selects two identical entangling boundaries as is shown in
the left panel of Fig.~\ref{fig: dimerized chain}(c).
On the other hand, choosing open boundary
conditions selects two identical physical boundaries as is shown in
the right panel of Fig.~\ref{fig: dimerized chain}(b),
while it selects one physical boundary and one entangling
boundary as is shown in
the right panel of Fig.~\ref{fig: dimerized chain}(c).

It has been proposed that the existence of protected single-particle states
that are localized at the entangling boundaries, i.e., protected
entangling boundary states, is a more refined signature for
topological band insulators than the existence of protected physical
boundary states. Indeed, it is argued that the existence
of protected physical boundary states implies the existence
of protected entangling boundary states for a suitable partition%
~\cite{Ryu2006,Turner2010,Fidkowski2010,Hughes2011}.
However, the converse is known not to hold. For example,
three-dimensional inversion symmetric topological insulators
have been constructed such that the physical boundary states
can be gapped out by inversion-symmetric perturbations,
while the entangling boundary states remain gapless.%
~\cite{Hughes2011,Turner2010}

\subsection{Spectral gap and locality of the equal-time 
one-point correlation matrix}

In this context, the following observations are crucial.
On the one hand, none of the results of 
Sec.~\ref{sec: Symmetries and entanglement Spectrum}
are sensitive to the presence or absence of a spectral gap
$\Delta$
separating the Slater determinant entering the definition%
~(\ref{eq: def matrix elements C}) 
of the equal-time one-point correlation matrix $\mathcal{C}$
from all excited states when periodic boundary conditions are imposed.
On the other hand, the stability analysis 
from Secs.~\ref{sec: example A}-\ref{sec: example D}
only makes sense if $\Delta>0$, since
the exponential decay of the boundary states away from the boundaries
is controlled by the characteristic length 
$\xi\propto1/\Delta$. Consequently, 
the overlap of two boundary states localized
on disconnected boundaries a distance $\propto N$ apart 
vanishes exponentially fast as the thermodynamic limit $N\to\infty$ is taken,
i.e., the matrix elements $\mathcal{C}^{\,}_{IJ}$
in Eq.~(\ref{eq: def matrix elements C}) 
also decay exponentially fast if the pair of repeat unit cells 
in the collective labels $I$ and $J$ are a distance larger than $\xi$ apart.
In particular, the matrix elements entering the upper-right block
$C^{\,}_{AB}$ in Eq.~(\ref{eq: partitioning single particles})
and those entering the transformation%
~(\ref{eq: def Gamma mathsf O})
are suppressed exponentially in magnitude
if they correspond to two repeat unit cells a distance larger than $\xi$ apart.
It is only because of the presence of a spectral gap above the ground state
of the Hamiltonian $\mathcal{H}$ defined in 
Eq.~(\ref{eq: def noninteractin fermion Hamiltonian})
that $\mathcal{C}$ defined in 
Eq.~(\ref{eq: partitioning single particles})
inherits from $\mathcal{H}$ its locality in the orbital basis.

\begin{figure}[t]
\centering
(a)
\includegraphics[width=0.4\textwidth]{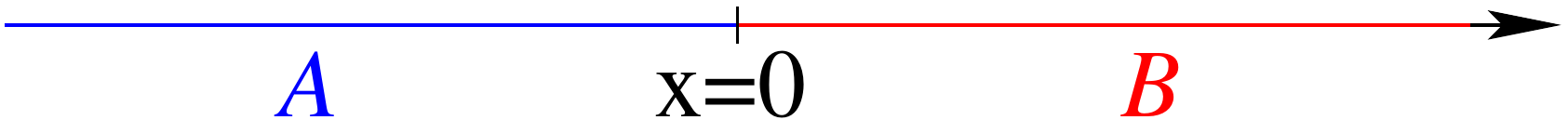}
%\medskip
(b)
\includegraphics[width=0.2\textwidth]{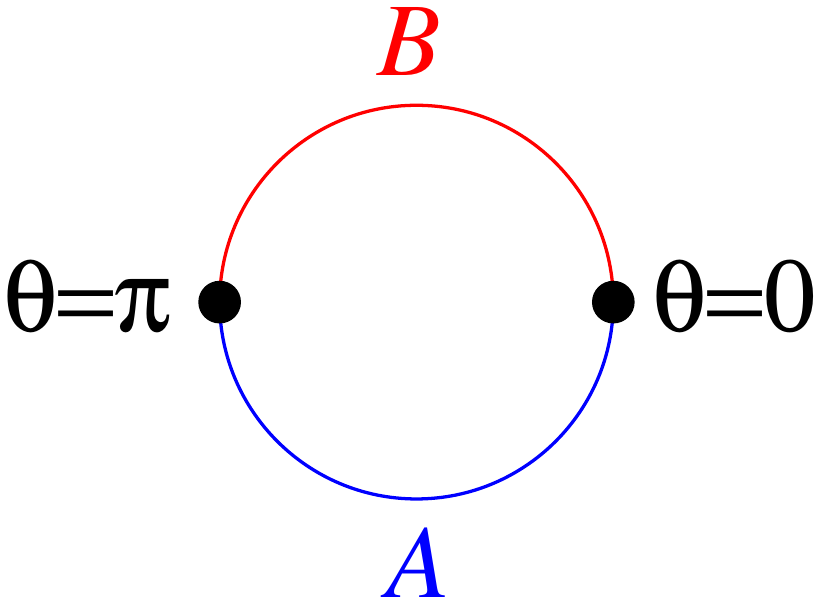}
\caption{%
(Color online)
(a)
The real line $\{x\in\mathbb{R}\}$
is partitioned into two open sets, 
the negative axis $A$
and the positive axis $B$.
The boundary between $A$ and $B$ is
the origin at $x=0$, a compact set.
This boundary is the entangling boundary of the real line,
a connected set.  
The real line has two disconnected physical boundaries at
$x=-\infty$ and $x=+\infty$. 
The inversion about the origin $x\mapsto-x$
is smooth,
exchanges $A$ and $B$,
and has the entangling boundary as its unique fixed point.
The map $x\mapsto-1/x$ is 
not smooth at the origin,
exchanges $A$ and $B$,
and has no fixed point.
The real line is the limit $r\to0$ 
of a cylinder with the radius $r$
embedded in three-dimensional Cartesian space.
(b)
The circle $\{\theta\in[0,2\pi[\}$
can be interpreted as the
compactification of the real line if the physical boundaries
at $x=-\infty$ and $x=+\infty$ are identified with the angle $\theta=\pi$
and the origin of the real line is identified as the angle $\theta=0$.
In doing so, the partition into the open sets $A$ and $B$ of the real line
acquires two entangling boundaries at $\theta=0$ and $\theta=\pi$,
respectively. These are two disconnected compact sets.
The inversion about $\theta=0$ defined by
$\theta\mapsto-\theta$
is smooth, 
exchanges $A$ and $B$,
and has the two distinct fixed points $\theta=0$ and $\theta=\pi$.
The inversion about $\theta=\pi$ defined by
$\theta\mapsto\pi-\theta$
is smooth, 
leaves $A$ ($B$) invariant as a set,
and exchanges $\theta=0$ and $\theta=\pi$.
The translation defined by
$\theta\mapsto\pi+\theta$,
the composition of the two previous inversions,
is smooth, 
exchanges $A$ and $B$,
and has no fixed points.
A circle of radius $R$ is the limit $r\to0$ of a 
ring torus obtained by revolving a circle of radius $r$
in three-dimensional Cartesian space 
about an axis coplanar with the circle a distance $R>r$
apart from the center of mass of the revolving circle.
         }
\label{fig: cylinder vs torus}
\end{figure}

\subsection{Spectral gap and locality of the 
spectral symmetry~(\ref{eq: def Gamma mathsf O})}
\label{subsec: Spectral gap and locality of the induced symmetry Gamma}

In the remaining of this paper, we shall be concerned with
exploring the effects of the spectral symmetry
~(\ref{eq: def Gamma mathsf O})
on the entanglement spectra
when a single-particle \textit{bulk} 
spectral gap $\Delta$ guarantees that the 
equal-time one-point correlation matrices
(\ref{eq: def matrix elements C})
and
(\ref{eq: def cal Q})
are local for bulk-like separations, 
i.e., their matrix elements in the position basis
can be bounded from above by
the exponential factor $a\,\exp(-b\,|\bm{r}|\,\Delta)$ where
$a$ and $b$ are some numerical factor of order unity and 
$|\bm{r}|\geq1/\Delta$ (in units with the Planck constant set to
$\hbar=1$ and the speed of light set to $c=1$)
is the bulk-like distance in space of the lattice degrees of freedom 
entering the bra and ket.
In other words, we seek to understand how the spectral properties of
the block $Q^{\,}_{A}$ of the equal-time one-point correlation matrix defined in
Eq.~(\ref{eq: def cal Q})
are affected by the assumption that a symmetry operation $\mathscr{O}$ 
conspires with the partition of the degrees of freedom of the single-particle
Hilbert space so as to obey Eq.~(\ref{eq: cal O off block diagonal}), i.e.,
the map
\begin{equation}
\Gamma_{\mathscr{O} A}:
\mathfrak{H}^{\,}_{A}\mapsto
\mathfrak{H}^{\,}_{A}
\end{equation}
defined by the matrix multiplication
$\Gamma^{\,}_{\mathscr{O} A}:=C^{\,}_{AB}\,O^{\dag}_{AB}$
anti-commutes with $Q^{\,}_{A}$.

Typically, the symmetry operation $\mathscr{O}$ is
a point-group symmetry, say an inversion or reflection symmetry.
Hence, we shall assume that the symmetry operation $\mathscr{O}$ 
is not local. The spectral symmetry of $Q^{\,}_{A}$ generated by
$\Gamma^{\,}_{\mathscr{O}\,A}$ 
allows to draw definitive conclusions 
on the existence of protected zero modes in the entanglement spectrum
$\sigma(Q^{\,}_{A})$
only if $\Gamma^{\,}_{\mathscr{O}\,A}$ 
is a local operation for bulk-like separations.
For this reason, we devote section%
~(\ref{subsec: Spectral gap and locality of the induced symmetry Gamma})
to studying the conditions under which
$\Gamma^{\,}_{\mathscr{O}\,A}$ is local for bulk-like separations.

We shall consider two geometries for simplicity.
Either we impose on $d$-dimensional space the geometry of 
a cylinder by imposing periodic boundary conditions in all $(d-1)$
directions in space  while imposing open boundary conditions 
for the last direction.
Or we impose on  $d$-dimensional space the geometry of a torus by imposing
periodic boundary conditions to all $d$ directions in space.
When space is one dimensional, the cases of cylindrical and torus geometry
are illustrated in Fig.~\ref{fig: cylinder vs torus}(a) 
and \ref{fig: cylinder vs torus}(b), respectively.

For a cylindrical geometry,
as illustrated in Fig.~\ref{fig: cylinder vs torus}(a),
there are two disconnected compact physical boundaries at $x=\pm\infty$
and one compact entangling boundary at $x=0$, where $x$ 
is the non-compact coordinate along the cylinder axis.
We choose the symmetry operation $\mathscr{O}$ to be
the mirror operation $x\mapsto-x$ that leaves the 
compact entangling boundary at $x=0$
point-wise invariant,
while it exchanges the two
disconnected compact physical boundaries at $x=\pm\infty$.
The matrix elements of $\Gamma^{\,}_{\mathscr{O}\,A}$
involving bra and kets with lattice degrees of freedom
a distance $|\bm{r}|\gg1/\Delta$ away from the 
entangling boundary at $x=0$
are exponentially suppressed by the factor
$\exp(-b|\bm{r}|\Delta)$
originating from $C^{\dag}_{AB}$.
Hence, $\Gamma^{\,}_{\mathscr{O}\,A}$ is a local operator
for bulk-like separations
that generates a spectral symmetry on
$Q^{\,}_{A}$
without mixing boundary states localized on disconnected
boundaries (whether physical or entangling).

For a torus geometry,
as illustrated in Fig.~\ref{fig: cylinder vs torus}(a),
there are two disconnected compact entangling boundaries at 
$\theta=0$ and $\theta=\pi$. On the one hand,
we may choose the symmetry operation $\mathscr{O}$ to be
the mirror operation $\theta\mapsto-\theta$ that leaves the 
compact entangling boundary at $\theta=0$ and $\theta=\pi$
point-wise invariant.
The matrix elements of $\Gamma^{\,}_{\mathscr{O}\,A}$
involving bra and kets with lattice degrees of freedom
a distance $|\bm{r}|\gg1/\Delta$ away from the 
entangling boundaries at $\theta=0$ and $\theta=\pi$
are exponentially suppressed by the factor
$\exp(-b|\bm{r}|\Delta)$
originating from $C^{\dag}_{AB}$.
Hence, $\Gamma^{\,}_{\mathscr{O}\,A}$ is again a local operator
for bulk-like separations
that generates a spectral symmetry on
$Q^{\,}_{A}$
without mixing boundary states localized on disconnected
entangling boundaries.
On the other hand,
we may choose the symmetry operation $\mathscr{O}$ to be
the mirror operation $\theta\mapsto\pi+\theta$ that exchanges 
the compact entangling boundary at $\theta=0$ and $\theta=\pi$.
Hence, $\Gamma^{\,}_{\mathscr{O}\,A}$ is not a local operator
for it mixes with an amplitude of order unity
boundary states localized on disconnected
entangling boundaries separated by an arbitrary distance.
Hence, even though $\Gamma^{\,}_{\mathscr{O}\,A}$
generates a spectral symmetry of the entanglement spectrum
of $Q^{\,}_{A}$, it cannot be used to deduce any stability
properties of the zero modes localized on the
boundaries.

Sections~\ref{sec: example A}-\ref{sec: example D}
are devoted to the stability analysis of entangling boundary states
by way of examples in one and two dimensions.
This stability analysis requires distinguishing
the nature of the boundaries in the partition%
~(\ref{eq: partitioning single particles}),
for whether these boundaries are physical or entangling
depends on the choice of the boundary conditions imposed
along the $d$ dimensions of space, as we have illustrated for the
case of $d=1$ in Fig.~\ref{fig: dimerized chain}.
This stability analysis also requires determining if
the spectral symmetry
$\Gamma^{\,}_{\mathscr{O}\,A}$ is local or not,
as we have illustrated for the
case of $d=1$ in Fig.~\ref{fig: cylinder vs torus}.

\section{Topological insulator protected by
reflection (inversion) symmetry in one dimension}
\label{sec: example A}

\subsection{Hamiltonian}
\label{subsubsec: Hamiltonian 1D example}

%%%%figure for example 1
\begin{figure*}[t]
\centering
(a)
\includegraphics[width=0.4\textwidth]{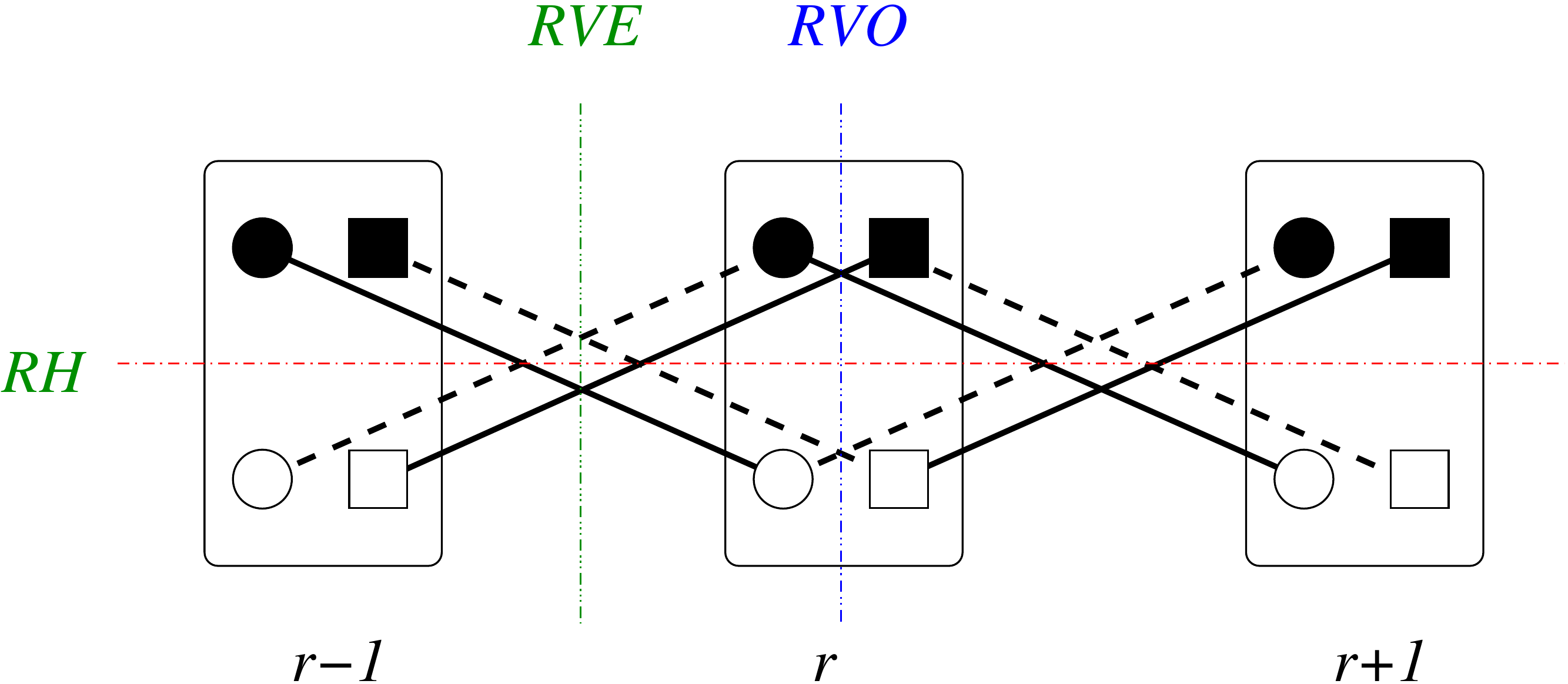}
%\\  
\medskip \medskip \medskip
(b) 
\includegraphics[width=0.4\textwidth]{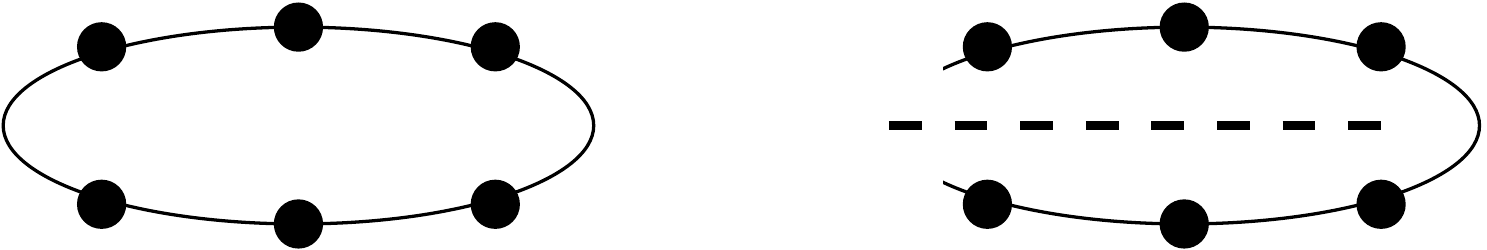}
\\ 
\medskip \medskip \medskip
%\, \, \, \,  
(c)
\includegraphics[width=0.4\textwidth]{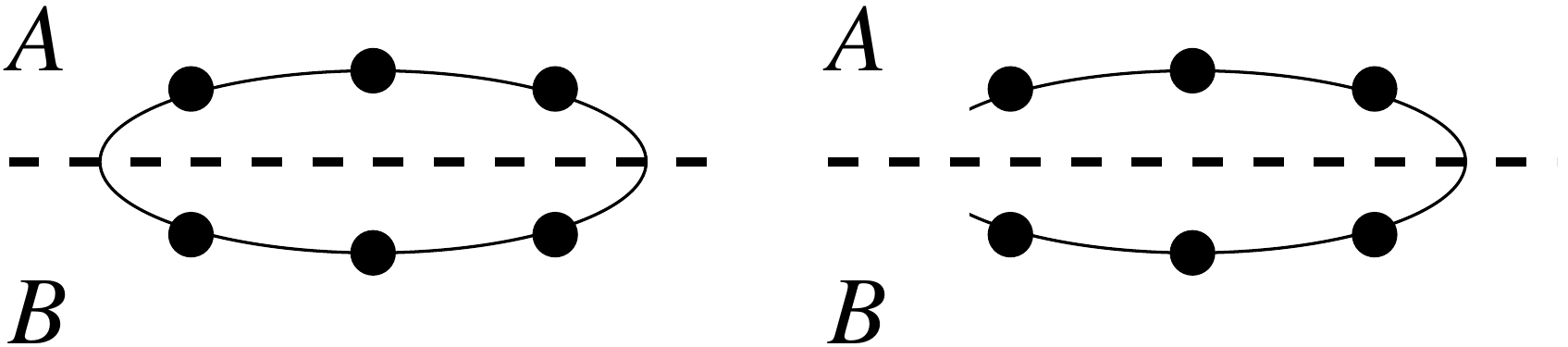}
\caption{%
(Color online)
(a)
Hoppings of fermions between three consecutive 
sites $r-1$, $r$, and $r+1$
along a one-dimensional ring with four
orbitals per site.
A repeat unit cell is labeled by the integer $r=1,\cdots, N$
and is pictured by a rounded rectangular frame.
A unit cell contains four orbitals that are pictured
by black or white discs or squares, respectively.
The hopping amplitude $t+\delta t\in\mathbb{R}$ is pictured
by a full connecting line.
The hopping amplitude $t-\delta t\in\mathbb{R}$ is pictured
by a dashed connecting line. Hopping is only possible
between orbitals of the same shape but distinct colors
belonging to nearest-neighbor repeat unit cell.
The figure is invariant under 
(i)
the composition of
the interchange of the full and dashed lines
with the interchange of the black and white filling colors
with a reflection about the horizontal dash-one-dot (red) line $RH$
and
(ii)
the composition of
the interchange of the circular and square shapes
with a reflection about the vertical dash-two-dots (blue) line $RVO$
if $N$ is odd or the vertical dash-three-dots (green) line $RVE$ if $N$ 
is even.
(b) Periodic boundary conditions are imposed (ring geometry) 
on the repeat unit cells represented by filled circles in the left panel,
whereas open boundary conditions are imposed (open line geometry)
on the repeat unit cells in the right panel.
There are two physical boundaries on either sides of the cut represented by
the dashed line a distance $N$ apart.
(c) Periodic boundary conditions are imposed (ring geometry) 
on the repeat unit cells represented by filled circles in the left panel,
whereas open boundary conditions are imposed (open line geometry)
on the repeat unit cells in the right panel. The partitions $A$ and $B$
are made of the unit cells above and below the dashed line, respectively.
There are two identical entangling boundaries an integer distance $N/2$ apart
in the left panel.
There are two identical physical boundaries a distance $N$ apart
in the right panel, each of which is an integer distance $N/2$ apart from
a single entangling boundary. 
         }
\label{fig: dimerized chain}
\end{figure*}

%%%%
%\begin{widetext}
Our first example is defined by choosing
$d=1$ and $N^{\,}_{\mathrm{orb}}=4$ in 
Eq.~(\ref{eq: def noninteractin fermion Hamiltonian}).
The lattice $\Lambda$ is one-dimensional
with the lattice spacing $2\mathfrak{a}$.
It is labeled by the integers $r=1,\cdots,N$.
To represent the single-particle Hamiltonian with the
matrix elements~(\ref{eq: def noninteractin fermion Hamiltonian d}),
we introduce two sets of Pauli matrices. 
We associate to the unit $2\times2$ matrix $\sigma^{\,}_{0}$
and the three Pauli matrices 
$\sigma^{\,}_{1}$, $\sigma^{\,}_{2}$, and $\sigma^{\,}_{3}$
two geometrical shapes, a square or a circle,
corresponding to the eigenvalues of $\sigma^{\,}_{3}$.
We associate to the unit $2\times2$ matrix $\tau^{\,}_{0}$
and the three Pauli matrices 
$\tau^{\,}_{1}$, $\tau^{\,}_{2}$, and $\tau^{\,}_{3}$
two colors, black or white,
corresponding to the eigenvalues of $\tau^{\,}_{3}$.
We choose the representation
\begin{subequations}
\label{eq: def 1D example}
\begin{equation}
\hat{\psi}^{\dag}_{r}\equiv
\begin{pmatrix}
\hat{\psi}^{\dag}_{\blacksquare;r}
&
\hat{\psi}^{\dag}_{\square;r}
&
\hat{\psi}^{\dag}_{\bullet;r}
&
\hat{\psi}^{\dag}_{\circ;r}.
\end{pmatrix}
\label{eq: def 1D example a} 
\end{equation}
and impose periodic boundary conditions
\begin{equation}
\hat{\psi}^{\dag}_{r+N}=
\hat{\psi}^{\dag}_{r},
\qquad 
r=1,\cdots,N.
\label{eq: def 1D example b}
\end{equation}
The non-interacting Hamiltonian is then defined by
\begin{equation}
\begin{split}
\hat{H}:=&\,
(t+\delta t)\,
\sum_{r=1}^{N}
\left(
\hat{\psi}^{\dag}_{\blacksquare;r+1}\,
\hat{\psi}^{\,}_{\square;r}
+
\hat{\psi}^{\dag}_{\circ;r+1}\,
\hat{\psi}^{\,}_{\bullet;r}
+
\hat{\psi}^{\dag}_{\square;r}\,
\hat{\psi}^{\,}_{\blacksquare;r+1}
+
\hat{\psi}^{\,}_{\bullet;r}\,
\hat{\psi}^{\dag}_{\circ;r+1}
\right)
\\
&\,
+
(t-\delta t)\,
\sum_{r=1}^{N}
\left(
\hat{\psi}^{\dag}_{\square;r+1}\,
\hat{\psi}^{\,}_{\blacksquare;r}
+
\hat{\psi}^{\dag}_{\bullet;r+1}\,
\hat{\psi}^{\,}_{\circ;r}
+
\hat{\psi}^{\,}_{\blacksquare;r}\,
\hat{\psi}^{\dag}_{\square;r+1}
+
\hat{\psi}^{\dag}_{\circ;r}\,
\hat{\psi}^{\,}_{\bullet;r+1}
\right)
\end{split}
\label{eq: def 1D example c}
\end{equation}
in second quantization.
It describes the nearest-neighbor hops of fermions with hopping amplitudes
$t\pm\delta t\in\mathbb{R}$
as is illustrated in
Fig.~\ref{fig: dimerized chain}. 
The corresponding single-particle Hamiltonian
can be written
[$\tau^{\,}_{\pm}:=(\tau^{\,}_{1}\pm\mathrm{i}\tau^{\,}_{2})/2$]
\begin{align}
\mathcal{H}^{\,}_{r,r'}(t,\delta t):=&\,
\delta^{\,}_{r',r-1}\,
\Bigg\{
(t+\delta t)\,
\left[
\frac{\sigma^{\,}_{0}+\sigma^{\,}_{3}}{2}
\otimes
\tau^{\,}_{+}
+
\frac{\sigma^{\,}_{0}-\sigma^{\,}_{3}}{2}
\otimes
\tau^{\,}_{-}
\right]
\\
&\,\hphantom{\delta^{\,}_{r',r-1}\,\Bigg\{}
+
(t-\delta t)\,
\left[
\frac{\sigma^{\,}_{0}+\sigma^{\,}_{3}}{2}
\otimes
\tau^{\,}_{-}
+
\frac{\sigma^{\,}_{0}-\sigma^{\,}_{3}}{2}
\otimes
\tau^{\,}_{+}
\right]
\Bigg\}
\nonumber\\
&\,
+
\delta^{\,}_{r',r+1}\,
\Bigg\{
(t+\delta t)\,
\left[
\frac{\sigma^{\,}_{0}+\sigma^{\,}_{3}}{2}
\otimes
\tau^{\,}_{-}
+
\frac{\sigma^{\,}_{0}-\sigma^{\,}_{3}}{2}
\otimes
\tau^{\,}_{+}
\right]
\nonumber\\
&\,\hphantom{\delta^{\,}_{r',r-1}\,\Bigg\{}
+
(t-\delta t)\,
\left[
\frac{\sigma^{\,}_{0}+\sigma^{\,}_{3}}{2}
\otimes
\tau^{\,}_{+}
+
\frac{\sigma^{\,}_{0}-\sigma^{\,}_{3}}{2}
\otimes
\tau^{\,}_{-}
\right]
\Bigg\}.
\label{eq: def 1D example d}
\end{align}
\end{subequations}

In the Bloch basis, the single-particle
Hamiltonian~(\ref{eq: def 1D example d})
becomes 
\begin{subequations}
\label{eq: def 1D example Bloch basis}
\begin{equation}
\mathcal{H}^{\,}_{k,k'}=
\mathcal{H}^{\,}_{k}\,
\delta^{\,}_{k,k'},
\qquad
\mathcal{H}^{\,}_{k}=
2\,t\,\cos k\,\sigma^{\,}_{0}\otimes\tau^{\,}_{1}
-
2\,\delta t\,\sin k\,\sigma^{\,}_{3}\otimes\tau^{\,}_{2},
\quad
k,k'=-\pi+\frac{2 \pi}{N},\cdots,\pi,
\label{eq: def 1D example Bloch basis a}
\end{equation}
with the single-particle spectrum consisting of the
two-fold degenerate pair of bands
\begin{equation}
\varepsilon^{\,}_{\pm;k}=
\pm
2\,|t|\,
\sqrt{
\cos^{2}k
+
(\delta t/t)^{2}\,
\sin^{2}k
     }.
\label{eq: def 1D example Bloch basis b}
\end{equation}
Consequently, the band gap
\begin{equation}
\Delta=
4\,|\delta\,t|
\label{eq: def 1D example Bloch basis c}
\end{equation}
\end{subequations}
opens at the boundary of the first Brillouin zone
for any non-vanishing $\delta t$.

%\vskip 20 true pt
%\end{widetext} 
%\vskip 20 true pt

\begin{table*}
\begin{center}
\resizebox{\columnwidth}{!}{%
\begin{tabular}{|c||cccc|cccc|cccc|cccc|cc|}
\hline 
\multicolumn{1}{|c||}{\multirow{2}{*}
{$\mathcal{X}^{\,}_{\mu\nu}\equiv\sigma^{\,}_{\mu}\otimes\tau^{\,}_{\nu}$}}
&
\multicolumn{4}{c|}{Symmetry under $\mathscr{P}$}
&
\multicolumn{4}{c|}{Symmetry under $\mathscr{T}$}
&
\multicolumn{4}{c|}{Symmetry under $\mathscr{C}$}
&
\multicolumn{4}{c|}{Symmetry under $\mathscr{S}$}
&
\multirow{2}{*}{$\sigma(\widetilde{\mathcal{H}}^{\,}_{\mu\nu})$}
&
\multirow{2}{*}{$\sigma(\widetilde{Q}^{\,}_{\mu\nu\,A})$}
\\
{}
&
$\mathcal{P}^{\,}_{01}$
&
$\mathcal{P}^{\,}_{10}$
&
$\mathcal{P}^{\,}_{20}$
&
$\mathcal{P}^{\,}_{31}$
&
$\mathcal{T}^{\,}_{00}$
&
$\mathcal{T}^{\,}_{11}$
&
$\mathcal{T}^{\,}_{21}$
&
$\mathcal{T}^{\,}_{30}$
&
$\mathcal{C}^{\,}_{03}$
&
$\mathcal{C}^{\,}_{12}$
&
$\mathcal{C}^{\,}_{22}$
&
$\mathcal{C}^{\,}_{33}$
&
$\mathcal{S}^{\,}_{03}$
&
$\mathcal{S}^{\,}_{12}$
&
$\mathcal{S}^{\,}_{22}$
&
$\mathcal{S}^{\,}_{33}$
&
{}
&
{}
\\
\hline
\hline
$\mathcal{X}^{\,}_{00}$
&
$\circ$
&
$\circ$
&
$\circ$
&
$\circ$  % END GROUP P FOR 00
&
$\circ$
&
$\circ$
&
$\circ$
&
$\circ$  % END GROUP T FOR 00
&
$\times$
&
$\times$
&
$\times$
&
$\times$  % END GROUP C FOR 00
&
$\times$
&
$\times$
&
$\times$
&
$\times$  % END GROUP S FOR 00
&
$\times$
&
$\circ$ % END SPECTRA FOR 00
\\
\hline
$\mathcal{X}^{\,}_{01}$
&
$\circ$
&
$\circ$
&
$\circ$
&
$\circ$  % END GROUP P FOR 01
&
$\circ$
&
$\circ$
&
$\circ$
&
$\circ$  % END GROUP T FOR 01
&
$\circ$
&
$\circ$
&
$\circ$
&
$\circ$  % END GROUP C FOR 01
&
$\circ$
&
$\circ$
&
$\circ$
&
$\circ$  % END GROUP S FOR 01
&
$\circ$
&
$\circ$ % END SPECTRA FOR 01
\\
\hline
$\mathcal{X}^{\,}_{02}$
&
$\times$
&
$\circ$
&
$\circ$
&
$\times$  % END GROUP P FOR 02
&
$\times$
&
$\circ$
&
$\circ$
&
$\times$  % END GROUP T FOR 02
&
$\times$
&
$\circ$
&
$\circ$
&
$\times$  % END GROUP C FOR 02
&
$\circ$
&
$\times$
&
$\times$
&
$\circ$  % END GROUP S FOR 02
&
$\circ$
&
$\circ$ % END SPECTRA FOR 02
\\
\hline
$\mathcal{X}^{\,}_{03}$
&
$\times$
&
$\circ$
&
$\circ$
&
$\times$  % END GROUP P FOR 03
&
$\circ$
&
$\times$
&
$\times$
&
$\circ$  % END GROUP T FOR 03
&
$\times$
&
$\circ$
&
$\circ$
&
$\times$  % END GROUP C FOR 03
&
$\times$
&
$\circ$
&
$\circ$
&
$\times$  % END GROUP S FOR 03
&
$\times$
&
$\times$ % END SPECTRA FOR 03
\\
\hline
$\mathcal{X}^{\,}_{10}$
&
$\circ$
&
$\circ$
&
$\times$
&
$\times$  % END GROUP P FOR 10
&
$\circ$
&
$\circ$
&
$\times$
&
$\times$  % END GROUP T FOR 10
&
$\times$
&
$\times$
&
$\circ$
&
$\circ$  % END GROUP C FOR 10
&
$\times$
&
$\times$
&
$\circ$
&
$\circ$  % END GROUP S FOR 10
&
$\circ$
&
$\circ$ % END SPECTRA FOR 10
\\
\hline
$\mathcal{X}^{\,}_{11}$
&
$\circ$
&
$\circ$
&
$\times$
&
$\times$  % END GROUP P FOR 11
&
$\circ$
&
$\circ$
&
$\times$
&
$\times$  % END GROUP T FOR 11
&
$\circ$
&
$\circ$
&
$\times$
&
$\times$  % END GROUP C FOR 11
&
$\circ$
&
$\circ$
&
$\times$
&
$\times$  % END GROUP S FOR 11
&
$\times$
&
$\circ$ % END SPECTRA FOR 11
\\
\hline
$\mathcal{X}^{\,}_{12}$
&
$\times$
&
$\circ$
&
$\times$
&
$\circ$  % END GROUP P FOR 12
&
$\times$
&
$\circ$
&
$\times$
&
$\circ$  % END GROUP T FOR 12
&
$\times$
&
$\circ$
&
$\times$
&
$\circ$  % END GROUP C FOR 12
&
$\circ$
&
$\times$
&
$\circ$
&
$\times$  % END GROUP S FOR 12
&
$\times$
&
$\times$ % END SPECTRA FOR 12
\\
\hline
$\mathcal{X}^{\,}_{13}$
&
$\times$
&
$\circ$
&
$\times$
&
$\circ$  % END GROUP P FOR 13
&
$\circ$
&
$\times$
&
$\circ$
&
$\times$  % END GROUP T FOR 13
&
$\times$
&
$\circ$
&
$\times$
&
$\circ$  % END GROUP C FOR 13
&
$\times$
&
$\circ$
&
$\times$
&
$\circ$  % END GROUP S FOR 13
&
$\circ$
&
$\circ$ % END SPECTRA FOR 13
\\%%%%%%%%(new rows added)
\hline
$\mathcal{X}^{\,}_{20}$
&
$\circ$
&
$\times$
&
$\circ$
&
$\times$  % END GROUP P FOR 20
&
$\times$
&
$\circ$
&
$\times$
&
$\circ$  % END GROUP T FOR 20
&
$\circ$
&
$\times$
&
$\circ$
&
$\times$  % END GROUP C FOR 20
&
$\times$
&
$\circ$
&
$\times$
&
$\circ$  % END GROUP S FOR 20
&
$\circ$
&
$\circ$ % END SPECTRA FOR 20
\\
\hline
$\mathcal{X}^{\,}_{21}$
&
$\circ$
&
$\times$
&
$\circ$
&
$\times$  % END GROUP P FOR 21
&
$\times$
&
$\circ$
&
$\times$
&
$\circ$  % END GROUP T FOR 21
&
$\times$
&
$\circ$
&
$\times$
&
$\circ$  % END GROUP C FOR 21
&
$\circ$
&
$\times$
&
$\circ$
&
$\times$  % END GROUP S FOR 21
&
$\times$
&
$\circ$ % END SPECTRA FOR 21
\\
\hline
$\mathcal{X}^{\,}_{22}$
&
$\times$
&
$\times$
&
$\circ$
&
$\circ$  % END GROUP P FOR 22
&
$\circ$
&
$\circ$
&
$\times$
&
$\times$  % END GROUP T FOR 22
&
$\circ$
&
$\circ$
&
$\times$
&
$\times$  % END GROUP C FOR 22
&
$\circ$
&
$\circ$
&
$\times$
&
$\times$  % END GROUP S FOR 22
&
$\times$
&
$\times$ % END SPECTRA FOR 22
\\
\hline
$\mathcal{X}^{\,}_{23}$
&
$\times$
&
$\times$
&
$\circ$
&
$\circ$  % END GROUP P FOR 23
&
$\times$
&
$\times$
&
$\circ$
&
$\circ$  % END GROUP T FOR 23
&
$\circ$
&
$\circ$
&
$\times$
&
$\times$  % END GROUP C FOR 23
&
$\times$
&
$\times$
&
$\circ$
&
$\circ$  % END GROUP S FOR 23
&
$\circ$
&
$\circ $ % END SPECTRA FOR 23
\\
\hline
$\mathcal{X}^{\,}_{30}$
&
$\circ$
&
$\times$
&
$\times$
&
$\circ$  % END GROUP P FOR 30
&
$\circ$
&
$\times$
&
$\times$
&
$\circ$  % END GROUP T FOR 30
&
$\times$
&
$\circ$
&
$\circ$
&
$\times$  % END GROUP C FOR 30
&
$\times$
&
$\circ$
&
$\circ$
&
$\times$  % END GROUP S FOR 30
&
$\times$
&
$\circ $ % END SPECTRA FOR 30
\\
\hline
$\mathcal{X}^{\,}_{31}$
&
$\circ$
&
$\times$
&
$\times$
&
$\circ$  % END GROUP P FOR 31
&
$\circ$
&
$\times$
&
$\times$
&
$\circ$  % END GROUP T FOR 31
&
$\circ$
&
$\times$
&
$\times$
&
$\circ$  % END GROUP C FOR 31
&
$\circ$
&
$\times$
&
$\times$
&
$\circ$  % END GROUP S FOR 31
&
$\circ$
&
$\circ $ % END SPECTRA FOR 31
\\
\hline\hline
$\mathcal{X}^{\,}_{32}$
&
$\times$
&
$\times$
&
$\times$
&
$\times$  % END GROUP P FOR 32
&
$\times$
&
$\times$
&
$\times$
&
$\times$  % END GROUP T FOR 32
&
$\times$
&
$\times$
&
$\times$
&
$\times$  % END GROUP C FOR 32
&
$\circ$
&
$\circ$
&
$\circ$
&
$\circ$  % END GROUP S FOR 32
&
$\circ$
&
$\circ$ % END SPECTRA FOR 32
\\
\hline
$\mathcal{X}^{\,}_{33}$
&
$\times$
&
$\times$
&
$\times$
&
$\times$  % END GROUP P FOR 33
&
$\circ$
&
$\circ$
&
$\circ$
&
$\circ$  % END GROUP T FOR 33
&
$\times$
&
$\times$
&
$\times$
&
$\times$  % END GROUP C FOR 33
&
$\times$
&
$\times$
&
$\times$
&
$\times$  % END GROUP S FOR 33
&
$\times$
&
$\times$ % END SPECTRA FOR 33
\\
\hline
\end{tabular}
}
\end{center}
\caption{%
The spectrum $\sigma(\widetilde{\mathcal{H}}^{\,}_{\mu\nu})$
of the single-particle Hamiltonian
$\widetilde{\mathcal{H}}^{\,}_{\mu\nu}$ 
defined by Eq.%
~(\ref{eq: def widetilde H for 1D example})
with open boundary conditions.
Hamiltonian $\widetilde{\mathcal{H}}^{\,}_{\mu\nu}$
is nothing but Hamiltonian $\mathcal{H}$ [defined in 
Eq.~(\ref{eq: def 1D example d})]
perturbed additively by the term
$\delta^{\,}_{r,r'}\,(t/10)\,\mathcal{X}^{\,}_{\mu\nu}$
with $\mathcal{X}^{\,}_{\mu\nu}\equiv\sigma^{\,}_{\mu}\otimes\tau^{\,}_{\nu}$.
The choices for $\mathcal{X}^{\,}_{\mu\nu}$
made in the first eight rows enumerate all 
perturbations $\mathcal{V}^{\,}_{\mu\nu}$
defined by Eq.~(\ref{eq: def widetilde H for 1D example b}) 
that enter 
Eq.~(\ref{eq: constant parity preserving perturbation in r space}),
i.e., that preserve the symmetry under the parity of $\mathcal{H}$
generated by $\mathcal{P}^{\,}_{10}$.
The last two rows are two examples of a perturbation
$\mathcal{V}^{\,}_{\mu\nu}$
that breaks parity.
The entanglement spectrum
$\sigma(\widetilde{Q}^{\,}_{\mu\nu A})$
defined by Eq.~(\ref{eq: def QA}) 
for the single-particle Hamiltonian
$\widetilde{\mathcal{H}}^{\,}_{\mu\nu}$ 
obeying periodic boundary conditions.
The entry
$\circ$ 
or
$\times$ 
denotes the presence or the absence, respectively,
of the symmetries under parity 
$\mathscr{P}$,
charge conjugation $\mathscr{C}$,
and time reversal $\mathscr{T}$
of the perturbation
$\delta^{\,}_{r,r'}\,(t/10)\,\mathcal{X}^{\,}_{\mu\nu}$
for the first sixteen columns.
In the last two columns, the entry
$\circ$ 
or
$\times$ 
denotes the presence or the absence, respectively,
of zero modes (mid-gap) states in the spectra 
$\sigma(\widetilde{\mathcal{H}}^{\,}_{\mu\nu})$
and
$\sigma(\widetilde{\mathcal{Q}}^{\,}_{\mu\nu})$ 
as determined by extrapolation to the
thermodynamic limit of exact diagonalization with $N=12$.
       }
\label{tab:1}
\end{table*}

\subsection{Symmetries}
\label{subsubsec: Symmetries 1D example}

The symmetries of the single-particle Hamiltonian defined by
Eq.~(\ref{eq: def 1D example d}) are the following, for
any pair of sites $r,r'=1,\cdots,N$.

[1]
Translation symmetry holds,
\begin{equation}
\mathcal{H}^{\,}_{r,r'}(t,\delta t)=
\mathcal{H}^{\,}_{r+n,r'+n}(t,\delta t),
\qquad 
\forall n\in\mathbb{Z}.
\end{equation}

[2] When the dimerization vanishes
\begin{equation}
\mathcal{H}^{\,}_{r,r'}(t,\delta t=0)=
t\,
\left(\delta^{\,}_{r',r-1}+\delta^{\,}_{r',r+1}\right)\,
\sigma^{\,}_{0}\otimes\tau^{\,}_{1},
\end{equation}
the discrete symmetries
\begin{subequations}
\begin{equation}
0=
\left[
\mathcal{H}^{\,}_{r,r'}(t,\delta t=0),\sigma^{\,}_{\mu}\otimes\tau^{\,}_{\nu}
\right]
\end{equation}
hold for 
$\mu=0,1,2,3$ and $\nu=0,1$, 
whereas the discrete spectral symmetries
\begin{equation}
0=
\left\{
\mathcal{H}^{\,}_{r,r'}(t,\delta t=0),\sigma^{\,}_{\mu}\otimes\tau^{\,}_{\mu}
\right\}
\end{equation}
\end{subequations}
hold for $\mu=0,1,2,3$ and $\nu=2,3$. 

[3] For any dimerization, the transformation laws
\begin{subequations}
\begin{equation}
\mathcal{H}^{\,}_{r,r'}(t,+\delta t)=
\sigma^{\,}_{\mu}
\otimes
\tau^{\,}_{1}\,
\mathcal{H}^{\,}_{r,r'}(t,-\delta t)\,
\sigma^{\,}_{\mu}
\otimes
\tau^{\,}_{1}
\end{equation}
with $\mu=0,3$
implement the symmetry of Fig.~\ref{fig: dimerized chain}
under the composition of
the interchange of the full and dashed lines
with the interchange of the black and white filling colors
with a reflection about the horizontal dash-one-dot (red) line $RH$,
whereas the transformation laws
\begin{equation}
\mathcal{H}^{\,}_{r,r'}(t,+\delta t)=
\sigma^{\,}_{\mu}
\otimes
\tau^{\,}_{1}\,
\mathcal{H}^{\,}_{r,r'}(t,+\delta t)\,
\sigma^{\,}_{\mu}
\otimes
\tau^{\,}_{1}
\end{equation}
\end{subequations}
hold for $\mu=1,2$ otherwise.

[4] Let $\mathscr{O}^{\,}_{\mathrm{R}}$ be the operation that interchanges
site $r^{\,}_{1}$ with site  $r^{\,}_{N}$,
site $r^{\,}_{2}$ with site  $r^{\,}_{N-1}$,
and so on, i.e., 
a reflection about the vertical dash-two-dots (blue) line $RVO$
if $N$ is odd 
or the vertical dash-three-dots (green) line $RVE$ 
if $N$ is even.
For any dimerization, the transformation laws
\begin{equation}
\mathcal{H}^{\,}_{r,r'}(t,\delta t)=
+
\sigma^{\,}_{\mu}
\otimes
\tau^{\,}_{0}\,
\mathcal{H}^{\,}_{\mathscr{O}^{\,}_{\mathrm{R}}r,\mathscr{O}^{\,}_{\mathrm{R}}r'}(t,\delta t)\,
\sigma^{\,}_{\mu}
\otimes
\tau^{\,}_{0}
\label{eq: two reflection symmetries about RV}
\end{equation}
with $\mu=1,2$ are unitary symmetries, whereas the transformation laws
\begin{equation}
\mathcal{H}^{\,}_{r,r'}(t,\delta t)=
-
\sigma^{\,}_{\mu}
\otimes
\tau^{\,}_{3}\,
\mathcal{H}^{\,}_{\mathscr{O}^{\,}_{\mathrm{R}}r,\mathscr{O}^{\,}_{\mathrm{R}}r'}(t,\delta t)\,
\sigma^{\,}_{\mu}
\otimes
\tau^{\,}_{3}
\end{equation}
with $\mu=1,2$ are unitary spectral symmetries. 

It is also instructive to
derive the symmetries and spectral symmetries
of the single-particle Hamiltonian%
~(\ref{eq: def 1D example Bloch basis})
for any 
$k=\pi/N,\cdots,\pi$
from the first Brillouin zone. 

To this end, it is convenient to introduce the more compact 
notation
\begin{subequations}
\label{eq: def X notation}
\begin{equation}
\mathcal{X}^{\,}_{\mu\nu}:=
\sigma^{\,}_{\mu}\otimes\tau^{\,}_{\nu},
\qquad
\mu,\nu=0,1,2,3,
\label{eq: def X notation a}
\end{equation}
for the sixteen linearly independent $4\times4$ Hermitian matrices that
generate the unitary group $U(4)$. In the Bloch basis%
~(\ref{eq: def 1D example Bloch basis a}),
\begin{equation}
\mathcal{H}^{\,}_{k,k'}=
\mathcal{H}^{\,}_{k}\,
\delta^{\,}_{k,k'},
\quad
\mathcal{H}^{\,}_{k}=
2\,t\,\cos k\,\mathcal{X}^{\,}_{01}
-
2\,\delta t\,\sin k\,\mathcal{X}^{\,}_{32},
\label{eq: def X notation b}
\end{equation}
\end{subequations}
for $k,k'=\pi/N,\cdots,\pi$.
We have taken advantage of the fact that
$\mathcal{X}^{\,}_{01}$ and $\mathcal{X}^{\,}_{32}$ anti-commute
to derive the band dispersions%
~(\ref{eq: def 1D example Bloch basis b}).

There are eight matrices $\mathcal{X}^{\,}_{\mu\nu}$
with $\mu=0,1,2,3$ and $\nu=0,1$ 
that commute with $\mathcal{X}^{\,}_{01}$,
there are eight matrices $\mathcal{X}^{\,}_{\mu\nu}$
with 
$\mu=1,2$ and $\nu=1,3$ 
or
$\mu=0,3$ and $\nu=0,2$ 
that commute with $\mathcal{X}^{\,}_{32}$.
This leaves the four matrices
$\mathcal{X}^{\,}_{00}$,
$\mathcal{X}^{\,}_{30}$, 
$\mathcal{X}^{\,}_{11}$, 
and
$\mathcal{X}^{\,}_{21}$ 
that commute with
$\mathcal{H}^{\,}_{k}$ 
for all $k$ in the Brillouin zone.

There are eight matrices $\mathcal{X}^{\,}_{\mu\nu}$
with $\mu=0,1,2,3$ and $\nu=2,3$ 
that anti-commute with $\mathcal{X}^{\,}_{01}$,
there are eight matrices $\mathcal{X}^{\,}_{\mu\nu}$
with 
$\mu=1,2$ and $\nu=0,2$ 
or
$\mu=0,3$ and $\nu=1,3$ 
that anti-commute with $\mathcal{X}^{\,}_{32}$.
This leaves the four matrices
$\mathcal{X}^{\,}_{12}$,
$\mathcal{X}^{\,}_{22}$, 
$\mathcal{X}^{\,}_{03}$, 
and
$\mathcal{X}^{\,}_{33}$ 
that anti-commute with
$\mathcal{H}^{\,}_{k}$ 
for all $k$ in the Brillouin zone.

The symmetries 
\begin{subequations}
\label{eq: four symmetries in 1D example}
\begin{align}
&
\mathcal{O}^{\dag}_{\mathscr{P}}\,
\mathcal{H}^{\,}_{-k}\,
\mathcal{O}^{\,}_{\mathscr{P}}=
\mathcal{H}^{\,}_{+k},
\label{eq: four symmetries in 1D example a}
\\
&
\mathcal{O}^{\dag}_{\mathscr{T}}\,
\mathcal{H}^{*}_{-k}\,
\mathcal{O}^{\,}_{\mathscr{T}}=
\mathcal{H}^{\,}_{+k},
\label{eq: four symmetries in 1D example b}
\end{align}
with
\begin{align}
&
\mathcal{O}^{\,}_{\mathscr{P}}\in
\{
\mathcal{X}^{\,}_{01},\
\mathcal{X}^{\,}_{10},\
\mathcal{X}^{\,}_{20},\
\mathcal{X}^{\,}_{31}
\},
\label{eq: four symmetries in 1D example c}
\\
&
\mathcal{O}^{\,}_{\mathscr{T}}\in
\{
\mathcal{X}^{\,}_{00},\
\mathcal{X}^{\,}_{11},\
\mathcal{X}^{\,}_{21},\
\mathcal{X}^{\,}_{30}
\},
\label{eq: four symmetries in 1D example d}
\end{align}
\end{subequations} 
whereas the spectral symmetries
\begin{subequations}
\begin{align}
&
\mathcal{O}^{\dag}_{\mathscr{C}}\,
\mathcal{H}^{\mathsf{T}}_{-k}\,
\mathcal{O}^{\,}_{\mathscr{C}}=
-
\mathcal{H}^{\,}_{k},
\\
&
\mathcal{O}^{\dag}_{\mathscr{S}}\,
\mathcal{H}^{\,}_{k}\,
\mathcal{O}^{\,}_{\mathscr{S}}=
-
\mathcal{H}^{\,}_{k},
\end{align}
with 
\begin{align}
&
\mathcal{O}^{\,}_{\mathscr{C}}\in
\{
\mathcal{X}^{\,}_{03},\
\mathcal{X}^{\,}_{12},\
\mathcal{X}^{\,}_{22},\
\mathcal{X}^{\,}_{33}
\},
\\
&
\mathcal{O}^{\,}_{\mathscr{S}}\in
\{
\mathcal{X}^{\,}_{03},\
\mathcal{X}^{\,}_{12},\
\mathcal{X}^{\,}_{22},\
\mathcal{X}^{\,}_{33}
\},
\end{align}
\end{subequations}
follow. As anticipated, the set of chiral-like spectral symmetries
is identical the set of particle-hole-like spectral symmetries in view
of the presence of four time-reversal-like symmetries.
We shall use the notation
$K$ for the anti-linear operation of complex conjugation.
We shall also introduce the notation
\begin{subequations}
\label{eq: def P C T S 1D example}
\begin{align}
&
\hbox{Parity:}
&
\mathcal{P}^{\,}_{10}:=
\sigma^{\,}_{1}\otimes\tau^{\,}_{0}=
+\mathcal{P}^{\mathsf{T}}_{10},
\\
&
\hbox{Charge Conjugation:}
&
\mathcal{C}^{\,}_{03}:=
\sigma^{\,}_{0}\otimes\tau^{\,}_{3}\,K=
+\mathcal{C}^{\mathsf{T}}_{03},
\\
&
\hbox{Time Reversal:}
&
\mathcal{T}^{\,}_{21}:=
\sigma^{\,}_{2}\otimes\tau^{\,}_{1}\,K=
-\mathcal{T}^{\mathsf{T}}_{21},
\\
&
\hbox{Chirality:}
&
\mathcal{S}^{\,}_{22}:=
\sigma^{\,}_{2}\otimes\tau^{\,}_{2}=
+\mathcal{S}^{\mathsf{T}}_{22},
\end{align}
\end{subequations}
say, to distinguish the operations of 
parity (reflection or inversion),
charge conjugation (particle hole interchange),
time reversal,
and chirality, respectively. 
The symmetry under parity of
$\mathcal{H}$ can be realized in four inequivalent ways.
Correspondingly, we define the $4\times4$ matrices
\begin{subequations}
\label{eq: def families of four P T C}
\begin{equation}
\hbox{Parity $\mathscr{P}$:}\qquad
\mathcal{P}^{\,}_{01},
\mathcal{P}^{\,}_{10},
\mathcal{P}^{\,}_{20},
\mathcal{P}^{\,}_{31},
\end{equation}
that realize the algebra of the unit $2\times2$ matrix $\rho^{\,}_{0}$
and of the three Pauli matrices 
$\rho^{\,}_{1}$,
$\rho^{\,}_{2}$,
and
$\rho^{\,}_{3}$. 
The symmetry under time reversal of
$\mathcal{H}$ can be realized in four inequivalent ways.
Correspondingly, we define the $4\times4$ matrices
\begin{equation}
\hbox{Time Reversal $\mathscr{T}$:}\qquad
\mathcal{T}^{\,}_{00},
\mathcal{T}^{\,}_{11},
\mathcal{T}^{\,}_{21},
\mathcal{T}^{\,}_{30},
\end{equation}
that realize the algebra of
$\rho^{\,}_{0}\,K$,
$\rho^{\,}_{1}\,K$,
$\rho^{\,}_{2}\,K$,
and
$\rho^{\,}_{3}\,K$. 
The symmetry under charge conjugation of
$\mathcal{H}$ can be realized in four inequivalent ways.
Correspondingly, we define the $4\times4$ matrices
\begin{equation}
\hbox{Charge Conjugation $\mathscr{C}$:}\qquad
\mathcal{C}^{\,}_{03},
\mathcal{C}^{\,}_{12},
\mathcal{C}^{\,}_{22},
\mathcal{C}^{\,}_{33},
\end{equation}
that realize the algebra of
$\rho^{\,}_{0}\,K$,
$\rho^{\,}_{1}\,K$,
$\rho^{\,}_{2}\,K$,
and
$\rho^{\,}_{3}\,K$. All the possible compositions of the operations for
charge conjugation and time reversal give four realizations
for the chiral symmetry of $\mathcal{H}$.
Correspondingly, we define the $4\times4$ matrices
\begin{equation}
\hbox{Chiral $\mathscr{S}$:}\qquad
\mathcal{S}^{\,}_{03},
\mathcal{S}^{\,}_{12},
\mathcal{S}^{\,}_{22},
\mathcal{S}^{\,}_{33},
\end{equation}
\end{subequations}
that realize the algebra of 
$\rho^{\,}_{0}$,
$\rho^{\,}_{1}$,
$\rho^{\,}_{2}$,
and
$\rho^{\,}_{3}$.
There are sixteen columns in Table~\ref{tab:1},
each of which correspond to one of these matrix operations.
We shall then select 12 triplets 
$(\mathcal{P}^{\,}_{\mu\nu},\mathcal{T}^{\,}_{\mu\nu},\mathcal{C}^{\,}_{\mu\nu})$ 
from Eqs.~(\ref{eq: def families of four P T C})
to built the rows of Table~\ref{tab:2}.

The single-particle Hamiltonian~$\mathcal{H}$ is extremely sparse. 
This is reflected by it obeying the symmetries
$(\mathcal{P}^{\,}_{\mu},\mathcal{T}^{\,}_{\mu},\mathcal{C}^{\,}_{\mu})$
with the pair $\mu$ and $\nu$ fixed by the columns from
Table~\ref{tab:1}. In particular,
$\mathcal{H}$ cannot be assigned in a unique way
the symmetry under time-reversal and the spectral symmetry
under charge conjugation without additional informations of microscopic
origin. Identifying the symmetric space generated by $\mathcal{H}$
is thus ambiguous.
Example 1,
$\mathcal{H}$ can be thought of as representative of 
the Cartan symmetry class CI if the choice
$\mathcal{T}^{\,}_{\mu\nu}$ obeying $\mathcal{T}^{2}_{\mu\nu}=+
%\openone
\mathbb{I}$
and 
$\mathcal{C}^{\,}_{\mu\nu}$ 
obeying $\mathcal{C}^{2}_{\mu\nu}=-\mathbb{I}$
for the symmetry under time-reversal and the spectral symmetry
under charge-conjugation
is dictated by a microscopic derivation of $\mathcal{H}$.
Example 2, $\mathcal{H}$
can be thought of as representative of the symmetry class DIII if 
the choice
$\mathcal{T}^{\,}_{\mu\nu}$ obeying $\mathcal{T}^{2}_{\mu\nu}=-%\openone
\mathbb{I}$
and 
$\mathcal{C}^{\,}_{\mu\nu}$ obeying $\mathcal{C}^{2}_{\mu\nu}=+%\openone
\mathbb{I}$
for the symmetry under time-reversal and the spectral symmetry
under charge-conjugation 
is dictated by a microscopic derivation of $\mathcal{H}$.
Example 3, $\mathcal{H}$
can be thought of as representative of the Cartan symmetry class BDI if 
the choice
$\mathcal{T}^{\,}_{\mu\nu}$ obeying $\mathcal{T}^{2}_{\mu\nu}=+%\openone
\mathbb{I}$
and 
$\mathcal{C}^{\,}_{\mu\nu}$ obeying $\mathcal{C}^{2}_{\mu\nu}=+%\openone
\mathbb{I}$
for the symmetry under time-reversal and the spectral symmetry
under charge-conjugation 
is dictated by a microscopic derivation of $\mathcal{H}$.
In fact, all complex Cartan symmetry classes 
AI, BDI, D, DIII, AII, CII, C, and CI
are obtained from perturbing
$\mathcal{H}$ under the condition that either a symmetry under
time reversal or a spectral symmetry under charge conjugation
is imposed by a microscopic derivation of $\mathcal{H}$.
Finally, if a microscopic derivation of $\mathcal{H}$ does not
prevent perturbations that break all the symmetries under
time reversal and all the spectral 
symmetries under charge conjugations
from Table~\ref{tab:1}, then the remaining two real Cartan symmetry classes
A and AIII are realized. 

\begin{table*}
\begin{center}
\resizebox{\columnwidth}{!}{%
\begin{tabular}{ | l | l | l | l | l | l }  \hline
Triplet $(\mathscr{P},\mathscr{T},\mathscr{C})$
of symmetries in class CI
& 
Generic perturbation
$\mathcal{V}^{\,}_{\mathscr{P};\mathscr{T};\mathscr{C}}$
& 
$(\eta^{\,}_{\mathscr{T}},\eta^{\,}_{\mathscr{C}})$ 
& 
Topological index 
&
$\sigma(Q^{\,}_{\mathscr{P};\mathscr{T};\mathscr{C}\, A})$  
\\ 
\hline 
\hline
$\mathcal{P}^{\,}_{01}$, 
$\mathcal{T}^{\,}_{00}$, 
$\mathcal{C}^{\,}_{12}$    
& 
$v^{\,}_{01}\mathcal{X}^{\,}_{01}
+v^{\,}_{11}\mathcal{X}^{\,}_{11}
+v^{\,}_{30}\mathcal{X}^{\,}_{30}$                                         
& 
$(+,-)$ 
& 
$\mathbb{Z}$  
& $\circ$  
\\ \hline
$\mathcal{P}^{\,}_{01}$, 
$\mathcal{T}^{\,}_{11}$, 
$\mathcal{C}^{\,}_{12}$    
& 
$v^{\,}_{01}\mathcal{X}^{\,}_{01}
+v^{\,}_{11}\mathcal{X}^{\,}_{11}
+v^{\,}_{21}\mathcal{X}^{\,}_{21}$                                         
& 
$(+,-)$ 
& 
$\mathbb{Z}$ 
& 
$\circ$  
\\ \hline
$\mathcal{P}^{\,}_{01}$, 
$\mathcal{T}^{\,}_{30}$, 
$\mathcal{C}^{\,}_{12}$    
& 
$v^{\,}_{01}\mathcal{X}^{\,}_{01}
+v^{\,}_{21}\mathcal{X}^{\,}_{21}
+v^{\,}_{30}\mathcal{X}^{\,}_{30}$                                        
& 
$(+,-)$ 
& 
$\mathbb{Z}$ 
& 
$\circ$  
\\ 
\hline
$\mathcal{P}^{\,}_{10}$, 
$\mathcal{T}^{\,}_{00}$, 
$\mathcal{C}^{\,}_{12}$    
& 
$v^{\,}_{01}\mathcal{X}^{\,}_{01}
+v^{\,}_{03}\mathcal{X}^{\,}_{03}
+v^{\,}_{11}\mathcal{X}^{\,}_{11}
+v^{\,}_{13}\mathcal{X}^{\,}_{13}$  
& 
$(+,+)$ 
& 
0                   
&   
$\times$    
\\ \hline
$\mathcal{P}^{\,}_{10}$, 
$\mathcal{T}^{\,}_{11}$, 
$\mathcal{C}^{\,}_{12}$    
& 
$v^{\,}_{01}\mathcal{X}^{\,}_{01}
+v^{\,}_{02}\mathcal{X}^{\,}_{02}
+v^{\,}_{11}\mathcal{X}^{\,}_{11}
+v^{\,}_{12}\mathcal{X}^{\,}_{12}$  
& 
$(+,+)$ 
& 
0                   
&   
$\times$   
\\ \hline
$\mathcal{P}^{\,}_{10}$, 
$\mathcal{T}^{\,}_{30}$, 
$\mathcal{C}^{\,}_{12}$    
& 
$v^{\,}_{01}\mathcal{X}^{\,}_{01}
+v^{\,}_{03}\mathcal{X}^{\,}_{03}
+v^{\,}_{12}\mathcal{X}^{\,}_{12}$                                         
& 
$(-,+)$ 
& 
0                   
&    
$\times$  
\\ \hline
$\mathcal{P}^{\,}_{20}$, 
$\mathcal{T}^{\,}_{00}$, 
$\mathcal{C}^{\,}_{12}$    
& 
$v^{\,}_{01}\mathcal{X}^{\,}_{01}
+v^{\,}_{03}\mathcal{X}^{\,}_{03}
+v^{\,}_{22}\mathcal{X}^{\,}_{22}$                                         
& 
$(-,+)$ 
& 
0                   
& 
$\times$   
\\ \hline
$\mathcal{P}^{\,}_{20}$, 
$\mathcal{T}^{\,}_{11}$, 
$\mathcal{C}^{\,}_{12}$    
& 
$v^{\,}_{01}\mathcal{X}^{\,}_{01}
+v^{\,}_{02}\mathcal{X}^{\,}_{02}
+v^{\,}_{21}\mathcal{X}^{\,}_{21}
+v^{\,}_{22}\mathcal{X}^{\,}_{22}$  
& 
$(+,+)$ 
& 
0                  
& 
$\times$   
\\ \hline
$\mathcal{P}^{\,}_{20}$, 
$\mathcal{T}^{\,}_{30}$, 
$\mathcal{C}^{\,}_{12}$    
& 
$v^{\,}_{01}\mathcal{X}^{\,}_{01}
+v^{\,}_{03}\mathcal{X}^{\,}_{03}
+v^{\,}_{21}\mathcal{X}^{\,}_{21}
+v^{\,}_{23}\mathcal{X}^{\,}_{23}$  
& 
$(+,+)$ 
& 
0                 
&
$\times$   
\\ \hline
$\mathcal{P}^{\,}_{31}$, 
$\mathcal{T}^{\,}_{00}$, 
$\mathcal{C}^{\,}_{12}$    
& 
$v^{\,}_{01}\mathcal{X}^{\,}_{01}
+v^{\,}_{13}\mathcal{X}^{\,}_{13}
+v^{\,}_{22}\mathcal{X}^{\,}_{22}
+v^{\,}_{30}\mathcal{X}^{\,}_{30}$  
& 
$(+,+)$ 
& 
0                 
& 
$\times$  
\\ \hline
$\mathcal{P}^{\,}_{31}$, 
$\mathcal{T}^{\,}_{11}$, 
$\mathcal{C}^{\,}_{12}$    
& 
$v^{\,}_{01}\mathcal{X}^{\,}_{01}
+v^{\,}_{12}\mathcal{X}^{\,}_{12}
+v^{\,}_{22}\mathcal{X}^{\,}_{22}$                                          
& 
$(-,+)$ 
& 
0                 
& 
$\times$   
\\ \hline
$\mathcal{P}^{\,}_{31}$, 
$\mathcal{T}^{\,}_{30}$, 
$\mathcal{C}^{\,}_{12}$    
& 
$v^{\,}_{01}\mathcal{X}^{\,}_{01}
+v^{\,}_{12}\mathcal{X}^{\,}_{12}
+v^{\,}_{23}\mathcal{X}^{\,}_{23}
+v^{\,}_{30}\mathcal{X}^{\,}_{30}$  
& 
$(+,+)$ 
& 
0                 
&
$\times$   
\\ \hline
\end{tabular}
}
\caption{%
The first column gives all possible
combinations for the triplet of symmetries 
$\mathcal{P}^{\,}_{\mu\nu}$, 
$\mathcal{T}^{\,}_{\mu\nu}$,
and $\mathcal{C}^{\,}_{\mu \nu}$,
from Table~\ref{tab:1}
that are compatible with the Cartan symmetry class CI defined by the conditions
$\mathcal{T}^{2}_{\mu\nu}=+1$ and $\mathcal{C}^{2}_{\mu\nu}=-1$.
The second column gives for each row the most general
perturbation $\mathcal{V}^{\,}_{\mathscr{P};\mathscr{T};\mathscr{C}}$
that obeys the triplet of symmetries 
$(\mathcal{P},\mathcal{T},\mathcal{C})$ on any given row.
The third column gives the  doublet
$(\eta^{\,}_{\mathscr{T}},\eta^{\,}_{\mathscr{C}})\in\{-,+\}\times\{-,+\}$ 
where the sign $\eta^{\,}_{\mathscr{T}}$ is defined by 
$\mathcal{P}\,\mathcal{T}\,\mathcal{P}=\eta^{\,}_{\mathscr{T}}\,\mathcal{T}$
and similarly for $\eta^{\,}_{\mathscr{C}}$.
The fourth column is an application
of the classification for the symmetry-protected topological band insulators
in one-dimensional space derived in Refs.%
~\cite{Chingkai2012} and \cite{Morimoto13a}
(Table VI from Ref.~\cite{Morimoto13a} was particularly useful).
The topological index $\mathbb{Z}$ and $0$ correspond to 
topologically nontrivial and trivial bulk phases, respectively. 
The entry $\circ$ or $\times$ in the last column denotes
the presence or absence of zero modes in the spectrum
$\sigma(\widetilde{Q}^{\,}_{\mu\nu\,A})$
as is explained in Sec.%
~\ref{subsubsec: Stability analysis of the zero modes 1D example}
and verified by numerics.
         }
\label{tab:2}
\end{center}
\end{table*}

\subsection{Partition, topological numbers, and zero modes}
\label{subsubsec: Partition, topological numbers, and zero modes 1D example}

It is time to turn our attention to the topological properties
of the single-particle Hamiltonian $\mathcal{H}$
defined by its matrix elements%
~(\ref{eq: def 1D example}) 
or%
~(\ref{eq: def 1D example Bloch basis})
in the orbital basis
or
in the Bloch basis, respectively.

The representation~(\ref{eq: def 1D example Bloch basis})
demonstrates that the single-particle Hamiltonian 
$\mathcal{H}$
is reducible for all $k$ in the Brillouin zone,
\begin{subequations}
\label{eq: diract sum decomposition 1D example}
\begin{align}
&
\mathcal{H}=
\bigoplus_{n=1}^{N}\mathcal{H}^{\,}_{\pi n/N},
\label{eq: diract sum decomposition 1D example aa}
\\
&
\mathcal{H}^{\,}_{k}=
\mathcal{H}^{(+)}_{k}
\oplus
\mathcal{H}^{(-)}_{k},
\label{eq: diract sum decomposition 1D example a}
\end{align}
where the $4\times4$ Hermitian matrices
$\mathcal{H}^{(+)}_{k}$
and 
$\mathcal{H}^{(-)}_{k}$
are isomorphic to the $2\times2$ Hermitian matrices
\begin{equation}
H^{(+)}_{k}:=
2\,t\,\cos k\,\tau^{\,}_{1}
-
2\,\delta t\sin k\,\tau^{\,}_{2}
\label{eq: diract sum decomposition 1D example b}
\end{equation}
and
\begin{equation}
H^{(-)}_{k}:=
2\,t\,\cos k\,\tau^{\,}_{1}
+
2\,\delta t\sin k\,\tau^{\,}_{2},
\label{eq: diract sum decomposition 1D example c}
\end{equation}
\end{subequations}
respectively. 

The reducibility%
~(\ref{eq: diract sum decomposition 1D example})
defines the partition%
~(\ref{eq: partitioning single particles})
for the one-dimensional example~(\ref{eq: def 1D example Bloch basis}).
For simplicity,
we take the thermodynamic limit $N\to\infty$ with $N=2M$ so that we may
define
\begin{subequations}
\label{eq: def frak HA and frak HB 1D example}
\begin{equation}
\mathfrak{H}^{\,}_{A}:=
\bigoplus_{n=1}^{M}
\bigoplus_{\mu,\nu=\pm}
|\chi^{(\mu)}_{\nu;\pi n/2M}\rangle
\langle\chi^{(\mu)}_{\nu;\pi n/2M}|
\label{eq: def frak HA and frak HB 1D example a}
\end{equation}
and
\begin{equation}
\mathfrak{H}^{\,}_{B}:=
\bigoplus_{n=M+1}^{2M}
\bigoplus_{\mu,\nu=\pm}
|\chi^{(\mu)}_{\nu;\pi n/2M}\rangle
\langle\chi^{(\mu)}_{\nu;\pi n/2M}|.
\label{eq: def frak HA and frak HB 1D example b}
\end{equation}
\end{subequations}
Here,
$\chi^{(+)}_{-;k}$  
and 
$\chi^{(+)}_{+;k}$ 
are the pair of eigenstates with eigenvalues
$\varepsilon^{(+)}_{-;k}\leqslant\varepsilon^{(+)}_{+;k}$ of $H^{(+)}_{k}$. 
Similarly, 
$\chi^{(-)}_{-,k}$,
$\chi^{(-)}_{+,k}$, 
and 
$\varepsilon^{(-)}_{-;k}\leqslant\varepsilon^{(-)}_{+;k}$ denote
the eigenstates and their eigenenergies from the lower and
upper bands of $H^{(-)}_{k}$. This partition satisfies
[recall Eq.~(\ref{eq: def P C T S 1D example})]
\begin{subequations}
\begin{align}
&
\mathscr{P}A=B,
&
\mathscr{P}B=A,
\\
&
\mathscr{C}A=A,
&
\mathscr{C}B=B,
\\
&
\mathscr{T}A=A,
&
\mathscr{T}B=B,
\\
&
\mathscr{S}A=A,
&
\mathscr{S}B=B.
\end{align}
\end{subequations}

Hamiltonian~(\ref{eq: diract sum decomposition 1D example b})
describes a single-particle that
hops between sites labeled by 
$\blacksquare$ and $\square$ in Fig.~\ref{fig: dimerized chain}
with the uniform hopping amplitude $t$
and the staggered hopping amplitude $\delta t$.
Hamiltonian~(\ref{eq: diract sum decomposition 1D example c})
describes a single-particle that
hops between sites labeled by $\bullet$ and $\circ$ 
in Fig.~\ref{fig: dimerized chain}
with the uniform hopping amplitude $t$
and the staggered hopping amplitude $\delta t$.

If we take the thermodynamic limit 
\begin{subequations}
\label{eq: def TL}
\begin{equation}
N,N^{\,}_{\mathrm{f}}\to\infty
\label{eq: def TL a}
\end{equation}
holding the fermion density 
\begin{equation}
N^{\,}_{\mathrm{f}}/N=1
\label{eq: def TL b}
\end{equation}
\end{subequations}
fixed, we find the non-vanishing winding numbers
\begin{equation}
W^{(+)}_{\mathrm{FS}}=
-W^{(-)}_{\mathrm{FS}}:=
\frac{\mathrm{i}}{2\pi}
\oint\mathrm{d}k\,
\chi^{(+)\dag}_{-}(k)
\left(\frac{\partial\chi^{(+)}_{-}}{\partial k}\right)(k)
\label{eq: def windings in A and B for 1d example}
\end{equation}
for any non-vanishing dimerization $\delta t$.
Owing to the reducibility of $\mathcal{H}^{\,}_{k}$,
the winding number $W^{\,}_{\mathrm{FS}}$ 
for the single-particle eigenstates
making up the Fermi sea of $\mathcal{H}^{\,}_{k}$ is
\begin{equation}
W^{\,}_{\mathrm{FS}}=
W^{(+)}_{\mathrm{FS}}
+ 
W^{(-)}_{\mathrm{FS}}=
0
\label{eq: def windings in A oplus B for 1d example}
\end{equation}
for any non-vanishing dimerization $\delta t$.
The non-vanishing values of the winding numbers
endow each of the single-particle Hamiltonians
$H^{(+)}$ and $H^{(-)}$
with a topological attribute.
The single-particle Hamiltonian $\mathcal{H}$
is topologically trivial as its winding number vanishes.

The very definition of the winding numbers 
(\ref{eq: def windings in A and B for 1d example})
and
(\ref{eq: def windings in A oplus B for 1d example})
requires twisted boundary conditions and a spectral gap
between the Fermi sea and all many-body excitations. 

With open boundary conditions,
two zero-dimensional boundaries at $r-1$ and $r$
follow from setting all hopping amplitudes
between site $r-1$ and $r$ in Fig.\ \ref{fig: dimerized chain}
to zero, i.e., erasing the connecting full and dashed lines
from Fig.\ \ref{fig: dimerized chain} 
that intersect the vertical line $RVE$.
In such an open geometry,
the winding numbers 
(\ref{eq: def windings in A and B for 1d example})
and
(\ref{eq: def windings in A oplus B for 1d example})
are ill defined. However, 
the bulk-edge correspondence implies the existence of mid-gap states
(zero modes) that are localized at the boundaries
(if the thermodynamic limit is taken with open boundary conditions), 
whenever the winding numbers 
(\ref{eq: def windings in A and B for 1d example})
are non-vanishing 
(if the thermodynamic limit is taken with twisted boundary conditions). 
There are four such zero modes,
a pair of zero modes for each boundary. On any
given boundary, one of the zero modes localized
on this boundary originates from
$H^{(+)}_{r,r'}$,
while the other originates from $H^{(-)}_{r,r'}$.
These four zero modes are eigenstates of either
$\mathcal{X}^{\,}_{03}+\mathcal{X}^{\,}_{33}$
or
$\mathcal{X}^{\,}_{03}-\mathcal{X}^{\,}_{33}$.
As a set, they are protected against any 
perturbation that anti-commutes with
$\mathcal{X}^{\,}_{03}$
and
$\mathcal{X}^{\,}_{33}$, i.e.,
they are protected against any 
linear combination of
$\mathcal{X}^{\,}_{01}$,
$\mathcal{X}^{\,}_{02}$,
$\mathcal{X}^{\,}_{31}$,
and
$\mathcal{X}^{\,}_{32}$.
[We have verified by exact diagonalization
that the zero modes in $\sigma(\mathcal{H})$ are indeed 
robust to the perturbations 
$\mathcal{X}^{\,}_{01}$,
$\mathcal{X}^{\,}_{02}$,
$\mathcal{X}^{\,}_{31}$,
or
$\mathcal{X}^{\,}_{32}$.]

It is believed that the existence of protected gapless boundary states, 
when taking the thermodynamic limit with open boundary conditions
for a topological band insulator with Hamiltonian $\mathcal{H}$,
implies the existence of protected gapless boundary states
in the entanglement spectrum of $Q^{\,}_{A}$ 
for a suitable partition of the form%
~(\ref{eq: partitioning single particles}),
when taking the thermodynamic limit with closed boundary conditions
for the equal-time one-point correlation matrix
$\mathcal{Q}$.%
~\cite{Ryu2006,Turner2010,Fidkowski2010,Hughes2011}

We have verified by exact diagonalization
that this is also the case
when we take the thermodynamic limit~(\ref{eq: def TL})
of $\mathcal{H}$ defined by Eq.%
~(\ref{eq: diract sum decomposition 1D example})
with an even number of sites $N=2\,M$ and
with the partition defined by Eq.%
~(\ref{eq: def frak HA and frak HB 1D example}).
The spectra for $\mathcal{H}$
and $Q^{\,}_{A}$ 
are shown in Fig.\ref{fig: EEES_{1}DChain}(a)
for $\delta t=t$ and $N=12$.
In both spectra, there are four zero modes within
an exponential accuracy resulting from finite-size corrections.
In the case of $\mathcal{H}$, a pair of zero modes
is exponentially localized at one physical boundary
a distance $2M$ apart from the second pair of zero modes localized
on the opposite boundary. 
In the case of $Q^{\,}_{A}$, a pair of zero modes 
is exponentially localized 
at the boundary with $B$ to the left of $A$
a distance $M$ apart from the second pair of zero modes 
exponentially localized
at the boundary with $A$ to the left of $B$.
The exponential decay of the zero modes away from their
boundary is inversely proportional to the band gap
of $\mathcal{H}$ when periodic boundary conditions are imposed.

We observe that the existence of zero modes in the
spectrum of $Q^{\,}_{A}$ does not imply the existence
of zero modes in the spectrum of $\mathcal{H}$.
For example, if we shift the spectrum of
$\mathcal{H}$ defined in Eq.%
~(\ref{eq: diract sum decomposition 1D example})
in a uniform way by adding a chemical potential
smaller than the band gap, we immediately lose
the zero modes. However, all single-particle
eigenstates are unperturbed by the chemical
potential, for it enters as a perturbation that
commutes with $\mathcal{H}$. Consequently,
neither $\mathcal{Q}$, nor $Q^{\,}_{A}$,
nor their spectra depend on the chemical potential.
In particular, the spectrum of $Q^{\,}_{A}$
with a non-vanishing chemical potential
contains the very same four zero modes
that are present when the chemical potential vanishes.

The question we are after is the following.
Are the four zero modes of $\mathcal{H}^{\,}_{r,r'}(t,\delta t)$,
that are localized at both ends of an open chain,
stable against perturbations that (i) 
commute with a reflection from Table~\ref{tab:1}
and (ii) have a characteristic energy 
that is small relative to the unperturbed band gap?
Similarly, how stable are the four zero modes of 
$Q^{\,}_{A}$?

\begin{figure}[t]
\centering
\includegraphics[height=8 cm]{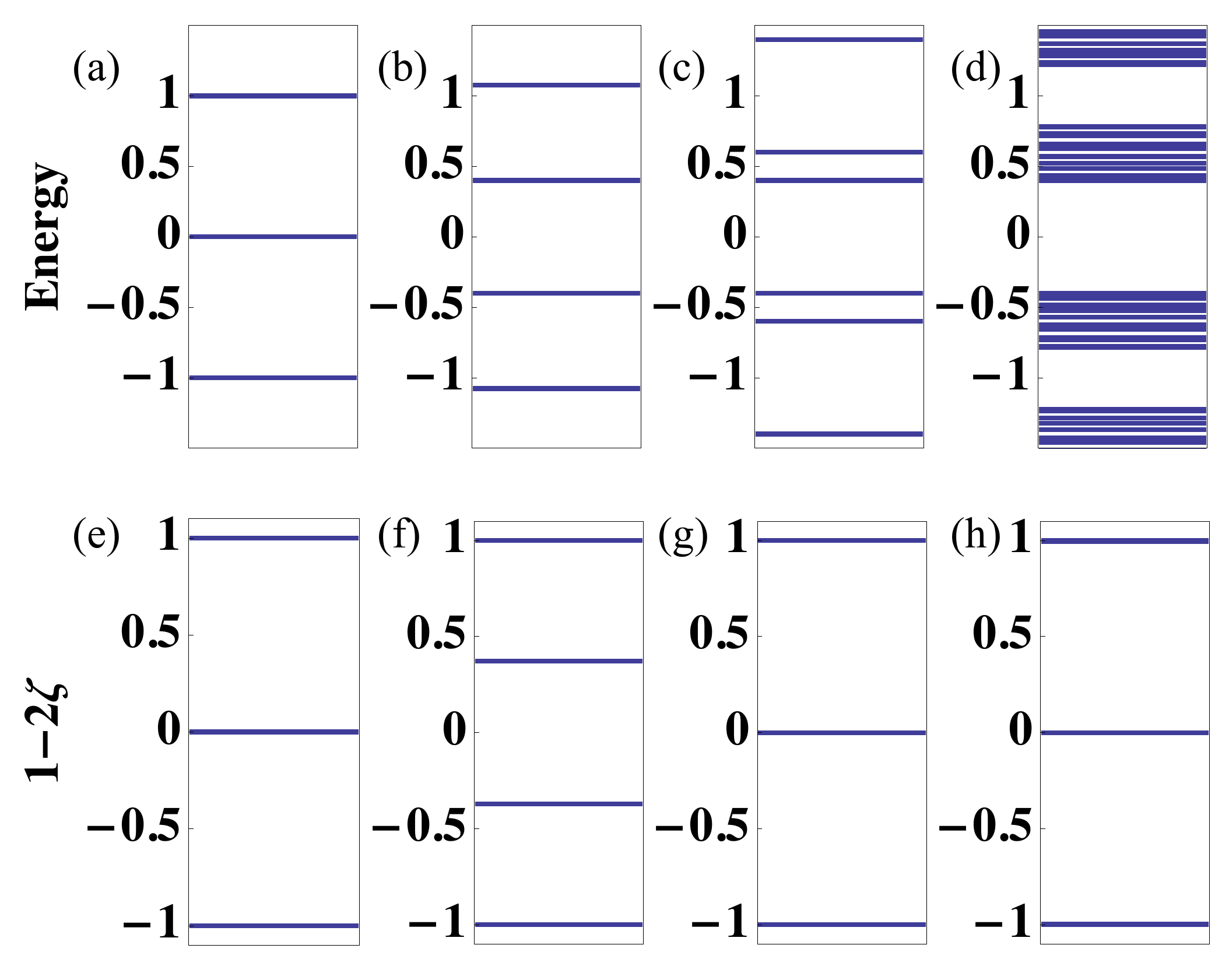}
\caption{
Energy spectra $\sigma(\widetilde{\mathcal{H}}^{\,}_{\mu\nu})$
in units of $2\,t$
with $\widetilde{\mathcal{H}}^{\,}_{\mu\nu}$ defined by 
Eq.~(\ref{eq: def widetilde H for 1D example})
obeying open boundary conditions 
are plotted in panels (a-d).
Entanglement spectra $\sigma(\widetilde{Q}^{\,}_{\mu\nu\,A})$
for the equal-time one-point correlation matrix%
~(\ref{eq: def cal Q})
derived from $\widetilde{\mathcal{H}}^{\,}_{\mu\nu}$ 
obeying periodic boundary conditions  
are plotted in panels (f-h).
The choices 
$\delta t=t$ and $\mathcal{V}^{\,}_{\mu\nu}=0$ 
for panels (a) and (e),
$\delta t=t$ and 
$\mathcal{V}^{\,}_{12 rr'}=\delta^{\,}_{r,r'}\,(t/10)\,\mathcal{X}^{\,}_{12}$
for panels (b) and (f),
$\delta t=t$ and 
$\mathcal{V}^{\,}_{11 rr'}=\delta^{\,}_{r,r'}\,(t/10)\,\mathcal{X}^{\,}_{11}$
for panels (c) and (g),
and
$\delta t=2\,t/3$ and 
$\mathcal{V}^{\,}_{11 rr'}=\delta^{\,}_{r,r'}\,(t/10)\,\mathcal{X}^{\,}_{11}$
for panels (d) and (h)
are made. The spectra in panels (b) and (f)
are unchanged if $\mathcal{V}^{\,}_{12}$ 
is replaced by either
$\mathcal{V}^{\,}_{03}$ or $\mathcal{V}^{\,}_{33}$.
        }
\label{fig: EEES_{1}DChain}
\end{figure}

\subsection{Stability analysis of the zero modes}
\label{subsubsec: Stability analysis of the zero modes 1D example}

The topological classification of non-interacting fermionic insulators 
based on the presence or absence of the discrete symmetries 
under the operations of
time reversal, charge conjugation, and chirality 
satisfies a bulk-edge (holography) correspondence principle. 
According to this principle a non-vanishing value for
a certain topological index defined for the bulk 
is equivalent to the existence of extended boundary states 
on physical boundaries. This correspondence is lost when 
crystalline symmetries such as inversion about a point 
or reflection about a mirror plane are also imposed. 
What remains is a symmetry-protected topological classification of 
non-interacting bulk fermionic insulators that
obey a crystalline symmetry\cite{Chingkai2012,Morimoto13a}.
This classification has two distinctive features.
First, a symmetry-protected topologically phase is not required
to support extended boundary states localized at the physical edges.
Second, a bulk-edge correspondence principle is nevertheless believed to
hold for the entanglement spectrum. 

We are going to verify how this correspondence principle 
for symmetry-protected topologically phases of non-interacting fermionic
insulating phases holds for the one-dimensional model 
with the elementary building block defined by Eq.~(\ref{eq: def 1D example})
and explain why. 
When $(\mu,\nu)=(0,0);(1,1);(2,1);(3,0)$,
we are going to show that
$\widetilde{\mathcal{H}}^{\,}_{\mu\nu}$ 
defined in Eq.~(\ref{eq: def widetilde H for 1D example})
fails to support robust edge states for an open geometry, 
whereas $\widetilde{\mathcal{Q}}^{\,}_{\mu\nu\,A}$
supports robust edge states at the entangling boundaries.

To this end, our strategy is going to be to first 
explain the rows of Table \ref{tab:1}.
We shall then choose a combination of symmetry under time reversal 
and spectral symmetry under charge conjugation that puts
$\widetilde{\mathcal{H}}^{\,}_{\mu\nu}$ in the Cartan symmetry class CI.
This symmetry class is a topologically trivial one, 
a generic perturbation of $\widetilde{\mathcal{H}}^{\,}_{\mu\nu}$
destroys any boundary state that $\mathcal{H}^{\,}_{\mu\nu}$
supports at a physical boundary. However, we shall show that
imposing a suitable symmetry under parity guarantees the existence of
protected boundary states in the entanglement spectrum 
$\sigma(\widetilde{\mathcal{Q}}^{\,}_{\mu\nu\,A})$.

If periodic boundary conditions are imposed, a perturbation
that commutes with the parity transformation generated by
$\mathcal{X}^{\,}_{10}$
is of the general form
\begin{equation}
\mathcal{V}^{\,}_{k}:=\,
\sum_{\nu=0}^{3}
\left[
\sum_{\mu=0,1}
f^{\,}_{\mu\nu;k}
+
\sum_{\mu=2,3}
g^{\,}_{\mu\nu;k}
\right]
\mathcal{X}^{\,}_{\mu\nu},
\label{eq: generic parity preserving perturbation}
\end{equation}
where the functions $f^{\,}_{\mu\nu;k}$
and $g^{\,}_{\mu\nu;k}$ are even and odd under 
the inversion $k\to-k$, respectively
[recall Eq.~(\ref{eq: four symmetries in 1D example b})].
For simplicity, the perturbation $\mathcal{V}^{\,}_{k}$
is taken independent of $k$, i.e.,
\begin{equation}
\mathcal{V}^{\,}_{k}:=\,
\sum_{\nu=0}^{3}
\sum_{\mu=0,1}
v^{\,}_{\mu\nu}
\mathcal{X}^{\,}_{\mu\nu}=
\mathcal{X}^{\,}_{10}\,
V^{\,}_{-k}\,
\mathcal{X}^{\,}_{10}
\label{eq: constant parity preserving perturbation in k space}
\end{equation}
with the eight parameters $v^{\,}_{\mu\nu}$ real valued.
In the orbital basis, we have
\begin{equation}
\begin{split}
\mathcal{V}^{\,}_{r,r'}=&\,
\delta^{\,}_{r,r'}\,
\left[
\underline{\underline{
v^{\,}_{01}\,\mathcal{X}^{\,}_{01}
          }          }
+
\underline{\underline{
v^{\,}_{02}\,\mathcal{X}^{\,}_{02}
          }          }
+
v^{\,}_{00}\,\mathcal{X}^{\,}_{00}
+
v^{\,}_{03}\,\mathcal{X}^{\,}_{03}
\right]
\\
&\,
+
\delta^{\,}_{r,r'}\,
\left[
\underline{
v^{\,}_{10}\,\mathcal{X}^{\,}_{10}
          }
+
\underline{
v^{\,}_{11}\,\mathcal{X}^{\,}_{11}
          }
+
\underline{
v^{\,}_{12}\,\mathcal{X}^{\,}_{12}
          }
+
\underline{
v^{\,}_{13}\,\mathcal{X}^{\,}_{13}
          }
\right].
\end{split}
\label{eq: constant parity preserving perturbation in r space}
\end{equation}
The terms that have been underlined twice anti-commute with
the two commuting matrices 
$\mathcal{X}^{\,}_{03}$
and
$\mathcal{X}^{\,}_{33}$
that share the zero modes as eigenstates.
Hence, the zero modes are protected as a set against the
parity-preserving perturbations that are linear combinations
of 
$\mathcal{X}^{\,}_{01}$
and
$\mathcal{X}^{\,}_{02}$.
All other terms in Eq.%
~(\ref{eq: constant parity preserving perturbation in r space})
fail to anti-commute with both
$\mathcal{X}^{\,}_{03}$
and
$\mathcal{X}^{\,}_{33}$.
The zero modes are not necessarily protected as a set under these 
parity-preserving perturbations. The terms in Eq.%
~(\ref{eq: constant parity preserving perturbation in r space})
that are underlined once either anti-commute with
$\mathcal{X}^{\,}_{03}$ 
or
$\mathcal{X}^{\,}_{33}$
but not with both. The terms in Eq.%
~(\ref{eq: constant parity preserving perturbation in r space})
that are not underlined fail to  anti-commute with both
$\mathcal{X}^{\,}_{03}$ 
and
$\mathcal{X}^{\,}_{33}$.

The same exercise can be repeated for
the parity transformations generated by
$\mathcal{X}^{\,}_{01}$,
$\mathcal{X}^{\,}_{20}$,
and
$\mathcal{X}^{\,}_{31}$.
This deliver the first fourteen rows in Table \ref{tab:1}.

For any $\mu,\nu=0,1,2,3$,
we define the single-particle Hamiltonian
$\widetilde{\mathcal{H}}^{\,}_{\mu\nu}\equiv
\mathcal{H}+\mathcal{V}^{\,}_{\mu\nu}$ 
by its matrix elements
\begin{subequations}
\label{eq: def widetilde H for 1D example}
\begin{align}
&
\widetilde{\mathcal{H}}^{\,}_{\mu\nu\,r,r'}:=
\mathcal{H}^{\,}_{r,r'}
+
\mathcal{V}^{\,}_{\mu\nu\,r,r'},
\\
&
\mathcal{V}^{\,}_{\mu\nu\,r,r'}:=
\delta^{\,}_{r,r'}\,
v^{\,}_{\mu\nu}\,
\mathcal{X}^{\,}_{\mu\nu}.
\label{eq: def widetilde H for 1D example b}
\end{align}
\end{subequations}
The single-particle Hamiltonian
$\mathcal{H}^{\,}_{r,r'}$
was defined in Eq.~(\ref{eq: def 1D example d})
for $r,r'=1,\cdots,N-1$. We choose between
imposing open boundary conditions by setting the hopping
amplitudes to zero between sites $N$ and $N+1$
or periodic boundary conditions. The perturbation strength
$v^{\,}_{\mu\nu}$ is real-valued. It will be set to $t/10$.
The corresponding equal-time one-point
correlation matrix defined by Eq.~(\ref{eq: def cal Q})
is denoted $\widetilde{\mathcal{Q}}^{\,}_{\mu\nu}$.

For illustrative purposes,
we plot in Figs.~\ref{fig: EEES_{1}DChain}(a-d)
the energy eigenvalue spectrum $\sigma(\widetilde{H}^{\,}_{\mu\nu})$
of $\widetilde{H}^{\,}_{\mu\nu}$ 
obeying open boundary conditions by an exact diagonalization with $N=12$
that can be extrapolated to the thermodynamic limit.
Energy eigenvalues are measured in units of $2t$.
In panels (a-c), $\delta t=t$ implies that
all energy eigenstates have wave-functions that are localized
on a pair of consecutive sites for which the hopping amplitude
is $t+\delta t=2t$.
In the remaining panel (d), a $\delta t\neq t$
delocalizes bulk energy eigenstates that acquire a dispersion,
i.e., a band width. The zero modes in panel (a) are four edge states. 
They are protected  by the chiral symmetry in that they are eigenstates
of either $\mathcal{X}^{\,}_{03}$ or $\mathcal{X}^{\,}_{33}$. 
For example, they are robust to 
changing the value of $\delta t$ away from $t$. 
However, these zero modes are shifted to non-vanishing energies 
by the perturbations 
$\mathcal{V}^{\,}_{12}$ 
for panel (b) and
$\mathcal{V}^{\,}_{11}$ 
for panels (c-d).
The spectrum in panel (b) is unchanged by
the substitutions
$\mathcal{V}^{\,}_{12}\to\mathcal{V}^{\,}_{03}\to\mathcal{V}^{\,}_{33}$.
The presence or absence of protected (against the perturbation
from the second column)
zero modes localized on the physical
boundaries has been verified in this way for all rows and is reported
in the penultimate column of
Table \ref{tab:1}.

We plot in Figs.~\ref{fig: EEES_{1}DChain}(e-h)
the eigenvalue spectrum $\sigma(\widetilde{Q}^{\,}_{\mu\nu A})$
of the upper-left block $\widetilde{Q}^{\,}_{\mu\nu A}$
in the equal-time one-point correlation matrix
$\widetilde{\mathcal{Q}}^{\,}_{\mu\nu}$ corresponding to 
$\widetilde{\mathcal{H}}^{\,}_{\mu\nu}$ obeying periodic boundary conditions
by an exact diagonalization with $N=12$
that can be extrapolated to the thermodynamic limit.
Panels (e-g) have $\delta t=t$.
Panel (h) has $\delta t=2\,t/3$.
There is no perturbation in panel (e),
in which case four zero modes
are present in the spectrum $\sigma(\widetilde{Q}^{\,}_{\mu\nu A})$.
The perturbation $\mathcal{V}^{\,}_{12}$ splits the four
zero modes into two pairs of degenerate eigenstates with
eigenvalues only differing by their sign in panel (f), 
as was the case for the Hamiltonian in panel (b).
The spectrum in panel (f) is unchanged by
the substitutions
$\mathcal{V}^{\,}_{12}\to\mathcal{V}^{\,}_{03}\to\mathcal{V}^{\,}_{33}$.
Unlike in panels (c) and (d) for the Hamiltonian, 
panels (g) and (h) show that
the four zero modes of 
$\widetilde{Q}^{\,}_{\mu\nu A}$ are robust to
the perturbation $\mathcal{V}^{\,}_{11}$.

The lesson from
Fig.~\ref{fig: EEES_{1}DChain}
and 
Table~\ref{tab:1}
is that of all the seven 
\begin{equation}
\mathcal{V}^{\,}_{00},
\mathcal{V}^{\,}_{03},
\mathcal{V}^{\,}_{11},
\mathcal{V}^{\,}_{12},
\mathcal{V}^{\,}_{21},
\mathcal{V}^{\,}_{22},
\mathcal{V}^{\,}_{30}
\end{equation}
out of fourteen parity-preserving perturbations
that gap the zero modes of 
$\mathcal{H}$,
only four, namely
\begin{equation}
\mathcal{V}^{\,}_{00},
\mathcal{V}^{\,}_{11}, 
\mathcal{V}^{\,}_{21}, 
\mathcal{V}^{\,}_{30}, 
\end{equation}
fail to also gap the zero modes of
$Q^{\,}_{A}$. 

To explain this observation, we rely on the reasoning
that delivers Table~\ref{tab:2}.
We assume that the underlying microscopic model
has the triplet of symmetries 
$(\mathscr{P};\mathscr{T};\mathscr{C})\sim
(\mathcal{P}^{\,}_{\mu\nu},\mathcal{T}^{\,}_{\mu\nu},\mathcal{C}^{\,}_{\mu\nu})$
where the doublet 
$(\mathscr{T},\mathscr{C})\sim
(\mathcal{T}^{\,}_{\mu\nu},\mathcal{C}^{\,}_{\mu\nu})$
defines the symmetry class CI, i.e.,
$\mathcal{T}^{2}_{\mu\nu}=+%\openone
\mathbb{I}$
and
$\mathcal{C}^{2}_{\mu\nu}=-%\openone
\mathbb{I}$.
We denote the most general perturbation 
that is compliant with the triplet of symmetries 
$(\mathscr{P};\mathscr{T};\mathscr{C})$
defining a given row of Table~\ref{tab:2} by
$\mathcal{V}^{\,}_{\mathscr{P};\mathscr{T};\mathscr{C}}$.
The explicit form of this perturbation
is to be found in the second column of
Table~\ref{tab:2}
as one varies $(\mathscr{P};\mathscr{T};\mathscr{C})$.
For a given row in Table~\ref{tab:2},
$\mathcal{V}^{\,}_{\mathscr{P};\mathscr{T};\mathscr{C}}$
is contained in the most general perturbation 
$\mathcal{V}^{\,}_{\mathscr{T};\mathscr{C}}$
that is compliant with the doublet of symmetries
$(\mathscr{T},\mathscr{C})$.
We define the single-particle Hamiltonian
$\mathcal{H}^{\,}_{\mathscr{P};\mathscr{T};\mathscr{C}}\equiv
\mathcal{H}+\mathcal{V}^{\,}_{\mathscr{P};\mathscr{T};\mathscr{C}}$ 
by its matrix elements
\begin{subequations}
\label{eq: def H PTC for 1D example}
\begin{align}
&
\mathcal{H}^{\,}_{\mathscr{P};\mathscr{T};\mathscr{C}\,r,r'}:=
\mathcal{H}^{\,}_{r,r'}
+
\mathcal{V}^{\,}_{\mathscr{P};\mathscr{T};\mathscr{C}\,r,r'},
\\
&
\mathcal{V}^{\,}_{\mathscr{P};\mathscr{T};\mathscr{C}\,r,r'}:=
\delta^{\,}_{r,r'}\,
\sum_{\mu,\nu\in\mathrm{row}}
v^{\,}_{\mu\nu}\,
\mathcal{X}^{\,}_{\mu\nu}.
\label{eq: def H PTC for 1D example b}
\end{align}
\end{subequations}
The corresponding equal-time one-point correlation matrix
is $\mathcal{Q}^{\,}_{\mathscr{P};\mathscr{T};\mathscr{C}}$
and its upper-left block is
$Q^{\,}_{\mathscr{P};\mathscr{T};\mathscr{C}\,A}$.
The third column in Table~\ref{tab:2}
provides two signs for each row. 
The first sign $\eta^{\,}_{\mathscr{T}}$
is positive if $\mathscr{P}$ commutes with $\mathscr{T}$
and negative if $\mathscr{P}$ anti-commutes with $\mathscr{T}$.
The second sign $\eta^{\,}_{\mathscr{C}}$
is positive if $\mathscr{P}$ commutes with $\mathscr{C}$
and negative if $\mathscr{P}$ anti-commutes with $\mathscr{C}$.
The information contained with the doublet 
$(\eta^{\,}_{\mathscr{T}},\eta^{\,}_{\mathscr{C}})$
is needed to read from Table VI of Ref.~\cite{Morimoto13a}
the bulk topological index of the single-particle Hamiltonian%
~(\ref{eq: def H PTC for 1D example}).
This topological index does not guarantee that
$\mathcal{H}^{\,}_{\mathscr{P};\mathscr{T};\mathscr{C}}$
supports boundary states in an open geometry.
In fact, $\mathcal{H}^{\,}_{\mathscr{P};\mathscr{T};\mathscr{C}}$
does not support boundary states in an open geometry,
for the physical boundaries are interchanged 
under the operation of parity $\mathscr{P}$.
On the other hand, $Q^{\,}_{\mathscr{P};\mathscr{T};\mathscr{C}\,A}$
supports boundary states on the entangling boundaries
under the following conditions.

Hamiltonian $\mathcal{H}^{\,}_{\mathscr{P};\mathscr{T};\mathscr{C}}$
is local by assumption and gaped if periodic boundary conditions
are imposed. As explained in Sec.%
~\ref{subsec: Spectral gap and locality of the induced symmetry Gamma},
$\mathcal{Q}^{\,}_{\mathscr{P};\mathscr{T};\mathscr{C}}$ and all its four blocks
inherit this locality. We have shown with Eq.%
~(\ref{eq: O operates on QA if block diagonal})
that the upper-left block $Q^{\,}_{\mathscr{P};\mathscr{T};\mathscr{C}\,A}$
inherits the symmetry $\mathcal{T}$ and the spectral symmetry $\mathcal{C}$ 
of $\mathcal{H}^{\,}_{\mathscr{P};\mathscr{T};\mathscr{C}}$.
We have also shown with Eq.%
~(\ref{eq: main result on how to get chiral sym})
that the symmetry $\mathcal{P}$ of
$\mathcal{H}^{\,}_{\mathscr{P};\mathscr{T};\mathscr{C}}$
is turned into a spectral symmetry of
$Q^{\,}_{\mathscr{P};\mathscr{T};\mathscr{C}\,A}$
under 
$\Gamma^{\ }_{\mathscr{P}A}:=
C^{\,}_{\mathscr{P};\mathscr{T};\mathscr{C}\,AB}\,P^{\dag}_{AB}$.
The unperturbed upper-left block $Q^{\,}_{A}$
has two zero modes per entangling boundary.
Because of locality, the perturbation
\begin{equation}
\delta Q^{\,}_{\mathscr{P};\mathscr{T};\mathscr{C}\,A}:=
Q^{\,}_{\mathscr{P};\mathscr{T};\mathscr{C}\,A}-Q^{\,}_{A}
\end{equation}
only mixes the two members of a doublet of
boundary states of 
$Q^{\,}_{A}$
on a given entangling boundary. Hence,
we may represent the effect of the perturbation
$\delta Q^{\,}_{\mathscr{P};\mathscr{T};\mathscr{C}\,A}$
by imposing on the Hermitian $2\times2$ matrix
\begin{equation}
\delta Q^{\,}_{\mathrm{boundary}}:=
\sum_{\mu=0}^{3}
a^{\,}_{\mu}\,\rho^{\,}_{\mu},
\qquad
a^{\,}_{\mu}\in\mathbb{R},
\end{equation}
($\rho^{\,}_{0}$ is the unit $2\times2$ matrix
and $\rho^{\,}_{1}$, $\rho^{\,}_{2}$, $\rho^{\,}_{3}$ are
the Pauli matrices)
the condition imposed by the symmetry $\mathscr{T}\sim\rho^{\,}_{0}\,K$ and
the spectral symmetries $\mathscr{C}\sim\rho^{\,}_{2}\,K$ 
and $\Gamma^{\ }_{\mathscr{P}A}$. The first two symmetries imply that
\begin{equation}
\delta Q^{\,}_{\mathrm{boundary}}=
a^{\,}_{1}\,\rho^{\,}_{1}
+
a^{\,}_{3}\,\rho^{\,}_{3}.
\label{eq: Q boundary if T and C} 
\end{equation}
A doublet of zero modes is thus protected if and only if
\begin{equation}
\Gamma^{\ }_{\mathscr{P}A}:=
C^{\,}_{\mathscr{P};\mathscr{T};\mathscr{C}\,AB}\,
P^{\dag}_{AB}\sim
\rho^{\,}_{0},
\label{eq: Q boundary if CI}
\end{equation}
for $\{\rho^{\,}_{0},\delta Q^{\,}_{\mathrm{boundary}}\}=0$
can then only be satisfied if $a^{\,}_{1}=a^{\,}_{3}=0$.
We now show that condition~(\ref{eq: Q boundary if CI})
is only met for the first three rows of Table~\ref{tab:2},
the only rows from Table~\ref{tab:2} with
$(\eta^{\,}_{\mathscr{T}},\eta^{\,}_{\mathscr{C}},)=(+,-)$,
i.e., the only choice for the triplet
$(\mathcal{P}^{\,}_{\mu\nu},\mathcal{T}^{\,}_{\mu\nu},\mathcal{C}^{\,}_{\mu\nu})$
for which
\begin{equation}
[\mathscr{P},\mathscr{T}]=0,
\qquad
\{\mathscr{P},\mathscr{C}\}=0.
\label{eq: identity A needed for proof}
\end{equation}
To see this,
we are going to combine 
Eq.~(\ref{eq: identity A needed for proof})
with
\begin{equation}
[C^{\,}_{\mathscr{P};\mathscr{T};\mathscr{C}\,AB},T^{\,}_{\mathscr{T}}]=0,
\qquad
\{C^{\,}_{\mathscr{P};\mathscr{T};\mathscr{C}\,AB},C^{\,}_{\mathscr{C}}\}=0,
\label{eq: identity B needed for proof}
\end{equation}
where $T^{\,}_{\mathscr{T}}$ and $C^{\,}_{\mathscr{C}}$
represent the actions of time reversal and charge conjugation
on the partition grading of the equal-time one-point correlation
matrix. If we use the algebraic identity
\begin{subequations}
\label{eq: identity C needed for proof}
\begin{align}
&[\Gamma^{\,}_{\mathscr{P}\,A},T^{\,}_{\mathscr{T}}]=
C^{\,}_{\mathscr{P};\mathscr{T};\mathscr{C}\,AB}\,
[P^{\dag}_{AB},T^{\,}_{\mathscr{T}}]
+
[C^{\,}_{\mathscr{P};\mathscr{T};\mathscr{C}\,AB},T^{\,}_{\mathscr{T}}]\,
P^{\dag}_{AB},
\\
&[\Gamma^{\,}_{\mathscr{P}\,A},C^{\,}_{\mathscr{C}}]=
C^{\,}_{\mathscr{P};\mathscr{T};\mathscr{C}\,AB}\,
\{P^{\dag}_{AB},C^{\,}_{\mathscr{C}}\}
-
\{C^{\,}_{\mathscr{P};\mathscr{T};\mathscr{C}\,AB},C^{\,}_{\mathscr{C}}\}\,
P^{\dag}_{AB},
\end{align}
\end{subequations}
Eq.~(\ref{eq: identity C needed for proof}),
when combined with Eqs.~(\ref{eq: identity B needed for proof})
and~(\ref{eq: identity A needed for proof}), simplifies to
\begin{equation}
[\Gamma^{\,}_{\mathscr{P}\,A},T^{\,}_{\mathscr{T}}]=
[\Gamma^{\,}_{\mathscr{P}\,A},C^{\,}_{\mathscr{C}}]=0.
\label{eq: identity D needed for proof}
\end{equation}
Equation~(\ref{eq: identity D needed for proof})
allows us to deduce that $\Gamma^{\,}_{\mathscr{P}\,A}$
must be represented by $\rho^{\,}_{0}$ on the two-dimensional
Hilbert space spanned by the boundary states on an entangling boundary,
\begin{equation}
\Gamma^{\,}_{\mathscr{P}\,A}\sim \rho^{\,}_{0}.
\label{eq: identity E needed for proof}
\end{equation}
Hence, the only 
$\delta Q^{\,}_{\mathrm{boundary}}$
in Eq.~(\ref{eq: Q boundary if T and C})
that anti-commutes with $\rho^{\,}_{0}$ is
$\delta Q^{\,}_{\mathrm{boundary}}=0$,
thereby proving the stability of the boundary states
on an entangling boundary for the first three rows
of Table~\ref{tab:2}.

\begin{figure*}[t]
\centering
\subfigure[] {
\label{1D0}
\includegraphics[height=2.2cm]{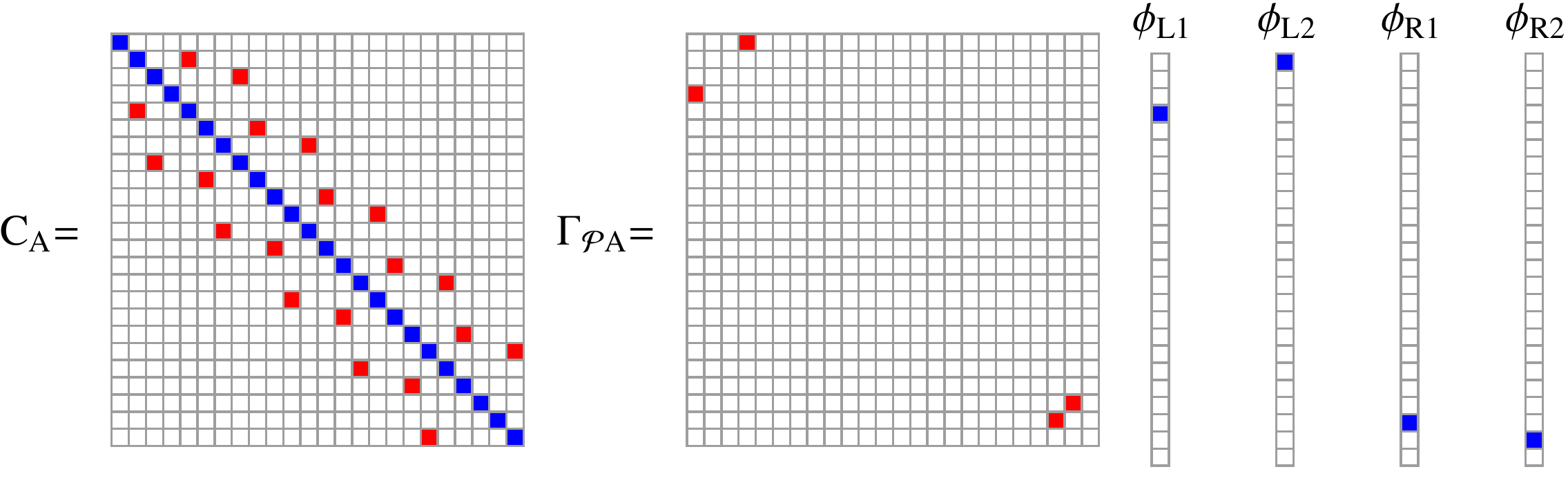}
}
\subfigure[] {
\label{1D1}
\includegraphics[height=2.2cm]{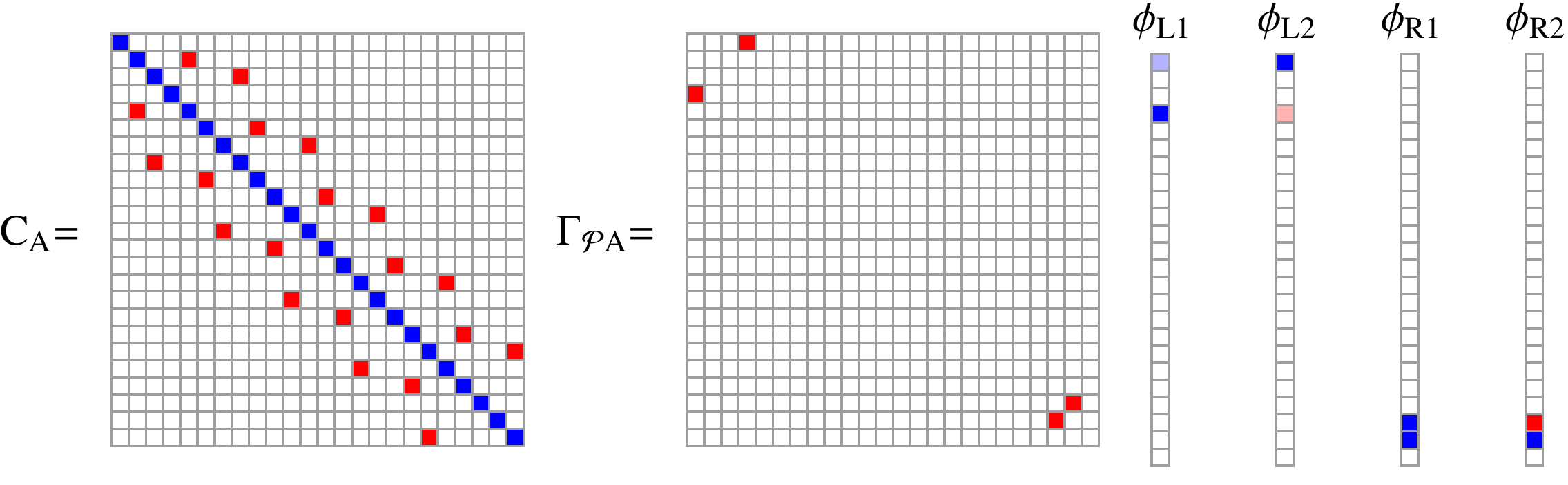}
}
\subfigure[] {
\label{1D2}
\includegraphics[height=2.2cm]{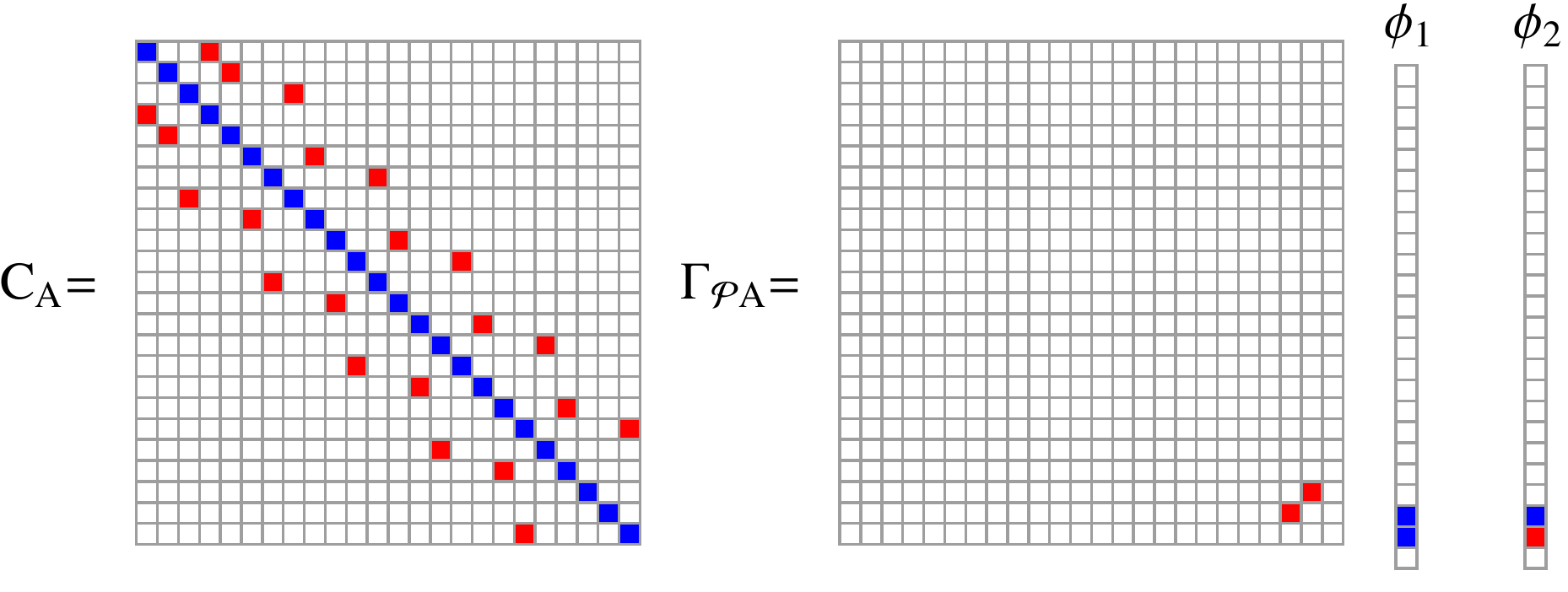}
}
\subfigure[] {
\label{1D3}
\includegraphics[height=2.2cm]{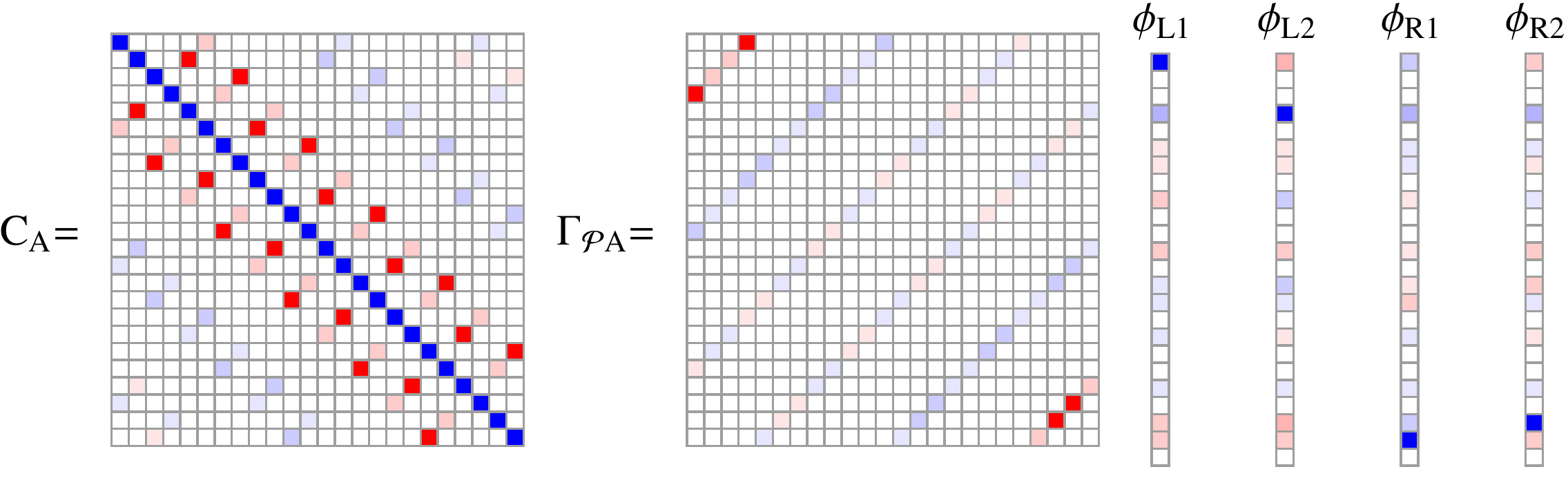}
}
\subfigure[] {
\label{1D4}
\includegraphics[height=2.2cm]{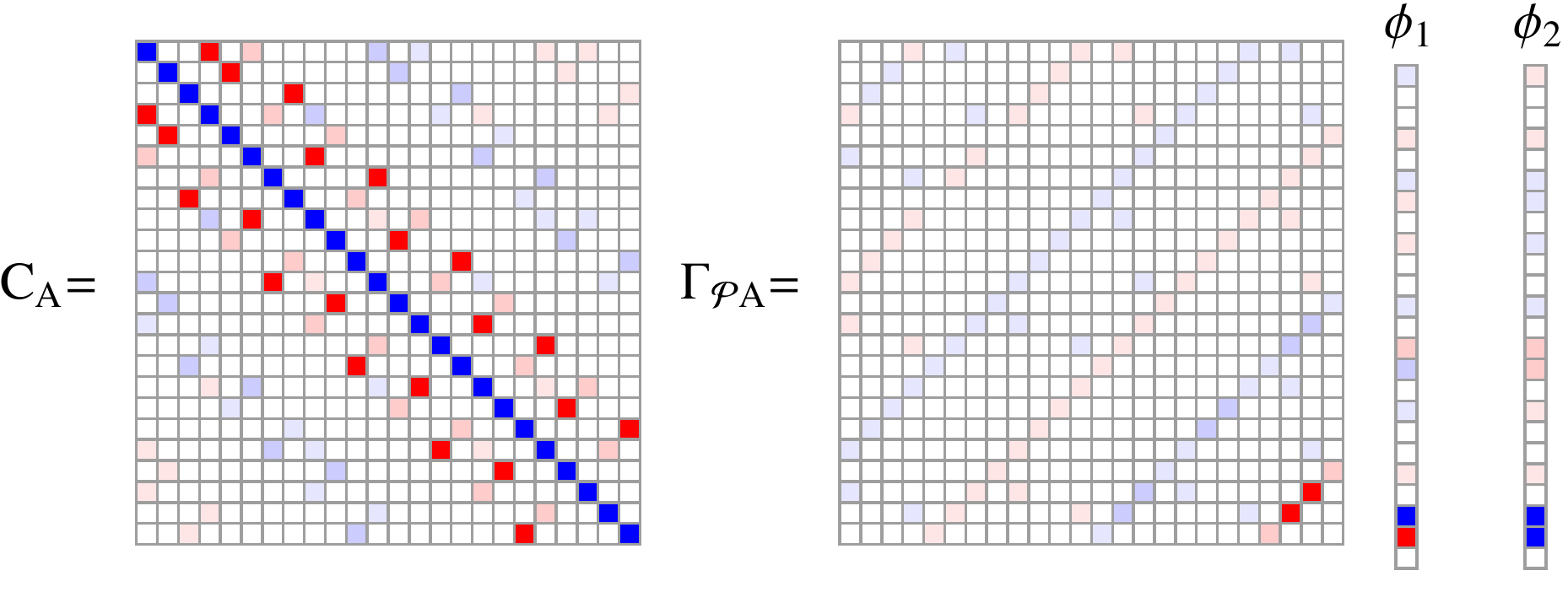}
}
\caption{
The one-dimensional lattice model is defined by Fig.%
~\ref{fig: dimerized chain}
with $N:=N^{\,}_{\mathrm{tot}}/N^{\,}_{\mathrm{orb}}=48/4=12$ repeat unit cells
with either a torus or a cylindrical geometry.
The partition is done by defining $A$ ($B$) to be the set of all the orbitals
localized to the left (right) of the dash-three-dots (green) line $RVE$ in
Fig.%
~\ref{fig: dimerized chain}(a).
The equal-time-correlation matrix $\mathcal{C}$ is defined
in Eq.~(\ref{eq: def matrix elements C}).
It is a $48\times48$ Hermitian matrix with a $24\times24$ Hermitian block defining
the matrix $C^{\,}_{A}$, see Eq.~(\ref{eq: partitioning single particles b}).
The symmetry operation is the parity transformation (reflection) $\mathscr{P}$
about the dash-three-dots (green) line $RVE$ in
Fig.%
~\ref{fig: dimerized chain}(a).
It interchanges $A$ and $B$ while leaving the entangling boundary
$RVE$ unchanged. Hence, $\mathscr{P}$ can be represented by
the $48\times48$ matrix $\mathcal{P}$ with the off-diagonal block
structure displayed in Eq.~(\ref{eq: cal O off block diagonal b}). 
There follows the existence of the $24\times24$ matrix
$\Gamma^{\,}_{\mathscr{P}\,A}$ defined in
Eq.~(\ref{eq: def Gamma mathsf O})
that anti-commutes with $Q^{\,}_{A}:=%\openone
\mathbb{I}-2\,C^{\,}_{A}$. 
The amplitudes of the matrix elements of $C^{\,}_{A}$ 
and $\Gamma^{\,}_{\mathscr{P}\,A}$
are represented by the coloring of the $24\times24$ elementary plaquettes
of a square lattice. The  blue (red) color of a plaquette 
determines the positive (negative) sign of the matrix element.
The lighter the color, the smaller the magnitude with white representing zero.
The darker the color, the larger the magnitude with $0.5$ the largest magnitude.
A $24\times1$ rectangular lattice represents as a column vector an eigenstate
of $Q^{\,}_{A}:=%\openone
\mathbb{I}-2\,C^{\,}_{A}$. 
(a) The case of Hamiltonian~(\ref{eq: def 1D example c})
obeying periodic boundary conditions for $\delta t=t$.
The four entangling zero modes are also plotted.
(b) The case of Hamiltonian~(\ref{eq: def 1D example c})
obeying periodic boundary conditions for $\delta t=t$ and perturbed by
$0.05\,t\,\mathcal{X}^{\,}_{11}$.
The four entangling zero modes are also plotted.
(c) The case of Hamiltonian~(\ref{eq: def 1D example c})
obeying open boundary conditions for $\delta t=t$ and perturbed by
$0.05\,t\,\mathcal{X}^{\,}_{11}$.
The two entangling zero modes are also plotted.
(d) The case of Hamiltonian~(\ref{eq: def 1D example c})
obeying periodic boundary conditions for $\delta t=9t/11$ and perturbed by
$0.05\,t\,\mathcal{X}^{\,}_{11}$.
The four entangling zero modes are also plotted.
(e) The case of Hamiltonian~(\ref{eq: def 1D example c})
obeying open boundary conditions for $\delta t=9t/11$ and perturbed by
$0.05\,t\,\mathcal{X}^{\,}_{11}$. 
The two entangling zero modes are also plotted.
        } 
\label{matrixplot}
\end{figure*}

\subsection{Numerical verification that $\Gamma^{\,}_{\mathscr{P}\,A}$ is local} 

For completeness, we verify numerically the prediction from
Sec.~\ref{subsec: Spectral gap and locality of the induced symmetry Gamma}
that $\Gamma^{\,}_{\mathscr{P}\,A}$ is local in that it 
does not mix zero modes localized
on boundaries separated by a bulk-like distance. 
To this end,
we consider Hamiltonian~(\ref{eq: def 1D example c})
with or without perturbations for $N^{\,}_{\mathrm{tot}}=48$ orbitals 
(twelve repeat unit cells with four orbitals per repeat unit cell, 
$N=12$ and $N^{\,}_{\mathrm{orb}}=4$).
Plotted in Fig.~\ref{matrixplot} as columnar vectors in the orbital basis
are the zero modes $\phi$ that are localized at the entangling boundaries.

When periodic boundary conditions are imposed, there are two entangling
boundaries separated by the bulk-like distance of order $N/2=6$. 
Correspondingly, there are two zero modes
$\phi^{\,}_{\mathrm{L}1}$ and $\phi^{\,}_{\mathrm{L}2}$ 
localized on the left entangling boundary and there are two zero modes
$\phi^{\,}_{\mathrm{R}1}$ and $\phi^{\,}_{\mathrm{R}2}$ localized
on the right entangling boundary. They are
represented in Figs.~\ref{matrixplot}(a),
\ref{matrixplot}(b), and \ref{matrixplot}(d)
as columnar vectors in the orbital basis.
One verifies that the overlap of any pair of zero modes
with one zero mode localized on the left
entangling boundary 
and the other zero mode localized on the right
entangling boundary
are exponentially suppressed in magnitude
by a factor of order $\exp(-b\,r\,\Delta)$
with $b$ a number or order unity and
the separation $r$ of order $N/2=6$.

When open boundary conditions are imposed, there are two physical
boundaries separated by the bulk-like distance of order 
$N/2=6$.
and one entangling boundary a distance of order 
$N/4=3$ away from either physical boundaries. The perturbation
$0.05\,t\,\mathcal{X}^{\,}_{11}$
has been added to the 
Hamiltonian~(\ref{eq: def 1D example c}).
According to Table%
~\ref{tab:1},
this perturbation gaps the zero modes
localized on the physical boundaries.
Correspondingly, there are two zero modes
$\phi^{\,}_{1}$ and $\phi^{\,}_{2}$ 
localized on the entangling boundary 
represented in Figs.~\ref{matrixplot}(c)
and \ref{matrixplot}(e)
by columnar vectors in the orbital basis.

\section{Topological insulator protected by
one reflection symmetry in two dimensions}
\label{sec: example B}

\subsection{Hamiltonian and topological quantum numbers}
\label{subsec: Hamiltonian example B}

Our second example is defined by choosing
$d=2$ and $N^{\,}_{\mathrm{orb}}=4$ in 
Eq.~(\ref{eq: def noninteractin fermion Hamiltonian}).
We represent the action on the orbital degrees of freedom
by the $4\times4$ matrices defined in 
Eq.~(\ref{eq: def X notation a}).
The Brillouin zone (BZ) is two-dimensional
and the single-particle Hamiltonian admits 
the direct-sum decomposition~\cite{Yao2012}
\begin{subequations}
\label{eq: direct sum decomposition example B}
\begin{align}
&
\mathcal{H}=
\bigoplus_{\bm{k}\in\mathrm{BZ}}
\mathcal{H}^{\,}_{\bm{k}},
\label{eq: diract sum decomposition example B a}
\\
&
\mathcal{H}^{\,}_{\bm{k}}:=
\left[
2\,t\,
\left(
\cos k^{\,}_{1}
+
\cos k^{\,}_{2}
\right)
-
\mu
\right]\,
\mathcal{X}^{\,}_{03}
%\nonumber\\
%&
%\hphantom{\mathcal{H}^{\,}_{\bm{k}}:=}
-
2\,\Delta\,
\sin k^{\,}_{1}\,
\mathcal{X}^{\,}_{31}
-
2\,\Delta\,
\sin k^{\,}_{2}\,
\mathcal{X}^{\,}_{02},
\label{eq: diract sum decomposition example B b}
\end{align}
\end{subequations}
with the real-valued characteristic energy scales 
$t$,
$\Delta$,
and
$\mu$.

Hamiltonian $\mathcal{H}$ can be interpreted as the direct sum 
\begin{subequations}
\begin{equation}
\mathcal{H}^{\,}_{\bm{k}}=
\mathcal{H}^{(+)}_{\bm{k}}
\oplus
\mathcal{H}^{(-)}_{\bm{k}}
\end{equation}
of the Bogoliubov-de-Gennes Hamiltonian
\begin{align}
\mathcal{H}^{(+)}_{\bm{k}}\sim%&\,
\left[
2\,t\,
\left(
\cos k^{\,}_{1}
+
\cos k^{\,}_{2}
\right)
-
\mu
\right]
\tau^{\,}_{3}
%\nonumber\\
%&\,
-
2\,\Delta\,
\left[
\sin k^{\,}_{1}\,
\tau^{\,}_{1}
+
\sin k^{\,}_{2}\,
\tau^{\,}_{2}
\right]
\end{align}
describing $p^{\,}_{1}+\mathrm{i}p^{\,}_{2}$ superconducting order
and the Bogoliubov-de-Gennes Hamiltonian
\begin{align}
\mathcal{H}^{(-)}_{\bm{k}}\sim%&\,
\left[
2\,t\,
\left(
\cos k^{\,}_{1}
+
\cos k^{\,}_{2}
\right)
-
\mu
\right]
\tau^{\,}_{3}
%\nonumber\\
%&\,
+
2\,\Delta\,
\left[
\sin k^{\,}_{1}\,
\tau^{\,}_{1}
-
\sin k^{\,}_{2}\,
\tau^{\,}_{2}
\right]
\end{align}
\end{subequations}
describing $p^{\,}_{1}-\mathrm{i}p^{\,}_{2}$ superconducting order.
A gap is present when $|\mu|<4\,t$, an assumption that is made throughout
in Sec.~\ref{sec: example B}.

The bundle of single-particle eigenstates obtained by collecting 
the lower band from the bundle 
$\{\mathcal{H}^{(+)}_{\bm{k}},\ \bm{k}\in\mathrm{BZ}\}$
has the opposite non-vanishing Chern number to that of
the bundle of single-particle eigenstates obtained by collecting 
the lower band from the bundle 
$\{\mathcal{H}^{(-)}_{\bm{k}},\ \bm{k}\in\mathrm{BZ}\}$.
The bundle of single-particle eigenstates obtained by collecting 
the lower band from the bundle 
$\{\mathcal{H}^{\,}_{\bm{k}},\ \bm{k}\in\mathrm{BZ}\}$
is topologically trivial.

\subsection{Symmetries}
\label{subsubsec: Symmetries example B}

The symmetries of the single-particle Hamiltonian defined by
Eq.~(\ref{eq: direct sum decomposition example B}) are the following.

[1] There are two symmetries of the inversion type.
If $\mathscr{I}$ denotes the inversion of space
\begin{subequations}
\begin{equation}
\mathscr{I}:\bm{r}\mapsto-\bm{r},
\end{equation}
then
\begin{equation}
\mathcal{O}^{\dag}_{\mathscr{I}}\,
\mathcal{H}^{\,}_{-\bm{k}}\,
\mathcal{O}^{\,}_{\mathscr{I}}=
\mathcal{H}^{\,}_{+\bm{k}}
\end{equation}
with
\begin{equation}
\mathcal{O}^{\,}_{\mathscr{I}}\in
\{
\mathcal{X}^{\,}_{03},
\mathcal{X}^{\,}_{33}
\}.
\end{equation}
\end{subequations}

[2] There are two symmetries of the reflection about
the horizontal axis $\bm{k}=(k^{\,}_{1},0)$ type.
If $\mathscr{R}^{\,}_{1}$ denotes the reflection
\begin{subequations}
\label{eq: horizontal reflection doublet symmetries}
\begin{equation}
\mathscr{R}^{\,}_{1}:r^{\,}_{1}\mapsto+r^{\,}_{1},
\qquad
\mathscr{R}^{\,}_{1}:r^{\,}_{2}\mapsto-r^{\,}_{2},
\end{equation}
then
\begin{equation}
\mathcal{O}^{\dag}_{\mathscr{R}^{\,}_{1}}\,
\mathcal{H}^{\,}_{+k^{\,}_{1},-k^{\,}_{2}}\,
\mathcal{O}^{\,}_{\mathscr{R}^{\,}_{1}}=
\mathcal{H}^{\,}_{+\bm{k}}
\label{eq: r1}
\end{equation}
with
\begin{equation}
\mathcal{O}^{\,}_{\mathscr{R}^{\,}_{1}}\in
\{
\mathcal{X}^{\,}_{13},
\mathcal{X}^{\,}_{23}
\}.
\end{equation}
\end{subequations}

[3] There are two symmetries of the reflection about
the vertical axis $\bm{k}=(0,k^{\,}_{2})$ type.
If $\mathscr{R}^{\,}_{2}$ denotes the reflection
\begin{subequations}
\label{eq: vertical reflection doublet symmetries}
\begin{equation}
\mathscr{R}^{\,}_{2}:r^{\,}_{1}\mapsto-r^{\,}_{1},
\qquad
\mathscr{R}^{\,}_{2}:r^{\,}_{2}\mapsto+r^{\,}_{2},
\end{equation}
then
\begin{equation}
\mathcal{O}^{\dag}_{\mathscr{R}^{\,}_{2}}\,
\mathcal{H}^{\,}_{-k^{\,}_{1},+k^{\,}_{2}}\,
\mathcal{O}^{\,}_{\mathscr{R}^{\,}_{2}}=
\mathcal{H}^{\,}_{+\bm{k}}
\end{equation}
with
\begin{equation}
\mathcal{O}^{\,}_{\mathscr{R}^{\,}_{2}}\in
\{
\mathcal{X}^{\,}_{10},
\mathcal{X}^{\,}_{20}
\}.
\end{equation}
\label{eq: r2}
\end{subequations}

[4] There are two symmetries of the time-reversal type.
If $\mathscr{T}$ denotes the reversal of time
\begin{subequations}
\begin{equation}
\mathscr{T}:t\mapsto-t,
\end{equation}
then
\begin{equation}
\mathcal{O}^{\dag}_{\mathscr{T}}\,
\mathcal{H}^{*}_{-\bm{k}}\,
\mathcal{O}^{\,}_{\mathscr{T}}=
\mathcal{H}^{\,}_{+\bm{k}}
\end{equation}
with
\begin{equation}
\mathcal{O}^{\,}_{\mathscr{T}}\in
\{
\mathcal{X}^{\,}_{10},
\mathcal{X}^{\,}_{20}
\}.
\end{equation}
\end{subequations}

[5] There are two spectral symmetries of the charge-conjugation type,
\begin{subequations}
\begin{equation}
\mathcal{O}^{\dag}_{\mathscr{C}}\,
\mathcal{H}^{\mathrm{T}}_{-\bm{k}}\,
\mathcal{O}^{\,}_{\mathscr{C}}=
-
\mathcal{H}^{\,}_{+\bm{k}}
\end{equation}
with
\begin{equation}
\mathcal{O}^{\,}_{\mathscr{C}}\in
\{
\mathcal{X}^{\,}_{01},
\mathcal{X}^{\,}_{31}
\}.
\end{equation}
\end{subequations}

[6] There are two spectral symmetries of the chiral type,
\begin{subequations}
\label{eq: chiral doublet symmetries}
\begin{equation}
\mathcal{O}^{\dag}_{\mathscr{S}}\,
\mathcal{H}^{\,}_{\bm{k}}\,
\mathcal{O}^{\,}_{\mathscr{S}}=
-
\mathcal{H}^{\,}_{\bm{k}}
\end{equation}
with
\begin{equation}
\mathcal{O}^{\,}_{\mathscr{C}}\in
\{
\mathcal{X}^{\,}_{11},
\mathcal{X}^{\,}_{21}
\}.
\end{equation}
\end{subequations}

\begin{figure*}[t]
\centering
\includegraphics[width=1.0 \textwidth]{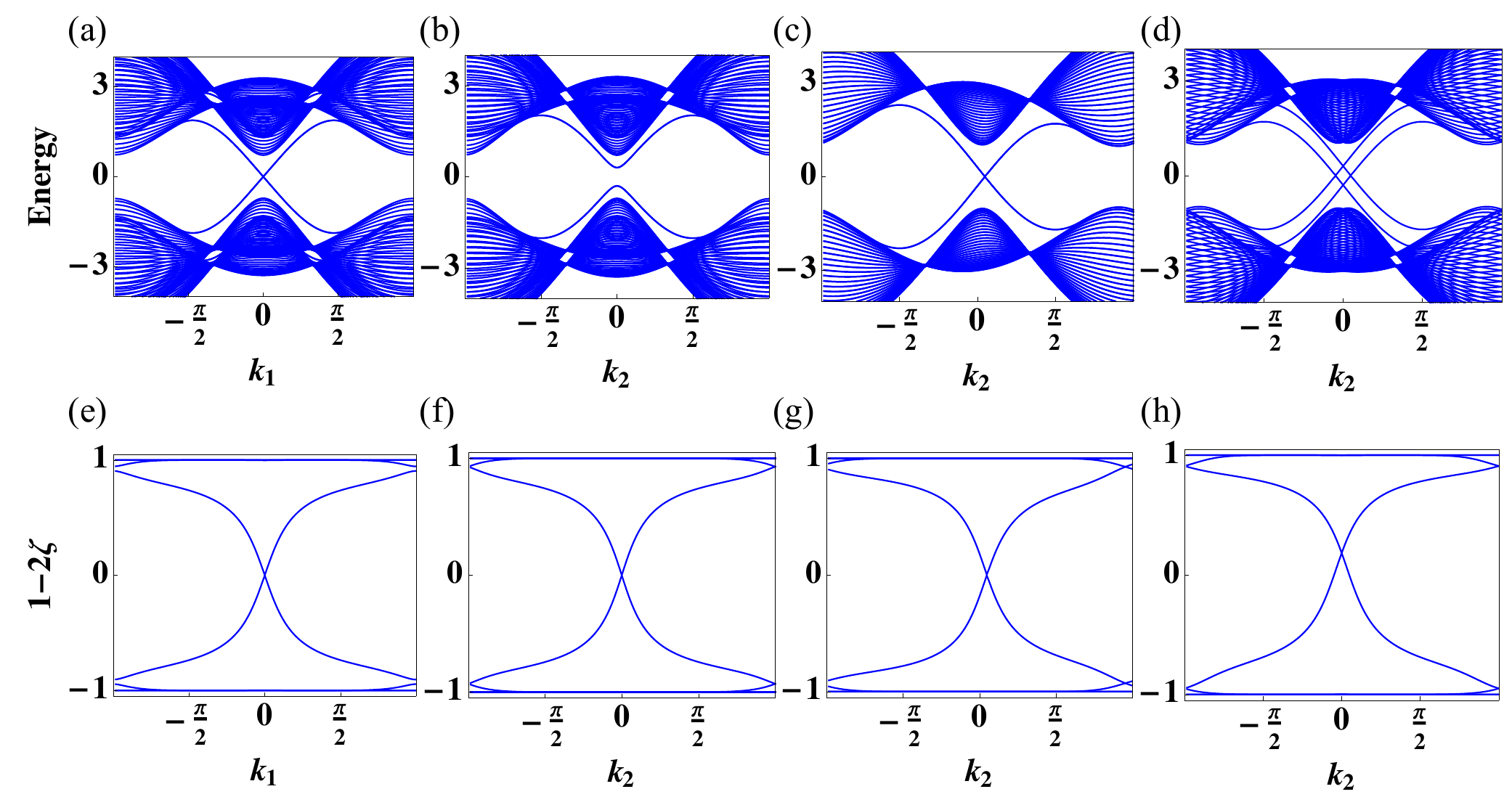}
\caption{
(Color online)
(a-d)
Single-particle energy spectra of 
$\widetilde{\mathcal{H}}^{\,}_{k^{\,}_{i}\,\mu\nu}$
defined in Eq.~(\ref{eq: def widetilde mathcal Hkimunu})
with $i=1,2$.
(e-f)
Single-particle spectra $1-2\,\zeta$ of
$\widetilde{\mathcal{Q}}^{\,}_{k^{\,}_{i}\,\mu\nu A^{\,}_{i+1}}$
with $i+1$ defined modulo 2.
%(i-l) 
%Spectra $\ln(1-\zeta)/\zeta$ extracted from
%$\widetilde{\mathcal{Q}}^{\,}_{k^{\,}_{i}\,\mu\nu A^{\,}_{i+1}}$
%with $i+1$ defined modulo 2.
The energy scales in
$\widetilde{\mathcal{H}}^{\,}_{k^{\,}_{i}\,\mu\nu}$
are chosen to be $t=\Delta=-\mu=1$, see
Eq.~(\ref{eq: diract sum decomposition example B b}).
The choice of $\mu=0,1,2,3$ and $\nu=0,1,2,3$ determines
the perturbation to Eq.~(\ref{eq: diract sum decomposition example B b}).
It is $0.3\,t\,\mathcal{X}^{\,}_{13}$ for panels (a, b, e, f).
It is $0.3\,t\,\mathcal{X}^{\,}_{02}$
for panels (c, g).
It is $0.3\,t\,\mathcal{X}_{32}$
for panels (d, h).
          }
\label{fig: EEES_{2}D}
\end{figure*}

\subsection{Partition and zero modes}
\label{subsubsec: Partition, and zero modes example B}

From now on, we adopt a cylindrical geometry instead
of the torus geometry from Sec.~\ref{subsec: Hamiltonian example B},
i.e., we compactify two-dimensional space 
$\{\bm{x}:=(x^{\,}_{1},x^{\,}_{2})\in\mathbb{R}^{2}\}$
only along one Cartesian coordinate. Consequently, only one component of the
two-dimensional momentum 
$\{\bm{k}:=(k^{\,}_{1},k^{\,}_{2})\in\mathbb{R}^{2}\}$
is chosen to be a good quantum number
in Eq.~(\ref{eq: direct sum decomposition example B}).

For any choice of the direction $i=1,2$ along which periodic boundary 
conditions are imposed while open boundary conditions are imposed along
the orthogonal direction $i+1$ modulo 2,
for any positive integer $M^{\,}_{i}$,
and for any good momentum quantum number
\begin{subequations}
\label{eq: def mathcal{H} ki;r(i+1),r'(i+1)}
\begin{equation}
k^{\,}_{i}=
\frac{2\pi}{2M^{\,}_{i}\,\mathfrak{a}}\,
n^{\,}_{i},
\qquad
n^{\,}_{i}=1,\cdots,2M^{\,}_{i},
\qquad
i=1,2,
\end{equation}
we do the following.
First, we denote by
$\mathcal{H}^{\,}_{k^{\,}_{i}}$
the $8M^{\,}_{i+1}\times8M^{\,}_{i+1}$ Hermitian
matrix such that the one-dimensional Fourier transform of 
\begin{equation}
\mathcal{H}^{\,}_{k^{\,}_{i}\,r^{\,}_{i+1},r^{\prime}_{i+1}}:=
\langle r^{\,}_{i+1}|
\mathcal{H}^{\,}_{k^{\,}_{i}}
|r^{\prime}_{i+1}\rangle,
\label{eq: def mathcal{H} ki;r(i+1),r'(i+1) a}
\end{equation}
delivers Eq.~(\ref{eq: diract sum decomposition example B b}).
Here,
\begin{equation}
r^{\,}_{i+1}=
n^{\,}_{i+1}\,\mathfrak{a},
\qquad 
n^{\,}_{i+1}=1,\cdots,2M^{\,}_{i+1},
\label{eq: def mathcal{H} ki;r(i+1),r'(i+1) b}
\end{equation}
and
\begin{equation}
r^{\prime}_{i+1}=n^{\prime}_{i+1}\,\mathfrak{a},
\qquad 
n^{\prime}_{i+1}=1,\cdots,2M^{\,}_{i+1},
\label{eq: def mathcal{H} ki;r(i+1),r'(i+1) c}
\end{equation}
\end{subequations}
are the lattice sites from an open chain 
along the direction $i+1$ (defined modulo 2) of a rectangular lattice
with the lattice spacing $\mathfrak{a}$ and the number of lattice sites 
$M^{\,}_{i+1}$ along the direction $i+1$ modulo 2.
Second, we define the $k^{\,}_{i}$-dependent partition
\begin{subequations}
\label{eq: def i-dependent partition example B}
\begin{align}
%&
\mathfrak{H}:=\bigoplus_{k^{\,}_{i}}
\mathfrak{H}^{\,}_{k^{\,}_{i}},
%\\
%&
\mathfrak{H}^{\,}_{k^{\,}_{i}}:=
\mathfrak{H}^{\,}_{A^{\,}_{i+1}}
\oplus
\mathfrak{H}^{\,}_{B^{\,}_{i+1}},
\end{align}
where
\begin{equation}
\mathfrak{H}^{\,}_{A^{\,}_{i+1}}:=
\bigoplus_{n^{\,}_{i+1}=1}^{M^{\,}_{i+1}}
\bigoplus_{\alpha=1}^{4}
|k^{\,}_{i},n^{\,}_{i+1},\alpha\rangle
\langle k^{\,}_{i},n^{\,}_{i+1},\alpha|,
\end{equation}
\begin{equation}
\mathfrak{H}^{\,}_{B^{\,}_{i+1}}:=
\bigoplus_{n^{\,}_{i+1}=M^{\,}_{i+1}+1}^{2M^{\,}_{i+1}}
\bigoplus_{\alpha=1}^{4}
|k^{\,}_{i},n^{\,}_{i+1},\alpha\rangle
\langle k^{\,}_{i},n^{\,}_{i+1},\alpha|,
\end{equation}
\end{subequations}
and the ket $|k^{\,}_{i},n^{\,}_{i+1},\alpha\rangle$
denotes the single-particle state with the Bloch index
$k^{\,}_{i}$, for it is extended along the $i=1,2$ direction,
the unit repeat cell index $n^{\,}_{i+1}$, for it is localized
at the site $n^{\,}_{i+1}\,\mathfrak{a}$ along the $i+1$ modulo 2 direction,
and the orbital index $\alpha=1,2,3,4$.
If we denote by $\mathscr{R}^{\,}_{i}$ the reflection that
leaves $k^{\,}_{i}$ unchanged but reverses the sign of
$k^{\,}_{i+1}$, i.e.,
\begin{equation}
\mathscr{R}^{\,}_{i}\,k^{\,}_{i}=+k^{\,}_{i},
\qquad
\mathscr{R}^{\,}_{i}\,k^{\,}_{i+1}=-k^{\,}_{i+1},
\end{equation}
we then have that
\begin{subequations}
\begin{align}
&
\mathscr{R}^{\,}_{i}A^{\,}_{i+1}=B^{\,}_{i+1},
&
\mathscr{R}^{\,}_{i}B^{\,}_{i+1}=A^{\,}_{i+1},
\\
&
\mathscr{C}A^{\,}_{i+1}=A^{\,}_{i+1},
&
\mathscr{C}B^{\,}_{i+1}=B^{\,}_{i+1},
\\
&
\mathscr{T}A^{\,}_{i+1}=A^{\,}_{i+1},
&
\mathscr{T}B^{\,}_{i+1}=B^{\,}_{i+1},
\\
&
\mathscr{S}A^{\,}_{i+1}=A^{\,}_{i+1},
&
\mathscr{S}B^{\,}_{i+1}=B^{\,}_{i+1}.
\end{align}
\end{subequations}

The thermodynamic limit is defined by
\begin{subequations}
\label{eq: def TL B}
\begin{equation}
N,N^{\,}_{\mathrm{f}}\to\infty,
\label{eq: def TL a B}
\end{equation}
with $N=(2M^{\,}_{1})\times(2M^{\,}_{2})$, holding the fermion density 
\begin{equation}
N^{\,}_{\mathrm{f}}/N=1
\label{eq: def TL b B}
\end{equation}
\end{subequations}
fixed. In the thermodynamic limit,
the single-particle energy eigenvalue spectrum
$\sigma(\mathcal{H}^{\,}_{i})$ of
\begin{subequations}
\label{eq: def mathcal{H}{i} and mathcal{Q}{Ai+1}}
\begin{equation}
\mathcal{H}^{\,}_{i}:=
\bigoplus_{k^{\,}_{i}}
\mathcal{H}^{\,}_{k^{\,}_{i}}
\label{eq: def mathcal{H}{i} and mathcal{Q}{Ai+1} a}
\end{equation}
supports two pairs of zero modes,
each of which are localized at the opposite (physical)
boundaries on the cylinder whose symmetry axis coincides 
with the direction $i+1$ (defined modulo 2) 
along which open boundary conditions have been imposed.
Similarly, the entanglement spectrum $\sigma(Q^{\,}_{A^{\,}_{i+1}})$
of the equal-time one-point correlation matrix
\begin{equation}
Q^{\,}_{A^{\,}_{i+1}}:=
%\openone
\mathbb{I}-
2\,
C^{\,}_{A^{\,}_{i+1}}
\label{eq: def mathcal{H}{i} and mathcal{Q}{Ai+1} b}
\end{equation}
\end{subequations}
supports a pair of zero modes localized at the entangling boundary.

In order not to confuse gapless boundary modes originating from the
physical boundaries with gapless boundary modes originating from the
entangling boundary, whenever the physical boundaries support gapless
boundary states in the energy spectrum,
we opt for a torus geometry when computing the entanglement spectrum.

Desired is a study of the stability of these zero modes under all generic
perturbations of $\mathcal{H}^{\,}_{i}$ that respect any one 
($\mathscr{S}$) 
of the two chiral symmetry from Eq.%
~(\ref{eq: chiral doublet symmetries})
and any one ($\mathscr{R}$)
of the four reflection symmetries from Eqs.%
~(\ref{eq: horizontal reflection doublet symmetries}) 
and (\ref{eq: vertical reflection doublet symmetries}).

\begin{table*}
\begin{center}
\resizebox{\columnwidth}{!}{%
\begin{tabular}{|c||cc|cc|cc|cc|cc|cc|cc|}
\hline 
\multicolumn{1}{|c||}{\multirow{2}{*}
{$\mathcal{X}^{\,}_{\mu\nu}\equiv\sigma^{\,}_{\mu}\otimes\tau^{\,}_{\nu}$}}
&
\multicolumn{2}{c|}{$\mathscr{R}_{1}$}
&
\multicolumn{2}{c|}{$\mathscr{R}_{2}$}
&
\multicolumn{2}{c|}{$\mathscr{T}$}
&
\multicolumn{2}{c|}{$\mathscr{C}$}
&
\multicolumn{2}{c|}{$\mathscr{S}$}
&
\multirow{2}{*}{$\sigma(\widetilde{\mathcal{H}}_{k^{\,}_{1}\,\mu\nu})$}
&
\multirow{2}{*}{$\sigma(\widetilde{Q}^{\,}_{k^{\,}_{1}\,\mu\nu\,A^{\,}_{2}})$}
&
\multirow{2}{*}{$\sigma(\widetilde{\mathcal{H}}_{k^{\,}_{2}\,\mu\nu})$}
&
\multirow{2}{*}{$\sigma(\widetilde{Q}^{\,}_{k^{\,}_{2}\,\mu\nu\,A^{\,}_{1}})$}
\\
&
$\mathcal{R}^{\,}_{13}$
&
$\mathcal{R}^{\,}_{23}$
&
$\mathcal{R}^{\,}_{10}$
&
$\mathcal{R}^{\,}_{20}$
&
$\mathcal{T}^{\,}_{10}$
&
$\mathcal{T}^{\,}_{20}$
&
$\mathcal{C}^{\,}_{01}$
&
$\mathcal{C}^{\,}_{31}$
&
$\mathcal{S}^{\,}_{11}$
&
$\mathcal{S}^{\,}_{21}$
&
{}
&
{}
&
{}
&
{}
\\
\hline
\hline
$\mathcal{X}^{\,}_{00}$
&
$\circ$
&
$\circ$
&
$\circ$
&
$\circ$  % END GROUP P FOR 00
&
$\circ$
&
$\circ$  % END GROUP T FOR 00
&
$\times$
&
$\times$  % END GROUP C FOR 00
&
$\times$
&
$\times$  % END GROUP S FOR 00
&
$\otimes$
&
$\circ$ % END SPECTRA FOR 00
&
$\otimes$
&
$\circ$ % END EESPECTRA FOR 00
\\
\hline
$\mathcal{X}^{\,}_{01}$
&
$\times$
&
$\times$
&
$\circ$
&
$\circ$  % END GROUP P FOR 01
&
$\circ$
&
$\circ$  % END GROUP T FOR 01
&
$\times$
&
$\times$  % END GROUP C FOR 01
&
$\times$
&
$\times$  % END GROUP S FOR 01
&
$\otimes$
&
$\otimes$ % END SPECTRA FOR 01
&
$\circ$
&
$\circ$ % END EESPECTRA FOR 01
\\
\hline
$\mathcal{X}^{\,}_{02}$
&
$\times$
&
$\times$
&
$\circ$
&
$\circ$  % END GROUP P FOR 02
&
$\times$
&
$\times$  % END GROUP T FOR 02
&
$\times$
&
$\times$  % END GROUP C FOR 02
&
$\circ$
&
$\circ$  % END GROUP S FOR 02
&
$\circ$
&
$\circ$ % END SPECTRA FOR 02
&
$\otimes$
&
$\otimes$ % END EESPECTRA FOR 02
\\
\hline
$\mathcal{X}^{\,}_{03}$
&
$\circ$
&
$\circ$
&
$\circ$
&
$\circ$  % END GROUP P FOR 03
&
$\circ$
&
$\circ$  % END GROUP T FOR 03
&
$\circ$
&
$\circ$  % END GROUP C FOR 03
&
$\circ$
&
$\circ$  % END GROUP S FOR 03
&
$\circ$
&
$\circ$ % END SPECTRA FOR 03
&
$\circ$
&
$\circ$ % END EESPECTRA FOR 03
\\
\hline
$\mathcal{X}^{\,}_{10}$
&
$\circ$
&
$\times$
&
$\circ$
&
$\times$  % END GROUP P FOR 10
&
$\circ$
&
$\times$  % END GROUP T FOR 10
&
$\times$
&
$\circ$  % END GROUP C FOR 10
&
$\times$
&
$\circ$  % END GROUP S FOR 10
&
$\times$
&
$\circ$ % END SPECTRA FOR 10
&
$\circ$
&
$\circ$ % END EESPECTRA FOR 10
\\
\hline
$\mathcal{X}^{\,}_{11}$
&
$\times$
&
$\circ$
&
$\circ$
&
$\times$  % END GROUP P FOR 11
&
$\circ$
&
$\times$  % END GROUP T FOR 11
&
$\times$
&
$\circ$  % END GROUP C FOR 11
&
$\times$
&
$\circ$  % END GROUP S FOR 11
&
$\times$
&
$\times$ % END SPECTRA FOR 11
&
$\times$
&
$\times$ % END EESPECTRA FOR 11

\\
\hline
$\mathcal{X}^{\,}_{12}$
&
$\times$
&
$\circ$
&
$\circ$
&
$\times$  % END GROUP P FOR 12
&
$\times$
&
$\circ$  % END GROUP T FOR 12
&
$\times$
&
$\circ$  % END GROUP C FOR 12
&
$\circ$
&
$\times$  % END GROUP S FOR 12
&
$\circ$
&
$\circ$ % END SPECTRA FOR 12
&
$\circ$
&
$\circ$ % END EESPECTRA FOR 12
\\
\hline
$\mathcal{X}^{\,}_{13}$
&
$\circ$
&
$\times$
&
$\circ$
&
$\times$  % END GROUP P FOR 13
&
$\circ$
&
$\times$  % END GROUP T FOR 13
&
$\circ$
&
$\times$  % END GROUP C FOR 13
&
$\circ$
&
$\times$  % END GROUP S FOR 13
&
$\circ$
&
$\circ$ % END SPECTRA FOR 13
&
$\times$
&
$\circ$ % END SPECTRA FOR 13
\\%%%%%%%%(new rows added)
\hline
$\mathcal{X}^{\,}_{20}$
&
$\times$
&
$\circ$
&
$\times$
&
$\circ$  % END GROUP P FOR 20
&
$\circ$
&
$\times$  % END GROUP T FOR 20
&
$\circ$
&
$\times$  % END GROUP C FOR 20
&
$\circ$
&
$\times$  % END GROUP S FOR 20
&
$\times$
&
$\circ$ % END SPECTRA FOR 20
&
$\circ$
&
$\circ$ % END EESPECTRA FOR 20
\\
\hline
$\mathcal{X}^{\,}_{21}$
&
$\circ$
&
$\times$
&
$\times$
&
$\circ$  % END GROUP P FOR 21
&
$\circ$
&
$\times$  % END GROUP T FOR 21
&
$\circ$
&
$\times$  % END GROUP C FOR 21
&
$\circ$
&
$\times$  % END GROUP S FOR 21
&
$\times$
&
$\times$ % END SPECTRA FOR 21
&
$\times$
&
$\times$ % END EESPECTRA FOR 21
\\
\hline
$\mathcal{X}^{\,}_{22}$
&
$\circ$
&
$\times$
&
$\times$
&
$\circ$  % END GROUP P FOR 22
&
$\times$
&
$\circ$  % END GROUP T FOR 22
&
$\circ$
&
$\times$  % END GROUP C FOR 22
&
$\times$
&
$\circ$  % END GROUP S FOR 22
&
$\circ$
&
$\circ$ % END SPECTRA FOR 22
&
$\circ$
&
$\circ$ % END SPECTRA FOR 22
\\
\hline
$\mathcal{X}^{\,}_{23}$
&
$\times$
&
$\circ$
&
$\times$
&
$\circ$  % END GROUP P FOR 23
&
$\circ$
&
$\times$  % END GROUP T FOR 23
&
$\times$
&
$\circ$  % END GROUP C FOR 23
&
$\times$
&
$\circ$  % END GROUP S FOR 23
&
$\circ$
&
$\circ$ % END SPECTRA FOR 23
&
$\times$
&
$\circ $ % END EESPECTRA FOR 23
\\
\hline
$\mathcal{X}^{\,}_{30}$
&
$\times$
&
$\times$
&
$\times$
&
$\times$  % END GROUP P FOR 30
&
$\times$
&
$\times$  % END GROUP T FOR 30
&
$\times$
&
$\times$  % END GROUP C FOR 30
&
$\circ$
&
$\circ$  % END GROUP S FOR 30
&
$\otimes$
&
$\circ $ % END SPECTRA FOR 30
&
$\otimes$
&
$\circ $ % END SPECTRA FOR 30
\\
\hline
$\mathcal{X}^{\,}_{31}$
&
$\circ$
&
$\circ$
&
$\times$
&
$\times$  % END GROUP P FOR 31
&
$\times$
&
$\times$  % END GROUP T FOR 31
&
$\times$
&
$\times$  % END GROUP C FOR 31
&
$\circ$
&
$\circ$  % END GROUP S FOR 31
&
$\otimes$
&
$\otimes$ % END SPECTRA FOR 31
&
$\circ$
&
$\circ $ % END EESPECTRA FOR 31
\\
\hline
$\mathcal{X}^{\,}_{32}$
&
$\circ$
&
$\circ$
&
$\times$
&
$\times$  % END GROUP P FOR 32
&
$\circ$
&
$\circ$  % END GROUP T FOR 32
&
$\times$
&
$\times$  % END GROUP C FOR 32
&
$\times$
&
$\times$  % END GROUP S FOR 32
&
$\circ$
&
$\circ$ % END SPECTRA FOR 32
&
$\otimes$
&
$\otimes$ % END EESPECTRA FOR 32
\\
\hline
$\mathcal{X}^{\,}_{33}$
&
$\times$
&
$\times$
&
$\times$
&
$\times$  % END GROUP P FOR 33
&
$\times$
&
$\times$  % END GROUP T FOR 33
&
$\circ$
&
$\circ$  % END GROUP C FOR 33
&
$\times$
&
$\times$  % END GROUP S FOR 33
&
$\circ$
&
$\circ$ % END SPECTRA FOR 33
&
$\circ$
&
$\circ$ % END EESPECTRA FOR 33
\\ 
\hline
\end{tabular}
}
\end{center}
\caption{%
The spectrum $\sigma(\widetilde{\mathcal{H}}^{\,}_{k^{\,}_{i}\,\mu\nu})$
of the single-particle Hamiltonian
$\widetilde{\mathcal{H}}^{\,}_{k^{\,}_{i}\,\mu\nu}$
defined by
Eq.~(\ref{eq: def widetilde mathcal Hkimunu}) 
and obeying periodic boundary conditions along the $i=1,2$ direction
and open boundary conditions along the $i+1$ (modulo 2) direction.
The entanglement spectrum
$\sigma(\widetilde{Q}^{\,}_{k^{\,}_{i}\,\mu\nu A^{\,}_{i+1}})$
defined by Eq.~(\ref{eq: def QA}) 
for the single-particle Hamiltonian
$\widetilde{\mathcal{H}}^{\,}_{k^{\,}_{i}\,\mu\nu}$
obeying periodic boundary conditions along 
both the $i=1,2$ and the $i+1$ (modulo 2) direction.
The entry
$\circ$ 
or
$\times$ 
in the second to sixth columns
denotes the presence or the absence, respectively,
of the symmetries under reflections about the directions 1
($\mathscr{R}^{\,}_{1}$)
and 2
($\mathscr{R}^{\,}_{2}$),
charge conjugation $\mathscr{C}$,
time reversal $\mathscr{T}$,
and chiral $\mathscr{S}$ 
of the perturbation
$\delta^{\,}_{r^{\,}_{i+1},r^{\prime}_{i+1}}\,0.3t\,\mathcal{X}^{\,}_{\mu\nu}$
for the sixteen rows. In the last two columns, the entries
$\circ$ and $\times$
denote the presence and the absence, respectively,
of non-propagating zero modes in the spectra 
$\sigma(\widetilde{\mathcal{H}}^{\,}_{k^{\,}_{i}\,\mu\nu})$
and
$\sigma(\widetilde{\mathcal{Q}}^{\,}_{k^{\,}_{i}\,\mu\nu A^{\,}_{i+1}})$
as determined by extrapolation to the
thermodynamic limit of exact diagonalization with
the open direction running over $32$ repeat unit cells
and the momentum along the compactified direction running over 
128 values. The entry $\otimes$ in the last two columns denotes
the existence of crossings between the mid-gap branches,
whereby the crossings are away from vanishing energy 
(entanglement eigenvalue)
and vanishing momentum in the spectra 
$\sigma(\widetilde{\mathcal{H}}^{\,}_{k^{\,}_{i}\,\mu\nu})$
and
$\sigma(\widetilde{\mathcal{Q}}^{\,}_{k^{\,}_{i}\,\mu\nu A^{\,}_{i+1}})$.
         }
\label{tab:3}
\end{table*}

\subsection{Stability analysis of the zero modes}
\label{subsubsec: Stability analysis of the zero modes example B}

\subsubsection{Definitions of $\widetilde{\mathcal{H}}^{\,}_{i\,\mu\nu}$
and $\widetilde{\mathcal{Q}}^{\,}_{k^{\,}_{i}\,\mu\nu A^{\,}_{i+1}}$}

To begin with the stability analysis, we choose $i=1,2$ and
perturb Hamiltonian~(\ref{eq: def mathcal{H} ki;r(i+1),r'(i+1) a})
by adding locally any one of the sixteen matrices
$\mathcal{X}^{\,}_{\mu\nu}\equiv\sigma^{\,}_{\mu}\otimes\tau^{\,}_{\nu}$
parametrized by $\mu,\nu=0,1,2,3$. Thus, we define
\begin{subequations}
\label{eq: def widetilde mathcal Hkimunu}
\begin{equation}
\widetilde{\mathcal{H}}^{\,}_{i\,\mu\nu}:=
\bigoplus_{k^{\,}_{i}}\widetilde{\mathcal{H}}^{\,}_{k^{\,}_{i}\,\mu\nu}\equiv
\bigoplus_{k^{\,}_{i}}
\left(\mathcal{H}^{\,}_{k^{\,}_{i}}+\mathcal{V}^{\,}_{\mu\nu}\right)
\end{equation}
by the matrix elements
\begin{align}
&
\widetilde{\mathcal{H}}^{\,}_{k^{\,}_{i}\,\mu\nu\,r^{\,}_{i+1},r^{\prime}_{i+1}}:=
\mathcal{H}^{\,}_{k^{\,}_{i}\,r^{\,}_{i+1},r^{\prime}_{i+1}}
+
\mathcal{V}^{\,}_{\mu\nu\,r^{\,}_{i+1},r^{\prime}_{i+1}},
\\
&
\mathcal{V}^{\,}_{\mu\nu\,r^{\,}_{i+1},r^{\prime}_{i+1}}:=
\delta^{\,}_{r^{\,}_{i+1},r^{\prime}_{i+1}}\,
v^{\,}_{\mu\nu}\,
\mathcal{X}^{\,}_{\mu\nu},
\end{align}
\end{subequations}
where the matrix elements of
$\mathcal{H}^{\,}_{k^{\,}_{i}}$ were defined in
Eq.~(\ref{eq: def mathcal{H} ki;r(i+1),r'(i+1)}) and
the perturbation strength $v^{\,}_{\mu\nu}$ is real valued.
Equipped with
$\widetilde{\mathcal{H}}^{\,}_{k^{\,}_{i}\,\mu\nu}$,
we define
$\widetilde{\mathcal{Q}}^{\,}_{k^{\,}_{i}\,\mu\nu A^{\,}_{i+1}}$
from Eq.~(\ref{eq: def QA})
with the caveat that periodic boundary conditions are imposed
along the direction $i+1$ (modulo 2).

\subsubsection{Spectra $\sigma(\widetilde{\mathcal{H}}^{\,}_{i\,\mu\nu})$}

For any $i=1,2$, the spectra
$\sigma(\widetilde{\mathcal{H}}^{\,}_{i\,\mu\nu})$
for $\mu,\nu=0,1,2,3$
were obtained from exact diagonalization
with the unperturbed energy scales $t=\Delta=-\mu=1$.
These spectra are characterized by two continua 
of single-particle excitations separated by an energy gap
as is illustrated in Figs.~\ref{fig: EEES_{2}D}(a-d).
A discrete number (four) of branches are seen to
peel off from these continua, 
some of which eventually cross the band gap. 
These mid-gap branches disperse along the physical boundaries
(edges) while they decay exponentially fast away from the edges.
They are thus called edge states.
Their crossings, if any, define degenerate edge states
with group velocities of opposite signs along the edges.
Their crossings at vanishing energy \textit{and} momentum, 
if any, define degenerate non-propagating zero modes
with group velocities of opposite signs along the edges.
Their crossings at vanishing energy \textit{and} non-vanishing momenta, 
if any, define degenerate propagating zero modes 
with group velocities of opposite signs along the edges.
Their crossings at non-vanishing energies \textit{and} vanishing momentum, 
if any, define degenerate non-propagating edge states
with group velocities of opposite signs along the edges.
The entries $\circ$ in the last two columns of
Table~\ref{tab:3} accounts for the existence of 
at least one crossing at vanishing energy and momentum.
The entries $\otimes$ in the last two columns of
Table~\ref{tab:3} account for the existence of 
crossings not necessarily at vanishing energy and momentum. 
The entries $\times$ in the last two column of
Table~\ref{tab:3} accounts for the absence of crossings.

The origin of the mid-gap branches 
in Figs.~\ref{fig: EEES_{2}D}(a-f)
is the following.
The single-particle Hamiltonian defined in 
Eq.~(\ref{eq: diract sum decomposition example B b})
for a torus geometry supports four boundary states 
for a cylindrical geometry
that are extended along the boundary but 
exponentially localized away from the boundary
due to the non-vanishing Chern numbers for 
each of $\mathcal{H}^{(+)}_{\bm{k}}$ and $\mathcal{H}^{(-)}_{\bm{k}}$.
These dispersing boundary states show up in the single-particle
energy spectrum as four branches of mid-gap states
that crosses at vanishing energy and momentum, thereby defining
four degenerate non-propagating zero modes.
In the presence of the perturbation $0.3\,t\,\mathcal{X}^{\,}_{13}$
when the boundary dictates that $k^{\,}_{1}$ is a good quantum number,
the energy spectrum shows a four-fold degenerate
crossing at vanishing energy and momentum in
Fig.~\ref{fig: EEES_{2}D}(a).
However, the very same perturbation,
when the boundary dictates that $k^{\,}_{2}$ is a good quantum number,
gaps the unperturbed crossing at vanishing energy and momentum
in Fig.~\ref{fig: EEES_{2}D}(b).
In the presence of the perturbation $0.3\,t\,\mathcal{X}^{\,}_{02}$ 
when the boundary dictates that $k^{\,}_{2}$ is a good quantum number,
the energy spectrum shows a four-fold-degenerate
crossing at vanishing energy but away from
$k^{\,}_{2}=0$ in Fig.~\ref{fig: EEES_{2}D}(c)].
In the presence of the perturbation $0.3\,t\,\mathcal{X}^{\,}_{32}$ 
when the boundary dictates that $k^{\,}_{2}$ is a good quantum number,
the energy spectrum shows four non-degenerate
crossings away from vanishing energy and 
vanishing momentum in Fig.~\ref{fig: EEES_{2}D}(d).
Even though all the crossings in the unperturbed 
entanglement spectra are robust to the perturbations
in Fig.~\ref{fig: EEES_{2}D}(e-h), they can be shifted away 
from vanishing energy as in Fig.~\ref{fig: EEES_{2}D}(h) 
or momentum as in Fig.~\ref{fig: EEES_{2}D}(g).

\begin{figure}[t] 
\centering     
\includegraphics[width=0.3\textwidth]{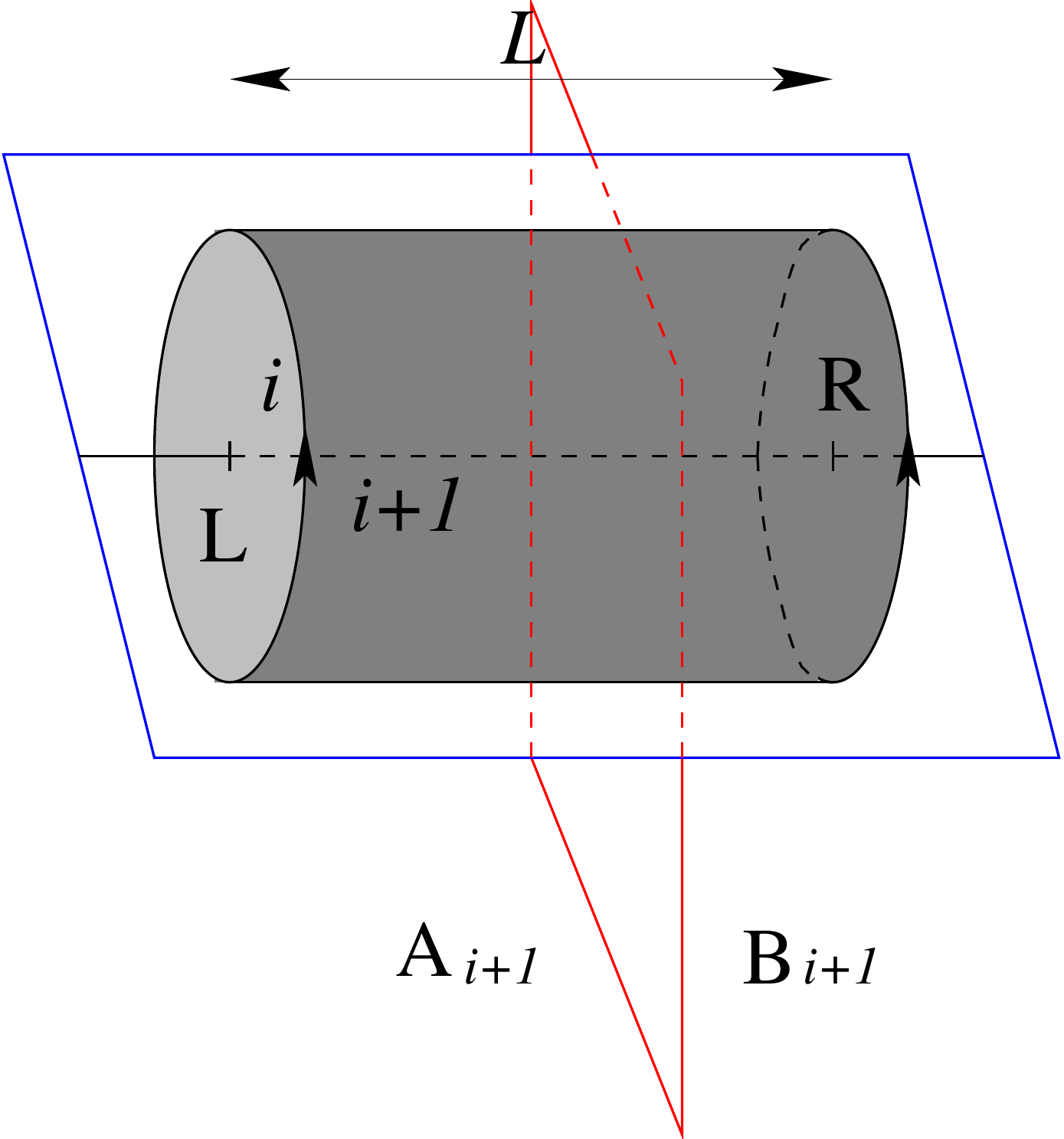}
\caption{
(Color online)
A cylinder of length $L$ along the $i+1$ modulo 2 direction,
while the coordinate $i=1,2$ has been compactified.
We define two reflection planes. The first is
defined by the blue frame that includes the cylinder axis. The second 
is defined by the red frame normal to the cylinder axis and intersecting
the cylinder axis at its mid-point. The circles centered at the point
L and R on the cylinder axis are the two disconnected boundaries of the
cylinder. The reflection about the plane framed in blue leaves
each circle invariant as a set.  
The reflection about the plane framed in red exchanges 
the circles centered on the cylinder axis at L and R.
This plane framed in red defines the entangling boundary
in a cylindrical geometry
as defined by the partition~(\ref{eq: def i-dependent partition example B}).
        }
\label{fig: reflection planes cylinder}
\end{figure}

To understand the effect of any one of the sixteen perturbations 
from Table~\ref{tab:3}
on the zero modes of Hamiltonian 
$\mathcal{H}^{\,}_{i}$
defined in Eq.~(\ref{eq: def mathcal{H}{i} and mathcal{Q}{Ai+1} a}), 
where $i=1,2$ is the choice made along the direction for which
periodic boundary conditions is imposed
[recall Eq.~(\ref{eq: def mathcal{H} ki;r(i+1),r'(i+1)})],
it is useful to consider 
Fig.~\ref{fig: reflection planes cylinder}.
The physical boundaries are the two circles centered
at L and R on the symmetry axis $i+1$ modulo 2 
of the cylinder in Fig.~\ref{fig: reflection planes cylinder}.
In the limit for which the band gap is taken to infinity,
the effective theory for the single-particle eigenstates of 
$\mathcal{H}^{\,}_{i}$ defined in 
Eq.~(\ref{eq: def mathcal{H}{i} and mathcal{Q}{Ai+1} a})
that are propagating along the edges $L$ and $R$ 
with the same group velocity but localized away from them
takes the form
\begin{subequations}
\label{eq: unperturbed edge theory example B}
\begin{equation}
\mathcal{H}^{\,}_{\mathrm{edges}\,k^{\,}_{i}}=
k^{\,}_{i}\,
Y^{\,}_{33}
\qquad
i=1,2.
\label{eq: unperturbed edge theory example B a}
\end{equation}
Here, we have set the group velocity to unity and
\begin{equation}
Y^{\,}_{\mu\nu}:=
\rho^{\,}_{\mu}
\otimes
\varrho^{\,}_{\nu},
\label{eq: unperturbed edge theory example B b}
\end{equation}
\end{subequations}
where we have introduced the two sets of unit $2\times2$
matrix and Pauli matrices $\rho^{\,}_{\mu}$ with $\mu=0,1,2,3$
and $\varrho^{\,}_{\nu}$ with $\nu=0,1,2,3$.
The eigenvalue $-1$ of
$\varrho^{\,}_{3}$
is interpreted as the left edge L 
in Fig.~\ref{fig: reflection planes cylinder}.
The eigenvalue $+1$ of
$\varrho^{\,}_{3}$
is interpreted as the right edge R
in Fig.~\ref{fig: reflection planes cylinder}.
Hence, The matrices $\varrho^{\,}_{1}$ and $\varrho^{\,}_{2}$
mix edge states localized on opposite edges. The eigenstates
of $\rho^{\,}_{3}$ describe right and left movers on a given edge.
The matrices $\rho^{\,}_{1}$ and $\rho^{\,}_{2}$ mix left and right
movers on a given edge.
Hamiltonian~(\ref{eq: unperturbed edge theory example B})
describes four single-particle states propagating along
the two edges with the momentum $\pm k^{\,}_{i}$.
These single-particle states are solely supported on the edges, 
reason for which we call them edge states.
Their direction of propagation originates from the
relative sign in $p^{\,}_{1}\pm\mathrm{i}p^{\,}_{2}$ 
if $\mathcal{H}^{\,}_{i}$ is interpreted as a Bogoliubov-de-Gennes 
superconductor.

A generic perturbation to the effective Hamiltonian%
~(\ref{eq: unperturbed edge theory example B})
that allows mixing between all four edge states is
\begin{equation}
\mathcal{V}^{\,}_{\mathrm{edges}}=
\sum_{\mu,\nu=0}^{3}
v^{\,}_{\mathrm{edges}\,\mu\nu}\,
Y^{\,}_{\mu\nu},
\qquad
v^{\,}_{\mathrm{edges}\,\mu\nu}\in\mathbb{R},
\label{eq: def mathcalV edges mu nu with mu nu summed}
\end{equation}
to lowest order in a gradient expansion.
For any given $\mu,\nu=0,1,2,3$,
the perturbation
\begin{equation}
\mathcal{V}^{\,}_{\mathrm{edges}\,\mu\nu}:=
v^{\,}_{\mathrm{edges}\,\mu\nu}\,
Y^{\,}_{\mu\nu}
\label{eq: def mathcalV edges mu nu}
\end{equation}
opens a gap in the spectrum of
\begin{equation}
\widetilde{\mathcal{H}}^{\,}_{\mathrm{edges}\, k^{\,}_{i}\,\mu\nu}:=
\mathcal{H}^{\,}_{\mathrm{edges}\, k^{\,}_{i}}
+
\mathcal{V}^{\,}_{\mathrm{edges}\,\mu\nu}
\end{equation}
if it anti-commutes with $Y^{\,}_{33}$, 
thereby forbidding the crossing of the bulk gap by
any one of the four mid-gap branches 
from the effective Hamiltonian%
~(\ref{eq: unperturbed edge theory example B}).
This is what happens in
Fig.~\ref{fig: EEES_{2}D}(b).
A perturbation
$\mathcal{V}^{\,}_{\mathrm{edges}\,\mu\nu}$
that commutes with $Y^{\,}_{33}$
can shift the location of the two crossings of the
four mid-gap branches from the effective Hamiltonian%
~(\ref{eq: unperturbed edge theory example B})
away from either vanishing energy or vanishing momentum.
For example, the perturbation $\mathcal{V}^{\,}_{\mathrm{edges}\,33}$
with $v^{\,}_{\mathrm{edges}\,33}<0$ 
moves the crossing in the effective Hamiltonian%
~(\ref{eq: unperturbed edge theory example B})
to the right, as happens in
Fig.\ref{fig: EEES_{2}D}(c). 
Figure~\ref{fig: EEES_{2}D}(d) 
is realized when only one of 
$v^{\,}_{\mathrm{edges}\,03}$ 
and
$v^{\,}_{\mathrm{edges}\,30}$
are non-vanishing in the perturbation%
~(\ref{eq: def mathcalV edges mu nu}).
Evidently,
Fig.\ref{fig: EEES_{2}D}(a)
suggests that not all perturbations parametrized by %
~(\ref{eq: def mathcalV edges mu nu with mu nu summed})
move the degenerate crossing of the effective Hamiltonian
Eq.~(\ref{eq: unperturbed edge theory example B}).

To understand Fig.\ref{fig: EEES_{2}D}(a),
we assume that the perturbation $\mathcal{V}^{\,}_{\mathrm{edges}}$
is local and that the ratio $\xi/L$, where $L$ is the length
of the cylinder while $\xi\propto1/\Delta$ 
is the bulk correlation length associated to the bulk gap $\Delta$,
is taken to zero. 
Hence, the effective edge theory for
\begin{equation}
\widetilde{\mathcal{H}}^{\,}_{i\,\mu\nu}:=
\mathcal{H}^{\,}_{i}+\mathcal{V}^{\,}_{\mu\nu}
\end{equation}
remains block diagonal, with one of the block given by
\begin{subequations}
\label{eq: perturbed edge theory example 2 by 2 B}
\begin{equation}
\widetilde{\mathcal{H}}^{\,}_{\mathrm{edge}\,k^{\,}_{i}}=
v^{\,}_{0}\,\rho^{\,}_{0}
+
v^{\,}_{1}\,\rho^{\,}_{1}
+
v^{\,}_{2}\,\rho^{\,}_{2}
+
\left(
k^{\,}_{i}
+
v^{\,}_{3}
\right)
\rho^{\,}_{3}
\label{eq: perturbed edge theory example 2 by 2 B a}
\end{equation}
for some effective real-valued couplings
$v^{\,}_{0}$,
$v^{\,}_{1}$,
$v^{\,}_{2}$,
and
$v^{\,}_{3}$.
As was the case for the bulk Hamiltonian%
~(\ref{eq: direct sum decomposition example B}),
\begin{equation}
\mathcal{H}^{\,}_{\mathrm{edge}\,k^{\,}_{i}}:=
k^{\,}_{i}\,\rho^{\,}_{3}=:
\widetilde{\mathcal{H}}^{\,}_{\mathrm{edge}\,k^{\,}_{i}\,\mu\nu}
-
\widetilde{\mathcal{V}}^{\,}_{\mathrm{edge}\,\mu\nu}
\label{eq: unperturbed edge theory example 2 by 2 B b}
\end{equation}
\end{subequations}
obeys two spectral chiral symmetries, for it anti-commutes with
$\rho^{\,}_{1}$ and $\rho^{\,}_{2}$. Moreover, 
$\mathcal{H}^{\,}_{\mathrm{edge}\,k^{\,}_{i}}$
commutes with the operation by which $k^{\,}_{i}\to-k^{\,}_{i}$
is composed with the conjugation of
$\mathcal{H}^{\,}_{\mathrm{edge}\,k^{\,}_{i}}$
with either $\rho^{\,}_{1}$ and $\rho^{\,}_{2}$.
Imposing simultaneously this pair of chiral symmetries and this pair
of reflection symmetries restricts the effective edge theory
to Eq.~(\ref{eq: unperturbed edge theory example 2 by 2 B b}). 
However, these symmetry constraints are redundant to
enforce the form%
~(\ref{eq: unperturbed edge theory example 2 by 2 B b})
as we now show.

The crossing at vanishing energy and momentum
of Hamiltonian~(\ref{eq: unperturbed edge theory example 2 by 2 B b}) 
is not generically robust to the perturbations in the
perturbed Hamiltonian% 
~(\ref{eq: perturbed edge theory example 2 by 2 B a}).
However, if we impose that the perturbed Hamiltonian%
~(\ref{eq: perturbed edge theory example 2 by 2 B a})
anti-commutes with either $\rho^{\,}_{1}$ or $\rho^{\,}_{2}$,
i.e., the effective perturbed Hamiltonian belongs to the symmetry class AIII, 
then the edge theory simplifies to the direct sum over blocks of the form
\begin{equation}
\mathcal{H}^{\mathrm{AIII}}_{\mathrm{edge}\,k^{\,}_{i}\,\mu\nu}=
v^{\,}_{2}\,\rho^{\,}_{2}
+
\left(
k^{\,}_{i}
+
v^{\,}_{3}
\right)
\rho^{\,}_{3}.
\label{eq: perturbed edge theory example AIII B}
\end{equation}
Here, we have chosen, without loss of generality, 
to implement the chiral symmetry by demanding that
$\mathcal{H}^{\mathrm{AIII}}_{\mathrm{edge}\,k^{\,}_{i}\,\mu\nu}$
anti-commutes with $\rho^{\,}_{1}$.
As is, the perturbations in Hamiltonian%
~(\ref{eq: perturbed edge theory example AIII B})
still gaps the unperturbed dispersion%
~(\ref{eq: unperturbed edge theory example 2 by 2 B b}).
However, if we impose the symmetry constraint
\begin{equation}
\rho^{\,}_{1}\,
\mathcal{H}^{\,}_{\mathrm{edge}\,(-k^{\,}_{i})\,\mu\nu\,\mathscr{S}\mathscr{R}}\,
\rho^{\,}_{1}=
\mathcal{H}^{\,}_{\mathrm{edge}\,(+k^{\,}_{i})\,\mu\nu\,\mathscr{S}\mathscr{R}}\,
\end{equation}
to define 
$\mathcal{H}^{\,}_{\mathrm{edge}\,k^{\,}_{i}\,\mu\nu\,\mathscr{S}\mathscr{R}}$,
we find that
\begin{equation}
\mathcal{H}^{\,}_{\mathrm{edge}\,k^{\,}_{i}\,\mu\nu\,\mathscr{S}\mathscr{R}}=
k^{\,}_{i}\,
\rho^{\,}_{3}.
\end{equation}
In other words, if we impose, in addition to the chiral symmetry,
the reflection symmetry generated by $\rho^{\,}_{1}$
and $k^{\,}_{i}\to-k^{\,}_{i}$, we find that
no perturbation can gap 
the unperturbed dispersion%
~(\ref{eq: unperturbed edge theory example 2 by 2 B b}).
Had we chosen to impose the reflection symmetry generated by $\rho^{\,}_{2}$
instead of $\rho^{\,}_{1}$, i.e., a realization of parity that
anti-commutes with the choice $\rho^{\,}_{1}$ we made to implement the
chiral transformation, then the perturbation $v^{\,}_{2}\,\rho^{\,}_{2}$
in Eq.~(\ref{eq: perturbed edge theory example AIII B})
would not be prevented
from removing the crossing of the edge states through a spectral gap.

Now, the reflection symmetry obeyed by
the effective edge Hamiltonian%
~(\ref{eq: unperturbed edge theory example 2 by 2 B b})
can only originate from a reflection about the plane frame
in blue that contains the cylinder axis in 
Fig.~\ref{fig: reflection planes cylinder}.
A reflection about the plane framed in red that
is orthogonal to the cylinder axis in 
Fig.~\ref{fig: reflection planes cylinder}
exchanges the edges $L$ and $R$.
As such, it can only be represented within
the representation defined by Eqs.%
~(\ref{eq: unperturbed edge theory example B})
and~(\ref{eq: def mathcalV edges mu nu with mu nu summed})
of the effective edge Hamiltonian that allows mixing
of the two opposite edges.

\subsubsection{Spectra $\sigma(\widetilde{\mathcal{Q}}^{\,}_{\mu\nu A^{\,}_{i+1}})$}

For any $i=1,2$, the spectra
$\sigma(\widetilde{\mathcal{Q}}^{\,}_{\mu\nu A^{\,}_{i+1}})$
for $\mu,\nu=0,1,2,3$
were also obtained from exact diagonalization
with the unperturbed energy scales $t=\Delta=-\mu=1$.
These spectra are characterized by two nearly flat bands at $\pm1$
from which pairs of mid-gap branches occasionally peel off
as is illustrated in Figs.~\ref{fig: EEES_{2}D}(e-f).
We observe in both Figs.~\ref{fig: EEES_{2}D}(e) and \ref{fig: EEES_{2}D}(f)
one crossing of the mid-gap branches 
at vanishing entanglement eigenvalue and momentum.
The crossing in \ref{fig: EEES_{2}D}(g) takes place 
at vanishing entanglement eigenvalue but non-vanishing momentum.
The crossing in \ref{fig: EEES_{2}D}(h) takes place 
at non-vanishing entanglement but vanishing momentum.
Crossings of the mid-gap branches taking place 
at vanishing entanglement eigenvalue and momentum 
are indicated by $\circ$ in the last two columns of 
Table~\ref{tab:3}.
The absence of any crossing of the mid-gap branches 
is indicated by $\times$ in the last two columns of 
Table~\ref{tab:3}.
Crossings of the mid-gap branches taking place away from
vanishing entanglement eigenvalue and momentum
are indicated by $\otimes$ in the last two columns of 
Table~\ref{tab:3}.

Comparison of 
Figs.~\ref{fig: EEES_{2}D}(a-d)
and
Figs.~\ref{fig: EEES_{2}D}(e-f)
suggests that the existence of crossings in the entanglement spectrum 
defined by Eq.~(\ref{eq: def mathcal{H}{i} and mathcal{Q}{Ai+1} b})
is more robust to perturbations than that for the Hamiltonian spectrum.
Choose $i=1,2$ ($i+1$ modulo 2), $\mu=0,1,2,3$, and $\nu=0,1,2,3$.
According to Table~\ref{tab:3} any crossing 
in the perturbed Hamiltonian spectrum
$\sigma(\widetilde{\mathcal{H}}^{\,}_{i\,\mu\nu})$
implies a crossing in the perturbed entanglement spectrum
$\sigma(\widetilde{\mathcal{Q}}^{\,}_{\,\mu\nu A^{\,}_{i+1}})$.
The converse is not true.
There are crossings in the perturbed entanglement spectrum 
$\sigma(\widetilde{\mathcal{Q}}^{\,}_{\,\mu\nu A^{\,}_{i+1}})$
but no crossings in the perturbed Hamiltonian spectrum
$\sigma(\widetilde{\mathcal{H}}^{\,}_{i\,\mu\nu})$,
as is shown explicitly when comparing 
Fig.~\ref{fig: EEES_{2}D}(f)
to
Fig.~\ref{fig: EEES_{2}D}(b).
In fact, Table~\ref{tab:3} shows the following difference between
the perturbed Hamiltonian  and entanglement spectra.
On the one hand,
the existence of crossings in
$\sigma(\widetilde{\mathcal{H}}^{\,}_{i\,\mu\nu})$
does not necessarily implies the existence of crossings in
$\sigma(\widetilde{\mathcal{H}}^{\,}_{(i+1)\,\mu\nu})$.
On the one hand,
the existence of crossings in
$\sigma(\widetilde{\mathcal{Q}}^{\,}_{\,\mu\nu A^{\,}_{i}})$
implies the existence of crossings in
$\sigma(\widetilde{\mathcal{Q}}^{\,}_{\,\mu\nu A^{\,}_{i+1}})$
and vice versa.

To understand the effect of any one of the sixteen perturbations 
from Table~\ref{tab:3}
on the zero modes of 
$\mathcal{Q}^{\,}_{A^{\,}_{i+1}}$
defined in Eq.~(\ref{eq: def mathcal{H}{i} and mathcal{Q}{Ai+1} b}), 
where $i+1$ modulo 2
is the choice made for the direction along which the partition
is made [recall Eq.~(\ref{eq: def mathcal{H} ki;r(i+1),r'(i+1)})],
it is useful to consider
the compactifcation along the cylinder axis of
Fig.~\ref{fig: reflection planes cylinder}
consisting in identifying the circles at L and R.
We thereby obtain a torus, whose intersection
with the plane framed in red from
Fig.~\ref{fig: reflection planes cylinder}
defines two entangling boundaries separated by
the distance $L/2$. The spectrum 
$\sigma(\mathcal{Q}^{\,}_{A^{\,}_{i+1}})$
supports four mid-gap branches, two per entangling boundaries.
In the limit for which the band gap is taken to infinity,
the effective theory for the eigenstates of 
$\mathcal{Q}^{\,}_{A^{\,}_{i+1}}$
that propagate along the entangling boundaries but are
localized away from them is given by
\begin{equation}
\mathcal{Q}^{\,}_{\mathrm{edges}\,k^{\,}_{i}\,A^{\,}_{i+1}}=
k^{\,}_{i}\, Y^{\,}_{33}.
\label{eq: unperturbed entangling theory example B}
\end{equation}
The interpretation of the
$4\times4$ matrices $Y^{\,}_{\mu\nu}$ with $\mu,\nu=0,1,2,3$
is the same as the one given below 
Eq.~(\ref{eq: unperturbed edge theory example B b}).

A generic perturbation to the effective Hamiltonian%
~(\ref{eq: unperturbed entangling theory example B})
that allows mixing between all four entangling states is
\begin{equation}
\delta\mathcal{Q}^{\,}_{\mathrm{edges}\,A^{\,}_{i+1}}=
\sum_{\mu,\nu=0}^{3}
v^{\,}_{\mathrm{edges}\,\mu\nu\,A^{\,}_{i+1}}\,
Y^{\,}_{\mu\nu},
\
v^{\,}_{\mathrm{edges}\,\mu\nu\,A^{\,}_{i+1}}\in\mathbb{R},
\label{eq: def mathcalV entangling mu nu with mu nu summed}
\end{equation}
to lowest order in a gradient expansion.
For any given $\mu,\nu=0,1,2,3$,
the perturbation
\begin{equation}
\delta\mathcal{Q}^{\,}_{\mathrm{edges}\,\mu\nu\,A^{\,}_{i+1}}:=
v^{\,}_{\mathrm{edges}\,\mu\nu\,A^{\,}_{i+1}}\,
Y^{\,}_{\mu\nu}
\label{eq: def mathcalV entangling mu nu}
\end{equation}
opens a gap in the spectrum of
\begin{equation}
\widetilde{\mathcal{Q}}^{\,}_{\mathrm{edges}\,k^{\,}_{i}\,\mu\nu\,A^{\,}_{i+1}}:=
\mathcal{Q}^{\,}_{\mathrm{edges}\,k^{\,}_{i}\,A^{\,}_{i+1}}
+
\delta\mathcal{Q}^{\,}_{\mathrm{edges}\,\mu\nu\,A^{\,}_{i+1}}
\end{equation}
if it anti-commutes with $Y^{\,}_{33}$, 
thereby forbidding the crossing of the bulk gap by
any one of the four mid-gap branches 
from the effective Hamiltonian%
~(\ref{eq: unperturbed entangling theory example B}).
A perturbation
$\mathcal{V}^{\,}_{\mathrm{edges}\,\mu\nu\,A^{\,}_{i+1}}$
that commutes with $Y^{\,}_{33}$
can shift the location of the two crossings of the
four mid-gap branches from the effective Hamiltonian%
~(\ref{eq: unperturbed entangling theory example B})
away from either vanishing energy or vanishing momentum.

To understand the robustness of 
the effective~(\ref{eq: unperturbed entangling theory example B})
to perturbations, we assume that the perturbation 
$\mathcal{V}^{\,}_{\mathrm{edges}\,A^{\,}_{i+1}}$
is local and that the ratio $\xi/L$, where $L$ is the length
of the cylinder while $\xi\propto1/\Delta$ 
is the bulk correlation length associated to the bulk gap $\Delta$,
is taken to zero. Hence, the effective edge theory for
\begin{equation}
\widetilde{\mathcal{Q}}^{\,}_{\mu\nu\,A^{\,}_{i+1}}:=
\mathcal{Q}^{\,}_{A^{\,}_{i+1}\,}
+
\delta\mathcal{Q}^{\,}_{\mu\nu\,A^{\,}_{i+1}}
\end{equation}
remains block diagonal, with one of the block given by
\begin{equation}
\widetilde{\mathcal{Q}}^{\,}_{\mathrm{edge}\,k^{\,}_{i}\,A^{\,}_{i+1}}=
v^{\,}_{0}\,\rho^{\,}_{0}
+
v^{\,}_{1}\,\rho^{\,}_{1}
+
v^{\,}_{2}\,\rho^{\,}_{2}
+
\left(
k^{\,}_{i}
+
v^{\,}_{3}
\right)
\rho^{\,}_{3}
\label{eq: perturbed entangling theory example 2 by 2 B a}
\end{equation}
for some effective real-valued couplings
$v^{\,}_{0}$,
$v^{\,}_{1}$,
$v^{\,}_{2}$,
and
$v^{\,}_{3}$.

To proceed, we recall that if
$\mathcal{S}$ is a unitary spectral symmetry of $\mathcal{H}$
in that $\mathcal{S}^{-1}\,\mathcal{H}\,\mathcal{S}=-\mathcal{H}$
and if $\mathcal{S}$ is block diagonal with respect to the
partition into $A$ and $B$, we then deduce from
\begin{equation}
\begin{split}
\mathcal{Q}=&\,
\mathcal{S}^{-1}\,(-1)\mathcal{Q}\,\mathcal{S}  
\\
=&\,
\begin{pmatrix}
S^{-1}_{A} 
& 
0
\\
0
& 
S^{-1}_{B}
\end{pmatrix} 
\begin{pmatrix}
-Q^{\,}_{A}
& 
+2C^{\,}_{AB}
\\
+2C^{\,}_{BA}
& 
-Q^{\,}_B
\end{pmatrix} 
\begin{pmatrix}
S^{\,}_{A}
& 
0
\\
0
& 
S^{\,}_{B}
\end{pmatrix} 
\end{split}
\end{equation}
that
\begin{subequations}
\begin{equation}
S^{-1}_{A} \,Q^{\,}_{A}\,S^{\,}_{A}=-Q^{\,}_{A},
\qquad
S^{-1}_{B} \,Q^{\,}_{B}\,S^{\,}_{B}=-Q^{\,}_{B},
\end{equation}
and
\begin{equation}
S^{-1}_{A} \,C^{\,}_{AB}\,S^{\,}_{B}=-C^{\,}_{AB},
\quad
S^{-1}_{B} \,C^{\,}_{BA}\,S^{\,}_{A}=-C^{\,}_{BA}.
\end{equation}
\end{subequations}

As was the case for the bulk Hamiltonian%
~(\ref{eq: direct sum decomposition example B}),
\begin{equation}
\mathcal{Q}^{\,}_{\mathrm{edge}\,k^{\,}_{i}\,A^{\,}_{i+1}}:=
k^{\,}_{i}\,\rho^{\,}_{3}
\label{eq: unperturbed entangling theory example 2 by 2 B b}
\end{equation}
obeys two spectral chiral symmetries, for it anti-commutes with
$\rho^{\,}_{1}$ and $\rho^{\,}_{2}$, and two spectral symmetries under
charge conjugation, for it anti-commutes with the composition
of $k^{\,}_{i}\to-k^{\,}_{i}$ with conjugation by $\rho^{\,}_{0}$
or $\rho^{\,}_{3}$.

The crossing at vanishing energy 
in Eq.~(\ref{eq: unperturbed entangling theory example 2 by 2 B b}) 
is not generically robust to the perturbations in
Eq.~(\ref{eq: perturbed entangling theory example 2 by 2 B a}).
However, if we impose that 
Eq.~(\ref{eq: perturbed entangling theory example 2 by 2 B a})
anti-commutes with either $\rho^{\,}_{1}$ or $\rho^{\,}_{2}$,
then the entangling theory simplifies to the direct sum over blocks of the form
\begin{equation}
\mathcal{Q}^{\mathrm{AIII}}_{\mathrm{edge}\,k^{\,}_{i}\,\mu\nu\,A^{\,}_{i+1}}=
v^{\,}_{2}\,\rho^{\,}_{2}
+
\left(
k^{\,}_{i}
+
v^{\,}_{3}
\right)
\rho^{\,}_{3}.
\label{eq: perturbed entangling theory example AIII B}
\end{equation}
Here and without loss of generality,  we have chosen
to implement the chiral symmetry by demanding that
$\mathcal{Q}^{\mathrm{AIII}}_{\mathrm{edge}\,k^{\,}_{i}\,\mu\nu\,A^{\,}_{i+1}}$
anti-commutes with $\rho^{\,}_{1}$.
As is, one perturbation in Hamiltonian%
~(\ref{eq: perturbed entangling  theory example AIII B})
still gaps the unperturbed dispersion%
~(\ref{eq: unperturbed entangling theory example 2 by 2 B b}).
However, if we impose the symmetry constraint
\begin{equation}
\rho^{\,}_{2}\,
\widetilde{\mathcal{Q}}^{\,}_
{\mathrm{edge}\,k^{\,}_{i}\,\mu\nu\,A^{\,}_{i+1}\,\mathscr{S}\mathscr{R}^{\,}_{i}}\,
\rho^{\,}_{2}=
\widetilde{\mathcal{Q}}^{\,}_
{\mathrm{edge}\,k^{\,}_{i}\,\mu\nu\,\,A^{\,}_{i+1}\mathscr{S}\mathscr{R}^{\,}_{i}}\,
\end{equation}
to define 
$\widetilde{\mathcal{Q}}^{\,}_
{\mathrm{edge}\,k^{\,}_{i}\,\mu\nu\,A^{\,}_{i+1}\,\mathscr{S}\mathscr{R}^{\,}_{i}}$,
we find that
\begin{equation}
\widetilde{\mathcal{Q}}^{\,}_
{\mathrm{edge}\,k^{\,}_{i}\,\mu\nu\,A^{\,}_{i+1}\,\mathscr{S}\mathscr{R}^{\,}_{i}}=
\left(k^{\,}_{i}+v^{\,}_{3}\right)
\rho^{\,}_{3}
\end{equation}
displays a crossing at vanishing energy and momentum
$k^{\,}_{i}=-v^{\,}_{3}$.
In other words, if we impose, in addition to the chiral symmetry,
the effective chiral symmetry generated by $\rho^{\,}_{2}$, we find that
no perturbation can gap 
the unperturbed dispersion%
~(\ref{eq: unperturbed entangling theory example 2 by 2 B b}),
although it can move the momentum of the zero mode away from
vanishing energy. As is implied by the notation, the effective
chiral symmetry generated by $\rho^{\,}_{2}$ originates from
protecting the symmetry class AIII by demanding that
reflection symmetry about the plane defining the entangling
boundaries, i.e., the plane framed in red in
Fig.~\ref{fig: reflection planes cylinder},
holds. According to Sec.%
~\ref{subsec: Spectral gap and locality of the induced symmetry Gamma}
the reflection about the plane framed in red in
Fig.~\ref{fig: reflection planes cylinder}
induces a local effective chiral symmetry of the form
given in Eq.~(\ref{eq: main result on how to get chiral sym}),
i.e., there exists a
\begin{equation}
\Gamma^{\,}_{\mathscr{R}^{\,}_{i}}:=
C^{\,}_{A^{\,}_{i+1}B^{\,}_{i+1}\,k^{\,}_{i}}\,
\mathcal{R}^{\,}_{i}
\end{equation}
such that
\begin{equation}
\left\{
\Gamma^{\,}_{\mathscr{R}^{\,}_{i}},
\widetilde{\mathcal{Q}}^{\,}_
{\mathrm{edge}\,k^{\,}_{i}\,\mu\nu\,A^{\,}_{i+1}\,\mathscr{S}\mathscr{R}^{\,}_{i}}
\right\}=0.
\end{equation}
The choice to represent
$\Gamma^{\,}_{\mathscr{R}^{\,}_{i}}$
by $\rho^{\,}_{2}$, given the choice to represent the chiral symmetry
$\mathscr{S}$ by $\rho^{\,}_{1}$, is a consequence of the following
assumption and the following fact.
First, we demand that $\mathscr{S}$ and $\mathscr{R}^{\,}_{i}$ commute,
i.e., we demand that
\begin{equation}
\left[
\mathcal{S},
\mathcal{R}^{\,}_{i}
\right]=0.
\end{equation}
Second, the identity
\begin{equation}
\begin{split}
\left\{
\mathcal{S},
\mathcal{R}^{\,}_{i}
\right\}=&\,
\left\{
\mathcal{S},
C^{\,}_{A^{\,}_{i+1}B^{\,}_{i+1}\,k^{\,}_{i}}\,
\mathcal{R}^{\,}_{i}
\right\}
\\
=&\,
\left\{
\mathcal{S},
C^{\,}_{A^{\,}_{i+1}B^{\,}_{i+1}\,k^{\,}_{i}}\,
\right\}
\mathcal{R}^{\,}_{i}
-
C^{\,}_{A^{\,}_{i+1}B^{\,}_{i+1}\,k^{\,}_{i}}\,
\left[
\mathcal{S},
\mathcal{R}^{\,}_{i}
\vphantom{C^{\,}_{A^{\,}_{i+1}B^{\,}_{i+1}\,k^{\,}_{i}}}
\right]
\end{split}
\end{equation}
holds.

\subsubsection{Spectra
$\sigma(\mathcal{H}^{\,}_{k^{\,}_{i}\,\mathscr{R};\mathscr{S}})$  
and
$\sigma(Q^{\,}_{k^{\,}_{i}\,\mathscr{R};\mathscr{S}\, A^{\,}_{i+1}})$}
          
It is time to present the reasoning
that delivers Table~\ref{tab:4}. Let $i=1,2$ and define $i+1$ modulo 2.
We assume that the underlying microscopic model
has the doublet of symmetries 
$(\mathscr{R};\mathscr{S})\sim
(\mathcal{R}^{\,}_{\mu\nu},\mathcal{S}^{\,}_{\mu\nu})$
where
$\mathscr{S}\sim
\mathcal{S}^{\,}_{\mu\nu}$
defines the symmetry class AIII.
We denote the most general perturbation 
that is compliant with the doublet of symmetries 
$(\mathscr{R};\mathscr{S})$
defining a given row of Table~\ref{tab:2} by
$\mathcal{V}^{\,}_{\mathscr{R};\mathscr{S}}$.
The explicit form of this perturbation
is to be found in the second column of
Table~\ref{tab:4}
as one varies $(\mathscr{R};\mathscr{S})$.
For a given row in Table~\ref{tab:4},
$\mathcal{V}^{\,}_{\mathscr{R};\mathscr{S}}$
is contained in the most general perturbation 
$\mathcal{V}^{\,}_{\mathscr{S}}$
that is compliant with the AIII symmetry.
We define the single-particle Hamiltonian
$\mathcal{H}^{\,}_{i\,\mathscr{R};\mathscr{S}}\equiv
\mathcal{H}^{\,}_{i}+\mathcal{V}^{\,}_{\mathscr{R};\mathscr{S}}$ 
by its matrix elements
\begin{subequations}
\label{eq: def H SR for 2D example B}
\begin{align}
&
\mathcal{H}^{\,}_{k^{\,}_{i}\mathscr{R};\mathscr{S}\,r^{\,}_{i+1},r^{\prime}_{i+1}}:=
\mathcal{H}^{\,}_{k^{\,}_{i}r^{\,}_{i+1},r^{\prime}_{i+1}}
+
\mathcal{V}^{\,}_{\mathscr{R};\mathscr{S}\,r^{\,}_{i+1},r^{\prime}_{i+1}},
\label{eq: def H SR for 2D example B a}
\\
&
\mathcal{V}^{\,}_{\mathscr{R};\mathscr{S}\,r^{\,}_{i+1},r^{\prime}_{i+1}}:=
\delta^{\,}_{r^{\,}_{i+1},r^{\prime}_{i+1}}\,
\sum_{\mu,\nu\in\mathrm{row}}
v^{\,}_{\mu\nu}\,
\mathcal{X}^{\,}_{\mu\nu}.
\label{eq: def H SR for 2D example B b}
\end{align}
\end{subequations}

The corresponding equal-time one-point correlation matrix
is $\mathcal{Q}^{\,}_{\mathscr{R};\mathscr{S}\,(i+1)}$
and its upper-left block is
$Q^{\,}_{\mathscr{R};\mathscr{S}\,A^{\,}_{i+1}}$.
The third column in Table~\ref{tab:4}
provides one sign for each row. 
The sign $\eta^{\,}_{\mathscr{S}}$
is positive if $\mathscr{R}$ commutes with $\mathscr{S}$
and negative if $\mathscr{R}$ anti-commutes with $\mathscr{S}$.
The information contained in $\eta^{\,}_{\mathscr{S}}$
is needed to read from Table VI of Ref.~\cite{Morimoto13a}
the bulk topological index of the single-particle Hamiltonian%
~(\ref{eq: def H SR for 2D example B}).
This topological index does not guarantee that
$\mathcal{H}^{\,}_{\mathscr{R};\mathscr{S}}$
supports boundary states in an open geometry.
On the other hand, $Q^{\,}_{\mathscr{R};\mathscr{S}\,A^{\,}_{i+1}}$
supports boundary states on the entangling boundaries.
The entries $\circ$ for the columns
$\sigma(H^{\,}_{k^{\,}_{i}\,\mathscr{R};\mathscr{S}})$
and
$\sigma(Q^{\,}_{k^{\,}_{i}\,\mathscr{R};\mathscr{S}\,A^{\,}_{i+1}})$
are a consequence of our stability analysis.
On the one hand,
Table~\ref{tab:4} demonstrates explicitly that
the presence of a reflection symmetry in addition to the chiral symmetry
does not guarantee that a non-vanishing bulk topological index
implies protected edge states in the spectrum
$\sigma(H^{\,}_{k^{\,}_{i}\,\mathscr{R};\mathscr{S}})$
for both edges $i=1,2$.
On the other hand,
Table~\ref{tab:4} demonstrates explicitly that
a non-vanishing bulk topological index
always implies protected entangling states in the spectrum
$\sigma(Q^{\,}_{k^{\,}_{i}\,\mathscr{R};\mathscr{S}\,A^{\,}_{i+1}})$
irrespective of the choice $i=1,2$ 
made for the entangling boundary.

\begin{table*}
\begin{center}
\resizebox{\columnwidth}{!}{%
\begin{tabular}{ | c | c |  c | c | c | c c | c c  | }  \hline
 \multicolumn{2}{|c|}{Symmetries in class AIII}
&
Generic perturbation
$\mathcal{V}^{\,}_{\mathscr{R};\mathscr{S}}$
& 
$\eta^{\,}_{\mathscr{S}}$ 
& 
Index
&
$\sigma(\mathcal{H}^{\,}_{k^{\,}_{1}\,\mathscr{R};\mathscr{S}})$  
&
$\sigma(Q^{\,}_{k^{\,}_{1}\,\mathscr{R};\mathscr{S}\, A^{\,}_{2}})$ 
&
$\sigma(\mathcal{H}^{\,}_{k^{\,}_{2}\,\mathscr{R};\mathscr{S}})$  
&
$\sigma(Q^{\,}_{k^{\,}_{2}\,\mathscr{R};\mathscr{S}\, A^{\,}_{1}})$   
\\ 
\hline 
\hline 
\multirow{4}{*}{$(\mathscr{R}^{\,}_{1},\mathscr{S})$}
&
$\mathcal{R}^{\,}_{13}$, $\mathcal{S}^{\,}_{11}$
&
$v^{\,}_{03}\mathcal{X}^{\,}_{03}
+v^{\,}_{13}\mathcal{X}^{\,}_{13}
+v^{\,}_{21}\mathcal{X}^{\,}_{21}
+v^{\,}_{31}\mathcal{X}^{\,}_{31}$                                         
& 
$-$ 
& 
$0$  
& 
$\times$  
&
$\times$  
&
$\times$  
&
$\times$  
\\ \cline{2-9}
&
$\mathcal{R}^{\,}_{13}$, $\mathcal{S}^{\,}_{21}$
& 
$v^{\,}_{03}\mathcal{X}^{\,}_{03}
+v^{\,}_{10}\mathcal{X}^{\,}_{10}
+v^{\,}_{22}\mathcal{X}^{\,}_{22}
+v^{\,}_{31}\mathcal{X}^{\,}_{31}$                                              
& 
$+$ 
& 
$\mathbb{Z}$  
& 
$\times$  
& 
$\odot$
& 
$\circ$  
& 
$\otimes$ 
\\  \cline{2-9}
&
$\mathcal{R}^{\,}_{23}$, $\mathcal{S}^{\,}_{11}$ 
& 
$v^{\,}_{03}\mathcal{X}^{\,}_{03}
+v^{\,}_{12}\mathcal{X}^{\,}_{12}
+v^{\,}_{20}\mathcal{X}^{\,}_{20}
+v^{\,}_{31}\mathcal{X}^{\,}_{31}$                                      
& 
$+$ 
& 
$\mathbb{Z}$ 
& 
$\times$  
& 
$\odot$
& 
$\circ$  
& 
$\otimes$  
\\  \cline{2-9}
&
$\mathcal{R}^{\,}_{23}$, $\mathcal{S}^{\,}_{21}$ 
& 
$v^{\,}_{03}\mathcal{X}^{\,}_{03}
+v^{\,}_{11}\mathcal{X}^{\,}_{11}
+v^{\,}_{23}\mathcal{X}^{\,}_{23}
+v^{\,}_{31}\mathcal{X}^{\,}_{31}$  
& 
$-$ 
& 
0                   
& 
$\times$  
&
$\times$  
&
$\times$  
&
$\times$  
\\ \hline \hline
\multirow{4}{*}{$(\mathscr{R}^{\,}_{2}, \mathscr{S})$}
&
 $\mathcal{R}^{\,}_{10}$, $\mathcal{S}^{\,}_{11}$ 
&
$v^{\,}_{02}\mathcal{X}^{\,}_{02}
+v^{\,}_{03}\mathcal{X}^{\,}_{03}
+v^{\,}_{12}\mathcal{X}^{\,}_{12}
+v^{\,}_{13}\mathcal{X}^{\,}_{13}$  
& 
$+$ 
& 
$\mathbb{Z}$                   
& 
$\circ$  
& 
$\otimes$
& 
$\times$  
& 
$\odot$
\\  \cline{2-9}

&
$\mathcal{R}^{\,}_{10}$, $\mathcal{S}^{\,}_{21}$ 
& 
$v^{\,}_{02}\mathcal{X}^{\,}_{02}
+v^{\,}_{03}\mathcal{X}^{\,}_{03}
+v^{\,}_{10}\mathcal{X}^{\,}_{10}
+v^{\,}_{11}\mathcal{X}^{\,}_{11}$                                      
& 
$-$ 
& 
0                   
&    
$\times$  
&
$\times$  
&
$\times$  
&
$\times$  
\\ \cline{2-9}
&
$\mathcal{R}^{\,}_{20}$, $\mathcal{S}^{\,}_{11}$    
& 
$v^{\,}_{02}\mathcal{X}^{\,}_{02}
+v^{\,}_{03}\mathcal{X}^{\,}_{03}
+v^{\,}_{20}\mathcal{X}^{\,}_{20}
+v^{\,}_{21}\mathcal{X}^{\,}_{21}$                                                
& 
$-$ 
& 
0                   
& 
$\times$  
&
$\times$  
&
$\times$  
&
$\times$ 
\\ \cline{2-9}
&
$\mathcal{R}^{\,}_{20}$, $\mathcal{S}^{\,}_{21}$    
& 
$v^{\,}_{02}\mathcal{X}^{\,}_{02}
+v^{\,}_{03}\mathcal{X}^{\,}_{03}
+v^{\,}_{22}\mathcal{X}^{\,}_{22}
+v^{\,}_{23}\mathcal{X}^{\,}_{23}$       
& 
$+$ 
& 
$\mathbb{Z}$                  
& 
$\circ$  
& 
$\otimes$
& 
$\times$  
& 
$\odot$
\\ \hline
\end{tabular}
}
\caption{%
The first four rows of the second column give all possible
doublets of generators consisting of a reflection about
the direction 1 and a chiral transformation.
The last four rows of the second column give all possible
doublets of generators consisting of a reflection about
the direction 2 and a chiral transformation.
The third column gives for each row the most general
perturbation $\mathcal{V}^{\,}_{\mathscr{R}; \mathscr{S}}$
that commutes with the operation of reflection
and anti-commutes with the operation of chirality.
The fourth column gives the sign $\eta^{\,}_{\mathscr{S}}$ defined by 
$\mathcal{R}\,\mathcal{S}\,\mathcal{R}=\eta^{\,}_{\mathscr{S}}\,\mathcal{S}$.
The fifth column is an application
of the classification for the symmetry-protected topological band insulators
in two-dimensional spaces derived in Refs.%
~\cite{Chingkai2012} and \cite{Morimoto13a}
(Table VI from Ref.~\cite{Morimoto13a} was particularly useful).
The topological indices $\mathbb{Z}$ and $0$ correspond to 
topologically nontrivial and trivial bulk phases, respectively. 
The entries $\circ$ or $\times$ in the last two columns denote
the presence or absence, respectively,  of zero modes in the spectra of
$\sigma(\mathcal{H}^{\,}_{k^{\,}_{1}\,\mathscr{R};\mathscr{S}})$,
on the one hand, and
$\sigma(\mathcal{H}^{\,}_{k^{\,}_{2}\,\mathscr{R};\mathscr{S}})$,
on the other hand. Whereas $\times$ denotes the absence of zero modes,
the entries $\odot$ or $\otimes$ in the last two columns denote
the presence of zero modes with or without spectral flow, respectively, 
in the spectra of 
$\sigma(\widetilde{Q}^{\,}_{k^{\,}_{1}\,\mathscr{R};\mathscr{S}\,A^{\,}_{2}})$,
on the one hand, and
$\sigma(\widetilde{Q}^{\,}_{k^{\,}_{2}\,\mathscr{R};\mathscr{S}\,A^{\,}_{1}})$,
on the other hand.
        }   
\label{tab:4}
\end{center}
\end{table*}

\subsection{Existence of spectral flows in the entanglement spectra}
\label{subsubsec: Spectral flow in the entanglement energy example B}

\begin{figure}[t] 
\centering     
\includegraphics[width=0.9 \textwidth]{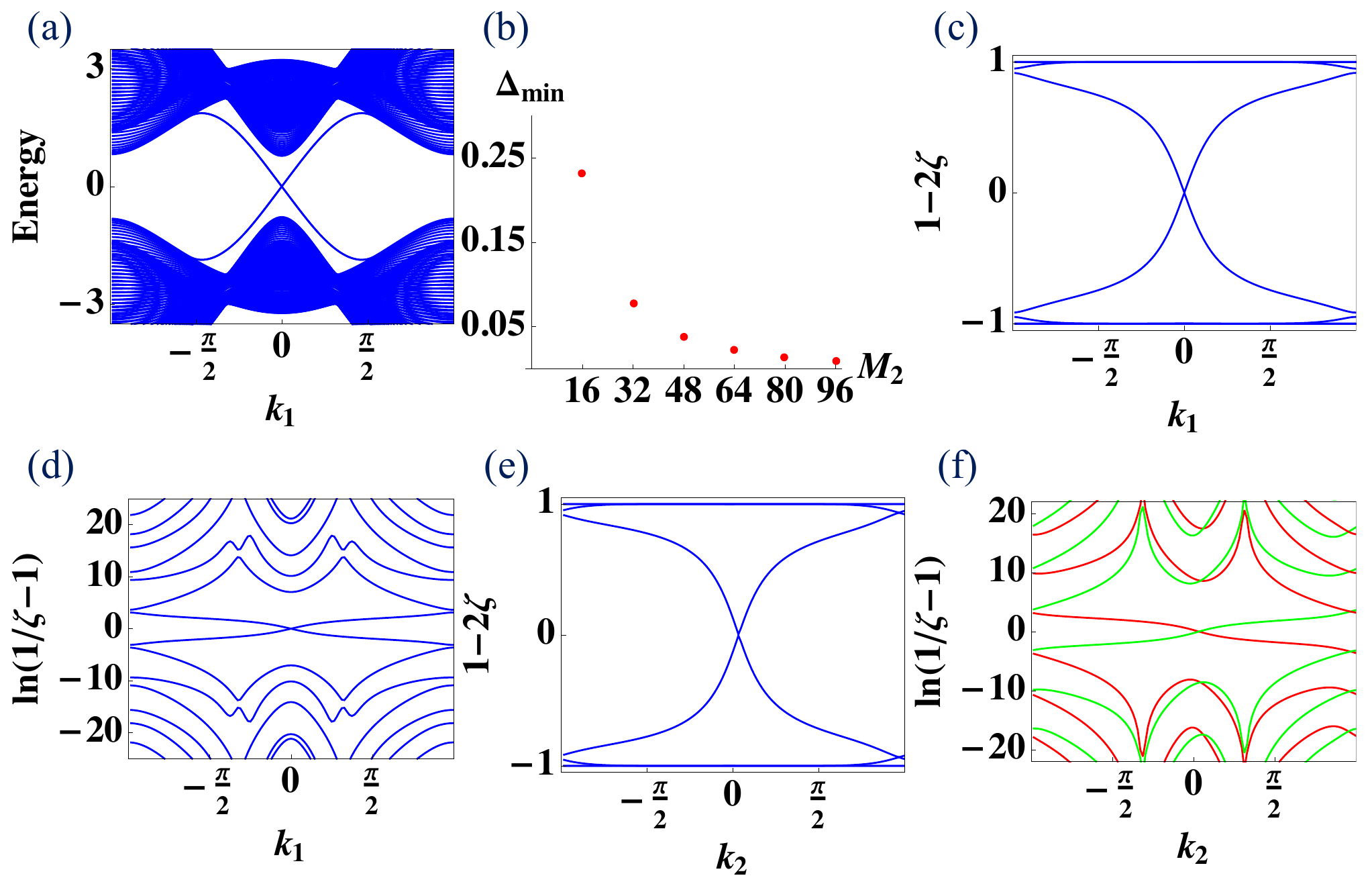}
\caption{
(Color online)
Spectra 
$\sigma(\mathcal{H}^{\,}_{k_i\,\mathscr{R}^{\,}_{2};\mathscr{S}})$,
$\sigma(Q^{\,}_{k_i\,\mathscr{R}^{\,}_{2};\mathscr{S}\,A^{\,}_{i+1}})$,
and
$\sigma\big(\ln(Q^{-1}_{k_i \mathscr{R}^{\,}_{2};\mathscr{S}\,A^{\,}_{i+1}}-1)\big)$
(with $i+1$ defined modulo 2)
for the single-particle Hamiltonian%
~$\mathcal{H}^{\,}_{k_i \,\mathscr{R}^{\,}_{2};\mathscr{S}}$
defined by the matrix elements~(\ref{eq: def H SR for 2D example B})
and the corresponding upper-left block%
~$Q^{\,}_{k_i \,\mathscr{R}^{\,}_{2};\mathscr{S}\,A^{\,}_{i+1}}$
from the equal-time one-point correlation matrix
with 
$(v^{\,}_{02},v^{\,}_{03},v^{\,}_{12},v^{\,}_{13})=(0.2,-0.1,0.05,0.3)$
taken from the fifth line from Table \ref{tab:4}.
(a) Spectrum 
$\sigma(\mathcal{H}^{\,}_{k_{1}\,\mathscr{R}^{\,}_{2};\mathscr{S}})$
for the linear sizes 
$M^{\,}_{1}=128$
and
$M^{\,}_{2}=64$.
(b) Scaling of the direct gap between the conduction bands
at the momentum $k^{\,}_{1}=\pi/2$
and the mid-gap branch with positive energy eigenvalue
at the momentum $k^{\,}_{1}=\pi/2$ as a function of increasing
$M^{\,}_{2}=16,32,64,80,96$
holding $M^{\,}_{1}=128$ fixed.
(c) Spectrum 
$Q^{\,}_{k^{\,}_{1}\,\mathscr{R}^{\,}_{2};\mathscr{S}\,A^{\,}_{2}}$
for the linear sizes 
$M^{\,}_{1}=128$
and
$M^{\,}_{2}=64$.
(d) Spectrum
$\sigma\big(\ln(Q^{-1}_{k^{\,}_{1}\,\mathscr{R}^{\,}_{2};\mathscr{S}\,A^{\,}_{2}}-1)\big)$
for the linear sizes 
$M^{\,}_{1}=128$
and
$M^{\,}_{2}=64$.
(e) Spectrum 
$Q^{\,}_{k^{\,}_{2}\,\mathscr{R}^{\,}_{2};\mathscr{S}\,A^{\,}_{1}}$
for the linear sizes 
$M^{\,}_{1}=64$
and
$M^{\,}_{2}=128$.
(f) Spectrum
$\sigma\big(\ln(Q^{-1}_{k^{\,}_{2}\,\mathscr{R}^{\,}_{2};\mathscr{S}\,A^{\,}_{1}}-1)\big)$
for the linear sizes 
$M^{\,}_{1}=64$
and
$M^{\,}_{2}=128$.
The coloring follows from the existence of
the operator~(\ref{eq: def Lambda example B})
that commutes with 
$Q^{\,}_{k^{\,}_{2}\,\mathscr{R}^{\,}_{2};\mathscr{S}\,A^{\,}_{1}}$.
The coloring demonstrates the existence of a spectral flow
that connects the valence to the conduction continua
through the mid-gap branches
in the thermodynamic limit 
$M^{\,}_{1},M^{\,}_{2}\to\infty$.
        }
\label{fig:spectral_flow}
\end{figure}

We are going to study the dependences of the
spectra 
$\sigma(\mathcal{H}^{\,}_{k_i \,\mathscr{R};\mathscr{S}})$
and
$\sigma(Q^{\,}_{k_i \, \mathscr{R};\mathscr{S}\,A^{\,}_{i+1}})$
(with $i+1$ defined modulo 2)
on the system sizes for the single-particle Hamiltonian%
~$\mathcal{H}^{\,}_{k_i \,\mathscr{R};\mathscr{S}}$
defined by the matrix elements~(\ref{eq: def H SR for 2D example B})
and the corresponding upper-left block%
~$Q^{\,}_{k_i\,\mathscr{R};\mathscr{S}\,A^{\,}_{i+1}}$
from the equal-time one-point correlation matrix. 
Without loss of generality,
we will present numerical results obtained by choosing the fifth row
in Table \ref{tab:4} to define 
$\mathcal{H}^{\,}_{k_i \,\mathscr{R}^{\,}_{2};\mathscr{S}}$
and
$Q^{\,}_{k_i \,\mathscr{R}^{\,}_{2};\mathscr{S}\,A^{\,}_{i+1}}$
with 
$(v^{\,}_{02},v^{\,}_{03},v^{\,}_{12},v^{\,}_{13})=(0.2,-0.1,0.05,0.3)$.

The question we want to address is the following.
On the one hand,
Figs.~\ref{fig: EEES_{2}D}(a-d)
suggest that the mid-gap branches merge into the continuum
when $\pi/2<|k^{\,}_{i}|<\pi$.
On the other hand,
the mid-gap branches in
Fig.~\ref{fig: EEES_{2}D}(e)
seem to be separated by a very small gap from all other bands
in the vicinity of $|k^{\,}_{1}|=\pi$,
while no gap is resolved in energy between the mid-gap branches
and all other branches in the vicinity of $|k^{\,}_{1}|=\pi$
for Figs.~\ref{fig: EEES_{2}D}(f-h).

The first question we are going to address is how to interpret
the ``peeling of'' the mig-gap branches from the continua
in Figs.~\ref{fig: EEES_{2}D}(a-d). To this end, we present
in Fig~\ref{fig:spectral_flow}(a)
the spectrum
$\sigma(\mathcal{H}^{\,}_{k_{1}\,\mathscr{R}^{\,}_{2};\mathscr{S}})$
obtained by imposing periodic boundary conditions along the direction
$i=1$ and open boundary conditions along the direction $i+1=2$
for the linear system sizes 
$M^{\,}_{1}=128$
and
$M^{\,}_{2}=64$.
We then compute the spectrum of
$\mathcal{H}^{\,}_{k_{1}\,\mathscr{R}^{\,}_{2};\mathscr{S}}$
for each value $M^{\,}_{2}=16,32,48,64,80,96$ 
of the cylinder height. At last,
we compute the minimal value 
$\Delta^{\,}_{\mathrm{min}}(\pi/2,M^{\,}_{2})>0$
taken by the difference in energy
between the positive energy eigenvalues at $k^{\,}_{1}=\pi/2$
from any one of the bulk bands and the mig-gap branch with
positive energy eigenvalue at $k^{\,}_{1}=\pi/2$.
The dependence of this direct gap
$\Delta^{\,}_{\mathrm{min}}(\pi/2,M^{\,}_{2})>0$
on $M^{\,}_{2}=16,32,48,64,80,96$ 
holding $M^{\,}_{1}=128$ fixed is plotted
in Fig~\ref{fig:spectral_flow}(b).
The fast decrease of
$\Delta^{\,}_{\mathrm{min}}(\pi/2,M^{\,}_{2})>0$
with increasing values of
$M^{\,}_{2}=16,32,48,64,80,96$ 
holding $M^{\,}_{1}=128$ fixed
is interpreted as the merging of the mid-gap branch into
the continuum of conduction bulk states in the 
quasi-one-dimensional limit $M^{\,}_{2}\to\infty$ 
holding $M^{\,}_{1}=128$ and
$\pi/2\leq k^{\,}_{1}\leq\pi$ fixed.
Similarly, one may verify that
the mig-gap branches merge into the valence and conduction
bulk continua in Figs.~\ref{fig: EEES_{2}D}(a-d)
for a non-vanishing interval of momenta $k^{\,}_{i}$ in the
quasi-one-dimensional limit
$M^{\,}_{i+1}\to\infty$ holding $M^{\,}_{i}$ fixed.
If we change the conserved momentum $k^{\,}_{i}$
adiabatically, say by imposing twisted boundary conditions
instead of periodic ones along the $i$ direction,
a charge can be transferred from the valence bulk continuum
to the conduction bulk continuum through the mid-gap branches.
This is an example of spectral flow induced by mid-gap states
crossing a bulk gap.

The same quasi-one-dimensional scaling analysis,
when performed on the spectra
$\sigma(\mathcal{H}^{\,}_{k_{2}\,\mathscr{R}^{\,}_{2};\mathscr{S}})$
and
$\sigma(Q^{\,}_{k^{\,}_{1}\,\mathscr{R}^{\,}_{2};\mathscr{S}\,A^{\,}_{2}})$,
delivers a very different result.
First, the mid-gap branches of $\mathcal{H}^{\,}_{k_{2}\,\mathscr{R}^{\,}_{2};\mathscr{S}}$
fail to cross. Second, the minimal value 
$\Delta^{\,}_{\mathrm{min}}(\pi,M^{\,}_{2})>0$
taken by the difference between the positive eigenvalues at $k^{\,}_{1}=\pi$
from any one of the bulk bands and the mig-gap branch with
positive energy eigenvalues at $k^{\,}_{1}=\pi$
converge to a non-vanishing value
upon increasing $M^{\,}_{2}=16,32,48,64,80,96$ 
holding $M^{\,}_{1}=128$ fixed 
in Fig~\ref{fig:spectral_flow}(c).
Instead of plotting the evidence for the saturation value
$\lim_{M^{\,}_{2}\to\infty}\Delta^{\,}_{\mathrm{min}}(\pi,M^{\,}_{2})>0$,
we plot in Fig~\ref{fig:spectral_flow}(d)
the entanglement spectrum 
\begin{equation}
\varpi^{\,}_{k^{\,}_{1}}:=
\ln\left(\frac{1}{\zeta^{\,}_{k^{\,}_{1}}}-1\right),
\end{equation}
which we already encountered in 
Eq.~(\ref{eq: relation sigma(CA) to sigma(H)}).

We now reproduce 
Figs~\ref{fig:spectral_flow}(c)
and
Fig~\ref{fig:spectral_flow}(d)
for $M^{\,}_{1}=64$ and $M^{\,}_{2}=128$ 
with the single change that we choose the geometry in which
it is $k^{\,}_{2}$ instead of $k^{\,}_{1}$ that is chosen to be
the good quantum number. There follows 
Figs~\ref{fig:spectral_flow}(e)
and
Fig~\ref{fig:spectral_flow}(f)
for
$\sigma(Q^{\,}_{k^{\,}_{2}\,\mathscr{R}^{\,}_{2};\mathscr{S}\,A^{\,}_{1}})$
and
$\sigma\big(\ln(Q^{-1}_{k^{\,}_{2}\,\mathscr{R}^{\,}_{2};\mathscr{S}\,A^{\,}_{1}}-1)\big)$,
respectively.
We have verified that, 
if we increase simultaneously 
$M^{\,}_{1}$ and $M^{\,}_{2}$, 
there is a level crossing between the mid-gap branches 
and the bulk branches in the neighborhood of $\pm\pi$.
The existence of this level crossing is implied by the
coloring of the bands in
Fig~\ref{fig:spectral_flow}(f).
This coloring has the following origin.

For any row from Table \ref{tab:4},
we may construct the operator
\begin{equation}
\Lambda^{\,}_{k^{\,}_{i}\,\mathscr{R}\,A^{\,}_{i+1}}:= 
\Gamma^{\,}_{k^{\,}_{i}\,\mathscr{R}}\, 
S^{\,}_{A^{\,}_{i+1}}
\label{eq: def Lambda example B}
\end{equation}
whenever the reflection $\mathscr{R}$ is block off-diagonal
with respect to the partition.
Here, $\Gamma^{\,}_{k^{\,}_{i}\,\mathscr{R}}$
was defined in Eq.~(\ref{eq: main result on how to get chiral sym}),
while the representation $\mathcal{S}$ of the chiral operation
$\mathscr{S}$ is diagonal in the partition with
the upper-left block $S^{\,}_{A^{\,}_{i+1}}$.
 
For the fifth row from Table \ref{tab:4},
$\Gamma^{\,}_{k^{\,}_{1}\,\mathscr{R}^{\,}_{2}}$ 
does not exist in the geometry that defines
$Q^{\,}_{k^{\,}_{1}\,\mathscr{R}^{\,}_{2};\mathscr{S}\,A^{\,}_{2}}$,
since $\mathscr{R}^{\,}_{2}$ is block diagonal for this partition.
For the fifth row from Table \ref{tab:4},
$\Gamma^{\,}_{k^{\,}_{2}\,\mathscr{R}^{\,}_{2}}$ 
exists in the geometry that defines
$Q^{\,}_{k^{\,}_{2}\,\mathscr{R}^{\,}_{2};\mathscr{S}\,A^{\,}_{1}}$
and anti-commutes with
$Q^{\,}_{k^{\,}_{2}\,\mathscr{R}^{\,}_{2};\mathscr{S}\,A^{\,}_{1}}$
according to Eq.~(\ref{eq: induced symmetry}),
since $\mathscr{R}^{\,}_{2}$ is block off-diagonal for this partition.
Hence, $\Lambda^{\,}_{k^{\,}_{2}\,\mathscr{R}^{\,}_{2}\,A^{\,}_{1}}$ commutes with
$Q^{\,}_{k^{\,}_{2}\,\mathscr{R}^{\,}_{2};\mathscr{S}\,A^{\,}_{1}}$
and we may associate to any eigenstate of
$Q^{\,}_{k^{\,}_{2}\,\mathscr{R}^{\,}_{2};\mathscr{S}\,A^{\,}_{1}}$
an eigenvalue of $\Lambda^{\,}_{k^{\,}_{2}\,\mathscr{R}^{\,}_{2}\,A^{\,}_{1}}$.
This eigenvalue of $\Lambda^{\,}_{k^{\,}_{2}\,\mathscr{R}^{\,}_{2}\,A^{\,}_{1}}$
is complex valued as 
$\Lambda^{\,}_{k^{\,}_{2}\,\mathscr{R}^{\,}_{2}\,A^{\,}_{1}}$
is not necessarily Hermitian. 
It turns out that 
$\Lambda^{\,}_{k^{\,}_{2}\,\mathscr{R}^{\,}_{2}\,A^{\,}_{1}}$
is purely imaginary according to our exact diagonalizations.
The coloring in Fig~\ref{fig:spectral_flow}(f)
corresponds to the sign of the imaginary eigenvalue of
$\Lambda^{\,}_{k^{\,}_{2}\,\mathscr{R}^{\,}_{2}\,A^{\,}_{1}}$.
This quantum number rules out an avoided level crossing 
of the mid-gap branches in the neighborhood of
$k^{\,}_{2}=0$ in Fig~\ref{fig:spectral_flow}(f).
This quantum number also rules out an avoided level crossing 
at the zone boundary $k^{\,}_{2}=\pi$. Inspection of the 
colored dispersions in Fig~\ref{fig:spectral_flow}(f)
suggests that twisting boundary conditions induces a spectral
flow from the valence to the conduction bands through the mid-gap
branches. This is the basis of the distinction between
crossings of mid-gap branches that are compatible with a spectral flow 
and are denoted by $\odot$
in Table \ref{tab:4}
and crossing of mid-gap branches that are not compatible with a spectral flow 
and are denoted by $\otimes$
in Table \ref{tab:4}.

In summary, we can understand the fifth row from Table \ref{tab:4}
(an all remaining rows with the same reasoning)
as follows. Because the reflection and chiral operations commute,
the topological index of a generic Hamiltonian
obeying these two symmetries belongs to $\mathbb{Z}$.
Protected edge states can but need not be present in the spectrum of
the Hamiltonian in a cylindrical geometry. 
Protected edge states arise if and only if the 
disconnected boundaries are not mixed by the reflection symmetry,
i.e., for $\mathcal{H}^{\,}_{k_{1}\,\mathscr{R}^{\,}_{2};\mathscr{S}}$
but not for $\mathcal{H}^{\,}_{k_{2}\,\mathscr{R}^{\,}_{2};\mathscr{S}}$.
On the other hand, the upper-left block
$Q^{\,}_{k^{\,}_{1}\,\mathscr{R}^{\,}_{2};\mathscr{S}\,A^{\,}_{2}}$
and
$Q^{\,}_{k^{\,}_{2}\,\mathscr{R}^{\,}_{2};\mathscr{S}\,A^{\,}_{1}}$
of the equal-time correlation matrix
both have robust mid-gap branches that cross.
However, it is only the mid-gap branches of
$Q^{\,}_{k^{\,}_{2}\,\mathscr{R}^{\,}_{2};\mathscr{S}\,A^{\,}_{1}}$
that support a spectral flow between the valence and conduction continua,
for it is only then that the reflection
$\mathscr{R}^{\,}_{2}$
is block off-diagonal in the partition defined by
$A^{\,}_{1}$. The fifth row from Table \ref{tab:4} demonstrates
that topological protection can be absent from the spectrum of the
Hamiltonian but present in the spectrum of the equal-time correlation matrix.

\section{Topological band insulator protected by two reflection symmetries
in two dimensions}
\label{sec: example C}

\subsection{Hamiltonian and topological quantum numbers}

Our third example is defined by choosing
$d=2$ and $N^{\,}_{\mathrm{orb}}=2$ in 
Eq.~(\ref{eq: def noninteractin fermion Hamiltonian}).
We consider the two-dimensional plane $z=0$ from
the three-dimensional Cartesian space
$\mathbb{R}^{3}:=\{x\,\bm{x}+y\,\bm{y}+z\,\bm{z}|x,y,z\in\mathbb{R}\}$
with the orthonormal basis $\bm{x}$, $\bm{y}$, and $\bm{z}$.
The single-particle Hamiltonian is defined by 
(the convention $e>0$ is used for the electron charge)
\begin{subequations}
\label{eq: def H example C}
\begin{equation}
\hat{\mathcal{H}}:=
\sum_{\sigma=\pm}
\frac{\hbar^{2}}{2m}
\left[
\hat{\bm{p}}\,\sigma^{\,}_{0}
-
\frac{e}{c}\,
\bm{A}(\hat{\bm{r}})\,\sigma^{\,}_{3}
\right]^{2}
\label{eq: def H example C a}
\end{equation}
where $\sigma^{\,}_{0}$ for the $2\times2$ unit matrix
and $\bm{\sigma}$ for the Pauli matrices are the usual suspects,  
\begin{equation}
\hat{\bm{p}}:=
\begin{pmatrix}
\hat{p}^{\,}_{x}
\\
\hat{p}^{\,}_{y}
\end{pmatrix}
\end{equation}
is the momentum operator in the plane $z=0$,
\begin{equation}
\hat{\bm{r}}:=
\begin{pmatrix}
\hat{r}^{\,}_{x}
\\
\hat{r}^{\,}_{y}
\end{pmatrix}
\end{equation}
is the position operator in the plane $z=0$,
and $\bm{A}(\hat{r})$ is the classical electromagnetic
vector potential such that
\begin{equation}
\left(
\bm{\nabla}\wedge\bm{A}
\right)(\bm{r})=
B\bm{z}
\end{equation}
\end{subequations}
is the uniform magnetic of magnitude $B>0$ pointing along $\bm{z}$.
This model describes an electron; with mass $m$, 
electric charge $-e$,
and the conserved spin quantum number along the quantization
axis defined by the eigenstates of the Pauli matrix $\sigma^{\,}_{3}$;
undergoing the spin quantum Hall effect generated by the matrix-valued
field $B\sigma^{\,}_{3}\bm{z}$.

The eigenvalue spectrum consists of 
Landau levels with the single-particle energy eigenvalues
\begin{subequations}
\begin{equation}
\varepsilon^{\,}_{n}=
\hbar\,\omega^{\,}_{\mathrm{c}}\,
\left(n+\frac{1}{2}\right),
\qquad 
n=0,1,2,\cdots,
\end{equation}
where
\begin{equation}
\omega^{\,}_{\mathrm{c}}:=
\frac{e\,B}{m\,c}
\end{equation}
\end{subequations}
is the cyclotron frequency. In addition to the usual orbital degeneracy
$\Phi/\Phi^{\,}_{0}$ of Landau levels with $\Phi$ the total magnetic flux
and $\Phi^{\,}_{0}:=h\,c/e$ the flux quantum, 
there is the Kramers' degeneracy under the composition
of complex conjugation and conjugation by $\sigma^{\,}_{2}$.

Each Landau level fully filled by 
$\Phi/\Phi^{\,}_{0}$
electrons with spin up carries the Chern number $C^{\,}_{+}=+1$.
Each Landau level fully filled by 
$\Phi/\Phi^{\,}_{0}$
electrons with spin down carries the Chern number $C^{\,}_{-}=-1$.
The contribution to the Hall conductivity from any completely filled
Landau level is therefore
\begin{subequations}
\begin{equation}
\sigma^{\,}_{\mathrm{QHE}}=
\left(
C^{\,}_{+}
+
C^{\,}_{-}
\right)
\frac{e^{2}}{h}=
0.
\end{equation}
The contribution to the spin Hall conductivity from any completely filled
Landau level is
\begin{equation}
\sigma^{\,}_{\mathrm{SQHE}}=
\frac{\hbar}{2e}
\left(
C^{\,}_{+}
-
C^{\,}_{-}
\right)
\frac{e^{2}}{h}=
\frac{e}{2\pi}.
\end{equation}
\end{subequations}

On an infinite strip along the $\bm{x}$ direction of width $L^{\,}_{y}$ 
along the $\bm{y}$ direction, we impose translation symmetry along the
$\bm{y}$ direction through periodic boundary conditions.
For this cylindrical geometry
(recall Fig.~\ref{fig: reflection planes cylinder}
with the identification $i=y$ and $i+1=x$),
it is convenient to choose the Landau gauge 
\begin{subequations}
\begin{equation}
\bm{A}:=
B
\begin{pmatrix}
0\\x
\end{pmatrix}
\end{equation}
for which normalized single-particle eigenstates 
$| n, \sigma, k^{\,}_{y} \rangle$
in the lowest Landau level $n=0$
can be represented by the normalized eigen-wavefunctions 
in the lowest ($n=0$) Landau levels
\begin{equation}
\begin{split}
\langle x,y|n=0,\sigma,k^{\,}_{y}\rangle\equiv%&\,
f^{(\sigma)}_{k^{\,}_{y}}(x,y)
%\\
=%&\,
\frac{
e^{+\mathrm{i}k^{\,}_{y}\,y}\, 
e^{-(x-\sigma\,\ell^{2}_{B}\,k^{\,}_{y})^{2}/(2\ell^{2}_{B})}
     }
     {
\left(\sqrt{\pi}\,\ell^{\,}_{B}\,L^{\,}_{y}\right)^{1/2}
     }
\end{split}
\end{equation}
for a spin $\sigma$ electron. Here,
$\sigma=\pm$ is the spin of the electron along the quantization axis, 
$\bm{r}=(x,y)\in\mathbb{R}^{2}$ is the coordinate of the electron, and
\begin{equation}
k^{\,}_{y}=
\frac{2\pi}{L^{\,}_{y}}\,m,
\qquad 
m=1,\cdots,\frac{\Phi}{\Phi^{\,}_{0}},
\end{equation}
is the quantized wave number arising from the compactification along
the width $L^{\,}_{y}$ of the strip and
\begin{equation}
\ell^{\,}_{B}:=
\sqrt{\frac{\hbar\,c}{e\,B}}
\end{equation}
\end{subequations}
is the magnetic length that we shall set to unity from now on.

\subsection{Symmetries}

The symmetries of the single-particle Hamiltonian defined by
Eq.~(\ref{eq: def H example C}) are the following.

\begin{enumerate}

\item
Let $\mathscr{R}^{\,}_{x}$ define the reflection about the 
horizontal axis $\bm{r}=(x,0)$ by which the quantum numbers
of the position and momentum operators transform according to
\begin{subequations}
\label{eq: def reflection x example C}
\begin{equation}
\begin{split}
&
\mathscr{R}^{\,}_{x}:x\mapsto+x,
\qquad
\mathscr{R}^{\,}_{x}:y\mapsto-y,
\\
&
\mathscr{R}^{\,}_{x}:k^{\,}_{x}\mapsto+k^{\,}_{x},
\qquad
\mathscr{R}^{\,}_{x}:k^{\,}_{y}\mapsto-k^{\,}_{y},
\end{split}
\label{eq: def reflection x example C a}
\end{equation}
respectively.
The symmetry
\begin{equation}
\hat{\mathcal{O}}^{\dag}_{\mathscr{R}^{\,}_{x}}\,
\hat{\mathcal{H}}\,
\hat{\mathcal{O}}^{\,}_{\mathscr{R}^{\,}_{x}}=
\hat{\mathcal{H}}
\label{eq: def reflection x example C b}
\end{equation}
then holds for the single-particle Hamiltonian~(\ref{eq: def H example C a})
if $\mathscr{R}^{\,}_{x}$ is represented by
the composition of $x\mapsto+x$, $y\mapsto-y$
with the conjugation by either
$\sigma^{\,}_{x}$ or $\sigma^{\,}_{y}$.
We choose the representation
\begin{equation}
\hat{\mathcal{O}}^{\,}_{\mathscr{R}^{\,}_{x}}\chi^{\,}_{k^{\,}_{y}}(x,y):=
\sigma^{\,}_{x}\,\chi^{\,}_{-k^{\,}_{y}}(x,-y)
\label{eq: def reflection x example C c}
\end{equation}
\end{subequations}
for any two-component spinor $\chi^{\,}_{k^{\,}_{y}}(x,y)\in\mathbb{C}^{2}$.

\item
Let $\mathscr{R}^{\,}_{y}$ define the reflection about the 
vertical axis $\bm{r}=(0,y)$ by which the quantum numbers
of the position and momentum operators transform according to
\begin{subequations}
\label{eq: def reflection y example C}
\begin{equation}
\begin{split}
&
\mathscr{R}^{\,}_{x}:x\mapsto-x,
\qquad
\mathscr{R}^{\,}_{x}:y\mapsto+y,
\\
&
\mathscr{R}^{\,}_{x}:k^{\,}_{x}\mapsto-k^{\,}_{x},
\qquad
\mathscr{R}^{\,}_{x}:k^{\,}_{y}\mapsto+k^{\,}_{y},
\end{split}
\label{eq: def reflection x example C a}
\end{equation}
respectively.
The symmetry
\begin{equation}
\hat{\mathcal{O}}^{\dag}_{\mathscr{R}^{\,}_{y}}\,
\hat{\mathcal{H}}\,
\hat{\mathcal{O}}^{\,}_{\mathscr{R}^{\,}_{y}}=
\hat{\mathcal{H}}
\label{eq: def reflection x example C b}
\end{equation}
then holds for the single-particle Hamiltonian~(\ref{eq: def H example C a})
if $\mathscr{R}^{\,}_{y}$ is represented by
the composition of $x\mapsto-x$, $y\mapsto+y$
with the conjugation by 
$\sigma^{\,}_{x}$, given the choice made in
Eq.~(\ref{eq: def reflection x example C c}),
i.e.,
\begin{equation}
\hat{\mathcal{O}}^{\,}_{\mathscr{R}^{\,}_{y}}\chi^{\,}_{k^{\,}_{y}}(x,y):=
\sigma^{\,}_{x}\,\chi^{\,}_{k^{\,}_{y}}(-x,y)
\label{eq: def reflection y example C c}
\end{equation}
\end{subequations}
for any two-component spinor $\chi^{\,}_{k^{\,}_{y}}(x,y)\in\mathbb{C}^{2}$.
With this choice, the commutation relation
$[\mathscr{R}^{\,}_{x},\mathscr{R}^{\,}_{y}]=0$
is faithfully represented, i.e.,
\begin{equation}
[\hat{\mathcal{O}}^{\,}_{\mathscr{R}^{\,}_{x}},
\hat{\mathcal{O}}^{\,}_{\mathscr{R}^{\,}_{y}}]=0.
\end{equation}

\item
Let $\mathscr{T}$ define reversal of time by which time and
the quantum numbers
of the position and momentum operators transform according to
\begin{subequations}
\label{eq: def TR example C}
\begin{equation}
\begin{split}
&
\mathscr{T}:t\mapsto-t,
\\
&
\mathscr{R}^{\,}_{x}:x\mapsto x,
\qquad
\mathscr{R}^{\,}_{x}:y\mapsto y,
\\
&
\mathscr{R}^{\,}_{x}:k^{\,}_{x}\mapsto-k^{\,}_{x},
\qquad
\mathscr{R}^{\,}_{x}:k^{\,}_{y}\mapsto-k^{\,}_{y},
\end{split}
\label{eq: def TR example C a}
\end{equation}
respectively.
The symmetry
\begin{equation}
\hat{\mathcal{O}}^{\dag}_{\mathscr{T}}\,
\hat{\mathcal{H}}\,
\hat{\mathcal{O}}^{\,}_{\mathscr{T}}=
\hat{\mathcal{H}}
\label{eq: def TR example C b}
\end{equation}
then holds for the single-particle Hamiltonian~(\ref{eq: def H example C a})
if $\mathscr{T}$ is represented by
the composition of charge conjugation
with the conjugation by
$\sigma^{\,}_{x}$,
i.e.,
\begin{equation}
\hat{\mathcal{O}}^{\,}_{\mathscr{T}}\chi^{\,}_{k^{\,}_{y}}(x,y):=
\sigma^{\,}_{x}\,\chi^{*}_{-k^{\,}_{y}}(x,y)
\label{eq: def TR example C c}
\end{equation}
\end{subequations}
for any two-component spinor $\chi^{\,}_{k^{\,}_{y}}(x,y)\in\mathbb{C}^{2}$.
With this choice, the commutation relations
$[\mathscr{R}^{\,}_{x},\mathscr{T}]=[\mathscr{R}^{\,}_{y},\mathscr{T}]=0$
are faithfully represented, i.e.,
\begin{equation}
[\hat{\mathcal{O}}^{\,}_{\mathscr{T}},
\hat{\mathcal{O}}^{\,}_{\mathscr{R}^{\,}_{x}}]=
[\hat{\mathcal{O}}^{\,}_{\mathscr{T}},
\hat{\mathcal{O}}^{\,}_{\mathscr{R}^{\,}_{y}}]=0.
\end{equation}
Observe that had we implemented the two reflections by choosing
$\sigma^{\,}_{y}$ instead of $\sigma^{\,}_{x}$
on the right-hand sides of 
Eqs.~(\ref{eq: def reflection x example C c})
and (\ref{eq: def reflection y example C c}), 
then the transformation
of reversal of time would anti-commute with both reflections.

\end{enumerate}

\begin{figure}[t]
\centering
\includegraphics[width=0.4 \textwidth]{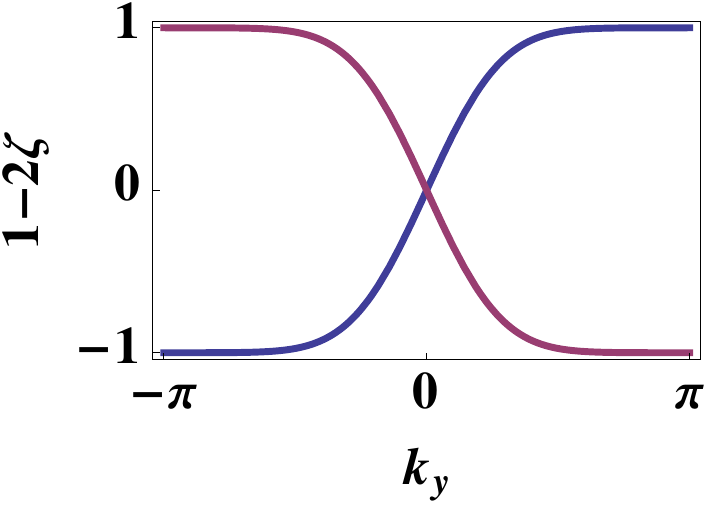}
\caption{(Color online)
The entanglement spectrum $\sigma(\hat{Q}^{\,}_{k^{\,}_{y}\,A})$
with $\hat{Q}^{\,}_{k^{\,}_{y}\,A}$ defined by 
Eqs.~(\ref{eq: zetapm(ky) exmaple C})
and~(\ref{eq: def hat Q ky A example C})
for the spin quantum Hall effect.
There are two branches of eigenstates with opposite chiralities
(these chiralities are denoted by the colors blue and red) 
that cross at vanishing energy and momentum.
        }
\label{Fig: SQHE}
\end{figure}

\subsection{Partition and zero modes}
\label{subsec: Partition and zero modes example C}

The counterpart to the projection of $\hat{\psi}^{\dag}_{I}\,\hat{\psi}^{\,}_{J}$
in the equal-time one-point correlation matrix%
~(\ref{eq: def matrix elements C})
onto the completely filled Landau level $n=0$
is the operator
\begin{subequations}
\label{eq: def hat mathcal C example C}
\begin{equation}
\hat{\mathcal{C}}=
\sum_{\sigma=\pm}
\sum_{L^{\,}_{y}\,k^{\,}_{y}/(2\pi)=1}^{\Phi/\Phi^{\,}_{0}}
|n=0,\sigma,k^{\,}_{y}\rangle\langle n=0,\sigma,k^{\,}_{y}|
\end{equation}
with the matrix element
\begin{equation}
\begin{split}
\mathcal{C}^{\,}_{\sigma,k^{\,}_{y}}(x,x')
\equiv%&\,
\langle n=0,\sigma,k^{\,}_{y},x|
\hat{\mathcal{C}}
|n=0,\sigma,k^{\,}_{y},x'\rangle
%\\
=%&\,
\frac{1}{\sqrt{\pi}}\,
e^{
-
\frac{1}{2}
\left[
(x-\sigma\,k^{\,}_{y})^{2}
+
(x'-\sigma\,k^{\,}_{y})^{2}
\right]
  }
\end{split}
\end{equation}
\end{subequations}
for any $x,x'\in\mathbb{R}$. 
One verifies that the $2\times2$ matrix
\begin{subequations}
\begin{equation}
\hat{\mathcal{C}}^{\,}_{k^{\,}_{y}}(x,x'):=
\mathcal{C}^{\,}_{+,k^{\,}_{y}}(x,x')\,
\sigma^{\,}_{11}
+
\mathcal{C}^{\,}_{-,k^{\,}_{y}}(x,x')\,
\sigma^{\,}_{22},
\end{equation}
where 
$\sigma^{\,}_{11}:=(\sigma^{\,}_{0}+\sigma^{\,}_{3})/2$
and
$\sigma^{\,}_{22}:=(\sigma^{\,}_{0}-\sigma^{\,}_{3})/2$,
obeys
\begin{align}
&
\hat{\mathcal{O}}^{\dag}_{\mathscr{R}^{\,}_{x}}\,
\hat{\mathcal{C}}^{\,}_{-k^{\,}_{y}}(x,x')\,
\hat{\mathcal{O}}^{\,}_{\mathscr{R}^{\,}_{x}}=
\hat{\mathcal{C}}^{\,}_{k^{\,}_{y}}(x,x'),
\label{eq: Rx hat cal C Rx example C}
\\
&
\hat{\mathcal{O}}^{\dag}_{\mathscr{R}^{\,}_{y}}\,
\hat{\mathcal{C}}^{\,}_{k^{\,}_{y}}(-x,-x')\,
\hat{\mathcal{O}}^{\,}_{\mathscr{R}^{\,}_{y}}=
\hat{\mathcal{C}}^{\,}_{k^{\,}_{y}}(x,x'),
\label{eq: Ry hat cal C Ry example C}\\
&
\hat{\mathcal{O}}^{\dag}_{\mathscr{T}}\,
\hat{\mathcal{C}}^{\,}_{-k^{\,}_{y}}(x,x')\,
\hat{\mathcal{O}}^{\,}_{\mathscr{T}}=
\hat{\mathcal{C}}^{\,}_{k^{\,}_{y}}(x,x').
\label{eq: T hat cal C T example C}
\end{align}
\end{subequations}

We define the partition
\begin{subequations}
\label{eq: def partition example C}
\begin{equation}
A:=
\left\{
\sigma,
k^{\,}_{y},
x
\left|
\sigma=\pm,
\,
k^{\,}_{y}=
\frac{2\pi}{L^{\,}_{y}},
\cdots,
\frac{2\pi\,\Phi}{L^{\,}_{y}\,\Phi^{\,}_{0}},
\,
x<0
\right.
\right\}
\end{equation}
and
\begin{equation}
B:=
\left\{
\sigma,
k^{\,}_{y},
x
\left|
\sigma=\pm,
\,
k^{\,}_{y}=
\frac{2\pi}{L^{\,}_{y}},
\cdots,
\frac{2\pi\,\Phi}{L^{\,}_{y}\,\Phi^{\,}_{0}},
\,
x>0
\right.
\right\}
\end{equation}
\end{subequations}
for the single-particle labels.
The restriction of the equal-time one-point correlation operator
defined by Eq.~(\ref{eq: def hat mathcal C example C})
to $x<0$ and $x'<0$ defines the operator
$\hat{C}^{\,}_{A}$
with its matrix elements
$C^{\,}_{\sigma,k^{\,}_{y}\,A}(x,x')$
and the $2\times2$ matrix
$\hat{C}^{\,}_{k^{\,}_{y}\,A}(x,x')$.
The restriction of the equal-time one-point correlation operator
defined by Eq.~(\ref{eq: def hat mathcal C example C})
to $x>0$ and $x'>0$ defines the operator
$\hat{C}^{\,}_{B}$
with its matrix elements
$C^{\,}_{\sigma,k^{\,}_{y}\,B}(x,x')$
and the $2\times2$ matrix
$\hat{C}^{\,}_{k^{\,}_{y}\,B}(x,x')$.
With this partition, for any $x,x'<0$,
\begin{subequations}
\begin{align}
&
\hat{\mathcal{O}}^{\dag}_{\mathscr{R}^{\,}_{x}}\,
\hat{C}^{\,}_{-k^{\,}_{y}\,A}(x,x')\,
\hat{\mathcal{O}}^{\,}_{\mathscr{R}^{\,}_{x}}=
\hat{C}^{\,}_{k^{\,}_{y}\,A}(x,x'),
\label{eq: Rx hat CA Rx example C}
\\
&
\hat{\mathcal{O}}^{\dag}_{\mathscr{R}^{\,}_{y}}\,
\hat{C}^{\,}_{k^{\,}_{y}\,A}(x,x')\,
\hat{\mathcal{O}}^{\,}_{\mathscr{R}^{\,}_{y}}=
\hat{C}^{\,}_{k^{\,}_{y}\,B}(-x,-x'),
\label{eq: Ry hat CA Ry example C}
\\
&
\hat{\mathcal{O}}^{\dag}_{\mathscr{T}}\,
\hat{C}^{\,}_{-k^{\,}_{y}\,A}(x,x')\,
\hat{\mathcal{O}}^{\,}_{\mathscr{T}}=
\hat{C}^{\,}_{k^{\,}_{y}\,A}(x,x').
\label{eq: T hat CA T example C}
\end{align}
\end{subequations}
The same equations hold if we do $x\to-x$, $x'\to-x'$,
$A\to B$, and $B\to A$. 
The restriction of the equal-time one-point correlation operator
defined by Eq.~(\ref{eq: def hat mathcal C example C})
to $x<0$ and $x'>0$ defines the operator
$\hat{C}^{\,}_{AB}$
with its matrix elements
$C^{\,}_{\sigma,k^{\,}_{y}\,AB}(x,x')$
and the $2\times2$ matrix
$\hat{C}^{\,}_{k^{\,}_{y}\,AB}(x,x')$.
Finally,
the restriction of the equal-time one-point correlation operator
defined by Eq.~(\ref{eq: def hat mathcal C example C})
to $x>0$ and $x'<0$ defines the operator
$\hat{C}^{\,}_{BA}=\hat{C}^{\dag}_{AB}$
with its matrix elements
$C^{\,}_{\sigma,k^{\,}_{y}\,BA}(x,x')=C^{\dag}_{\sigma,k^{\,}_{y}\,AB}(x',x)$
and the $2\times2$ matrix
$\hat{C}^{\,}_{k^{\,}_{y}\,BA}(x,x')=\hat{C}^{\dag}_{k^{\,}_{y}\,AB}(x',x)$.

The normalizable eigenstates of $\hat{C}^{\,}_{k^{\,}_{y}\,A}$ 
are represented by 
\begin{equation}
\phi^{\,}_{+,k^{\,}_{y}\,}(x)=
\frac{1}{\sqrt{2}\,\pi^{1/4}}\,
\begin{pmatrix}
e^{-\frac{1}{2}(x-k^{\,}_{y})^{2}}      
\\
0
\end{pmatrix}
\end{equation}
and
\begin{equation}
\phi^{\,}_{-,k^{\,}_{y}\,}(x)=
\frac{1}{\sqrt{2}\,\pi^{1/4}}\,
\begin{pmatrix}
0
\\
e^{-\frac{1}{2}(x+k^{\,}_{y})^{2}}      
\end{pmatrix}
\end{equation}
for any $-\infty<x\leq0$.
In the thermodynamic limit
$L^{\,}_{y}\to\infty$ 
holding the electron density fixed, 
the corresponding eigenvalues are given by
\begin{align}
\zeta^{\,}_{\pm}(k^{\,}_{y})=&\,
\frac{1}{\sqrt{\pi}}\,
\int\limits_{-\infty}^{0}\mathrm{d}x\,
e^{-(x\mp k^{\,}_{y})^{2}}
=:
\frac{1}{2}
\left[
1
-
\mathrm{erf}(\pm k^{\,}_{y})
\right].
\label{eq: zetapm(ky) exmaple C}
\end{align}
The spectrum 
$1-2\,\zeta^{\,}_{\pm}(k^{\,}_{y})$
of 
\begin{equation}
\hat{Q}^{\,}_{k^{\,}_{y}\,A}:=
\hat{%\openone
\mathbb{I}}-2\,\hat{C}^{\,}_{k^{\,}_{y}\,A}
\label{eq: def hat Q ky A example C}
\end{equation}
is shown in
Fig.\ \ref{Fig: SQHE}
Two chiral edge modes of opposite chirality
are seen crossing the spectrum 
with a crossing at vanishing entanglement eigenvalue. 

On the one hand,
Eqs.~(\ref{eq: Rx hat CA Rx example C})
and~(\ref{eq: T hat CA T example C})
imply that the reflection about the horizontal axis%
~(\ref{eq: def reflection x example C}) 
and reversal of time%
~(\ref{eq: def TR example C})
are represented by the operators
$\hat{\mathcal{O}}^{\,}_{\mathscr{R}^{\,}_{x}}$
and
$\hat{\mathcal{O}}^{\,}_{\mathscr{T}}$,
respectively,
that are block diagonal with respect to the partition
of the single-particle labels into the sets $A$ and $B$ defined by
Eq.~(\ref{eq: def partition example C}).
As explained in Eqs.~(\ref{eq: assumption cal O block diagonal}) 
and (\ref{eq: O operates on QA if block diagonal}),
the upper-left block~(\ref{eq: def hat Q ky A example C})
inherits these symmetries from $\hat{\mathcal{H}}$
defined in Eq.~(\ref{eq: def H example C}).

On the one hand,
Eq.~(\ref{eq: Ry hat CA Ry example C})
implies that
$\hat{\mathcal{O}}^{\,}_{\mathscr{R}^{\,}_{y}}$
is block off diagonal with respect to the partition
of the single-particle labels into the sets $A$ and $B$ defined by
Eq.~(\ref{eq: def partition example C}).
The counterparts to
Eqs.~(\ref{eq: cal O off block diagonal})
and~(\ref{eq: condition cal O dag cal C cal O= cal c})
then hold,
so that we may apply the counterpart to the spectral symmetry%
~(\ref{eq: main result on how to get chiral sym}).

To proceed, 
for any $x,x'<0$ and 
$k^{\,}_{y},k^{\prime}_{y}=
2\pi/ L^{\,}_{y},\cdots,(2\pi \Phi)/(L^{\,}_{y}\Phi^{\,}_{0})$,
we make use of the representations
\begin{equation}
\begin{split}
\hat{O}^{\,}_{k^{\,}_{y},k^{\prime}_{y}\,\mathscr{R}^{\,}_{x}\,A}(x,x'):=%&\,
\langle k^{\,}_{y},x|
\hat{\mathcal{O}}^{\,}_{\mathscr{R}^{\,}_{x}}
|k^{\prime}_{y},x'\rangle
%\\
=%&\,
\delta^{\,}_{k^{\,}_{y}+k^{\prime}_{y},0}\,
\delta(x-x')\,
\sigma^{\,}_{x},
\end{split}
\end{equation}
for the reflection about the horizontal axis,
\begin{equation}
\begin{split}
\hat{O}^{\,}_{k^{\,}_{y},k^{\prime}_{y}\,\mathscr{R}^{\,}_{y}\,A}(x,x'):=%&\,
\langle k^{\,}_{y},x|
\hat{\mathcal{O}}^{\,}_{\mathscr{R}^{\,}_{y}}
|k^{\prime}_{y},x'\rangle
%\\
=%&\,
\delta^{\,}_{k^{\,}_{y},k^{\prime}_{y}}\,
\delta(x+x')\,
\sigma^{\,}_{x},
\end{split}
\end{equation}
for the reflection about the vertical axis, and
\begin{equation}
\begin{split}
\hat{O}^{\,}_{k^{\,}_{y},k^{\prime}_{y}\,\mathscr{T}\,A}(x,x'):=\,
\langle k^{\,}_{y},x|
\hat{\mathcal{O}}^{\,}_{\mathscr{T}}
|k^{\prime}_{y},x'\rangle
=
\delta^{\,}_{k^{\,}_{y}+k^{\prime}_{y},0}\,
\delta(x-x')\,
\sigma^{\,}_{y}\,K,
\end{split}
\end{equation}
for the reversal of time ($K$ denotes complex conjugation).
%\begin{widetext}
Consequently,
the upper-left block~(\ref{eq: def hat Q ky A example C})
obeys the symmetries
\begin{subequations}
\label{eq: symmetry Q A example C}
\begin{align}
\hat{Q}^{\,}_{k^{\,}_{y}\,A}(x,x')=&\,
\sum_{k^{\prime\prime}_{y}}
\sum_{k^{\prime\prime\prime}_{y}}
\int\limits_{-\infty}^{0}\mathrm{d}x''
\int\limits_{-\infty}^{0}\mathrm{d}x'''\,
\nonumber\\
&\,
\hat{O}^{\dag}_{k^{\,}_{y},k^{\prime\prime}_{y}\,\mathscr{R}^{\,}_{x}\,A}(x,x'')\,
\delta^{\,}_{k^{\prime\prime}_{y},k^{\prime\prime\prime}_{y}}\,
\hat{Q}^{\,}_{k^{\prime\prime\prime}_{y}\,A}(x'',x''')
\hat{O}^{\,}_{k^{\prime\prime\prime}_{y},k^{\prime}_{y}\,\mathscr{R}^{\,}_{x}\,A}(x''',x')
\nonumber\\
=&\,
\sigma^{\,}_{x}\,\hat{Q}^{\,}_{-k^{\,}_{y}\,A}(x,x')\,\sigma^{\,}_{x}
\label{eq: symmetry Q A example C a}
\end{align}
under reflection about the horizontal axis and
\begin{align}
\hat{Q}^{\,}_{k^{\,}_{y}\,A}(x,x')=&\,
\sum_{k^{\prime\prime}_{y}}
\sum_{k^{\prime\prime\prime}_{y}}
\int\limits_{-\infty}^{0}\mathrm{d}x''
\int\limits_{-\infty}^{0}\mathrm{d}x'''\,
\nonumber\\
&\,
\hat{O}^{\dag}_{k^{\,}_{y},k^{\prime\prime}_{y}\,\mathscr{T}\,A}(x,x'')\,
\delta^{\,}_{k^{\prime\prime}_{y},k^{\prime\prime\prime}_{y}}\,
\hat{Q}^{\,}_{k^{\prime\prime\prime}_{y}\,A}(x'',x''')
\hat{O}^{\,}_{k^{\prime\prime\prime}_{y},k^{\prime}_{y}\,\mathscr{T}\,A}(x''',x')
\nonumber\\
=&\,
\sigma^{\,}_{y}\,\hat{Q}^{*}_{-k^{\,}_{y}\,A}(x,x')\,\sigma^{\,}_{y}
\label{eq: symmetry Q A example C b}
\end{align}
\end{subequations}
under reversal of time. 
The symmetry of the Hamiltonian~(\ref{eq: def H example C a})
under reflection about the vertical axis is turned into the spectral
symmetry
\begin{subequations}
\label{eq: spectral symmetry Q A example C}
\begin{align}
-\hat{Q}^{\,}_{k^{\,}_{y}\,A}(x,x')=&\,
\sum_{k^{\prime\prime}_{y}}
\sum_{k^{\prime\prime\prime}_{y}}
\int\limits_{-\infty}^{0}\mathrm{d}x''
\int\limits_{-\infty}^{0}\mathrm{d}x'''\,
\nonumber\\
&\,
\hat{\Gamma}^{\dag}_{k^{\,}_{y},k^{\prime\prime}_{y}\,\mathscr{R}^{\,}_{y}\,A}(x,x'')\,
\delta^{\,}_{k^{\prime\prime}_{y},k^{\prime\prime\prime}_{y}}\,
\hat{Q}^{\,}_{k^{\prime\prime\prime}_{y}\,A}(x'',x''')\,
\hat{\Gamma}^{\,}_{k^{\prime\prime\prime}_{y},k^{\prime}_{y}\,\mathscr{R}^{\,}_{y}\,A}(x''',x'),
\label{eq: spectral symmetry Q A example C a}
\end{align}
where
\begin{equation}
\begin{split}
\hat{\Gamma}^{\,}_{k^{\,}_{y},k^{\prime}_{y}\,\mathscr{R}^{\,}_{y}\,A}(x,x')=&\,
\sum_{k^{\prime\prime}_{y}}
\int\limits_{0}^{+\infty}\mathrm{d}x''\,
\delta^{\,}_{k^{\,}_{y},k^{\prime\prime}_{y}}\,
\hat{C}^{\,}_{k^{\,}_{y}\,AB}(x,x'')\,
\hat{O}^{\dag}_{k^{\prime\prime}_{y},k^{\prime}_{y}\,\mathscr{R}^{\,}_{y}\,A}(x'',x')
\\
=&\,
\delta^{\,}_{k^{\,}_{y},k^{\prime}_{y}}\,
\frac{1}{\sqrt{\pi}}
\left[
e^{
-
\frac{1}{2}
\left[
(x-k^{\,}_{y})^{2}
+
(x'+k^{\,}_{y})^{2}
\right]
  }\,
\sigma^{\,}_{12}
+
e^{
-
\frac{1}{2}
\left[
(x+k^{\,}_{y})^{2}
+
(x'-k^{\,}_{y})^{2}
\right]
  }\,
\sigma^{\,}_{21}
\right]
\end{split}
\label{eq: spectral symmetry Q A example C b}
\end{equation}
\end{subequations}
with 
$\sigma^{\,}_{12}:=(\sigma^{\,}_{1}+\mathrm{i}\sigma^{\,}_{2})/2$
and
$\sigma^{\,}_{21}:=(\sigma^{\,}_{1}-\mathrm{i}\sigma^{\,}_{2})/2$.
For any $x<0$ and 
$k^{\,}_{y}=
2\pi/ L^{\,}_{y},\cdots,(2\pi \Phi)/(L^{\,}_{y}\Phi^{\,}_{0})$,
one verifies the eigenvalue equation
\begin{equation}
\sum_{k^{\prime}_{y}}
\int\limits_{-\infty}^{0}\mathrm{d}x'\,
\hat{\Gamma}^{\,}_{k^{\,}_{y},k^{\prime}_{y}\,\mathscr{R}^{\,}_{y}\,A}(x,x')\,
\phi^{\,}_{\pm,k^{\prime}_{y}}(x')  
=
\frac{1}{2}\,
[1-\mathrm{erf}(k^{\,}_{y})]\,
\phi^{\,}_{\mp,k^{\,}_{y}}(x).
\end{equation}
%\end{widetext}

\subsection{Stability analysis of the zero modes}

We are after the stability of the crossings at vanishing energy and momentum
that characterizes the edge states in the Hamiltonian 
defined by Eq.~(\ref{eq: def H example C}) for a cylindrical geometry
and the states in the entanglement spectra localized 
on the entangling boundary at $x=0$ defined in
Sec.~\ref{subsec: Partition and zero modes example C}. 
We are going to show that the reflection symmetries 
$\mathscr{R}^{\,}_{x}$
and
$\mathscr{R}^{\,}_{y}$
do not protect the edge states
on the physical boundaries, but they protect the edge states on the
entangling boundary.

\subsubsection{Hamiltonian spectrum}
If we linearize the edge state spectrum close to vanishing energy and momentum,
we get the effective Hamiltonian (the group velocity has been set to one)
\begin{equation}
\hat{\mathcal{H}}^{\,}_{\mathrm{edges}}(k^{\,}_{y})\approx
k^{\,}_{y}\, Y^{\,}_{33}
\end{equation}
in the cylindrical geometry of 
Fig.~\ref{fig: reflection planes cylinder}
($i=y$ and $i+1=x$), 
where we have introduced a second set of $2\times2$ matrices 
generated by the unit $2\times2$ matrix $\rho^{\,}_{0}$ an the Pauli
matrices $\rho^{\,}_{1}$, $\rho^{\,}_{2}$, and $\rho^{\,}_{3}$
when defining 
\begin{equation}
Y^{\,}_{\mu\nu}:=\sigma^{\,}_{\mu}\otimes\rho^{\,}_{\nu}
\end{equation}
for $\mu,\nu=0,1,2,3$.
The matrices $\rho^{\,}_{0}$,
$\rho^{\,}_{1}$, 
$\rho^{\,}_{2}$, 
and $\rho^{\,}_{3}$
encode the mixing of edge states 
localized on opposite edges of the cylinder.
Reflection about the $x$ axis,
reflection about the $y$ axis,
and reversal of time are represented by
\begin{subequations}
\label{eq: def edges symmetries Hamiltonian example C}
\begin{align}
&
\hat{\mathcal{O}}^{\,}_{\mathrm{edges}\,\mathscr{R}^{\,}_{x}}
(k^{\,}_{y},k^{\prime}_{y}):=
\delta(k^{\,}_{y}+k^{\prime}_{y})\,
Y^{\,}_{10},
\\
&
\hat{\mathcal{O}}^{\,}_{\mathrm{edges}\,\mathscr{R}^{\,}_{y}}
(k^{\,}_{y},k^{\prime}_{y}):=
\delta(k^{\,}_{y}-k^{\prime}_{y})\,
Y^{\,}_{11},
\\
&
\hat{\mathcal{O}}^{\,}_{\mathrm{edges}\,\mathscr{T}}
(k^{\,}_{y},k^{\prime}_{y}):=
\delta(k^{\,}_{y}+k^{\prime}_{y})\,
Y^{\,}_{20}\,K,
\end{align}
\end{subequations}
where $K$ denotes charge conjugation. In a gradient expansion,
any perturbation of the form
\begin{equation}
\hat{\mathcal{V}}^{\,}_{\mathrm{edges}\,\mu\nu}=
v^{\,}_{\mu\nu}\,
Y^{\,}_{\mu\nu}
\end{equation}
for some energy scale $v^{\,}_{\mu\nu}\in\mathbb{R}$ that
is smaller than the cyclotron energy,
commutes with the three transformations%
~(\ref{eq: def edges symmetries Hamiltonian example C}),
and anti-commutes with $Y^{\,}_{33}$, 
say $\hat{\mathcal{V}}^{\,}_{\mathrm{edges}\,01}$,
opens a spectral gap on the edges.

In the thermodynamic limit by which the length of the cylinder
is taken to infinity keeping the density of electrons fixed,
there is no mixing between edge states localized on opposite boundaries
of the cylinder. The effective Hamiltonian on a single edge becomes
\begin{equation}
\hat{\mathcal{H}}^{\,}_{\mathrm{edge}}(k^{\,}_{y})\approx
k^{\,}_{y}\, \sigma^{\,}_{3}.
\end{equation}
It commutes with
\begin{subequations}
\label{eq: def edge symmetries Hamiltonian example C}
\begin{align}
&
\hat{\mathcal{O}}^{\,}_{\mathrm{edge}\,\mathscr{R}^{\,}_{x}}
(k^{\,}_{y},k^{\prime}_{y}):=
\delta(k^{\,}_{y}+k^{\prime}_{y})\,
\sigma^{\,}_{1},
\label{eq: def edge symmetries Hamiltonian example C a}
\\
&
\hat{\mathcal{O}}^{\,}_{\mathrm{edge}\,\mathscr{T}}
(k^{\,}_{y},k^{\prime}_{y}):=
\delta(k^{\,}_{y}+k^{\prime}_{y})\,
\sigma^{\,}_{2}\,K,
\label{eq: def edge symmetries Hamiltonian example C b}
\end{align}
\end{subequations}
where $K$ denotes charge conjugation.
In a gradient expansion,
a generic perturbation on this single edge is of the form
\begin{equation}
\hat{\mathcal{V}}^{\,}_{\mathrm{edge}}=
\sum_{\mu=0}^{3}
v^{\,}_{\mu}\,
\sigma^{\,}_{\mu}
\label{eq: mathcal V edge example C}
\end{equation}
with any $v^{\,}_{\mu}\in\mathbb{R}$ much smaller than the bulk gap.

Imposing reflection symmetry
enforces the conditions $v^{\,}_{2}=v^{\,}_{3}=0$ on the perturbation
$\hat{\mathcal{V}}^{\,}_{\mathrm{edge}}$, i.e., 
a reflection symmetric perturbation can open a gap at the crossing
in the unperturbed spectrum. The crossing in the 
unperturbed spectrum of a single edge is thus not protected by having
two reflection symmetries in the bulk Hamiltonian.
This is not true anymore for the entanglement spectrum.

Imposing time-reversal symmetry
enforces the conditions $v^{\,}_{1}=v^{\,}_{2}=v^{\,}_{3}=0$ on the perturbation
$\hat{\mathcal{V}}^{\,}_{\mathrm{edge}}$, i.e., 
a time-reversal symmetric perturbation does not 
destroy the crossing, 
it merely shifts the crossing in the unperturbed spectrum
at vanishing momentum to a non-vanishing energy.

\subsubsection{Entanglement spectrum}

In the thermodynamic limit by which the length of the cylinder
is taken to infinity keeping the density of electrons fixed,
there is no mixing between states localized on the physical boundaries
at $x=\pm\infty$ and the entangling boundary at $x=0$.
The entangling edge states close to the crossing at vanishing
entangling eigenvalue and momentum of Fig.~\ref{Fig: SQHE}
are governed by the effective Hamiltonian
(the group velocity has been set to one)
\begin{equation}
\hat{\mathcal{Q}}^{\,}_{\mathrm{edge}\,A}(k^{\,}_{y})\approx
k^{\,}_{y}\, \sigma^{\,}_{3}.
\end{equation}
The representations of the two symmetry transformations%
~(\ref{eq: def edge symmetries Hamiltonian example C a})
and
(\ref{eq: def edge symmetries Hamiltonian example C b})
remain valid. In addition, the spectral symmetry transformation%
~(\ref{eq: spectral symmetry Q A example C}) 
takes the form of an anti-commutation with
\begin{equation}
\hat{\Gamma}^{\,}_{\mathrm{edge}\,\mathscr{R}^{\,}_{y}\,A}(k^{\,}_{y},k^{\prime}_{y})\propto
\delta(k^{\,}_{y}-k^{\prime}_{y})\,
\sigma^{\,}_{1}
\label{eq: effective spectral symmetry generator example C}
\end{equation}
when $x=x'=0$ and $k^{\,}_{y},k^{\prime}_{y}\approx0$.
A perturbation of the form~(\ref{eq: mathcal V edge example C})
is then restricted to the conditions
$v^{\,}_{2}=v^{\,}_{3}=0$
if it commutes with the generator%
~(\ref{eq: def edge symmetries Hamiltonian example C a})
for reflection about the horizontal axis.
A perturbation of the form~(\ref{eq: mathcal V edge example C})
is restricted to the conditions 
$v^{\,}_{0}=v^{\,}_{1}=0$
if it anti-commutes with the generator%
~(\ref{eq: effective spectral symmetry generator example C})
for reflection about the vertical axis.
Imposing both reflection symmetries on the Hamiltonian
thus protects the crossing at vanishing entangling eigenvalue and momentum
in Fig.~\ref{Fig: SQHE}
from any perturbation, 
unlike for the crossing at vanishing energy and momentum 
of the edge states for the Hamiltonian. 

Finally,
a perturbation of the form~(\ref{eq: mathcal V edge example C})
is restricted to the conditions
$v^{\,}_{1}=v^{\,}_{2}=v^{\,}_{3}=0$
if it commutes with the generator%
~(\ref{eq: def edge symmetries Hamiltonian example C b})
for reversal of time.
Imposing time-reversal symmetry on the Hamiltonian
thus protects the existence of a crossing at vanishing momentum
in Fig.~\ref{Fig: SQHE}
from any perturbation.

\section{
Graphene with Kekule order as an 
inversion-symmetric topological insulator in two dimensions}
\label{sec: example D}

\subsection{Introduction}

Graphene has two single-particle bands that touch
linearly at two inequivalent points 
$\bm{K}$ and $\bm{K}'$
from the first Brillouin zone of the triangular lattice
(BZT). These two inequivalent points from 
the BZT of graphene are called Dirac points.
If graphene is modeled by a tight-binding model
with two bands, whereby spinless electrons can only hop
with a uniform amplitude $t$ between
the nearest-neighbor sites of the honeycomb lattice, 
then the linear band touching
is located at the two inequivalent 
corners of the hexagonal BZT of graphene.%
~\cite{Novoselov2004,Novoselov2005}
Graphene is a planar semi-metal with a low-energy 
and long-wave length electronic
structure that can be modeled by
a $4\times4$ massless Dirac Hamiltonian
in $(2+1)$-dimensional space and time.
The rank four of the Dirac matrices 
entering this Dirac Hamiltonian arises
because only the two bands with a linear 
touching are kept in the low-energy sector
of graphene and after linearization 
of the single-particle spectrum about
the two Dirac points $\bm{K}$ and $\bm{K'}$.

\begin{figure}[t] 
\centering
\subfigure[]{
\label{fig: Kekule+armchair a} 
\includegraphics[height=5.cm]{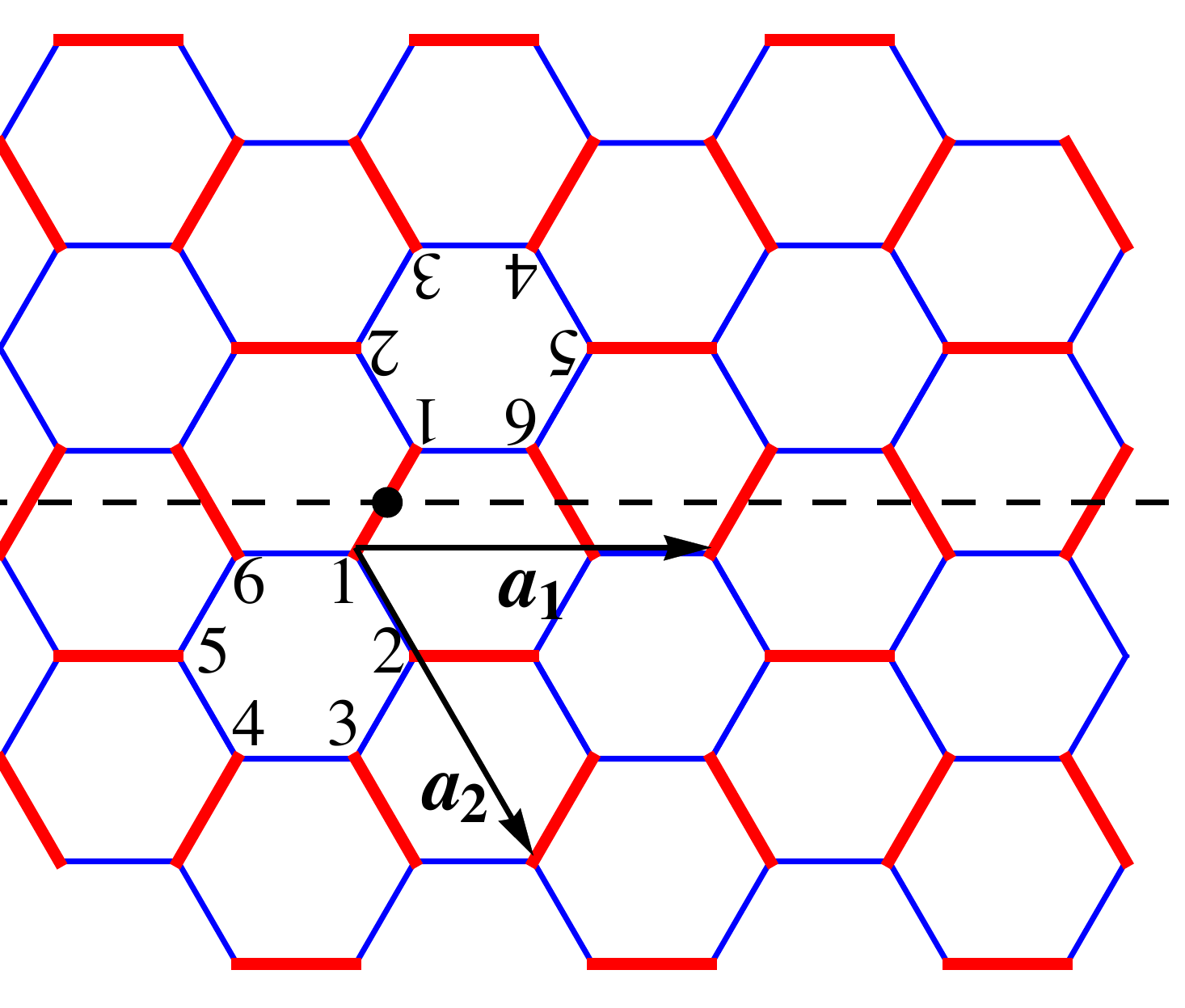}
            }
\hspace{1.cm}
\subfigure[]{
\label{fig: Kekule+armchair b}
\includegraphics[height=4.5 cm]{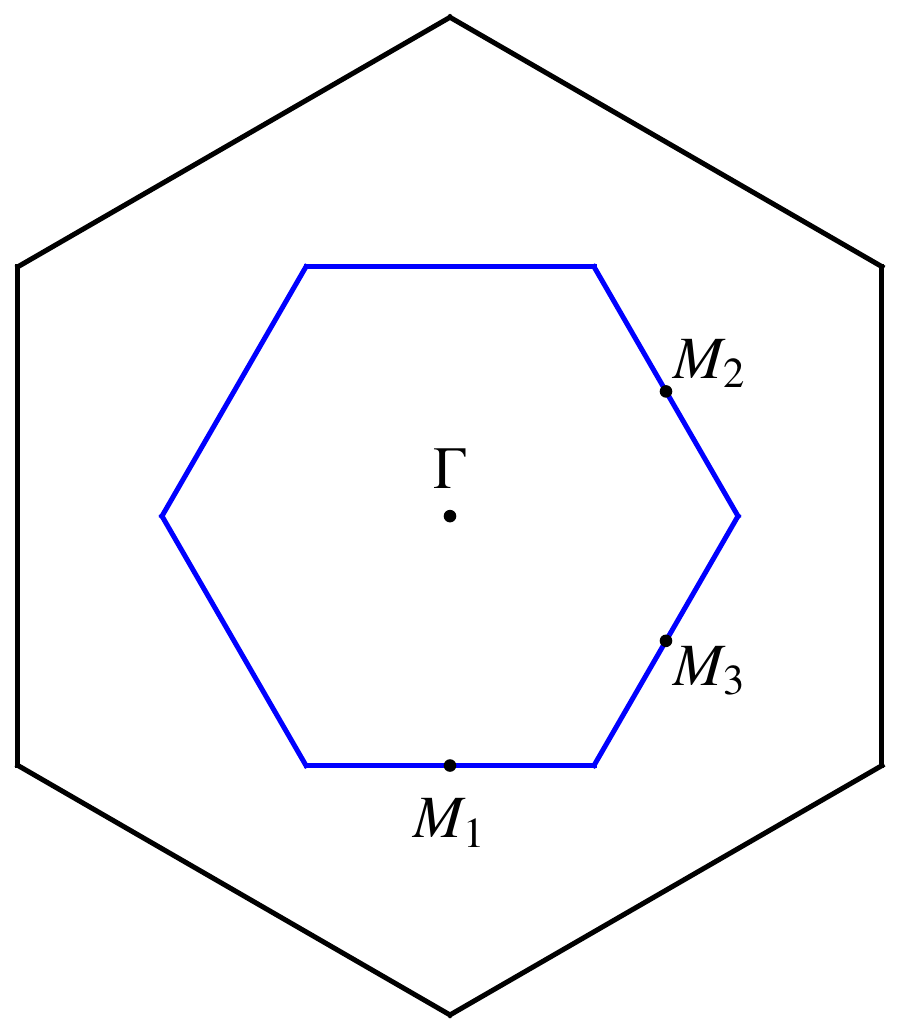}
            }
\caption{
(Color online)
(a) The simplest tight-binding model for graphene with a
Kekule distortion is defined by allowing spinless electrons to hop
between any two nearest-neighbor sites of the honeycomb lattice
with the real-valued modulated amplitudes
$t^{\,}_{1}$
and
$t^{\,}_{2}$
if a nearest-neighbor bond is colored in red or blue, respectively. 
The repeat unit cell of the strong and weak bonds associated to the
Kekule distortion can be chosen to be made of the following three
hexagons below the dashed line. The first hexagon is colored in blue 
and has six vertices numbered clockwise from 1 to 6.
The second hexagon shares the blue 
bond  $\langle 12\rangle$ with the first one.
The third hexagon shares the blue bond $\langle 23 \rangle$ to the first one.
This unit cell comprised of these three hexagons is three
time as large as the repeat unit cell of the honeycomb lattice with
all nearest-neighbor bonds colored in black (the limiting case when
$t^{\,}_{1}=t^{\,}_{2}$). This enlarged repeat unit cell
has 6 inequivalent sites. The spanning vectors of the honeycomb lattice
with the Kekule coloring of nearest-neighbor bonds are 
$\bm{a}^{\,}_{1}$
and
$\bm{a}^{\,}_{2}$.
(b) The large hexagon colored in black defines the first Brillouin zone
of the triangular lattice. The small hexagon colored in blue 
defines the first Brillouin zone
of the honeycomb lattice
with the Kekule coloring of nearest-neighbor bonds.
The ratio of the area of the black hexagon to the area of the 
blue hexagon is three to one. 
If the honeycomb lattice is cut along the horizontal dashed line,
an armchair edge is obtained. 
The point denoted by $\bullet$ at the mid-point where the dashed line intersects 
the nearest-neighbor bond coming out of vertex 1 below the dashed line
defines the inversion center. The points 
$\Gamma$, $M^{\,}_{1}$, $M^{\,}_{2}$ and $M^{\,}_{3}$ 
from the first Brillouin zone
of the honeycomb lattice
with the Kekule coloring of nearest-neighbor bonds
are invariant (fixed) under this inversion.
}
\label{fig: Kekule+armchair}
\end{figure}

The Kekule distortion is a pattern of symmetry breaking
on the honeycomb lattice by which
the nearest-neighbor hopping amplitude takes 
the two distinct real values 
$t^{\,}_{1}$
and
$t^{\,}_{2}$, 
respectively.
This pattern of symmetry breaking
is depicted in Fig.~\ref{fig: Kekule+armchair a}
through the coloring of the nearest-neighbor bonds of 
the honeycomb lattice. 
The Kekule distortion is weak if 
$|t^{\,}_{1}-t^{\,}_{2}|\ll|t^{\,}_{1}+t^{\,}_{2}|/2$, 
in which case a single-particle gap
$\Delta^{\,}_{\mathrm{K}}\propto|t^{\,}_{1}-t^{\,}_{2}|$
that is much smaller than the band width 
($\propto|t^{\,}_{1}+t^{\,}_{2}|/2$)
opens up a the Dirac points. A Kekule distortion can be
induced by fine-tuning of sufficiently 
large repulsive interactions.%
~\cite{Hou2007,Castro}
A Kekule distortion is also favored in the presence
of a sufficiently large magnetic field by some phonons.%
~\cite{Nomura2009,Hou2010}
A Kekule distortion  may also arise locally at the core of 
a vortex in the (proximity-induced) superconducting phase
of graphene.%
~\cite{Pouyan2010}
Finally, a Kekule distortion has been observed in
artificial graphene (molecular graphene)
obtained by patterning carbon monoxide
molecules on the 111 surface of copper.%
\cite{Gomes}

A distinctive feature of graphene is a density of states
that vanishes linearly if single-particle energies are 
measured relative to the energy
of the Dirac points at sufficiently small energies.
This semi-metallic behavior is turned into a semiconducting one
if a Kekule distortion opens a gap at the Dirac points that is larger
than the chemical potential measured relative to the energy
at the Dirac points. Graphene with a Kekule distortion is thus
a band insulator. From the point of view of the ten-fold classification
of band insulators in two-dimensional space,%
~\cite{Schnyder2008,Schnyder2009,Kitaev2009,Ryu2010,Stone2011,Ryu12} 
graphene with a Kekule distortion is topologically trivial 
in that it does not support gapless edge states in an open geometry 
that are robust to the breaking
of translation invariance by an on-site real-valued potential, say. 
Nevertheless, graphene with a Kekule distortion
supports unusual quantum numbers if the Kekule distortion is
defective. For example, a point defect in the Kekule distortion
binds locally a fractional value of the electron charge.%
~\cite{Hou2007,Hou2010,Shinsei2009,Claudio2008}

The goal of Sec.~\ref{sec: example D} is to study the entanglement
spectrum of graphene with a Kekule distortion and show that graphene
with a Kekule distortion is another example of a symmetry protected
topological phase of matter.

\subsection{Hamiltonian}
\label{sec:armchair H}
We start from a honeycomb lattice $\Lambda$,
whose sites are denoted with the symbol $\bm{i}$.
We then color the nearest-neighbor bonds of 
$\Lambda$ with two colors, say red and blue,
as is done in Fig.\ \ref{fig: Kekule+armchair a}.
A pair of nearest-neighbor sites 
of the honeycomb lattice is denoted by
$\langle\bm{i}^{\,}_{1}\bm{j}^{\,}_{1}\rangle$
($\langle\bm{i}^{\,}_{2}\bm{j}^{\,}_{2}\rangle$)
if they are covered by a bond colored in red (blue)
in Fig.\ \ref{fig: Kekule+armchair a}.

To each site $\bm{i}$,
we assign the anti-commuting pair 
$\hat{c}^{\dag}_{\bm{i}}$
and
$\hat{c}^{\,}_{\bm{i}}$
of creation and annihilation operators, respectively.
We then model graphene with a Kekule distortion 
on which spinless fermions hop by the Hamiltonian
\begin{subequations}
\label{eq: def hat H graphene with kekule}
\begin{equation}
\hat{H}^{\,}_{\mathrm{K}}:=
\sum_{n=1,2}
t^{\,}_{n}
\sum_{\langle\bm{i}^{\,}_{n}\bm{j}^{\,}_{n}\rangle}
\left(
\hat{c}^{\dag}_{\bm{i}^{\,}_{n}}\,
\hat{c}^{\,}_{\bm{j}^{\,}_{n}}
+
\mathrm{H.c.}
\right),
\end{equation}
where
\begin{equation}
t^{\,}_{n}:=
-
t
+
\frac{2\,\cos\left((n-1)\frac{2\pi}{3}\right)}{3}\,
\Delta^{\,}_{\mathrm{K}}
\end{equation}
\end{subequations}
with $t$ and $\Delta^{\,}_{\mathrm{K}}$ both real-valued.
The gapless spectrum of graphene with the uniform hopping amplitude 
$t$ is recovered when $\Delta^{\,}_{\mathrm{K}}=0$.

The energy scale $|\Delta^{\,}_{\mathrm{K}}|\ll|t|$ 
breaks the point-group symmetry of 
the honeycomb lattice $\Lambda$
as is implied by the colors blue and red
of inequivalent nearest-neighbor bonds
in Fig.\ \ref{fig: Kekule+armchair a}. The repeat unit cell
of the honeycomb lattice $\Lambda$
decorated by weak and strong bonds
as is implied by the colorings in Fig.\ \ref{fig: Kekule+armchair a}
can be chosen to be the three hexagons sharing
pairwise the blue bond $\langle 12\rangle$
and the blue bond $\langle 23\rangle$
in Fig.\ \ref{fig: Kekule+armchair a}.
The corresponding spanning vectors of the honeycomb lattice
are then $\bm{a}^{\,}_{1}$ and $\bm{a}^{\,}_{2}$.
We use the convention whereby we label 
the 6 sites of the blue hexagon as is done
below the dashed line in Fig.\ \ref{fig: Kekule+armchair a}.
The site 2 belongs to one repeat unit cell.
The sites 1, 3, 5 and their three images by translations 
belong to two repeat unit cells.
The sites 4 and 6 and their four images by translations
belong to three repeat unit cells.
Hence, there are a total of $1+3+2=6$ inequivalent sites
in the repeat unit cell for the honeycomb lattice
with the Kekule distortion. We label the site 2 from
the repeat unit cell for the honeycomb lattice
with the Kekule distortion by $\bm{I}$. 
We may then introduce the spinor
\begin{subequations}
\begin{equation}
\hat{\psi}^{\dag}_{\bm{I}}:=
\begin{pmatrix}
\hat{\psi}^{\dag}_{\bm{I}1}
&
\hat{\psi}^{\dag}_{\bm{I}2}
&
\hat{\psi}^{\dag}_{\bm{I}3}
&
\hat{\psi}^{\dag}_{\bm{I}4}
&
\hat{\psi}^{\dag}_{\bm{I}5}
&
\hat{\psi}^{\dag}_{\bm{I}6}
\end{pmatrix}
\end{equation}
obeying the fermion algebra
\begin{equation}
\{
\hat{\psi}^{\,}_{\bm{I}a},
\hat{\psi}^{\dag}_{\bm{I}'a'}
\}=
\delta^{\,}_{\bm{I},\bm{I}'}
\delta^{\,}_{a,a'},
\qquad
\{
\hat{\psi}^{\dag}_{\bm{I}a},
\hat{\psi}^{\dag}_{\bm{I}'a'}
\}=
0,
\end{equation}
\end{subequations}
and whose components create a spinless fermions
on any one of the six inequivalent sites
from the repeat unit cell $\bm{I}$
for the honeycomb lattice
with the Kekule distortion.
By performing a Fourier transformation
to reciprocal space with momenta $\bm{k}$
restricted to one third of the first Brillouin zone
of graphene,  
we may rewrite Hamiltonian%
~(\ref{eq: def hat H graphene with kekule}) 
as
\begin{subequations}
\begin{equation}
\hat{H}^{\,}_{\mathrm{K}}=
\sum_{\bm{k}\in\mathrm{BZK}}
\hat{\psi}^{\dag}_{\bm{k}}\,
\mathcal{H}^{\,}_{\bm{k}}\,
\hat{\psi}^{\,}_{\bm{k}},
\end{equation}
where the $6\times6$ Hermitian matrix 
$\mathcal{H}^{\,}_{\bm{k}}$
is given by
%\begin{widetext}
\begin{equation}
\label{eq: H of kekule}
\mathcal{H}^{\,}_{\bm{k}}=
\begin{pmatrix}
0 & t^{\,}_{2} & 0 & t^{\,}_{1} e^{-\mathrm{i} (k^{\,}_{2}-k^{\,}_{1})} & 0 & t^{\,}_{2} 
\\
t^{\,}_{2} & 0 & t^{\,}_{2} & 0 & t^{\,}_{1} e^{+\mathrm{i} k^{\,}_{1}} & 0 
\\
0 & t^{\,}_{2} & 0 & t^{\,}_{2} & 0 & t^{\,}_{1} e^{+\mathrm{i}k^{\,}_{2}} 
\\
t^{\,}_{1} e^{+\mathrm{i}(k^{\,}_{2}-k^{\,}_{1})} & 0 & t^{\,}_{2} & 0 & t^{\,}_{2} & 0 
\\
0 & t^{\,}_{1} e^{-\mathrm{i}k^{\,}_{1}} & 0 & t^{\,}_{2} & 0 & t^{\,}_{2} 
\\
t^{\,}_{2} & 0 & t^{\,}_{1} e^{-\mathrm{i} k^{\,}_{2}} & 0 & t^{\,}_{2} & 0
\end{pmatrix}.
\end{equation}
%\end{widetext}
\end{subequations}
The acronym BZK stands for the
reduced Brillouin zone 
in Fig.~\ref{fig: Kekule+armchair b}
associated to the repeat unit cell of the honeycomb lattice with a Kekule
distortion. It covers one third of the area of the BZT.

\subsection{Symmetries}
\label{subsec: Symmetries}

The symmetry
\begin{equation}
\mathcal{H}^{*}_{-\bm{k}}=
\mathcal{H}^{\,}_{+\bm{k}}
\label{eq: TRS for Kekule}
\end{equation}
implements time-reversal symmetry for spinless fermions.

The spectral symmetry
\begin{subequations}
\label{eq: SLS for Kekule}
\begin{equation}
\mathcal{S}^{-1}\,\mathcal{H}^{\,}_{\bm{k}}\,\mathcal{S}=
-\mathcal{H}^{\,}_{\bm{k}},
\end{equation}
where
\begin{equation}
\label{eq: chiral kekule}
\mathcal{S}:=
\mathrm{diag}\, 
\left(
1,-1, 1, -1,1,-1
\right)
%\begin{pmatrix}
%1 & 0 & 0 & 0 & 0 & 0 
%\\
%0 & -1 & 0 & 0 & 0 & 0 
%\\
%0 & 0 & 1 & 0 & 0 & 0 
%\\
%0 & 0 & 0 & -1 & 0 & 0 
%\\
%0 & 0 & 0 & 0 & 1 & 0 
%\\
%0 & 0 & 0 & 0 & 0 & -1
%\end{pmatrix}
\end{equation}
\end{subequations}
implements the chiral (sublattice) spectral symmetry.

The symmetry
\begin{subequations}
\label{eq: inversion symmetry Kekule}
\begin{equation}
\mathcal{P}^{-1}\,\mathcal{H}^{\,}_{-\bm{k}}\,\mathcal{P}=
\mathcal{H}^{\,}_{\bm{k}},
\label{eq: inversion symmetry Kekule a}
\end{equation}
where
\begin{equation}
\mathcal{P}:=
%\begin{pmatrix}
%0 & 0 & 0 & 1 & 0 & 0 
%\\
%0 & 0 & 0 & 0 & 1 & 0 
%\\
%0 & 0 & 0 & 0 & 0 & 1 
%\\
%1 & 0 & 0 & 0 & 0 & 0 
%\\
%0 & 1 & 0 & 0 & 0 & 0 
%\\
%0 & 0 & 1 & 0 & 0 & 0
%\end{pmatrix}
\begin{pmatrix}
 0 & \mathbb{I}_3  
\\
\mathbb{I}_3 & 0 
\end{pmatrix}
\label{eq: inversion symmetry Kekule b}
\end{equation}
\end{subequations}
implements the inversion symmetry defined with the help
of Fig.~\ref{fig: Kekule+armchair a}. To define the inversion symmetry,
we first draw the dashed line in Fig.~\ref{fig: Kekule+armchair a}.
A cut along this dashed line defines an arm-chair boundary.
We then select the intersection of the dashed line
with the mid-point of the bond emerging from the site 1
of the enlarged repeat unit cell below the dashed line
in Fig.~\ref{fig: Kekule+armchair a}. This mid-point, represented by
a filled circle in Fig.~\ref{fig: Kekule+armchair a}, 
defines the inversion center.
Performing an inversion about this point
maps the Kekule pattern below the dashed line into
the Kekule pattern above the dashed line.
The two patterns are identical, hence the inversion symmetry.
On the one hand, the labels
in the enlarged repeat unit cell below the dashed line becomes those
above the dashed line in Fig.~\ref{fig: Kekule+armchair a}
under this inversion. On the other hand,
if the convention for the labels of the enlarged
repeat unit cell are identical below and above the 
dashed line, the representation%
~(\ref{eq: inversion symmetry Kekule b})
follows.

Observe that 
\begin{equation}
\{\mathcal{S},\mathcal{P}\}=0.
\label{eq: mathcal{S} mathcal{P} anticommute for kekule}
\end{equation} 
We will make use of this anti-commutator in Sec.%
~(\ref{subsubsec: Stability analysis of the zero modes}).

\begin{figure}[!ht]
\centering
\includegraphics[height=8.5 cm]{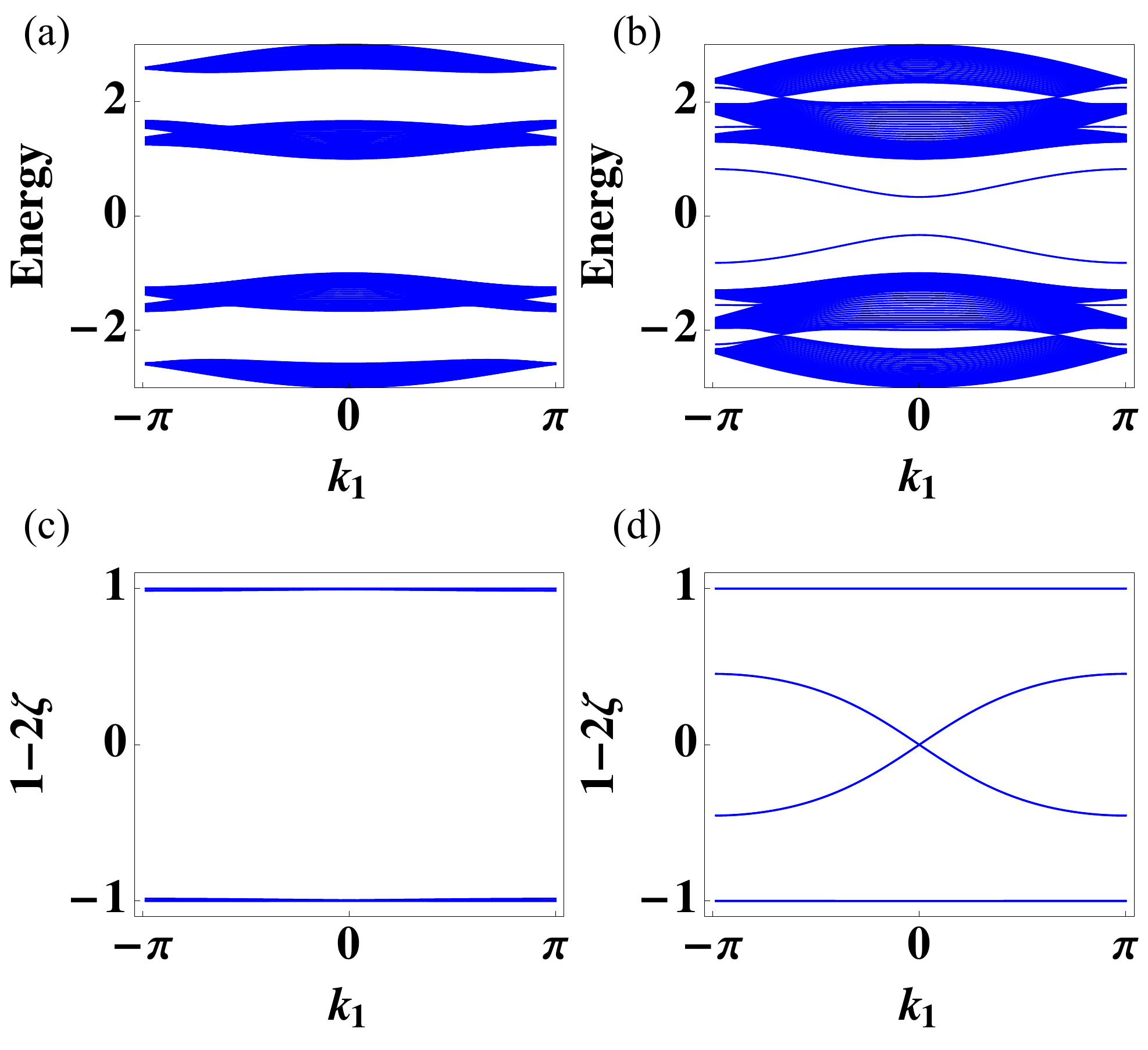}
\caption{
(Color online)
Energy spectrum 
of Hamiltonian~(\ref{eq: def hat H graphene with kekule})
with armchair edges for 
(a) 
$(t^{\,}_{1},t^{\,}_{2})=({1}/{3},{4}/{3})$ 
and 
(b) 
$(t^{\,}_{1},t^{\,}_{2})=({5}/{3},{2}/{3})$. 
Entanglement spectrum with armchair entangling edges for 
(c) 
$(t^{\,}_{1},t^{\,}_{2})=({1}/{3},{4}/{3})$ 
and  
(d)
$(t^{\,}_{1},t^{\,}_{2})=({5}/{3},{2}/{3})$.
The dimensions of the lattice are given by 
$(N^{\,}_{1},N^{\,}_{2})=(128,32)$,
where $N^{\,}_{i}$ 
is the number of the repeat unit cell from
Fig.~\ref{fig: Kekule+armchair a}
along the direction of the spanning vector
$\bm{a}^{\,}_{i}$ ($i=1,2$)
and in units for which the spanning vectors
$\bm{a}^{\,}_{1}$
and
$\bm{a}^{\,}_{2}$
are of unit length.
}
\label{fig: graphene spectrum 3}
\end{figure}

\subsection{Partition} 
\label{subsec: gaphene partition}

A slab geometry is cut from Fig.~\ref{fig: Kekule+armchair a}
by choosing two armchair edges
running parallel to the $\bm{a}^{\,}_{1}$ direction 
of the two-dimensional embedding Euclidean space. 
Periodic boundary conditions are imposed along the
$\bm{a}^{\,}_{1}$ direction, 
open ones along the $\bm{a}^{\,}_{2}$ direction. 
Hence, the momentum $k^{\,}_{1}$ is a good quantum number. 

This slab geometry with the choice of mixed periodic and
open boundary conditions is identical to the cylindrical geometry 
shown in Fig.~\ref{fig: reflection planes cylinder}.
The dashed line in Fig.~\ref{fig: Kekule+armchair a}
can be identified with the intersection of the red plane
with the cylinder in Fig.~\ref{fig: reflection planes cylinder}.
The dashed line in Fig.~\ref{fig: Kekule+armchair a} will shortly
be identified with an entangling
boundary that is invariant under the inversion symmetry
about the inversion center in Fig.~\ref{fig: Kekule+armchair a}.

We use the partition introduced in 
Sec.~\ref{subsubsec: Partition, and zero modes example B}
with the convention that the good quantum number is $k^{\,}_{1}$,
i.e., with $i=1$ in Sec.~\ref{subsubsec: Partition, and zero modes example B}.
The single-particle Hilbert space is the direct sum
\begin{subequations}
\begin{equation}
\mathfrak{H}=
\bigoplus_{k^{\,}_{1}}
\mathfrak{H}^{\,}_{k^{\,}_{1}}.
\end{equation}
For any good momentum quantum
number in the one-dimensional Brillouin zone
\begin{equation}
k^{\,}_{1}:=
\frac{2\pi}{2M^{\,}_{1}\,|\bm{a}^{\,}_{1}|}\,n^{\,}_{1}, 
\qquad
n^{\,}_{1}=1,\cdots,2M^{\,}_{1},
\end{equation}
where $N^{\,}_{1}=2M^{\,}_{1}$ 
is the number of enlarged unit cells along the $\bm{a}^{\,}_{1}$ direction,
the subspace $\mathfrak{H}^{\,}_{k^{\,}_{1}}$ is spanned by the
orthonormal single-particle states 
\begin{equation}
\left\{
|k^{\,}_{1},n^{\,}_{2},\alpha\rangle|
n^{\,}_{2}=1,\cdots,2M^{\,}_{2},
\qquad
\alpha=1,\cdots,6
\right\}
\end{equation}
\end{subequations}
with $n^{\,}_{2}$ labeling the $N^{\,}_{2}=2M^{\,}_{2}$
enlarged unit cells in the direction
$\bm{a}^{\,}_{2}$ and $\alpha=1,\cdots,6$ labeling the inequivalent sites
within an enlarged unit cell. The single-particle Hamiltonian
$\mathcal{H}^{\,}_{k^{\,}_{1}}$ has matrix elements of the form given in
Eq.~\ref{eq: def mathcal{H} ki;r(i+1),r'(i+1) a}.
The partition for any given good quantum number
$k^{\,}_{1}$ is then
\begin{subequations}
\label{eq: def partition kekule}
\begin{align}
&
\mathfrak{H}^{\,}_{k^{\,}_{1}}:=
\mathfrak{H}^{\,}_{A^{\,}_{2}}
\oplus
\mathfrak{H}^{\,}_{B^{\,}_{2}},
\end{align}
where
\begin{equation}
\mathfrak{H}^{\,}_{A^{\,}_{2}}:=
\bigoplus_{n^{\,}_{2}=1}^{M^{\,}_{2}}
\bigoplus_{\alpha=1}^{6}
|k^{\,}_{1},n^{\,}_{2},\alpha\rangle
\langle k^{\,}_{1},n^{\,}_{2},\alpha|
\end{equation}
and
\begin{equation}
\mathfrak{H}^{\,}_{B^{\,}_{2}}:=
\bigoplus_{n^{\,}_{2}=M^{\,}_{2}+1}^{2M^{\,}_{2}}
\bigoplus_{\alpha=1}^{4}
|k^{\,}_{1},n^{\,}_{2},\alpha\rangle
\langle k^{\,}_{1},n^{\,}_{2},\alpha|.
\end{equation}
\end{subequations}
If we denote by $\mathscr{P}$ the inversion 
about the inversion center in Fig.~\ref{fig: Kekule+armchair a}
that reverses the sign of the good quantum number
$k^{\,}_{1}$, i.e.,
\begin{equation}
\mathscr{P}\,k^{\,}_{1}=-k^{\,}_{1},
\end{equation}
we then have that
\begin{equation}
\mathscr{P}^{\,}A^{\,}_{2}=B^{\,}_{2},
\end{equation}
i.e., $\mathscr{P}$ interchanges the physical boundaries while
it leaves the entangling boundary between $A^{\,}_{2}$ and $B^{\,}_{2}$
invariant as a set.

\subsection{Kekule with armchair edges}
\label{sec:spectra of kekule armchair}

Both the energy and entanglement spectra are obtained 
by exact diagonalization with $(N^{\,}_{1},N^{\,}_{2})=(128,32)$ 
and presented in Fig.~\ref{fig: graphene spectrum 3}
for different values of $t^{\,}_{1}$ and $t^{\,}_{2}$
in Eq.~(\ref{eq: def hat H graphene with kekule}).

The energy spectrum is bulk-like when $t^{\,}_{1}<t^{\,}_{2}$
as is illustrated with Fig.~\ref{fig: graphene spectrum 3}(a).
Edge modes are not present when $t^{\,}_{1}<t^{\,}_{2}$
in the energy spectrum,
Fig.~\ref{fig: graphene spectrum 3}(a) being an example of this 
observation. For each armchair boundary, 
The energy spectrum supports a single pair of right- and left-moving
edge states
when $t^{\,}_{1}>t^{\,}_{2}$,
as is illustrated with Fig.~\ref{fig: graphene spectrum 3}(b).
These edge states do not cross, they are gaped at the
band center.

The entanglement spectrum is bulk-like when $t^{\,}_{1}<t^{\,}_{2}$
as is illustrated with Figs.~\ref{fig: graphene spectrum 3}(c).
Edge modes are not present when $t^{\,}_{1}<t^{\,}_{2}$
in the entanglement spectrum,
Fig.~\ref{fig: graphene spectrum 3}(c) being an example of this 
observation. For each armchair boundary, 
the entanglement spectrum supports 
a single pair of right- and left-moving edge states
when $t^{\,}_{1}>t^{\,}_{2}$,
as is illustrated with Fig.~\ref{fig: graphene spectrum 3}(d).
These edge states cross at the band center.

\begin{figure}[!ht]
\centering
\includegraphics[height=8.5 cm]{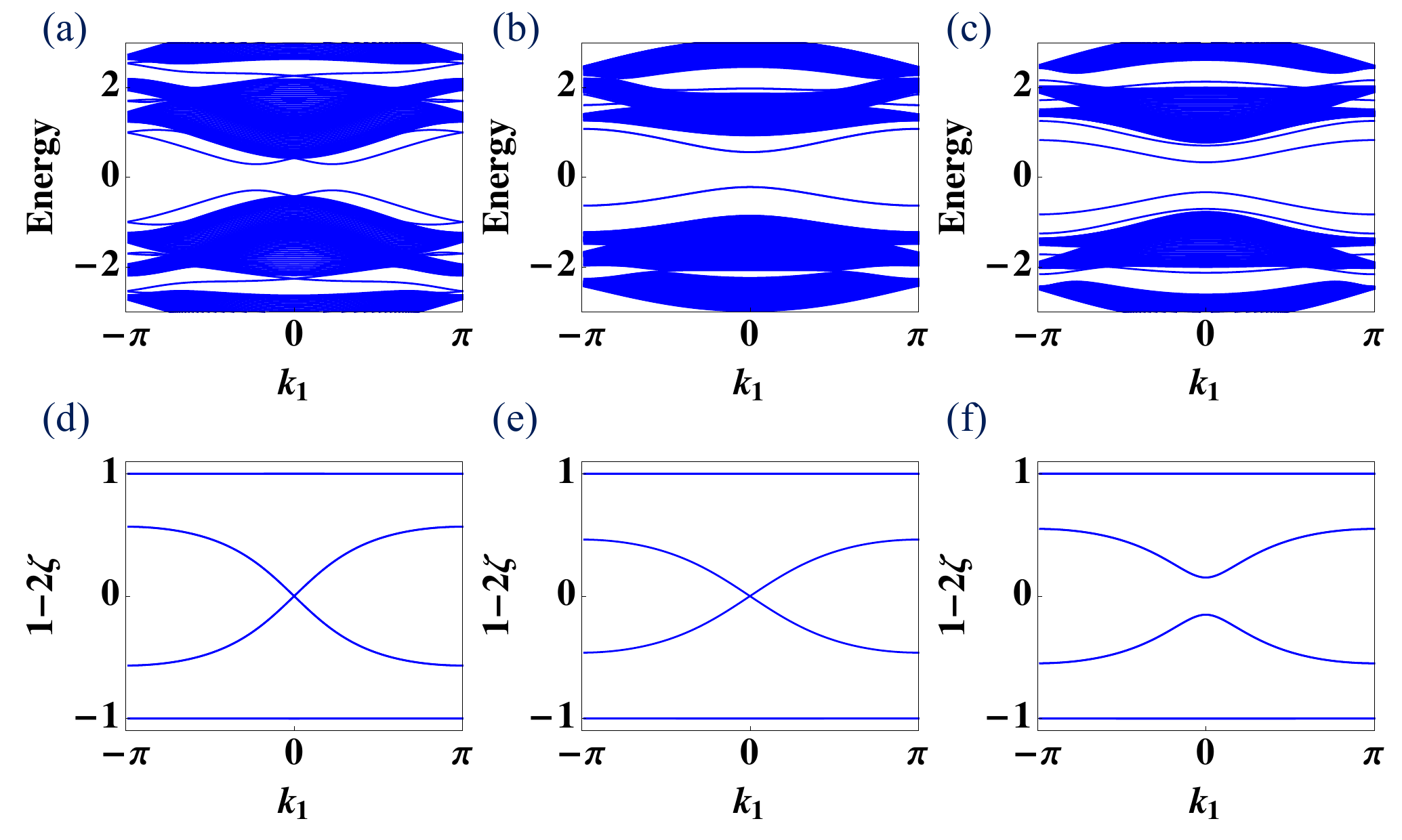}
\caption{
(Color online)
Energy spectra are presented in the left column,
entanglement spectra are presented in the right column.
In both cases, the geometry is that of a slab with armchair edges as in
Fig.~\ref{fig: Kekule+armchair a}
for $(t^{\,}_{1},t^{\,}_{2})=({5}/{3},{2}/{3})$
and
$(N^{\,}_{1},N^{\,}_{2})=(128,32)$.
The spectra (a) and (d) are obtained by choosing the
time-reversal-breaking perturbation%
~(\ref{eq: def mathcal H' TRSB})
in Hamiltonian~(\ref{eq: def perturbed HK}).
The spectra (b) and (e) are obtained by choosing the
chiral-symmetry-breaking perturbation%
~(\ref{eq: def mathcal H' chSB a})
with~(\ref{eq: def mathcal H' chSB c})
in Hamiltonian~(\ref{eq: def perturbed HK}).
The spectra (c) and (f) are obtained by choosing the
inversion-symmetry-breaking perturbation% 
~(\ref{eq: def mathcal H' ISB})
to Hamiltonian~(\ref{eq: def perturbed HK}).
}
\label{fig:4}
\end{figure}

In the following, we shall choose
$(t^{\,}_{1},t^{\,}_{2})=({5}/{3},{2}/{3})$
and study the robustness of the crossing of the edge states in
the entanglement spectrum in the presence of three symmetry-breaking
perturbations. We either break time-reversal symmetry,
chiral symmetry, or inversion symmetry once at a time.
Spectra obtained by exact diagonalization are presented
in Fig.~\ref{fig:4}, i.e., we study the spectra of
\begin{equation}
\hat{H}:=
\hat{H}^{\,}_{\mathrm{K}}
+
\hat{H}'
\label{eq: def perturbed HK}
\end{equation}
with $\hat{H}'$ a one-body perturbation that breaks
either time-reversal symmetry, chiral symmetry, or
inversion symmetry.

\subsubsection{Time-reversal symmetry breaking}
We choose in Eq.~(\ref{eq: def perturbed HK})
the perturbation defined by
\begin{subequations}
\label{eq: def mathcal H' TRSB}
\begin{equation}
\mathcal{H}':=
l
\begin{pmatrix}
0 & e^{\mathrm{i}\phi} & 0 & 0 & 0  & e^{-\mathrm{i}\phi} 
\\
e^{-\mathrm{i}\phi} & 0 & e^{\mathrm{i}\phi} & 0 & 0 & 0  
\\
0  & e^{-\mathrm{i}\phi} & 0 & e^{\mathrm{i}\phi} & 0  & 0 
\\
0 & 0 & e^{-\mathrm{i}\phi} & 0 & e^{\mathrm{i}\phi} & 0  
\\
0 & 0 & 0  & e^{-\mathrm{i}\phi} & 0 & e^{\mathrm{i}\phi} 
\\
e^{\mathrm{i}\phi} & 0 & 0 & 0  & e^{-\mathrm{i}\phi} & 0
\end{pmatrix},
\label{eq: def mathcal H' TRSB a}
\end{equation}
where the real number $l$ is a uniform hopping amplitude and 
$\phi$ is a uniform phase that breaks time-reversal symmetry if
not equal to $0$ or $\pi$. 
Figures~\ref{fig:4}(a)
and~\ref{fig:4}(b) 
give the physical energy and entanglement spectra, respectively,
for 
\begin{equation}
l=0.3,
\qquad
\phi=\pi/4.
\label{eq: def mathcal H' TRSB b}
\end{equation}
\end{subequations}
Both the chiral and the inversion symmetries are present 
for this perturbation that breaks time-reversal symmetry
and the spectra in 
Figs.~\ref{fig:4}(a)
and~\ref{fig:4}(d) 
follow. According to Fig.~\ref{fig:4}(a) 
the edge states are gaped in the energy spectrum.
According to \ref{fig:4}(d),
the existence and location of the crossing of the edge states 
in the entanglement spectrum is seen to be robust to this perturbation.

\subsubsection{Sublattice symmetry breaking}
The chiral (sublattice) spectral symmetry is broken by any on-site potential.
Any distribution of on-site potentials within the repeat unit cell
of Fig.~\ref{fig: Kekule+armchair a} that is unchanged under any linear combination
of the spanning vectors $\bm{a}^{\,}_{1}$ and $\bm{a}^{\,}_{2}$ 
preserves the translation symmetry of the honeycomb lattice
decorated by the colors of the Kekule strong and weak bonds.
Hence, we choose in Eq.~(\ref{eq: def perturbed HK}) the perturbation
\begin{equation}
\mathcal{H}':=
\mathrm{diag}\,
\left(
\mu_1, \mu_2, \mu_3,\mu_4,\mu_5,\mu_6
\right), 
%\begin{pmatrix}
%\mu^{\,}_{1} & 0 & 0 & 0 & 0 & 0 
%\\
%0 & \mu^{\,}_{2} & 0 & 0 & 0 & 0 
%\\
%0 & 0 & \mu^{\,}_{3} & 0 & 0 & 0 
%\\
%0 & 0 & 0 & \mu^{\,}_{4} & 0 & 0 
%\\
%0 & 0 & 0 & 0 & \mu^{\,}_{5} & 0 
%\\
%0 & 0 & 0 & 0 & 0 & \mu^{\,}_{6}
%\end{pmatrix},
\label{eq: def mathcal H' chSB a}
\end{equation}
where we demand that
\begin{equation}
\mu^{\,}_{1}=\mu^{\,}_{4},
\qquad
\mu^{\,}_{2}=\mu^{\,}_{5},
\qquad
\mu^{\,}_{3}=\mu^{\,}_{6},
\label{eq: def mathcal H' chSB b}
\end{equation}
breaks the chiral (sublattice) spectral symmetry
for any non-vanishing value of
$\mu^{\,}_{n}$ with $n=1,\cdots,3$
belonging to the repeat unit cell,
but preserves the reduced first Brillouin zone
from Fig.~\ref{fig: Kekule+armchair b}
with the inversion symmetric points
$\Gamma$, $M^{\,}_{1}$, $M^{\,}_{2}$, and $M^{\,}_{3}$.
Since an on-site potential
is represented by a real-valued matrix, 
$\mathcal{H}'$  does not break time-reversal symmetry.
If we use a slab geometry with two parallel armchair edges
running along the $x$ axis in the embedding two-dimensional Euclidean space,
we need only demand that $k^{\,}_{1}$ is a good quantum number, i.e.,
invariance under translation by $\bm{a}^{\,}_{1}$ only. If so,
we can make the choice
\begin{equation}
\mu^{\,}_{1}=1/2,
\quad
\mu^{\,}_{2}=-1/4,
\quad
\mu^{\,}_{3}=\cdots=\mu^{\,}_{6}=0,
\label{eq: def mathcal H' chSB c}
\end{equation}
for the real-valued chemical potentials within the repeat unit cell.
With this choice, $k^{\,}_{1}$ is a good quantum number,
the time-reversal and inversion symmetries are present,
but the chiral sublattice symmetry is broken,
and the spectra in 
Figs.~\ref{fig:4}(b)
and~\ref{fig:4}(e) 
follow. According to Fig.~\ref{fig:4}(b) 
the edge states are gaped in the energy spectrum.
According to \ref{fig:4}(e),
the existence and location of the crossing of the edge states 
in the entanglement spectrum is seen to be robust to this perturbation.

\subsubsection{Inversion symmetry breaking}

We choose in Eq.~(\ref{eq: def perturbed HK}) the perturbation
\begin{subequations}
\label{eq: def mathcal H' ISB}
\begin{equation}
\mathcal{H}'=
l\,
\begin{pmatrix}
0 & 0 & 0 &0& 0 & 1 
\\
0 & 0 & 0 & 0 & 0 & 0 
\\
0 & 0 & 0 & 0 & 0 & 0 
\\
0 & 0 & 0 & 0 & 0 & 0 
\\
0 & 0 & 0 & 0 & 0 & 0 
\\
1 & 0 & 0 & 0 & 0 & 0
\end{pmatrix},
\label{eq: def mathcal H' ISB a}
\end{equation}
where the real-valued 
\begin{equation}
l=0.5
\label{eq: def mathcal H' ISB b}
\end{equation}
\end{subequations}
in Figs.~\ref{fig:4}(c) and \ref{fig:4}(f).
Since this perturbation is a real-valued nearest-neighbor hopping
between site 1 and 6 from the repeat unit cell of 
Fig.~\ref{fig: Kekule+armchair a},
time-reversal and chiral (sublattice) symmetries are present,
but inversion symmetry is broken.
According to Figs.~\ref{fig:4}(c) and \ref{fig:4}(f),
both the energy and entanglement spectra are gaped. 

\subsubsection{Stability analysis of the zero modes}
\label{subsubsec: Stability analysis of the zero modes}

The two disconnected physical boundaries 
in the cylindrical geometry used to do the exact diagonalization
presented in Figs.~\ref{fig: graphene spectrum 3} and~\ref{fig:4} 
are interchanged by the inversion symmetry.
The single entangling boundary for the cylindrical geometry considered
in Figs.~\ref{fig: graphene spectrum 3} and~\ref{fig:4} 
is invariant as a set under the inversion symmetry. 
This difference explains why there are crossings
of gapless edge states in the entanglement spectra provided 
$\hat{H}$ defined by Eq.~(\ref{eq: def hat H graphene with kekule})
with $t^{\,}_{1}>t^{\,}_{2}$
respects the inversion (parity) symmetry $\mathscr{P}$
(and irrespectively of the presence or absence of
time reversal symmetry and spectral chiral symmetry),
while there are no crossing of edge states in the energy spectrum
under the same assumptions.

To explain this empirical observation deduced from
exact diagonalization, we assume that the existence of
a pair of left- and right-moving edge states along any one
of the three armchair boundaries, namely any one of the
two physical armchair boundaries and the entangling 
armchair boundary. We consider first the case of
a single physical armchair boundary and then the case
of the single entangling armchair boundary.
In both cases, our effective single-particle Hamiltonian
for the single pair of left and right movers on an edge is given by
\begin{subequations}
\label{eq: def single armchair edge}
\begin{equation}
\mathcal{H}^{\,}_{\mathrm{edge}}=
\bigoplus_{k}
\mathcal{H}^{\,}_{\mathrm{edge}\,k},
\label{eq: def single armchair edge a}
\end{equation}
where the momentum along the edge is denoted by $k$ and
\begin{equation}
\mathcal{H}^{\,}_{\mathrm{edge}\,k}=
v^{\,}_{0}\,\sigma^{\,}_{0}
+
v^{\,}_{1}\,\sigma^{\,}_{1}
+
v^{\,}_{2}\,\sigma^{\,}_{2}
+
k\,\sigma^{\,}_{3}
\label{eq: def single armchair edge b}
\end{equation}
\end{subequations}
to leading order in a gradient expansion.
Here, we have introduced the usual suspects, 
namely the
$2\times2$ unit matrix
$\sigma^{\,}_{0}$ 
and the three Pauli matrices
$\sigma^{\,}_{1}$,
$\sigma^{\,}_{2}$,
and
$\sigma^{\,}_{3}$
to which we associate three energy scales
through the real numbers 
$v^{\,}_{0}$,
$v^{\,}_{1}$,
$v^{\,}_{2}$,
and one group velocity $v^{\,}_{3}$ that we have
set to unity ($\hbar=1$ as well), respectively. 

\paragraph{Energy spectrum}
The effective Hamiltonian%
~(\ref{eq: def single armchair edge})
for a single physical armchair boundary
inherits two symmetries from Hamiltonian%
~(\ref{eq: def hat H graphene with kekule})
(the case $t^{\,}_{1}>t^{\,}_{2}$ is assumed to have edge states).
There is the symmetry  
$\mathscr{T}^{\,}_{\mathrm{edge}}$ under reversal of time.
There is the spectral symmetry $\mathscr{S}^{\,}_{\mathrm{edge}}$
under multiplication by a minus
sign of all the single-particle states in the position basis
on one and only one of the two triangular sublattices 
of the honeycomb lattice. There is no inversion symmetry
generated by a putative operator
$\mathscr{P}^{\,}_{\mathrm{edge}}$
for a single physical armchair boundary.
Because we have been considering spinless fermions,
a representation of $\mathscr{T}^{\,}_{\mathrm{edge}}$ 
is uniquely fixed by
demanding that reversal of time interchanges left and right movers
while squaring to unity as an insanitary operator, i.e.,
\begin{equation}
\mathcal{T}^{\,}_{\mathrm{edge}}:=
\sigma^{\,}_{1}\,
K,
\end{equation}
where $K$ denotes complex conjugation.
Symmetry under reversal of time is the condition
\begin{equation}
\mathcal{H}^{\,}_{\mathrm{edge}\,+k}=
\sigma^{\,}_{1}\,
\mathcal{H}^{*}_{\mathrm{edge}\,-k}\,
\sigma^{\,}_{1}
\end{equation}
that is met for any 
$v^{\,}_{0}$,
$v^{\,}_{1}$,
and
$v^{\,}_{2}$
in Eq.~(\ref{eq: def single armchair edge}).
Hence, it is only the sublattice symmetry that
restricts the allowed values of
$v^{\,}_{0}$,
$v^{\,}_{1}$,
and
$v^{\,}_{2}$
in Eq.~(\ref{eq: def single armchair edge}).
There are two possible choices to represent
$\mathscr{S}^{\,}_{\mathrm{edge}}$ by a unitary matrix 
$\mathcal{S}^{\,}_{\mathrm{edge}}$ such that
\begin{subequations}
\label{eq: SLS symmetry armchair edge with Kekule}
\begin{equation}
\mathcal{H}^{\,}_{\mathrm{edge}\,k}=
-
\mathcal{S}^{\,}_{\mathrm{edge}}\,
\mathcal{H}^{\,}_{\mathrm{edge}\,k}\,
\mathcal{S}^{-1}_{\mathrm{edge}},
\label{eq: SLS symmetry armchair edge with Kekule a}
\end{equation}
namely 
\begin{equation}
\mathcal{S}^{\,}_{\mathrm{edge}}=\sigma^{\,}_{1}
\label{eq: SLS symmetry armchair edge with Kekule b}
\end{equation}
or 
\begin{equation}
\mathcal{S}^{\,}_{\mathrm{edge}}=\sigma^{\,}_{2}.
\label{eq: SLS symmetry armchair edge with Kekule c}
\end{equation}
\end{subequations}
In both cases, the spectral symmetry%
~(\ref{eq: SLS symmetry armchair edge with Kekule})
fixes the chemical potential to the value $v^{\,}_{0}=0$.
In the former case, the spectral symmetry with the generator%
~(\ref{eq: SLS symmetry armchair edge with Kekule b})
fixes $v^{\,}_{1}=0$ but leaves $v^{\,}_{2}$
arbitrary so that a gap opens up as soon as $v^{\,}_{2}\neq0$
in the energy spectrum of the edge states.
In the latter case, the spectral symmetry with the generator%
~(\ref{eq: SLS symmetry armchair edge with Kekule c})
fixes  $v^{\,}_{2}=0$ but leaves $v^{\,}_{1}$
arbitrary so that a gap opens up as soon as $v^{\,}_{1}\neq0$
in the energy spectrum of the edge states.

\paragraph{Entanglement spectrum}
We now assume that the single pair of
left and right movers propagating along the entangling boundary
is governed by the effective
\begin{equation}
\mathcal{Q}^{\,}_{\mathrm{edge}\,A\,k}
\equiv
\mathcal{H}^{\,}_{\mathrm{edge}\,k}
\end{equation}
with $\mathcal{H}^{\,}_{\mathrm{edge}\,k}$ defined in
Eq.~(\ref{eq: def single armchair edge}).
The symmetries obeyed by
$\mathcal{Q}^{\,}_{\mathrm{edge}\,A\, k}$
are
\begin{subequations}
\begin{align}
&
\mathcal{Q}^{\,}_{\mathrm{edge}\,A\,+k}=
\mathcal{T}^{\,}_{\mathrm{edge}}\,
\mathcal{Q}^{*}_{\mathrm{edge}\,A\,-k}\,
\mathcal{T}^{-1}_{\mathrm{edge}},
\\
&
\mathcal{Q}^{\,}_{\mathrm{edge}\,A\,k},=
-
\mathcal{S}^{\,}_{\mathrm{edge}}\,
\mathcal{Q}^{\,}_{\mathrm{edge}\,A\,k}\,
\mathcal{S}^{-1}_{\mathrm{edge}},
\\
&
\mathcal{Q}^{\,}_{\mathrm{edge}\,A\,+k}=
-
\Gamma^{\,}_{\mathcal{P}^{\,}_{\mathrm{edge}}}\,
\mathcal{Q}^{\,}_{\mathrm{edge}\,A\,-k}\,
\Gamma^{-1}_{\mathcal{P}^{\,}_{\mathrm{edge}}},
\end{align}
\end{subequations}
[Recall Eq.~(\ref{eq: induced symmetry b})].
Needed is a representation of the unitary generator
$\Gamma^{\,}_{\mathcal{P}^{\,}_{\mathrm{edge}}}$
for inversion on the entangling boundary.
Since
\begin{equation}
k\,\sigma^{\,}_{3}=
-
\sigma^{\,}_{\mu}\,(-k\,\sigma^{\,}_{3})\,\sigma^{\,}_{\mu}
\end{equation}
for both $\mu=0$ or $\mu=3$, there seems to be an ambiguity
when defining $\Gamma^{\,}_{\mathcal{P}^{\,}_{\mathrm{edge}}}$.
On the one hand, choosing
\begin{subequations}
\label{eq: I symmetry armchair edge with Kekule}
\begin{equation}
\Gamma^{\,}_{\mathcal{P}^{\,}_{\mathrm{edge}}}=\sigma^{\,}_{0}
\label{eq: I symmetry armchair edge with Kekule a}
\end{equation} 
implies that $\Gamma^{\,}_{\mathcal{P}^{\,}_{\mathrm{edge}}}$ commutes 
with either choices 
(\ref{eq: SLS symmetry armchair edge with Kekule b})
or
(\ref{eq: SLS symmetry armchair edge with Kekule c}).
On the other hand, choosing
\begin{equation}
\Gamma^{\,}_{\mathcal{P}^{\,}_{\mathrm{edge}}}=\sigma^{\,}_{3}
\label{eq: I symmetry armchair edge with Kekule b}
\end{equation}
\end{subequations} 
implies that $\Gamma^{\,}_{\mathcal{P}^{\,}_{\mathrm{edge}}}$ anti-commutes 
with either choices 
(\ref{eq: SLS symmetry armchair edge with Kekule b})
or
(\ref{eq: SLS symmetry armchair edge with Kekule c}).
Now,
it is only when $\Gamma^{\,}_{\mathcal{P}^{\,}_{\mathrm{edge}}}$
commutes with $\mathcal{S}^{\,}_{\mathrm{edge}}$ that inversion symmetry   
protects the crossing of the pair of edge states
at the band center by fixing $v^{\,}_{0}=v^{\,}_{1}=v^{\,}_{2}=0$
irrespectively of whether sublattice symmetry holds or not!
The ambiguity in choosing between the representations
(\ref{eq: I symmetry armchair edge with Kekule a})
and
(\ref{eq: I symmetry armchair edge with Kekule b})
is spurious, however. We must choose
the representation
(\ref{eq: I symmetry armchair edge with Kekule a})
as we now demonstrate.

We are now going to show that irrespective of the choice
made to represent $\mathscr{S}^{\,}_{\mathrm{edge}}$, we must choose
to represent $\mathscr{P}^{\,}_{\mathrm{edge}}$ such that 
\begin{equation}
[\mathcal{S}^{\,}_{\mathrm{edge}},\Gamma^{\,}_{\mathcal{P}^{\,}_{\mathrm{edge}}}]=0.
\end{equation}
To this end, we use the fact that the chiral transformation representing
the spectral sublattice symmetry and that representing the inversion
symmetry in the bulk, recall Eq.~(\ref{eq: SLS for Kekule}) 
and~(\ref{eq: inversion symmetry Kekule}),
anti-commute according to Eq.%
~(\ref{eq: mathcal{S} mathcal{P} anticommute for kekule}).

In the presence of inversion symmetry, there is a spectral symmetry
$\Gamma^{\,}_{\mathscr{P}}$ in the entanglement spectrum. This spectral
symmetry was defined by
$\Gamma^{\,}_{\mathscr{P}}=
C^{\,}_{AB\,k}\, 
P$
in Eq.~(\ref{eq: def Gamma mathsf O})
(we are using the conventions of
Sec.~\ref{Chiral symmetry of the entanglement spectrum}.
for the choice of the fonts of the symmetry generators).
In the presence of the spectral symmetry
\begin{align}
S\,
Q^{\,}_{A\,k}\,
S^{-1}=
-Q^{\,}_{A\,k}
\end{align} 
and
\begin{equation}
S\,
C^{\,}_{AB\,k}\,
S^{-1}=
-C^{\,}_{AB\,k}
\end{equation}
with the anti-commutator
$\{S,P\}=0$
in agreement with
Eq.~(\ref{eq: mathcal{S} mathcal{P} anticommute for kekule}).
We then have
\begin{align}
[
S,
\Gamma^{\,}_{\mathscr{P}}
]=&\,
[
S,
C^{\,}_{AB\,k}\,
P
] 
\notag\\
=&\,
\{
S,C^{\,}_{AB\,k}
\}\,
P
-
C^{\,}_{AB\,k}\,
\{
S,
P
\}
\notag\\
=&\,
0.
\label{eq: emergent tcommutator S and P for Kekule}
\end{align} 
We have constructed the explicit representations of 
$S$
and
$\Gamma^{\,}_{\mathscr{P}}$
obtained after exact diagonalization of
Eq.~(\ref{eq: def perturbed HK})
and verified that
Eq.~(\ref{eq: emergent tcommutator S and P for Kekule})
holds.

\begin{figure}[t!] 
\centering
\subfigure[]{
\includegraphics[width=0.4\textwidth]{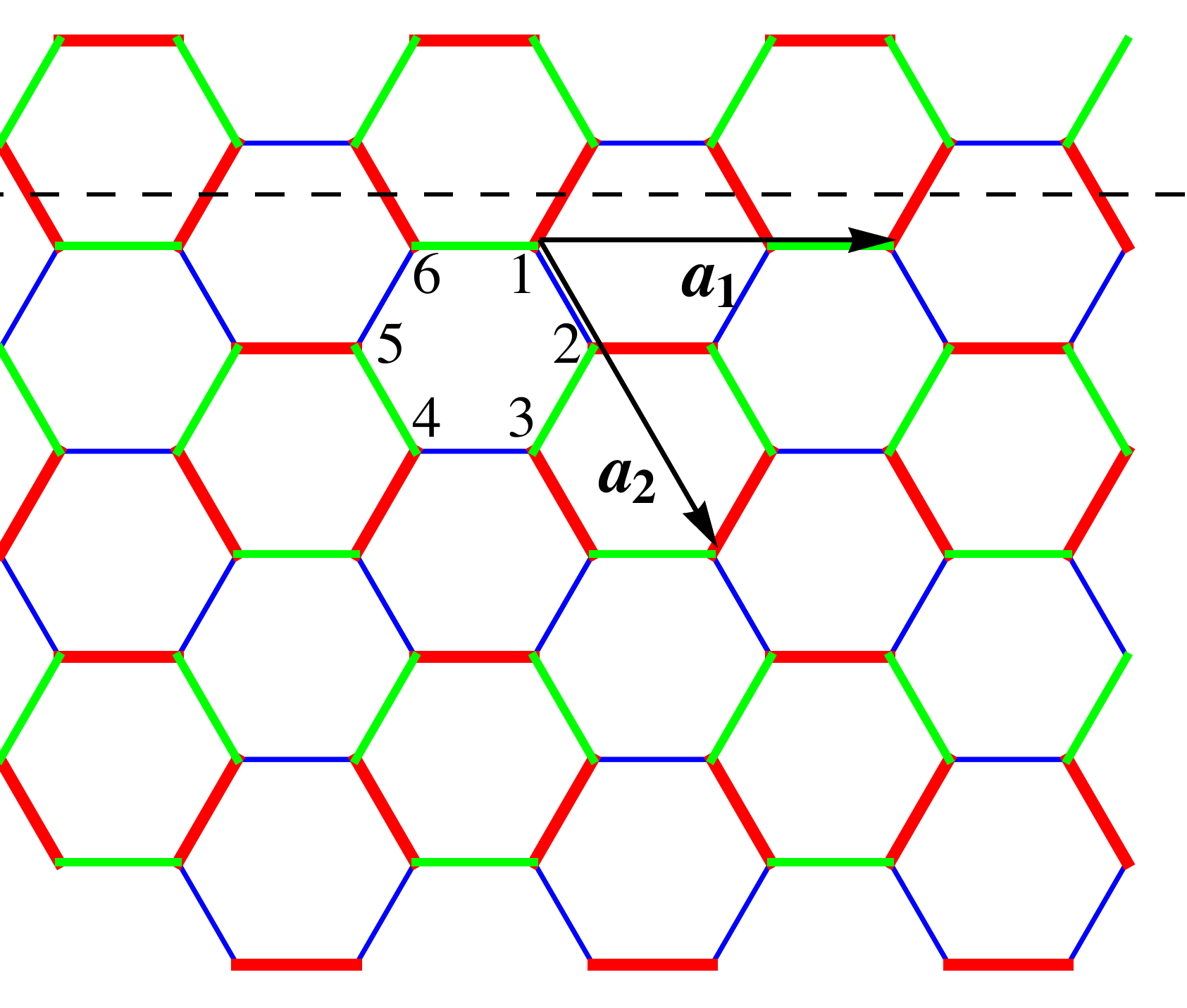}
\label{fig: Kekule+zigzag a}
}
\subfigure[]{
\includegraphics[width=0.1 \textwidth]{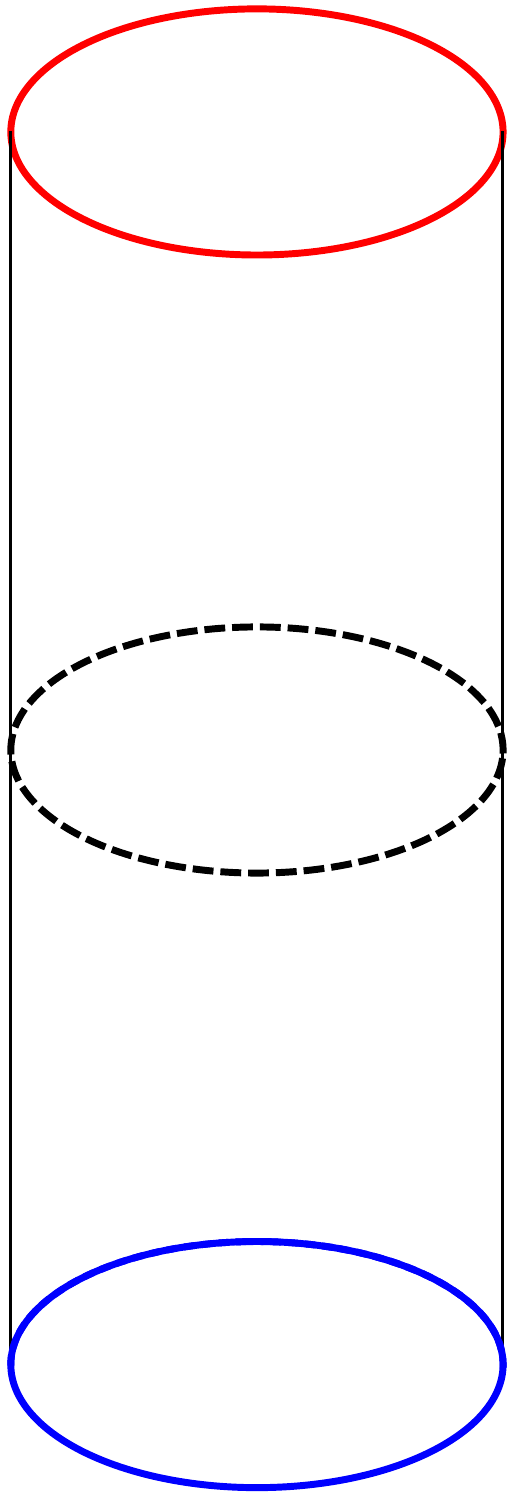}
\label{fig: Kekule+zigzag b}
}
\caption{(Color online)
(a) The nearest-neighbor bonds of the honeycomb lattice are colored
in red, blue, and green as depicted. The colors
red, blue, and green correspond to the values
$t^{\,}_{1}$, 
$t^{\,}_{2}$, 
and $t^{\,}_{3}$ 
taken by the nearest-neighbor hopping amplitudes
for spinless fermions hopping on the honeycomb lattice
with Hamiltonian%
~(\ref{eq: def hat H graphene with rotating kekule texture}), 
respectively. The repeat unit cell with its spanning vectors 
$\bm{a}^{\,}_{1}$ and $\bm{a}^{\,}_{2}$
was defined in Fig.~\ref{fig: Kekule+armchair a}. 
An armchair entangling edge is obtained 
by opening the honeycomb lattice through the dashed line. 
(b) The cylindrical geometry with the two 
armchair edges differing by their colors is selected
by imposing periodic boundary conditions along
the $\bm{a}^{\,}_{1}$ direction and open ones along the $\bm{a}^{\,}_{2}$
direction.
The top armchair edge denoted by a red ellipse has fermions hopping
along it with the consecutive hopping amplitudes 
$t^{\,}_{1}$, $t^{\,}_{2}$, $t^{\,}_{3}$, and $t^{\,}_{2}$.
The bottom armchair edge denoted by the blue ellipse has fermions hopping
along it with the consecutive hopping amplitudes 
$t^{\,}_{1}$, $t^{\,}_{3}$, $t^{\,}_{2}$, and $t^{\,}_{3}$.
[Note that these are not the armchair boundaries shown in panel (a).]
       }
\label{fig: Kekule+zigzag}
\end{figure}

\subsection{Rotated Kekule with armchair edges}
\label{Rotated Kekule with armchair edges}

The honeycomb lattice is unchanged under rotations by
$\pi/3$ about the center of an elementary hexagon.
The Kekule order breaks this point-group symmetry
down to rotations by $2\pi/3$ about the center of 
an elementary hexagon. This pattern of symmetry breaking
is that of $C^{\,}_{6}\to C^{\,}_{3}$, 
where $C^{\,}_{n}$ is the $n$-fold rotation symmetry group. 
Rotations by arbitrary angles about a point in a plane form a group, 
the Abelian group $U(1)$. 

This suggests a connection between the Kekule order parameter
and the spontaneous breaking of an internal 
symmetry group $U(1)$. This connection becomes 
precise in the approximation by which the spectrum of
graphene is linearized about the Dirac points.
The actions of rotations
about the center of an hexagon in graphene
involve, in the Dirac approximation,
a mixing of the components of the Dirac spinors
through the action of one of the Dirac matrices
denoted $\gamma^{\,}_{5}$.
In the terminology of high-energy physics,
$\gamma^{\,}_{5}$ is associated to
a pseudoscalar charge called the axial charge.
For graphene, this pseudoscalar charge is the
difference in the local density of electrons
associated to the two valleys of graphene.

In the Dirac limit,
the pattern of symmetry breaking induced by a
Kekule distortion becomes
the spontaneous breaking of a continuous $U(1)$
symmetry generated by the axial gauge charge.

The Kekule distortion can support a point defect
at which three Kekule distortions differing pairwise
by a global axial phase of either 
$2\pi/3$ or $4\pi/3$ (mod $2\pi$) 
meet. With open boundary conditions,
such a point defect was shown two support two localized
zero modes.
There is one zero mode localized around the point defect.
There is one zero mode localized somewhere on the boundary.
The location on the boundary of the latter zero mode 
depends on the value taken by the global axial phase
of the defective Kekule distortion.

Our purpose is to study the influence of the choice made
for the global axial phase of a uniform Kekule distortion
on the spectrum of the Kekule Hamiltonian
with armchair open boundary conditions.

\begin{figure}[t]
\centering
\includegraphics[width=0.7 \textwidth]{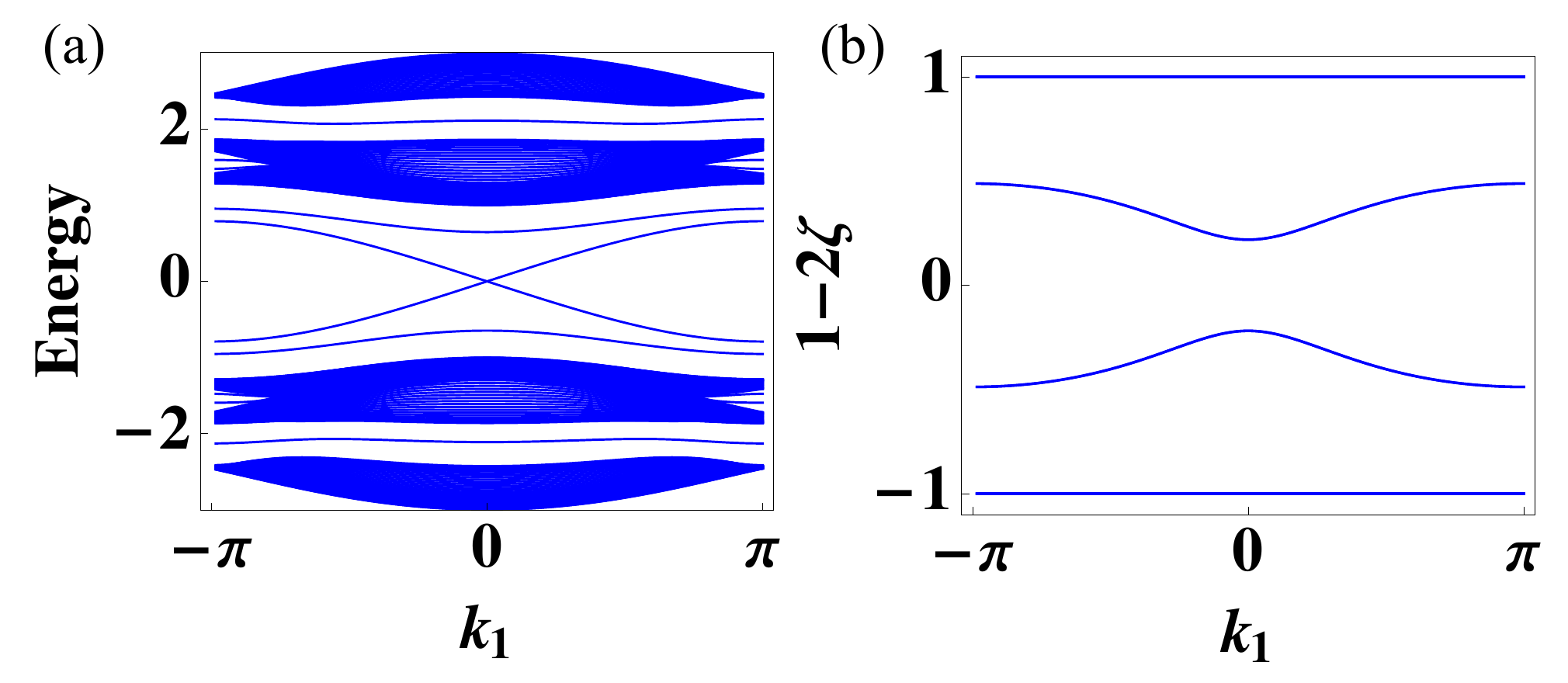}
\caption{(Color online) 
(a)
Energy spectrum with two armchair edges 
in the cylindrical geometry of Fig.~\ref{fig: Kekule+zigzag b}.
(b) Entanglement spectrum 
with two entangling armchair boundaries  
in a toroidal geometry.
The energy scales are
$\Delta^{\,}_0=1$ and $t=-1$.
The axial phase is
$\alpha \cong 5.927$.
The lattice size is $(N^{\,}_{1}, N^{\,}_{2})=(128,32)$. 
The number of unit cells along $\bm{a}^{\,}_{i}$
is $N^{\,}_{i}$ for $i=1,2$. 
       }
\label{ees2}
\end{figure}

\subsubsection{Rotated Kekule distortion} 
\label{subsubsec: Rotated Kekule distortion} 

To parametrize a global rotation of a Kekule distortion,
we define
\begin{subequations}
\label{eq: def hat H graphene with rotating kekule texture}
\begin{equation}
\hat{H}^{\,}_{\mathrm{K}}(\alpha):=
\sum_{n=1,2,3}
t^{\,}_{n}(\alpha)\,
\sum_{\langle\bm{i}^{\,}_{n}\bm{j}^{\,}_{n}\rangle}
\left(
\hat{c}^{\dag}_{\bm{i}^{\,}_{n}}\,
\hat{c}^{\,  }_{\bm{j}^{\,}_{n}}
+
\mathrm{H.c.}
\right),
\label{eq: def hat H graphene with rotating kekule texture a}
\end{equation}
where
\begin{align}
t^{\,}_{n}(\alpha):=&\,
-
t
+
\frac{1}{3}\,
\Big(
\Delta^{\,}_{\mathrm{K}}(\alpha)\,e^{i (n-1) 2\pi/3}
%\notag\\
%&\,
+
\Delta^{*}_{\mathrm{K}}(\alpha)\,e^{-i (n-1) 2\pi/3}
\Big),
\label{eq: def hat H graphene with rotating kekule texture b}
\end{align}
and
\begin{equation}
\Delta^{\,}_{\mathrm{K}}(\alpha):= 
\Delta^{\,}_{\mathrm{K}0}\,
e^{+\mathrm{i}\alpha}.
\label{eq: def hat H graphene with rotating kekule texture c}
\end{equation}
\end{subequations} 
Here, $t$ is real-valued and the Kekule distortion has the amplitude 
$\Delta^{\,}_{\mathrm{K}0}\geq0$ and global axial phase $0\leq\alpha<2\pi$.
Moreover, a pair of nearest-neighbor sites  
of the honeycomb lattice is denoted by
$\langle\bm{i}^{\,}_{i}\bm{j}^{\,}_{i}\rangle$
if they are connected by a bond colored in red ($i=1$), blue ($i=2$), and
green ($i=3$), respectively as is depicted in Fig.\ \ref{fig: Kekule+zigzag a}.

On a torus, two-dimensional momentum 
$\bm{k}\equiv(k^{\,}_{1},k^{\,}_{2})$
is a good quantum number and 
the single-particle Hamiltonian
with a rotated Kekule distortion becomes
(the $\alpha$ dependence is implicit)
%\begin{widetext}
\begin{equation}
\label{eq: H of rotation kekule}
\mathcal{H}^{\,}_{\bm{k}}=
\begin{pmatrix}
0 & 
t^{\,}_{2} & 
0 & 
t^{\,}_{1}\,e^{-\mathrm{i} (k^{\,}_{2}-k^{\,}_{1})} & 
0 & 
t^{\,}_{3} 
\\
t^{\,}_{2} & 
0 & 
t^{\,}_{3} & 
0 & 
t^{\,}_{1}\,e^{+\mathrm{i} k^{\,}_{1}} & 
0 
\\
0 & 
t^{\,}_{3} & 
0 & 
t^{\,}_{2} & 
0 & 
t^{\,}_{1}\,e^{+\mathrm{i}k^{\,}_{2}} 
\\
t^{\,}_{1}\,e^{+\mathrm{i}(k^{\,}_{2}-k^{\,}_{1})} & 
0 & 
t^{\,}_{2} & 
0 & 
t^{\,}_{3} & 
0 
\\
0 & 
t^{\,}_{1}\,e^{-\mathrm{i}k^{\,}_{1}} & 
0 & 
t^{\,}_{3} & 
0 & 
t^{\,}_{2}
\\
t^{\,}_{3} & 
0 & 
t^{\,}_{1}\,e^{-\mathrm{i} k^{\,}_{2}} & 
0 & 
t^{\,}_{2} & 
0
\end{pmatrix}.
\end{equation}
%\end{widetext}  
For vanishing global axial phase, 
Hamiltonian (\ref{eq: H of rotation kekule})
reduces to Hamiltonian (\ref{eq: H of kekule}).
Hamiltonian (\ref{eq: H of rotation kekule})
is invariant under the same operation for time-reversal
as was the case for Hamiltonian~(\ref{eq: H of kekule}).
The spectrum of Hamiltonian~(\ref{eq: H of rotation kekule})
is invariant under the same chiral (sublattice) operation  
as was the case for Hamiltonian (\ref{eq: H of kekule}).
Hamiltonian~(\ref{eq: H of rotation kekule}) breaks
the inversion symmetry enjoyed by
Hamiltonian (\ref{eq: H of kekule})
for any $\alpha\neq0$ mod $2\pi$.
This can be seen by inspection of the armchair boundaries
in Fig.~\ref{fig: Kekule+zigzag a}
and is indicated 
in Fig.~\ref{fig: Kekule+zigzag b} 
by the distinct colors used
to denote the two edges if a cylindrical geometry is selected
by the choice of boundary conditions with the direction
$\bm{a}^{\,}_{2}$ the open direction.

\subsubsection{Spectra for rotated Kekule}
\label{subsubsec: Spectra for rotated Kekule}

\begin{figure}[t]
\centering
\includegraphics[height=5.5 cm]{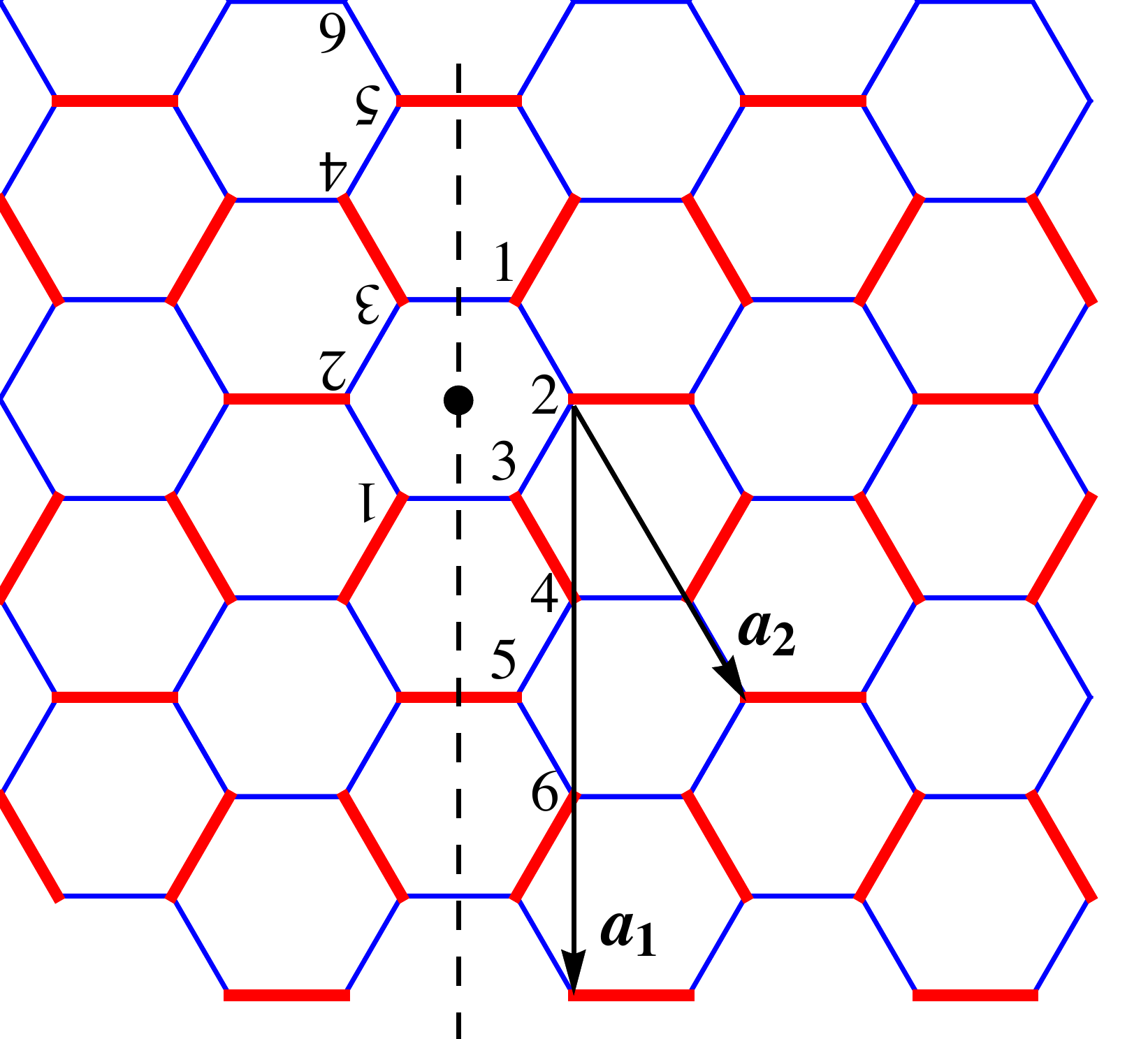}
\caption{(Color online) 
The nearest-neighbor bonds of the honeycomb lattice are colored
in red and blue as depicted. The colors red and blue correspond to the values
$t^{\,}_{1}$
and
$t^{\,}_{2}$
taken by the nearest-neighbor hopping amplitudes
for spinless fermions hopping on the honeycomb lattice
with Hamiltonian%
~(\ref{eq: H zigzag}), 
respectively. 
A Kekule distortion follows from choosing 
$t^{\,}_{1}\neq t^{\,}_{2}$.
A physical zigzag edge is constructed 
from cutting through the dashed line.
The dashed line also defines a zigzag entangling edge.
The symbol $\bullet$ denotes an inversion center. 
One repeat unit cell contains three hexagons 
defined as follows.
The first hexagon from the repeat unit cell has two sites numbered
1 and 2, whereby site 1 is connected by a blue bond to
site 2.
The second hexagon from the repeat unit cell has three sites numbered
2, 3, and 4, whereby site 2 is connected by a blue bond to site 3,
while site 3 is connected by a red bond to site 4.
The third hexagon from the repeat unit cell has all
six edges colored in blue with the vertices numbered
4, 5, and 6.
The spanning vectors corresponding to this unit cell are
$\bm{a}^{\,}_{1}$ and $\bm{a}^{\,}_{2}$.
The image of the repeat unit cell under inversion about 
the point $\bullet$ is has its three hexagons labeled with
the numbers $1$ to $6$ written upside down.
        }
\label{fig:zig}
\end{figure}

The energy spectrum as a function of the good momentum quantum number
$k^{\,}_{1}$ is computed by diagonalizing the Hamiltonian%
~(\ref{eq: def hat H graphene with rotating kekule texture}) 
in the cylindrical geometry of Fig.~\ref{fig: Kekule+zigzag b}.
For any $\alpha\neq0$ modulo $2\pi$, inversion symmetry is broken
so that the quantum dynamics on the opposite edges of the cylinder
in Fig.~\ref{fig: Kekule+zigzag b} differ. Remarkably, at a critical value
of the axial angle $\alpha^{\,}_{\mathrm{c}}$, the gap for the single pair of
left- and right-movers on one of the edges closes, while it
does not for the single pair of left- and right-movers from the other edge.
This property of the energy spectrum of the rotated Kekule distortion
is illustrated in Fig.~\ref{ees2}(a).
The critical value of the axial angle $\alpha^{\,}_{\mathrm{c}}$ is
$\alpha^{\,}_{\mathrm{c}}\approx 5.927$ when $(\Delta^{\,}_{\mathrm{K}0},t)=(1,-1)$,
as we now show. To this end, we use the $4\times4$ Hamiltonians
\begin{align}
\mathcal{H}^{\mathrm{edge}}_{\mathrm{top}\,k^{\,}_{1}}(\alpha):=
\begin{pmatrix}
0 & t^{\,}_{1}(\alpha) & 0 & t^{\,}_{2}(\alpha)\,e^{+\mathrm{i}k^{\,}_{1}} \\
t^{\,}_{1}(\alpha) & 0 & t^{\,}_{2}(\alpha) & 0 \\
0 & t^{\,}_{2}(\alpha) & 0 & t^{\,}_{3}(\alpha) \\
t^{\,}_{2}(\alpha)\,e^{-\mathrm{i}k^{\,}_{1}} & 0 & t^{\,}_{3}(\alpha) & 0
\end{pmatrix}
\end{align}
and
\begin{align}
\mathcal{H}^{\mathrm{edge}}_{\mathrm{bot}\,k^{\,}_{1}}(\alpha):=
\begin{pmatrix}
0 & t^{\,}_{1}(\alpha) & 0 & t^{\,}_{3}(\alpha)\,e^{+\mathrm{i}k^{\,}_{1}} \\
t^{\,}_{1}(\alpha) & 0 & t^{\,}_{3}(\alpha) & 0 \\
0 & t^{\,}_{3}(\alpha) & 0 & t^{\,}_{2}(\alpha) \\
t^{\,}_{3}(\alpha)\,e^{-\mathrm{i} k^{\,}_{1}} & 0 & t^{\,}_{2}(\alpha) & 0
\end{pmatrix}
\end{align}
to model hopping restricted to the top and bottom armchair boundaries
defined in Fig.~\ref{fig:zig}, respectively. For the top armchair edge,
we are using four orbitals per repeat unit cell with the conventions that
orbital 1 hops to orbital 2 with the amplitude $t^{\,}_{1}(\alpha)$,
orbital 2 hops to orbital 3 with the amplitude $t^{\,}_{2}(\alpha)$,
orbital 3 hops to orbital 4 with the amplitude $t^{\,}_{3}(\alpha)$,
and orbital 4 hops to orbital 1 in the neighboring repeat unit cell
with the amplitude $t^{\,}_{2}(\alpha)$.
For the bottom armchair edge, we are using the conventions 
that follow from those for the top armchair edge obtained 
by exchanging $t^{\,}_{2}(\alpha)$ and  $t^{\,}_{3}(\alpha)$.
Eigenstates of Hamiltonian $\mathcal{H}^{\mathrm{edge}}_{\mathrm{top}\,k^{\,}_{1}}(\alpha)$
at $k^{\,}_{1}=0$  with zero energy eigenvalue satisfy
$\mathrm{det}\,[\mathcal{H}^{\mathrm{edge}}_{\mathrm{top}\,k^{\,}_{1}=0}(\alpha)]=0$.
This condition gives 
$\alpha^{\,}_{\mathrm{c}}\approx 5.927$ 
when $(\Delta^{\,}_{\mathrm{K}0},t)=(1,-1)$
for the critical axial angle.

We have also calculated the entanglement spectrum 
using the same partition as the one used in 
Sec.~\ref{subsec: gaphene partition},
with the proviso that we are now using the dashed line
shown in Fig.~\ref{fig: Kekule+zigzag a}. To avoid contamination in the entanglement
spectrum arising from the gap closing along one of the physical edges
when $\alpha=\alpha^{\,}_{\mathrm{c}}$, we choose to impose full
periodic boundary conditions, i.e., the geometry of a torus.
We expect no closing of the entanglement spectrum because
of the breaking of inversion symmetry by any $\alpha\neq0$ modulo $2\pi$
and indeed, this is what is observed from exact diagonalization and
illustrated with the help of Fig.~\ref{ees2}(b).

\subsection{Kekule with zigzag edges}

\subsubsection{Hamiltonian}
The energy spectrum of semi-infinite graphene modeled by
a single nearest-neighbor hopping amplitude on the honeycomb lattice
with a zigzag edge shows flat (dispersionless) bands
connecting the two Dirac points.%
~\cite{Mitsutaka96} 
These zero-energy flat bands are protected by chiral symmetry,
see Ref. \ \cite{Shinsei2002}. 
These flat bands can become dispersive 
by tuning on when perturbed by a one-body potential 
that breaks the inversion symmetry, 
as demonstrated in Ref.~\cite{Yao2009}.
We consider the energy and entanglement spectra of graphene 
with a Kekule distortion in the presence of 
physical and entangling zigzag edges.
The repeat unit cell and the spanning vectors are defined
in Fig~\ref{fig:zig}. With the conventions of
Fig~\ref{fig:zig},
the BZ and four inversion symmetric momenta are identical to the
BZ and four inversion symmetric momenta from Fig.~\ref{fig: Kekule+armchair b}
if we impose periodic boundary conditions.

%\begin{widetext}
The single particle Hamiltonian in momentum space is 
\begin{equation}
\label{eq: H zigzag}
\mathcal{H}^{\,}_{\mathrm{zig}\,\bm{k}}=
\begin{pmatrix}
0 & 
t^{\,}_{2} & 
0 & 
t^{\,}_{2}\,e^{-\mathrm{i} k^{\,}_2} & 
0 & 
t^{\,}_{1}\,e^{-\mathrm{i} k^{\,}_{1}}
\\
t^{\,}_{2} & 
0 & 
t^{\,}_{2} & 
0 & 
t^{\,}_{1}\,e^{+\mathrm{i} (k^{\,}_{2}-k_1)} & 
0 
\\
0 & 
t^{\,}_{2} & 
0 & 
t^{\,}_{1} & 
0 & 
t^{\,}_{2}\,e^{-\mathrm{i}k^{\,}_{2}} 
\\
t^{\,}_{2}\,e^{+\mathrm{i}k^{\,}_{2}} & 
0 & 
t^{\,}_{1} & 
0 & 
t^{\,}_{2} & 
0 
\\
0 & 
t^{\,}_{1}\,e^{-\mathrm{i}(k^{\,}_{2}-k_1)} & 
0 & 
t^{\,}_{2} & 
0 & 
t^{\,}_{2} 
\\
t^{\,}_{1}\,e^{+\mathrm{i} k^{\,}_{1}} & 
0 & 
t^{\,}_{2}\,e^{+\mathrm{i} k^{\,}_{2}} & 
0 & 
t^{\,}_{2} & 
0
\end{pmatrix}.
\end{equation}
%\end{widetext}

\begin{figure}[t]
\centering
\includegraphics[height=9 cm]{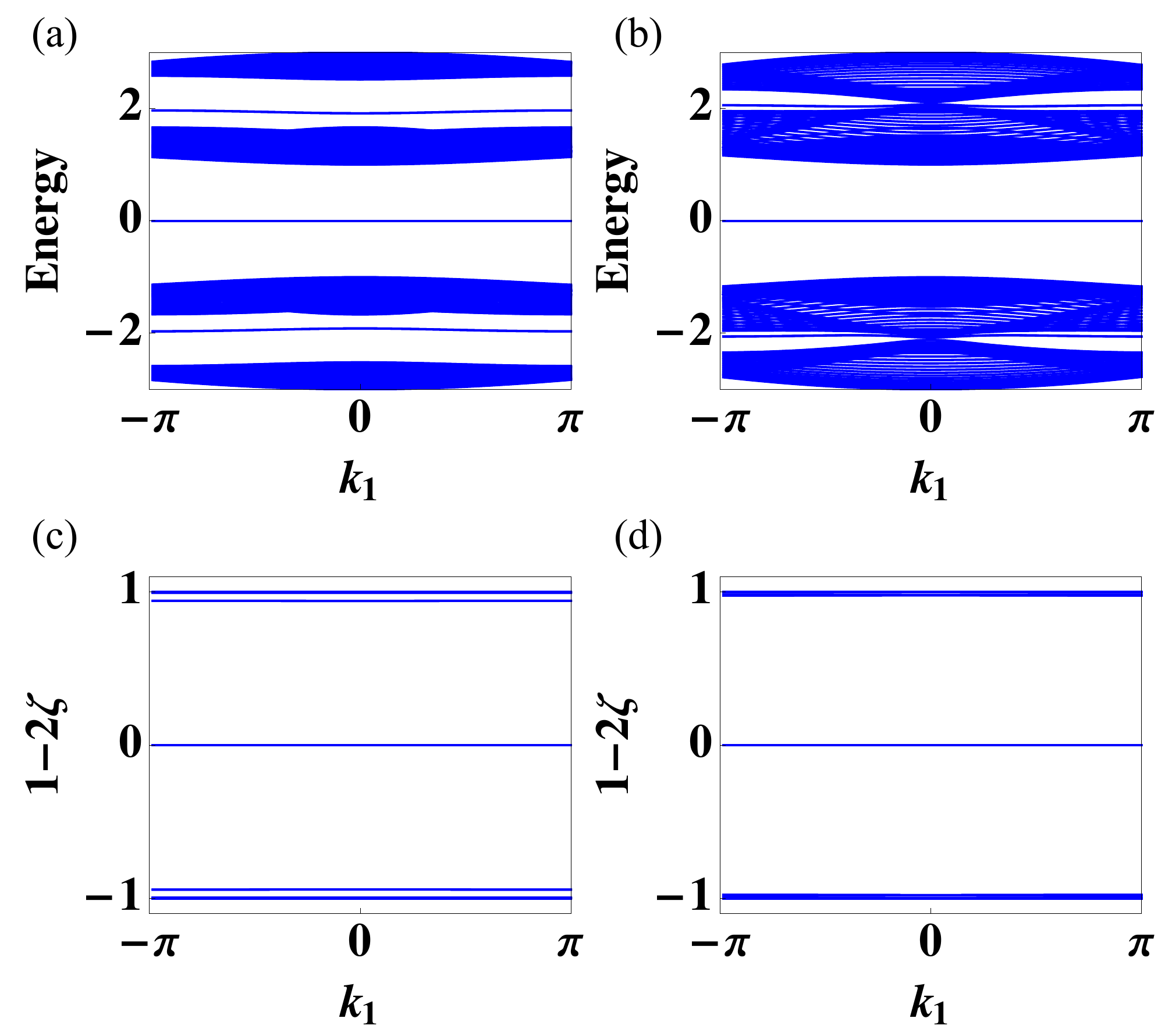}
\caption{(Color online)
Energy spectra 
of $\mathcal{H}^{\,}_{\mathrm{zig}\,k^{\,}_{1}}$ 
in Eq.~(\ref{eq: def perturbed HK zigzag})
with zigzag edges in a cylinder geometry for
(a) 
$(t^{\,}_{1},t^{\,}_{2})=({1}/{3},{4}/{3})$ 
and 
(b) 
$(t^{\,}_{1},t^{\,}_{2})=({5}/{3},{2}/{3})$. 
Entanglement spectra 
of $\mathcal{H}^{\,}_{\mathrm{zig}\,k^{\,}_{1}}$ 
in Eq.~(\ref{eq: def perturbed HK zigzag})
with zigzag entangling edges in a torus geometry for
(c) 
$(t^{\,}_{1},t^{\,}_{2})=({1}/{3},{4}/{3})$ 
and  
(d)
$(t^{\,}_{1},t^{\,}_{2})=({5}/{3},{2}/{3})$.
The dimensions of the lattice are given by 
$(N^{\,}_{1},N^{\,}_{2})=(128,32)$,
where $N^{\,}_{i}$ 
is the number of the repeat unit cell from
Fig.~\ref{fig:zig}
along the direction of the spanning vector
$\bm{a}^{\,}_{i}$ ($i=1,2$)
and in units for which the spanning vectors
$\bm{a}^{\,}_{1}$
and
$\bm{a}^{\,}_{2}$
are of unit length.}
\label{fig:zigzag}
\end{figure}

\subsubsection{Symmetries}

The symmetry
\begin{equation}
\mathcal{H}^{*}_{\mathrm{zig}\,-\bm{k}}=
\mathcal{H}^{\,}_{\mathrm{zig}\,+\bm{k}}
\label{eq: TRS for Kekule zigzag}
\end{equation}
implements time-reversal symmetry for spinless fermions.

The spectral symmetry
\begin{subequations}
\label{eq: SLS for Kekule zigzag}
\begin{equation}
\mathcal{S}^{-1}\,\mathcal{H}^{\,}_{\mathrm{zig}\,\bm{k}}\,\mathcal{S}=
-\mathcal{H}^{\,}_{\mathrm{zig}\,\bm{k}},
\end{equation}
where
\begin{equation}
\label{eq: chiral kekule zigzag}
\mathcal{S}:=
\mathrm{diag}\,
\left(
1, -1, 1, -1,1,-1
\right)
%\begin{pmatrix}
%1 & 0 & 0 & 0 & 0 & 0 
%\\
%0 & -1 & 0 & 0 & 0 & 0 
%\\
%0 & 0 & 1 & 0 & 0 & 0 
%\\
%0 & 0 & 0 & -1 & 0 & 0 
%\\
%0 & 0 & 0 & 0 & 1 & 0 
%\\
%0 & 0 & 0 & 0 & 0 & -1
%\end{pmatrix}
\end{equation}
\end{subequations}
implements the chiral (sublattice) spectral symmetry.
The symmetry
\begin{subequations}
\label{eq: inversion symmetry Kekule zigzag}
\begin{equation}
\mathcal{P}^{-1}_{\mathrm{zig}}\,
\mathcal{H}^{\,}_{\mathrm{zig}\,-\bm{k}}\,
\mathcal{P}^{\,}_{\mathrm{zig}}=
\mathcal{H}^{\,}_{\mathrm{zig}\,+\bm{k}},
\label{eq: inversion symmetry Kekule zigzag a}
\end{equation}
where
\begin{equation}
\mathcal{P}^{\,}_{\mathrm{zig}}:=
\begin{pmatrix}
0 & 0 & 0 & 0 & 0 & 1 
\\
0 & 0 & 0 & 0 & 1 & 0 
\\
0 & 0 & 0 & 1 & 0 & 0
\\
0 & 0& 1 & 0 & 0 & 0 
\\
0 & 1 & 0 & 0 & 0 & 0
\\
1 & 0 & 0 & 0 & 0 & 0
\end{pmatrix},
\label{eq: inversion symmetry Kekule zigzag b}
\end{equation}
\end{subequations}
implements the inversion symmetry defined with the help
of Fig.~\ref{fig:zig}. To define the inversion symmetry,
we first draw the dashed line in Fig.~\ref{fig:zig}.
A cut along this dashed line defines a zigzag boundary.
We then select the intersection of the dashed line
with the mid-point of a hexagon
in Fig.~\ref{fig:zig}. This mid-point, represented by
a filled circle in Fig.~\ref{fig:zig}, 
defines the inversion center.
Performing an inversion about this point
maps the Kekule pattern on the right of the dashed line into
the Kekule pattern on the left of the dashed line.
The two patterns are identical, hence the inversion symmetry.
On the one hand, the labels
in the enlarged repeat unit cell on the right of the dashed line becomes those
on the left the dashed line in Fig.~\ref{fig:zig}
under this inversion. On the other hand,
if the convention for the labels of the enlarged
repeat unit cell are identical on the right and left of the 
dashed line, the representation
~(\ref{eq: inversion symmetry Kekule zigzag})
follows.

Observe that 
\begin{equation}
\{\mathcal{S},\mathcal{P}^{\,}_{\mathrm{zig}}\}=0.
\label{eq: mathcal{S} mathcal{P} anticommute for kekule zigzag}
\end{equation} 

\subsubsection{Partition}

The partition is defined with respect to the dashed line
in Fig.~\ref{fig:zigzag}
as was done in Sec.~\ref{subsec: gaphene partition}.

\subsubsection{Hamiltonian and entanglement spectra}

Let
\begin{align}
\mathcal{H}^{\,}_{k^{\,}_{1}}:=
\mathcal{H}^{\,}_{\mathrm{zig}\,k^{\,}_{1}}+\mathcal{H}',
\label{eq: def perturbed HK zigzag}
\end{align}
where 
$\mathcal{H}'$ is a one-body perturbation that breaks either 
the chiral symmetry, the inversion symmetry, or both,
and we have imposed periodic boundary conditions
along the $\bm{a}^{\,}_{1}$ direction from Fig.~\ref{fig:zigzag}
so that $k^{\,}_{1}$ is a good quantum number.
Open boundary conditions are imposed 
along the $\bm{a}^{\,}_{2}$ direction from Fig.~\ref{fig:zigzag}
when computing energy spectra.
Periodic boundary conditions are imposed 
along the $\bm{a}^{\,}_{2}$ direction from Fig.~\ref{fig:zigzag}
when computing entanglement spectra in order to avoid a spectral
contamination of the entanglement boundary states
arising from the zero modes from the physical boundaries.

We have studied by exact diagonalization
both the energy and entanglement spectra
of Hamiltonian~(\ref{eq: def perturbed HK zigzag})
in a cylinder and torus geometry, respectively,

In the absence of the perturbation $\mathcal{H}'$, 
non-dispersing edge states at zero energy
are present for any Kekule distortion, i.e., as soon as 
$t^{\,}_{1}\neq t^{\,}_{2}$, 
as is illustrated in 
Figs.~\ref{fig:zigzag}(a) and \ref{fig:zigzag}(b).
Non-dispersing zero modes localized on the entangling boundary
are also found in the entanglement spectrum 
for any Kekule distortion, i.e., as soon as 
$t^{\,}_{1}\neq t^{\,}_{2}$, 
as is illustrated in 
in Figs.~\ref{fig:zigzag}(c) and \ref{fig:zigzag}(d).

In the following,
the number of repeat unit cells are
$(N^{\,}_{1},N^{\,}_{2})=(128,32)$
and we set $(t^{\,}_{1},t^{\,}_{2})=(5/3,2/3)$
when studying the robustness of the flat (entangling) edge states
in the presence of three distinct $\mathcal{H}'$.

\begin{figure}[t]
\centering
\includegraphics[height=8 cm]{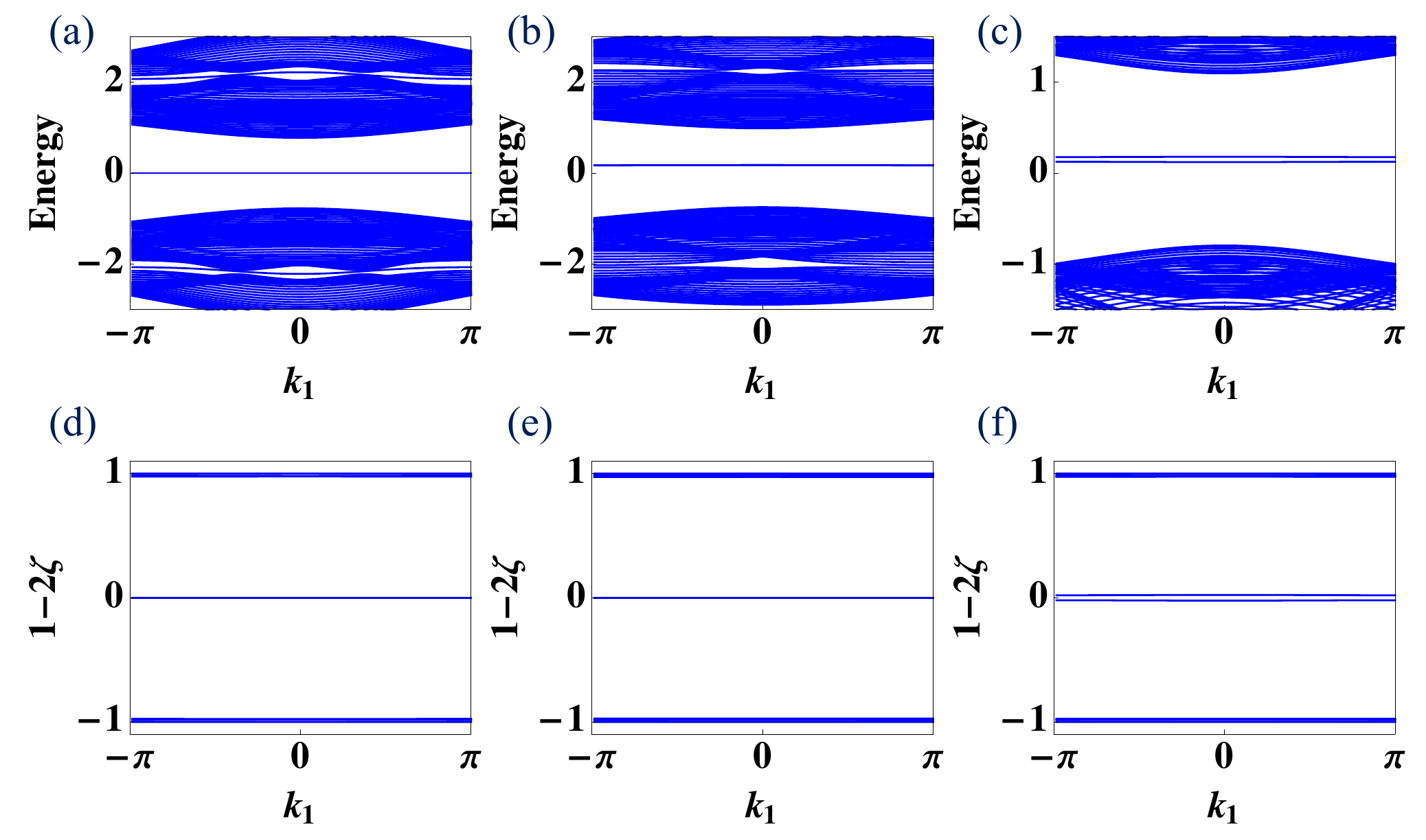}
\caption{(Color online)
Energy spectra of 
$\mathcal{H}^{\,}_{k^{\,}_{1}}$ in Eq.~(\ref{eq: def perturbed HK zigzag})
with zigzag edges in a cylindrical geometry are presented in the left column.
Entanglement spectra of 
$\mathcal{H}^{\,}_{k^{\,}_{1}}$ in Eq.~(\ref{eq: def perturbed HK zigzag})
with zigzag entangling edges in a torus geometry
are presented in the right column.
In both cases, we set 
$(t^{\,}_{1},t^{\,}_{2})=({5}/{3},{2}/{3})$
and
$(N^{\,}_{1},N^{\,}_{2})=(128,32)$.
The spectra (a) and (d) are obtained by choosing the
inversion-breaking perturbation%
~(\ref{eq: def mathcal H' zig IB})
in Hamiltonian~(\ref{eq: def perturbed HK zigzag}).
There are mid-gap flat bands that are two-fold degenerate.
The spectra (b) and (e) are obtained by choosing the
chiral-symmetry-breaking perturbation%
~(\ref{eq: def mathcal H'  zig chSB})
in Hamiltonian~(\ref{eq: def perturbed HK zigzag}).
The flat bands remain two-fold degenerate but are shifted
away from zero energy in panel (b).
The spectra (c) and (f) are obtained by choosing both 
the inversion-symmetry-breaking and chiral-symmetry-breaking perturbation% 
~(\ref{eq: def mathcal H' zig ISB})
to Hamiltonian~(\ref{eq: def perturbed HK zigzag}).
The two-fold degeneracy of the flat bands is lifted
and all flat bands are shifted away from the eigenvalue zero
in panels (c) and (f).
         }
\label{fig:zigzag2}
\end{figure}

\paragraph{Inversion symmetry breaking}
The inversion-symmetry-breaking perturbation is chosen 
in Eq.~\ref{eq: def perturbed HK zigzag} to be
\begin{align}
\label{eq: def mathcal H' zig IB}
\mathcal{H}':=
\left(\begin{array}{cccccc}
0 & v'_{1}\,e^{+\mathrm{i}k^{\,}_{1}} & 0 &0& 0 & 0 \\
 v'_{1}\,e^{-\mathrm{i}\,k^{\,}_{1}} & 0 &0 & 0 & 0 & 0 \\
0 & 0 & 0 & 0 & 0 & 0 \\
0 & 0 & 0 & 0 & 0 & 0 \\
0 & 0 & 0 & 0 & 0 &  0 \\ 
0 & 0 & 0 & 0 &  0 & 0
\end{array}\right),
\end{align}
where $v'_{1}=0.5$.
The flat bands when $v'_{1}=0.5$
are robust in both the energy and the entanglement spectra 
under this inversion-symmetry-breaking but
chiral-symmetry-preserving perturbation
as is shown in Fig. ~\ref{fig:zigzag2}(a) and \ref{fig:zigzag2}(d),
respectively.

\paragraph{Chiral symmetry breaking}
The chiral-symmetry-breaking perturbation 
is chosen in Eq.~(\ref{eq: def perturbed HK zigzag}) 
to be
\begin{align}
\label{eq: def mathcal H'  zig chSB}
\mathcal{H}':=
\mathrm{diag}\,\left(
o^{\,}_{1}, 
o^{\,}_{2}, 
o^{\,}_{3}, 
o^{\,}_{3}, 
o^{\,}_{2}, 
o^{\,}_{1} 
\right), 
\end{align}
where $o^{\,}_{1}=-0.13$, $o^{\,}_{2}=0.2$, and $o^{\,}_{3}=0.3$.
The flat bands when $o^{\,}_{1}=o^{\,}_{2}=o^{\,}_{3}=0$ 
are shifted away from zero energy in
the energy spectrum shown in Fig. ~\ref{fig:zigzag2}(b).
However, the flat bands when $o^{\,}_{1}=o^{\,}_{2}=o^{\,}_{3}=0$ 
are unchanged by $\mathcal{H}'$
in the entanglement spectrum 
shown in Fig. ~\ref{fig:zigzag2}(e). 

\paragraph{Chiral symmetry and inversion symmetry breaking}
The inversion-symmetry-breaking and 
chiral-symmetry-breaking perturbation 
is chosen in Eq.~(\ref{eq: def perturbed HK zigzag}) 
to be
\begin{align}
\label{eq: def mathcal H' zig ISB}
\mathcal{H}':=
\mathrm{diag}\,
\left(
o^{\,}_{1}, o^{\,}_{2},
o^{\,}_{3}, o^{\,}_{4},
o^{\,}_{5}, o^{\,}_{6}
\right), 
\end{align}
where 
$(o^{\,}_{1},o^{\,}_{2},o^{\,}_{3},o^{\,}_{4},o^{\,}_{5},o^{\,}_{6})=
(0, 0.1,0.2,0.3,0.2,0.1)$
in Figs.~\ref{fig:zigzag2}(c) and \ref{fig:zigzag2}(f).
This  chiral-symmetry-breaking and 
inversion-symmetry-breaking perturbation gaps out 
the flat bands when
$(o^{\,}_{1},o^{\,}_{2},o^{\,}_{3},o^{\,}_{4},o^{\,}_{5},o^{\,}_{6})=
(0,0,0,0,0,0)$ 
both in the energy and entanglement spectra shown in
Fig.~\ref{fig:zigzag2}(c) and \ref{fig:zigzag2}(f),
respectively. 

\subsubsection{Stability analysis of the zero modes}
\label{sec:zigzag stability}

We have observed numerically that flat bands in
$\sigma(\mathcal{H}^{\,}_{\mathrm{zig}\,k^{\,}_{1}})$ and
$\sigma(Q^{\,}_{A\,k^{\,}_{1}})$ 
can only be gaped when both the chiral and inversion symmetries are broken. 
To understand this result, we proceed in two steps.

First, we observe numerically that 
(i) there is one edge state per momentum and per edge 
(with a wavefunction that decays exponentially
fast as a function of the distance away from the edge)
in both the energy and entanglement spectra
in all panels from Fig.~\ref{fig:zigzag2},
(ii) this edge state is non-dispersive 
(independent of the good momentum quantum number $k^{\,}_{1}$) 
in all panels from Fig.~\ref{fig:zigzag2},
and (iii)
with vanishing eigenvalue 
in panels (a), (d), and (e) from Fig.~\ref{fig:zigzag2}.
Properties (i) and (ii) imply that the effective edge theory 
on an isolated single 
zigzag edge of the Hamiltonian 
$\mathcal{H}^{\,}_{\mathrm{zig}\,k^{\,}_{1}}$
is the momentum-resolved $1\times1$ matrix
\begin{subequations}
\begin{equation}
\mathcal{H}^{\,}_{\mathrm{edge}\, k^{\,}_{1}}=
m^{\,}_{\mathrm{energy}}\in\mathbb{R},
\end{equation}
whereas that on an isolated single
zigzag edge of the entangling operator
$Q^{\,}_{A\,k^{\,}_{1}}$ 
is the momentum-resolved $1\times1$ matrix
\begin{equation}
Q^{\,}_{\mathrm{edge}\,A\, k^{\,}_{1}}=
m^{\,}_{\mathrm{entanglement}}\in\mathbb{R}.
\end{equation}
\end{subequations}
Chiral symmetry imposes the constraints
\begin{subequations}
\begin{equation}
\mathcal{S}^{-1}_{\mathrm{edge}}\,
\mathcal{H}^{\,}_{\mathrm{edge}\, k^{\,}_{1}}\,
\mathcal{S}^{\,}_{\mathrm{edge}}=
-\mathcal{H}^{\,}_{\mathrm{edge}\, k^{\,}_{1}}
\end{equation}
and
\begin{equation}
\mathcal{S}^{-1}_{\mathrm{edge}}\,
Q^{\,}_{\mathrm{edge}\,A\, k^{\,}_{1}}\,
\mathcal{S}^{\,}_{\mathrm{edge}}=
-Q^{\,}_{\mathrm{edge}\,A\, k^{\,}_{1}}.
\end{equation}
\end{subequations}
These constraints can only be met if
\begin{subequations}
\begin{equation}
\mathcal{H}^{\,}_{\mathrm{edge}\,k^{\,}_{1}}=0
\end{equation}
and
\begin{equation}
Q^{\,}_{\mathrm{edge}\,A\, k^{\,}_{1}}=0.
\end{equation}
\end{subequations}
Thus, it is the spectral chiral symmetry~(\ref{eq: SLS for Kekule zigzag}) 
that explains the presence of the mid-gap flat bands in panels (a) and (d)
from Fig.~\ref{fig:zigzag2}. The breaking of 
the spectral chiral symmetry~(\ref{eq: SLS for Kekule zigzag}) 
by the perturbation~(\ref{eq: def mathcal H'  zig chSB})
in panels (b) and (e) is manifest in the fact that 
the flat bands in panel (b)
are not to be found anymore at the energy eigenvalue zero.
The fact that the flat bands in panel (e) remains at the eigenvalue zero
must be attributed to another protecting symmetry.

Second, in the presence of the inversion symmetry%
~(\ref{eq: inversion symmetry Kekule zigzag}), 
there exists the operator 
$\Gamma^{\,}_{\mathscr{P}^{\,}_{\mathrm{zig}}}$ such that
[recall Eq.~(\ref{eq: induced symmetry})]
\begin{equation}
\Gamma^{\,}_{\mathscr{P}^{\,}_{\mathrm{zig}}}\,
Q^{\,}_{\mathrm{edge}\,A\,+k^{\,}_{1}}=
-Q^{\,}_{\mathrm{edge}\,A\,-k^{\,}_{1}}\,
\Gamma^{\,}_{\mathscr{P}^{\,}_{\mathrm{zig}}}.
\end{equation}
This effective chiral symmetry of the entanglement spectrum is
the reason why panel (e) from Fig.~\ref{fig:zigzag2}
displays a two-fold degenerate flat band at the eigenvalue zero, whereas
the two-fold degenerate flat band in panel (b) from Fig.~\ref{fig:zigzag2}
is at a non-vanishing energy eigenvalue that is 
determined by the amount of breaking of the spectral chiral symmetry
by the perturbation~(\ref{eq: def mathcal H'  zig chSB}). 
As soon as the inversion symmetry and the spectral chiral symmetry
are simultaneously broken,
as it is in panels (c) and (f) from Fig.~\ref{fig:zigzag2}
by the perturbation~(\ref{eq: def mathcal H' zig ISB}),
the flat bands are to be found at non-vanishing distances
from the eigenvalue zero in both the energy and the entanglement spectra,
while their degeneracy has been lifted.

\subsection{Counting the mid-gap states protected by inversion symmetry}

The stability analysis of the zero modes that we have conducted so
far relied on the number of edges states determined numerically
before the introduction of perturbations.
However, this number of zero modes can be determined analytically as follows.

It was shown in Ref.~\cite{Hughes2011} that
a two-dimensional topological band insulator protected by inversion symmetry is 
characterized by the number of zero modes 
$n^{\mathrm{zero}}_{A,k^{\,}_{i},k^{\star}_{i}}$ 
in the entanglement spectrum $\sigma(Q^{\,}_{A\, k^{\,}_{i}})$ 
defined with periodic boundary conditions (torus geometry). 
Here, we recall that the entangling boundary defined by the partition $A$
is characterized by the good momentum quantum number $k^{\,}_{i}$ where
$i=1,2$ while $i+1$ is defined modulo 2 and that
$\bm{k}^{\star}=(k^{\star}_{1},k^{\star}_{2})$
is any momentum from the Brillouin zone
that is unchanged modulo the addition of a momentum from the
reciprocal lattice with the two spanning vectors 
$\bm{Q}^{\,}_{1}$
and
$\bm{Q}^{\,}_{2}$
under the operation of inversion.
In turn,
it was shown in Refs.~\cite{Hughes2011} and~\cite{Turner2010} that 
\begin{align}
n^{\mathrm{zero}}_{A,k^{\,}_{i},k^{\star}_{i}}=
2
\left|
n^{\,}_{\bm{k}^{\star}}
-
n^{\,}_{\bm{k}^{\star}+(\bm{Q}^{\,}_{i+1}/2)}
\right|,
\label{eq: counting I}
\end{align}
where
$n^{\,}_{\bm{k}^{\star}}$
is the number of occupied Bloch eigenstates at the inversion symmetric momentum 
$\bm{k}^{\star}$
of the single-particle Hamiltonian (defined with periodic boundary conditions)
that are simultaneous eigenstates of the inversion operator.
The counting formula~(\ref{eq: counting I})
is here meaningful because (i)
$\bm{k}^{\star}$ and $\bm{k}^{\star}+(\bm{Q}^{\,}_{i+1}/2)$
are both invariant under the operation of inversion
modulo the addition of a reciprocal momentum
and (ii) it is possible to simultaneously
diagonalize the Bloch Hamiltonian at any inversion symmetric momentum point
and the operator that represents the operation of inversion.
Equation~(\ref{eq: counting I}) is remarkable in that it relates a property
from the entangling boundary, the integer
$n^{\mathrm{zero}}_{A,k^{\,}_{i},k^{\star}_{i}}$,
to a property of the bulk, the integer
$
n^{\,}_{\bm{k}^{\star}}
-
n^{\,}_{\bm{k}^{\star}+(\bm{Q}^{\,}_{i+1}/2)}
$.
Equation~(\ref{eq: counting I}) is thus
an example of a bulk-boundary correspondence.
We choose the Fermi energy to be zero and apply Eq.~(\ref{eq: counting I})
to graphene with a Kekule distortion.

Graphene with a Kekule distortion,
in its simplest incarnation,
has six bands that are related by the chiral operation consisting in 
changing the sign of the wavefunction on all the sites of one
of the two triangular sublattices of the honeycomb lattice.
This spectrum of graphene with a Kekule distortion is thus
chiral symmetric. Hence, there are three occupied bands with
strictly negative energy eigenvalues when the Fermi energy is 
vanishing, i.e., coincides with the mid-gap single-particle energy.

We are going to apply the counting formula~(\ref{eq: counting I})
to graphene with the Kekule distortion and at a vanishing Fermi energy
by choosing the inversion
point to be either along an armchair cut as in Fig.~\ref{fig: Kekule+armchair}
or along a zigzag cut as in Fig.~\ref{fig: Kekule+zigzag}.
The inversion symmetric momenta from the Brillouin zone are then 
\begin{equation}
\bm{k}^{\star}\in
\left\{
(0,0)^{\mathsf{T}},
(\pi,0)^{\mathsf{T}},
(0,\pi)^{\mathsf{T}},
(\pi,\pi)^{\mathsf{T}}
\right\},
\label{eq: def k star}
\end{equation}
where we have chosen units such that the spanning vectors of the
reciprocal lattice are 
$\bm{Q}^{\,}_{1}=(2\pi,0)^{\mathsf{T}}$
and
$\bm{Q}^{\,}_{2}=(0,2\pi)^{\mathsf{T}}$.

\paragraph{Armchair cut}
For the armchair case with Hamiltonian (\ref{eq: H of kekule}),
we assign to each of the four inversion-symmetric momenta%
~(\ref{eq: def k star}) 
the row vector consisting of the three parities under the operation of inversion
of the three occupied Bloch states according to
\begin{align}
\begin{array}{ll}
\Gamma(k_{1}=0, k_{2}=0):&  (-,-,-), \\
M_{1}(k_{1}=\pi,k_{2}=0):&   (+,-,+), \\
M_{2} (k_{1}=0,k_{2}=\pi):&  (+,-,+), \\
M_{3}(k_{1}=\pi,k_{2}=\pi):& (+,-,+).
\end{array}
\end{align}
Hence, the number~(\ref{eq: counting I})  of zero modes is four 
(two per entangling edge in a torus geometry) 
at $k^{\,}_{1}=0$ and is zero at $k^{\,}_{1}=\pi$
in agreement with our numerics.

\paragraph{Zigzag cut}
For the zigzag case with Hamiltonian (\ref{eq: H zigzag}),
we assign to each of the four inversion-symmetric momenta%
~(\ref{eq: def k star}) 
the row vector consisting of the three parities under the operation of inversion
of the three occupied Bloch states according to
\begin{align}
\begin{array}{ll}
\Gamma(k_{1}=0, k_{2}=0):&(-,+,+),\\
M_{1}(k_{1}=\pi,k_{2}=0):& (-,-,+),\\
M_{2} (k_{1}=0,k_{2}=\pi):&(+,-,-),\\
M_{3}(k_{1}=\pi,k_{2}=\pi):&(+,-,+).
\end{array}
\end{align}
Hence, the number~(\ref{eq: counting I}) of zero modes is 
two (one per entangling edge in a torus geometry) 
at both $k^{\,}_{1}=0$ and $k^{\,}_{1}=\pi$
in agreement with our numerics.

\section{Conclusion}
\label{sec:conclusion}

The main focus of this paper has been on fermionic single-particle
local Hamiltonians obeying three conditions.
First, the many-body ground state is non-degenerate and incompressible
if periodic boundary conditions are chosen. In short, 
the ground state is that of a band insulator.
Second, point-group symmetries generated by non-local transformations
such as a reflection or inversion must hold. Third, certain
boundary conditions that are compatible with the point-group symmetries
must be imposed on the entanglement spectrum through the choice of
entangling boundaries.
We have then constructed several examples of model Hamiltonians
obeying all three conditions in one- and two-dimensional space
with the following two properties. First, each model
supports gapless boundary states in the entanglement spectrum
that are localized on an isolated entangling boundary, 
even though no gapless boundary states can be found in the energy spectrum 
on an isolated physical boundary. Second,
the stability under (one-body) perturbations 
of the gapless boundary states in the entanglement spectrum
is guaranteed by the point-group symmetries. 
Common to all these examples is the fact
that the non-local point-group symmetries in the energy spectrum become 
local spectral symmetries in the entanglement spectrum, 
as we have shown.
The existence of these symmetry-protected gapless boundary states 
in the entanglement spectrum is a signature of a topological character, 
for it is dependent on the choice of boundary conditions.
Whereas counting them relies explicitly on the point-group
symmetries, as in Eq.~(\ref{eq: counting I}) say,%
~\cite{Hughes2011,Turner2010}
our main results~(\ref{eq: main result on how to get chiral sym})
and~(\ref{eq: induced symmetry})
offer a complementary understanding to their stability.

\noindent
\ack

We thank 
C. Chamon, 
C.-K. Chiu,  
M. J. Gilbert, 
C.-Y. Hou,
C.-T. Hsieh,
T. Morimoto, 
T. H. Taylor, 
and 
J. C. Y.  Teo 
for
insightful discussions.

\section*{References}

\def\urlprefix{}
\def\url#1{}
\bibliographystyle{iopart-num}
\bibliography{combined}

\end{document}